\documentclass[12pt]{book}
\usepackage{graphicx}
\usepackage{amsmath}
\usepackage{amssymb}
\newcommand{\sat}{\markboth{V.P.Berezovoj, Yu.L.Bolotin, V.A.Cherkaskiy,
G.I.Ivashkevich}{Regular and Chaotic Classical and Quantum
Dynamics}}
\title{Regular and Chaotic Classical and Quantum Dynamics
in Multi-Well Potentials}
\author{V.P.Berezovoj \and Yu.L.Bolotin\and V.A.Cherkaskiy\and G.I.Ivashkevich}
\date{}
\begin{document}
\maketitle
\sat\tableofcontents\sat
\chapter{Introduction}\sat Currently we can consider as a
rigorously established fact the existence of dynamical systems with
a small number of degrees of freedom ($N\ge 2$ --- in autonomous
case, $N\ge 1$ --- for non-autonomous systems) for which under
certain conditions classical motion cannot be distinguished from
random motion \cite{zaslavsky,haake,gutzwiller,reichl}. Typical
features of these systems are nonlinearity and the absence of any
external source of randomness. Thus, using such synonyms for the
term "random" as "chaotic", "stochastic", "irregular", we can state
that there are nonlinear deterministic systems for which these
notions express adequately internal fundamental properties that
comprise important and interesting subjects for investigation. For
the last 30-40 years examples of chaotic motion have been detected
in every field of natural science, and their number continues to
grow.

The mechanism which is responsible for the existence of chaotic
regimes in purely deterministic systems is based on local
instability. It leads to exponential divergence of initially close
trajectories
\[d(t)=d(0)e^{ht}\]
where $d$ is the distance between two points in phase space that
belong to different trajectories. It could be shown \cite{krylov}
that local instability leads to mixing i.e. splitting of time
correlations
\[\lim_{t\rightarrow\infty}R(f,\varphi)\sim e^{-h_c t}\]
for arbitrary functions in phase space
$f(z),\varphi(z)\left(z(t)=[q(t),p(t)]\right)$. A principally new
result that was established by N.Krylov \cite{krylov} consisted in
the understanding that the averaged increment of local instability
$\langle h\rangle$ determines the time of correlations splitting
\[h_c=\langle h\rangle.\]
This equation exactly establishes the connection between the
dynamics of the system and its statistical properties. In other
words, we must understand stochasticity as a rise of statistical
features in the system as a result of local instability.

Local instability of trajectories makes the problem of assignment of
initial conditions in classical mechanics as fundamental as the
uncertainty principle makes the problem of measurement in quantum
mechanics. From the traditional viewpoint finite precision is
limited by the imperfection of the measuring instrument. But the new
viewpoint is different: we could not record the result of the
measurement with infinite precision even if the ideal measuring
instrument were constructed --- there is not enough energy, time and
paper. Uncertainty of measurement could not be completely eliminated
even as a theoretical idea because inaccuracy in the data-line of a
purely deterministic system is not produced by external randomness
or noise but is connected with the finite precision of initial
conditions assignment instead. In particular, at the beginning of
the last century Poincar\'e showed \cite{poincare} that for some
astronomical systems a tiny imprecision in the initial conditions
would grow in time at an enormous rate. Thus two
nearly-indistinguishable sets of initial conditions for the same
system would result in two vastly different final predictions. In
the presence of local instability this inaccuracy leads to very
peculiar behavior of deterministic systems. For most physicists
"dynamical instability" and "chaos" became convertible terms.

Fundamental progress in the understanding of classical nonlinear
dynamics caused numerous attempts to integrate the concept of
stochasticity into quantum mechanics. The core of the problem lies
in the fact that energy spectrum of every quantum system is discrete
and thus motion is quasiperiodic if the system demonstrates finite
motion. On the other hand, the correspondence principle demands the
possibility of transition to classical mechanics which demonstrates
not only regular regimes but chaotic too. Several important results
that clarify this contradiction were established. However, long
before complete solution of the problem there has been interest in
its reduced variant –-- the search for peculiarities in the behavior
of quantum systems which have chaotic classical analogues. These
peculiarities are called Quantum manifestations of classical
stochasticity (QMCS).

Many decades would pass before dynamical chaos ideology was realized
by the science community as a whole. According to the old ideology:

\begin{itemize}
\item Chaos is an attribute of a compound system.
\item In any compound system it is possible to find out the elements of chaos.
\item Useful information is contained in those few places in which chaos
is absent.
\item Physicists must look for non-chaos.
\end{itemize} The new ideology changed the situation principally:
\begin{itemize}
\item Chaos is universal inalienable
property of simple deterministic systems.
\item Chaotic dynamics is
the most general way of evolution of an arbitrary nonlinear system
\item New interesting information is contained exactly in those
branches of natural sciences where chaos is present.
\item Chaos is the major object of study.
\item The physicist must look for chaos!
\cite{barranger}. \end{itemize}

The basis of the present report is the new chaos ideology. In the
context of this approach a general investigation of arbitrary
nonlinear dynamical system involves the following steps:

\begin{enumerate}
\item Investigation of the classical phase space, detection of chaotic
regimes by numerical and analytical analysis of the classical
equations of motion.
\item Analytical estimation of the critical
energy for the onset of global stochasticity.
\item Test for QMCS in
the energy spectra, eigenfunctions and wave packet dynamics.
\item Consideration of the interrelationship between stochastic
dynamics and concrete physical effects.
\end{enumerate}

The basic subject of the current report is to realize the outlined
program for two-dimensional Hamiltonian systems with potential
energy surface which has several local minima, i.e. multi-well
potentials.

Despite the huge number of papers concerning chaotic dynamics,
Hamiltonian systems with multi-well potentials have been somewhat
neglected. Nevertheless the Hamiltonian system with multi-well
potential energy surface (PES) represents a realistic model,
describing the dynamics of transition between different equilibrium
states, including such important cases as chemical and nuclear
reactions, nuclear fission, phase transitions, string landscape and
many others. Being substantially nonlinear, such systems represent a
potentially important object, both for the study of classic chaos
and QMCS.
\chapter{Specifics of Classical Dynamics in Multi-Well Potentials --- Mixed State\label{ms}}\sat
Let us consider the characteristics of classical finite motion in
multi-well potentials. They are more complicated than in single-well
potentials and allow the existence of several critical energies even
for a fixed set of potential parameters. This fact results in the
so-called mixed state in such potentials \cite{bolotin-87}: at the
same energy there are different dynamical regimes in different
wells, either regular or chaotic. It is important to note that mixed
state is a general feature of the Hamiltonians with nontrivial PES.
For the first example, let us demonstrate the existence of mixed
state for nuclear quadrupole oscillations Hamiltonian
\cite{eizenberg}. It can be shown that, using only transformation
properties of the interaction, the deformation potential of surface
quadrupole oscillations of nuclei takes the form
\cite{mozel_greiner}:
\begin{equation}\label{uqo}U(a_0,a_2)=\sum\limits_{m,n}C_{mn}(a_0^2+2a_2^2)^m a_0^n
(6a_2^2-a_0^2)^n\end{equation} where $a_0$ and $a_2$ are internal
coordinates of the nuclear surface during the quadrupole
oscillations:
\[R(\theta,\varphi)=R_0\{1+a_0 Y_{2,0}(\theta,\varphi) + a_2[Y_{2,2}(\theta,\varphi)+Y_{2,-2}(\theta,\varphi)]\}\]
Constants $C_{mn}$ can be considered as phenomenological parameters.
Restricting to the terms of the fourth degree in the deformation and
assuming the equality of mass parameters for two independent
directions, we get $C_{3v}$-symmetric Hamiltonian:
\begin{equation}\label{qo_ham}H=\frac{p_x^2+p_y^2}{2m}+U_{QO}(x,y;a,b,c)\end{equation} where
\[\begin{array}{c}U_{QO}(x,y;a,b,c)=\frac a 2 (x^2+y^2) + b\left(xy^2-\frac{x^3}{3}\right) + c(x^2+y^2)^2\\
x= a_0,\ y=\sqrt2 a_2,\ a=2C_{10},\ b=3C_{01},\
c=C_{20}\end{array}\] This potential  is a generalization of the
well-known H\'enon-Heiles potential \cite{henon_heiles}, which
became a traditional object for examination of new ideas and methods
in investigations of stochasticity in Hamiltonian systems. It is
essential that, in contrast to H\'enon–-Heiles potential, motion in
(\ref{qo_ham}) is finite for all energies, assuring the existence of
stationary states in the quantum case. Hamiltonian (\ref{qo_ham})
and corresponding equations of motion depend only on parameter
$W=b^2/(ac)$, the unique dimensionless quantity we can build from
parameters $a,b,c$. The same parameter determines the geometry of
PES \begin{equation}\label{u_qo}U_{QO}(x,y;W)=\frac{1}{2W} (x^2+y^2)
+ xy^2-\frac{x^3}{3} + (x^2+y^2)^2\end{equation}

\begin{figure}
\includegraphics[width=\textwidth,draft=false]{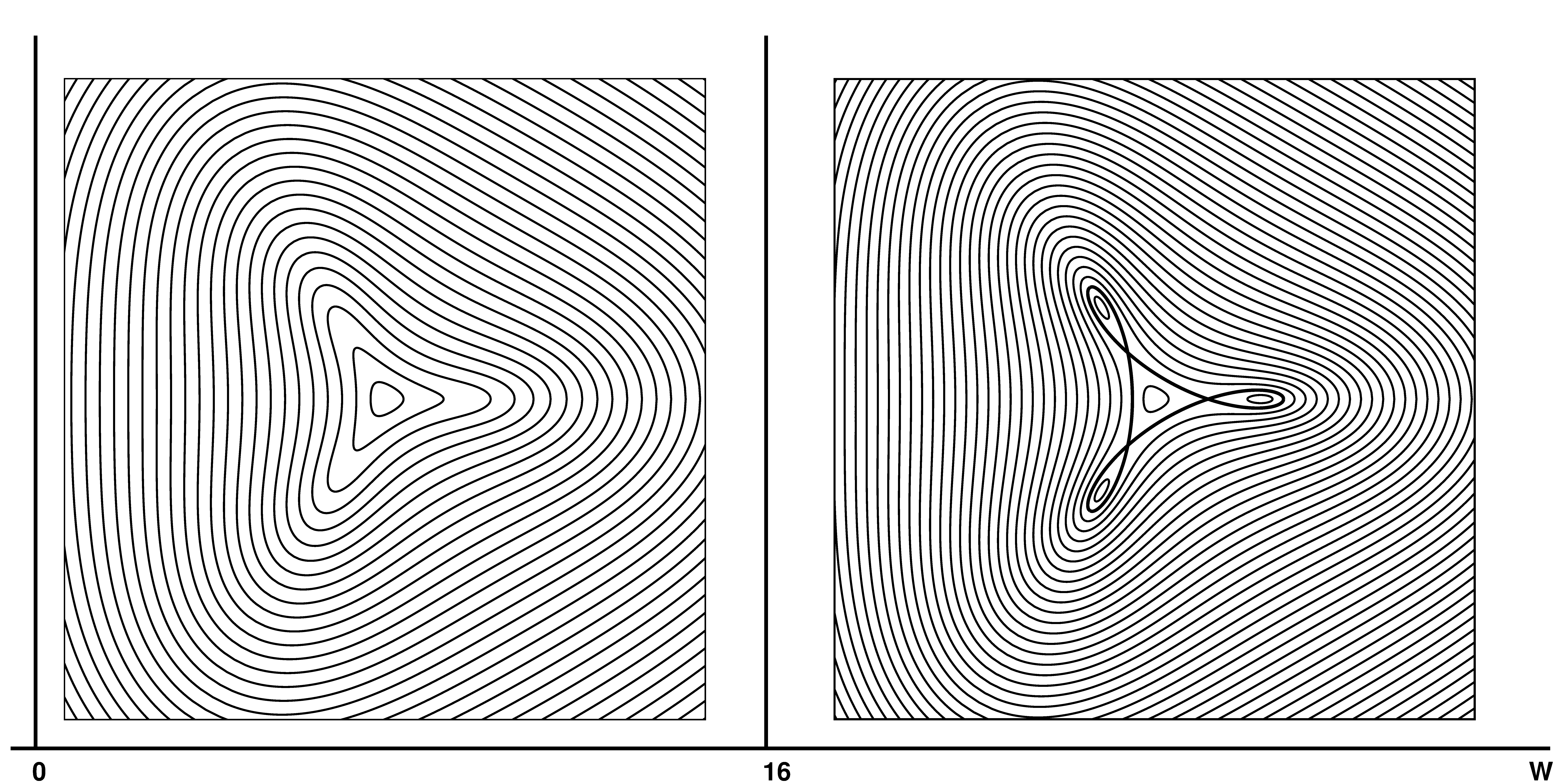}
\caption{The level lines of the PES (\ref{u_qo}) for different
structurally stable domains.\label{qo_ll}}
\end{figure}

\begin{figure}
\includegraphics[width=0.3\textwidth,draft=false]{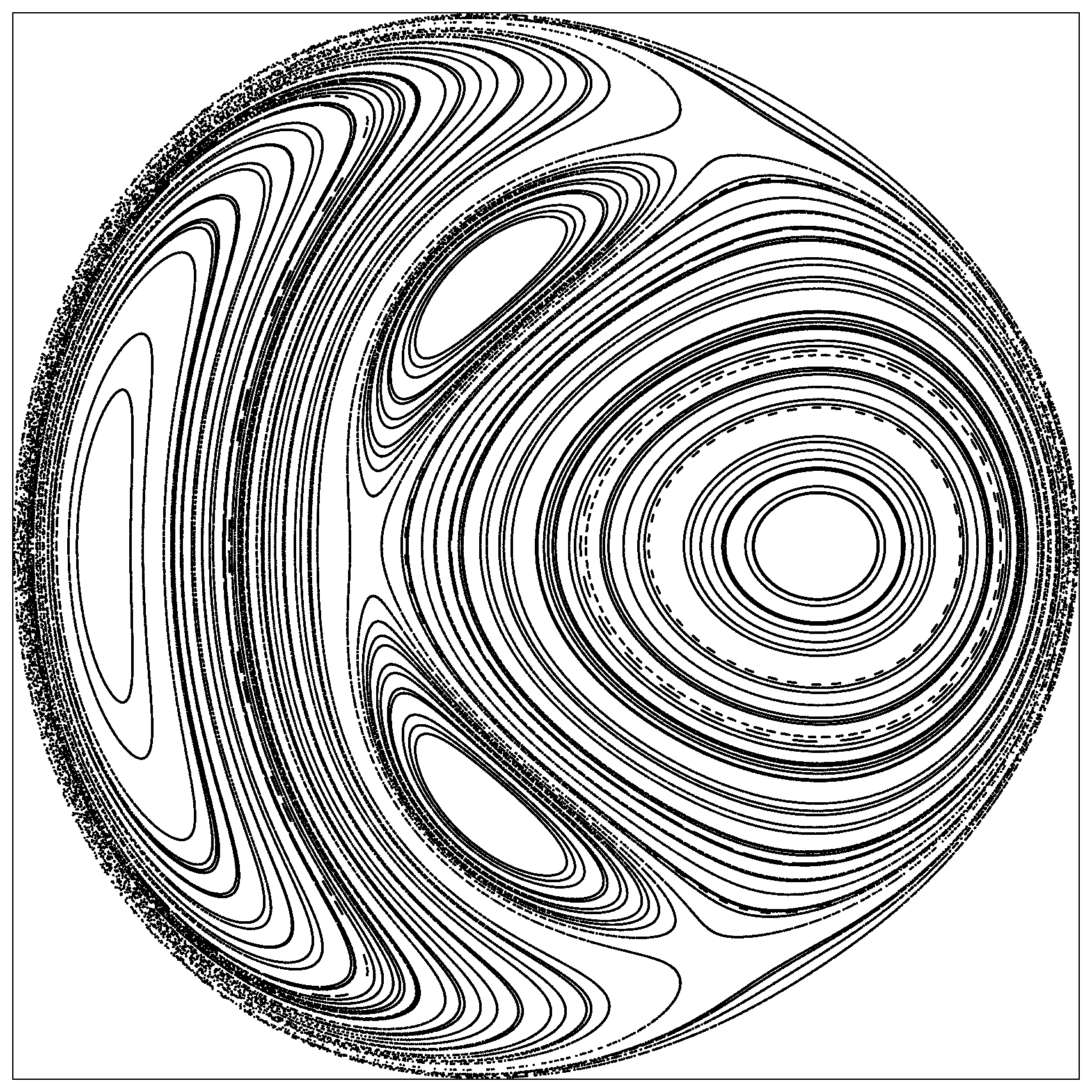}
\includegraphics[width=0.3\textwidth,draft=false]{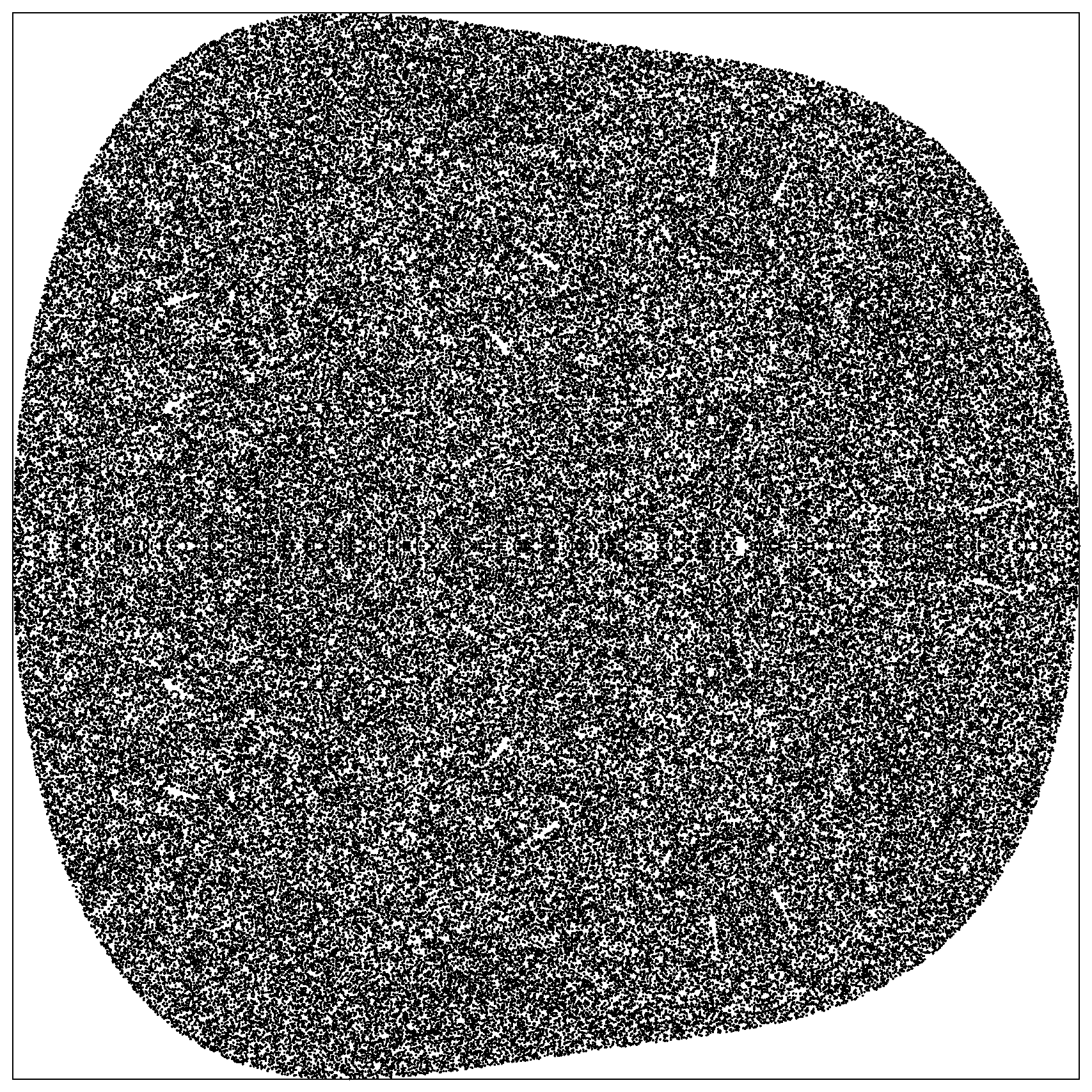}
\includegraphics[width=0.3\textwidth,draft=false]{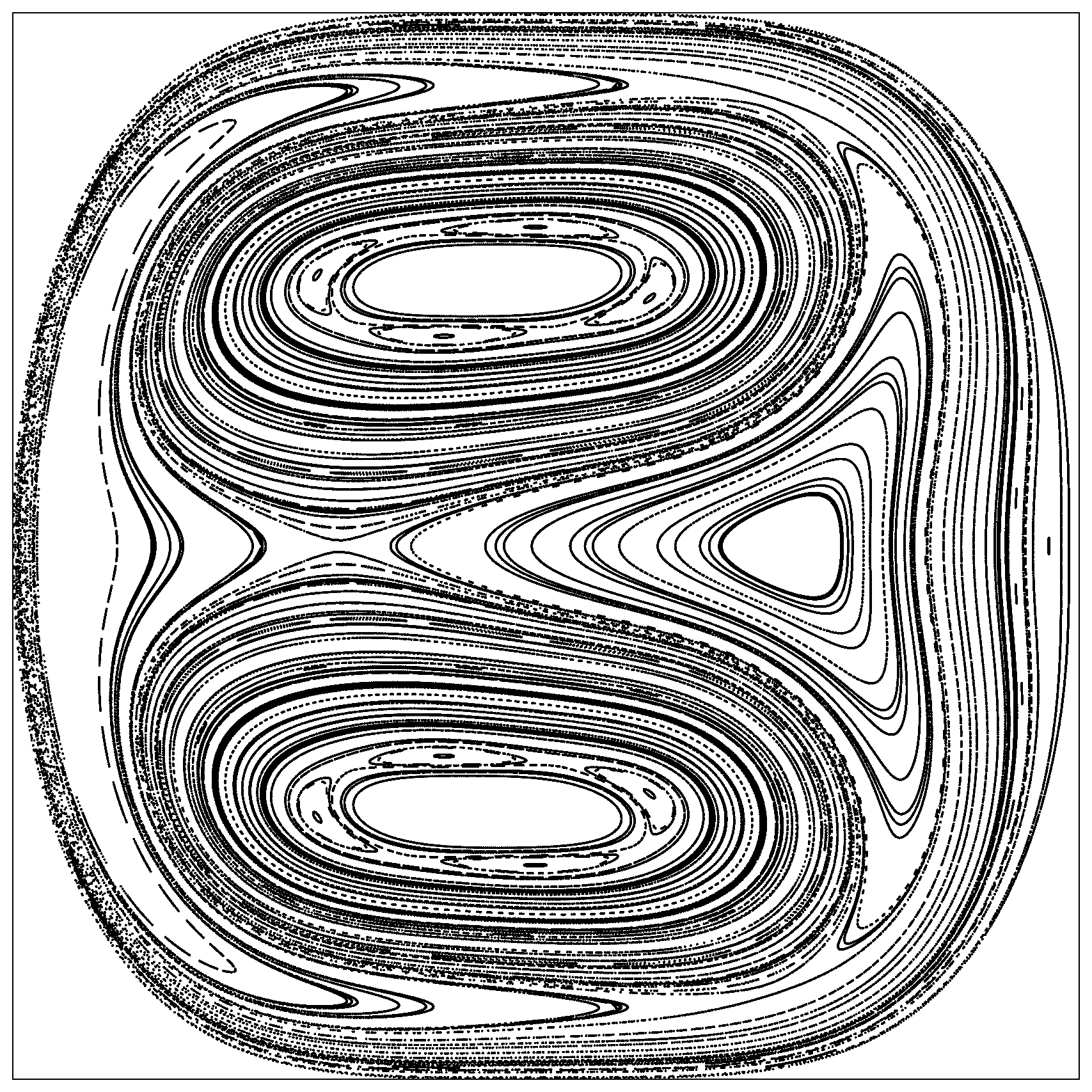}
\caption{The R-C-R transition in the quadrupole oscillations
potential (\ref{u_qo}).\label{rcr0}}
\end{figure}

Interval $0<W<16$ includes potentials with a single extremum
--- minimum in the origin that corresponds to the spherically symmetric
shape of the nucleus. In the interval $W>16$ PES of $U_{QO}$
contains seven extrema: four minima (one central, placed in the
origin, and three peripheral, which correspond to deformed states of
nuclei) and three saddles, which separate the peripheral minima from
the central one (Fig.\ref{qo_ll}).
Numerical calculations of equations of motion in the region $0<W<16$
(region of single-well potentials) indicate
regularity--chaos--regularity (R-C-R) transition: gradual transition
from regular to chaotic motion when energy increasesá and
restoration of regular motion for high energies (fig.\ref{rcr0}). In
the next section we will discuss in details possible stochastization
scenarios and methods of critical energy calculation.

\begin{figure}
\includegraphics[width=0.5\textwidth,draft=false]{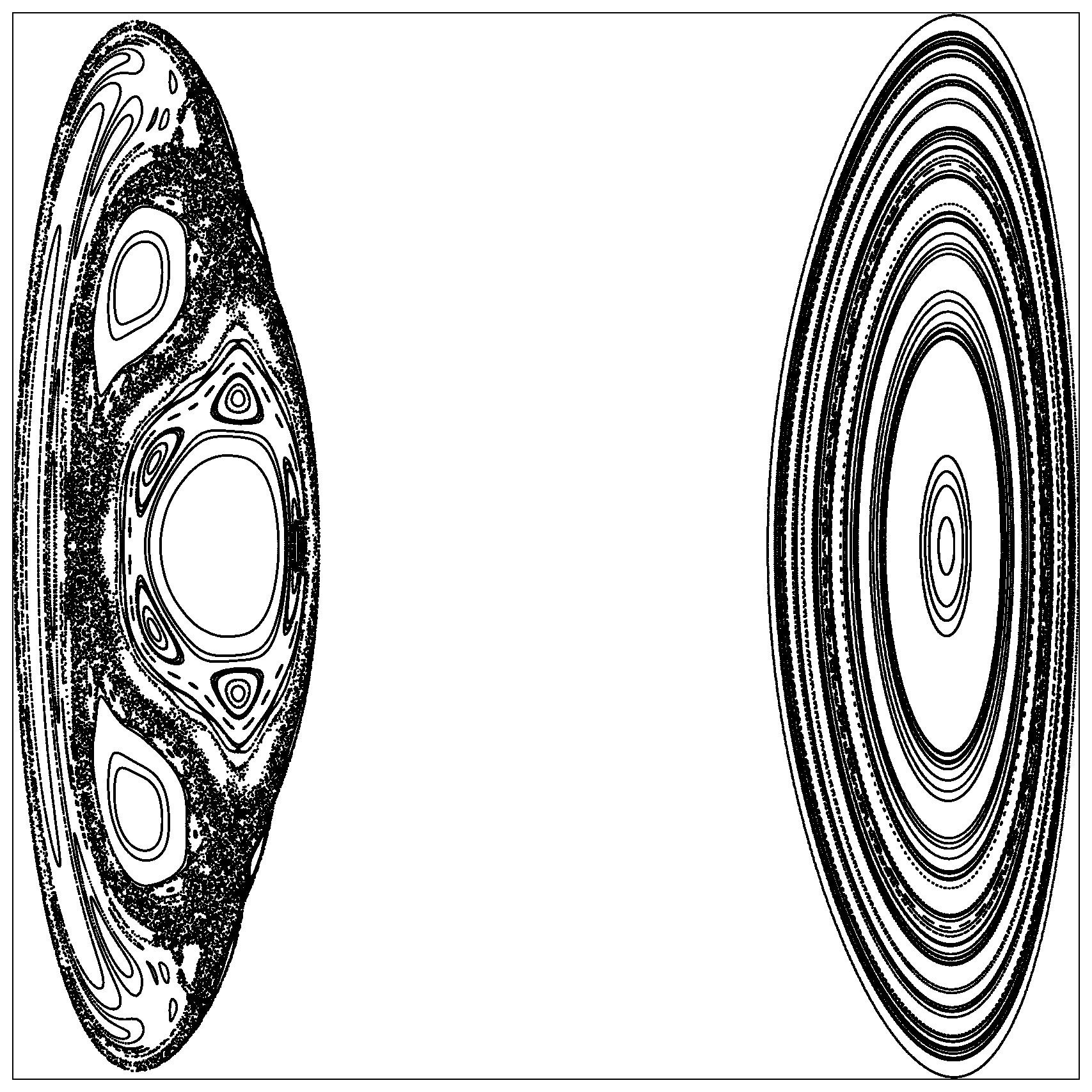}
\includegraphics[width=0.5\textwidth,draft=false]{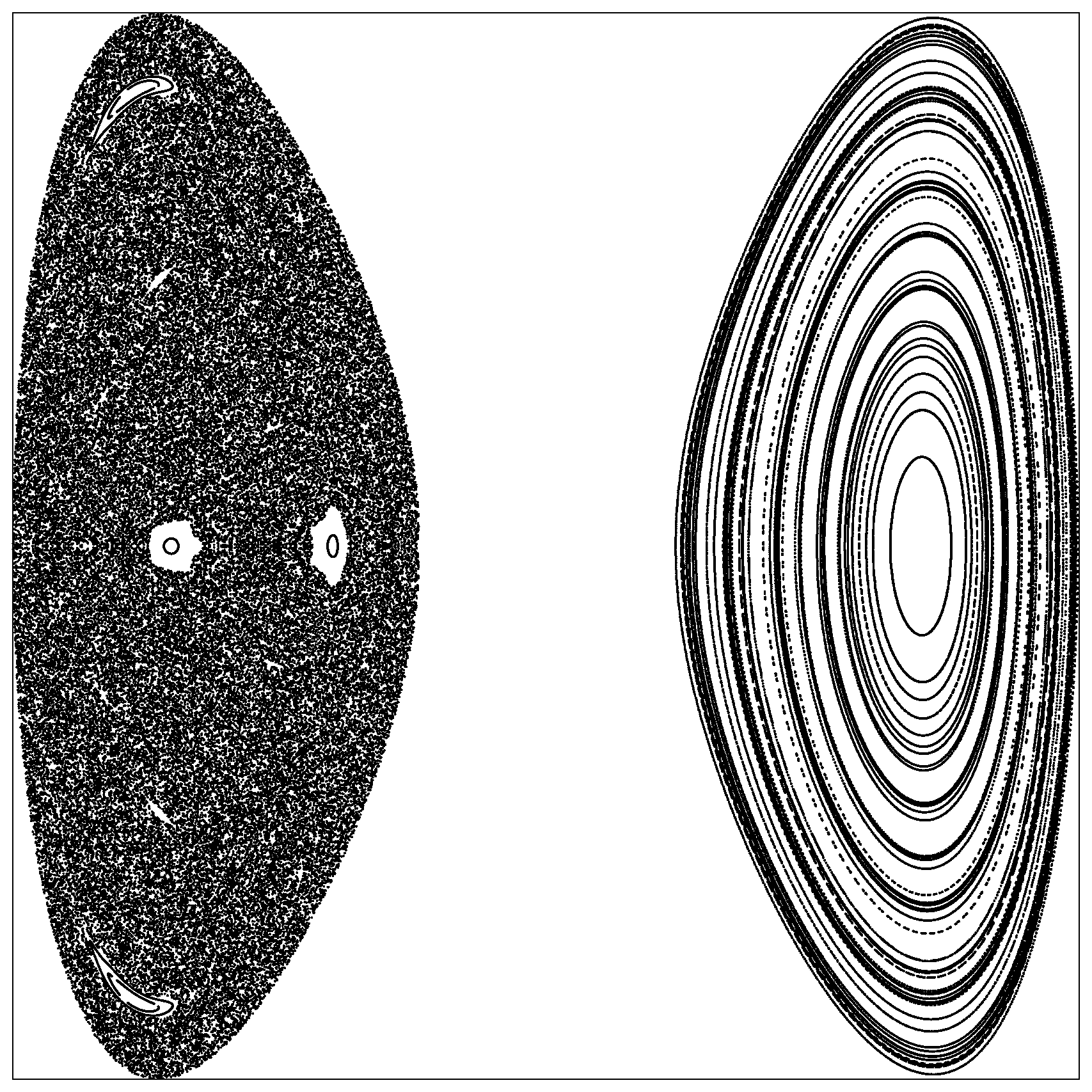}
\includegraphics[width=0.5\textwidth,draft=false]{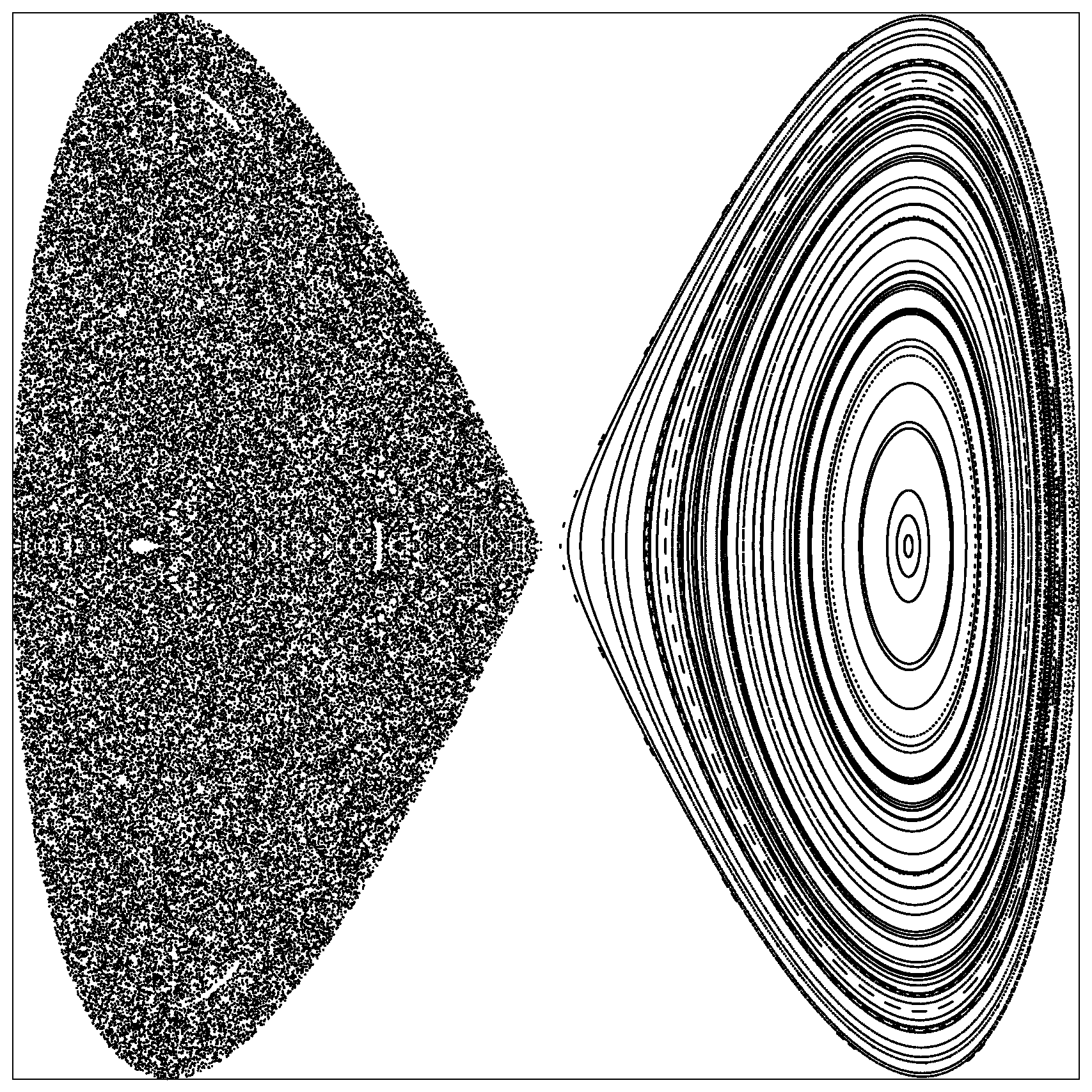}
\includegraphics[width=0.5\textwidth,draft=false]{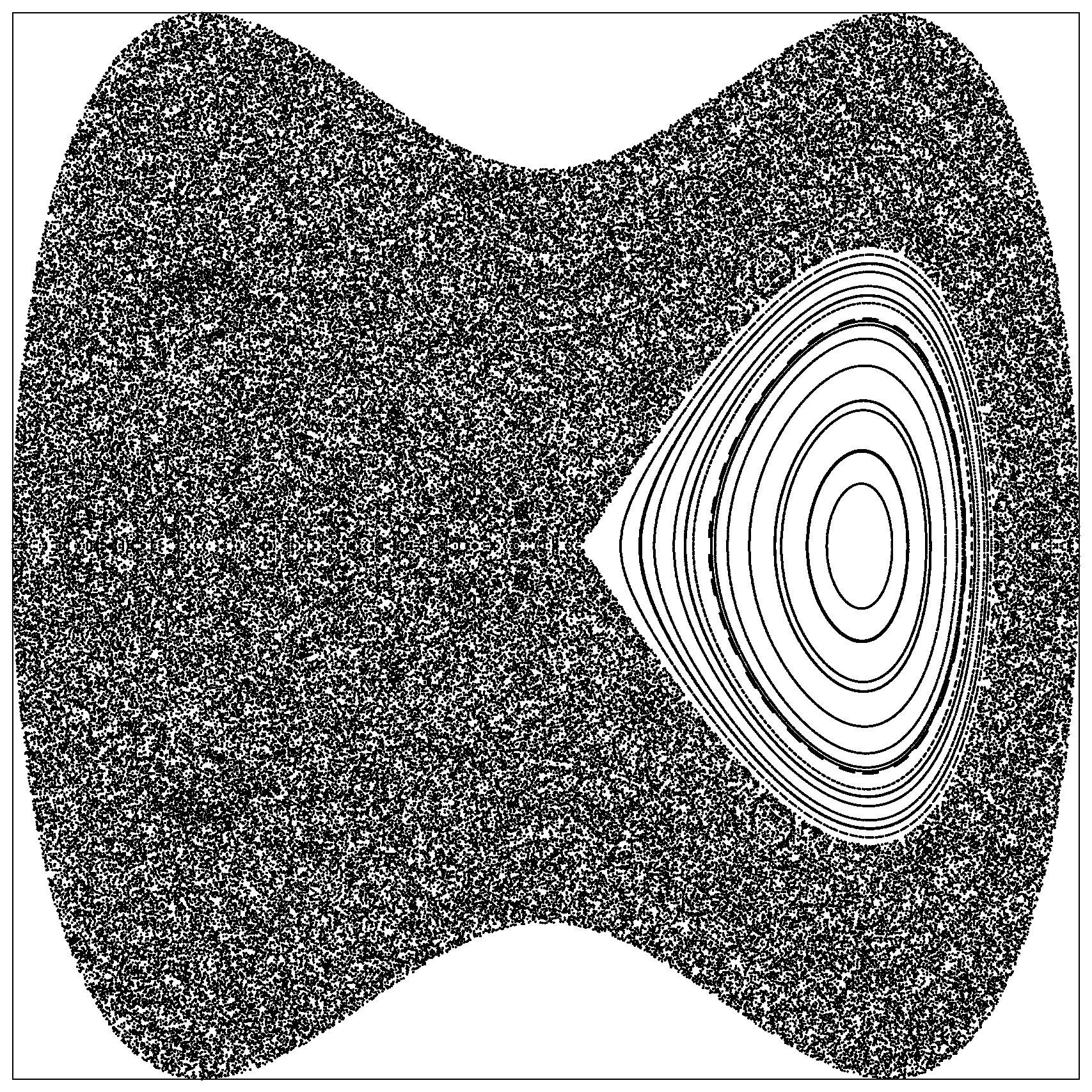}
\caption{PSS for motion in the potential $U_{QO}$ (\ref{u_qo}) with
$W=18$ at different energies:
$E/E_S=\{0.5,0.8,1,2\}$.\label{ms_pss}}
\end{figure}

In the region $W>16$ (multi-well potentials) we confront with a
substantially more complicated situation. In Figure \ref{ms_pss}
there are presented Poincar\'e surfaces of section (PSS) for
different energies. They demonstrate evolution of dynamics in
central and peripheral minima of QO potential $U_{QO}$ (with $W=18$,
when depths of central and peripheral minima are equal). At low
energies motion is clearly quasi-periodic for both minima. Let us
pay attention to the difference in topology of PSS. In the central
minimum, PSS structure is complicated and has fixed points of
different types, while in peripheral minima PSS has trivial
structure with only one elliptical fixed point.

When energy increases, gradual transition to chaos is observed, but
changes in character of motion are totally different in different
minima. In the central minimum already at energy equal to a half of
the saddle energy $E_S$, a sizeable part of the trajectories is
chaotic, and at the saddle energy there are almost no regular
trajectories at all. At the same conditions, in the peripheral
minimum motion remains quasiperiodic. Furthermore, even at energies
higher than the saddle energy there is a substantial part of the
phase space occupied by quasiperiodic motion. In other words
dynamics above the barrier has some kind of "memory": the structure
of phase space at energies greater than the saddle one is determined
by the character of the motion in the local minima.

Thus, in the energy region $E_S/2<E<E_S$, classical dynamics is
clearly chaotic in the central minimum and remains regular in
peripheral minima. This type of dynamics, when chaoticity measured
at fixed energy significantly differs in different local minima,
represents the common case situation in multi-well potentials and is
called the mixed state.

As an example of the concrete realization of the (\ref{uqo})
potential, let us consider deformational potential, that describes
quadrupole oscillations of Krypton isotopes. Seiwert, Ramayya and
Maruhn \cite{seivert} restored the parameters of the deformation
potential of the quadrupole oscillations, including the sixth degree
terms in deformation for isotopes $Kr^{74,76,78,80}$
\[\begin{array}{c}U_{QO}(x,y;a,b,c)=\frac a 2 (x^2+y^2) + b\left(xy^2-\frac{x^3}{3}\right) + c(x^2+y^2)^2 +\\
+ b\left(xy^2-\frac{x^3}{3}\right)(x^2+y^2) +
e\left(xy^2-\frac{x^3}{3}\right)^2 + f(x^2+y^2)^3.
\end{array}\]

\begin{figure}
\includegraphics[width=\textwidth,draft=false]{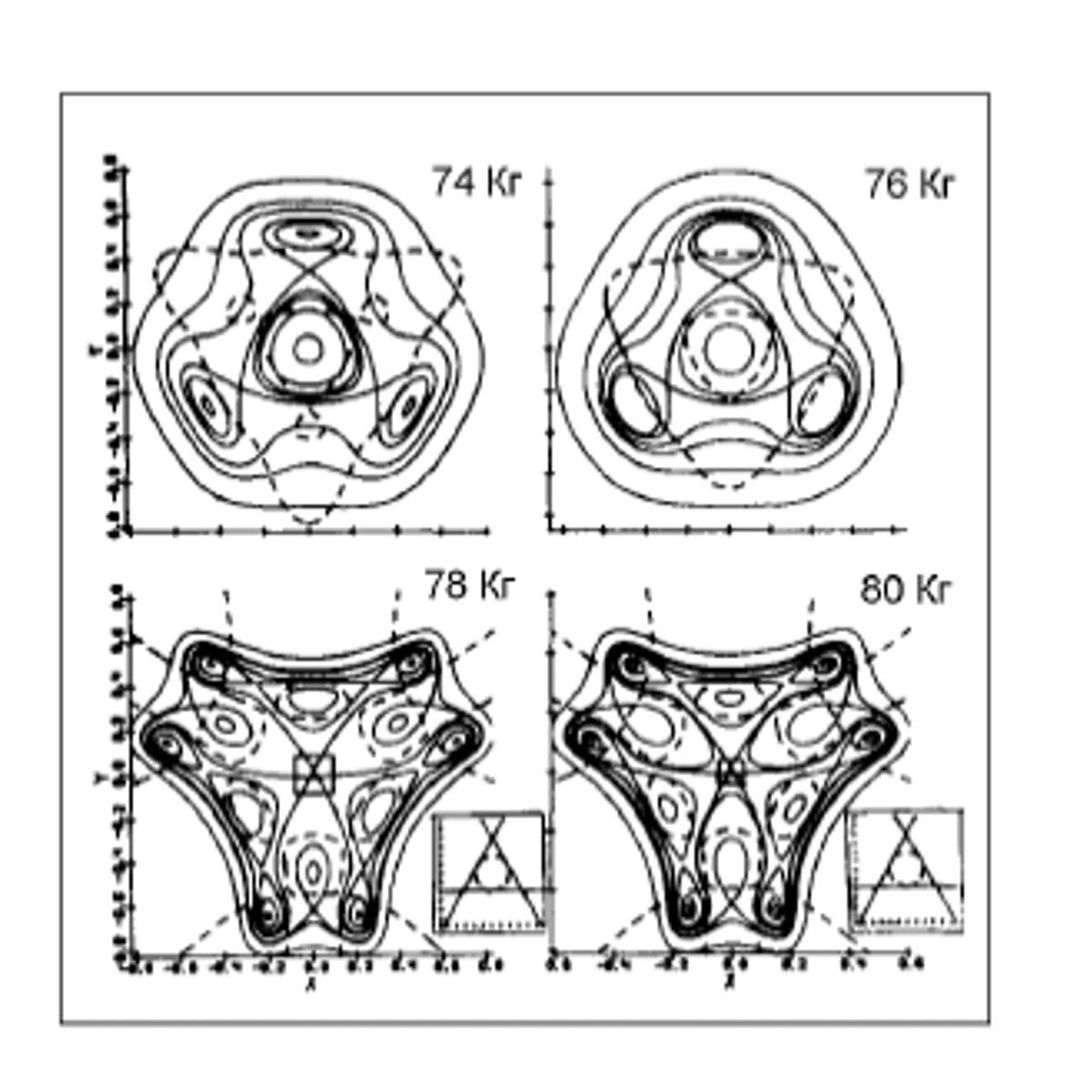}
\caption{The PES of Krypton isotopes.\label{kr}}
\end{figure}

The big experimental values of energy of the first $2^+$-states for
nuclei $Kr^{74,76}$ indicate a spherical shape of nucleus surface,
while the probabilities of the electromagnetic transitions
$2^+\rightarrow2^0$ and very low energies of first rotational states
imply the possibility of superdeformation. The nonlinear effects,
which are connected with the geometry of PES must be exhibited in
the superdeformed nuclei at relatively low energies of excitation.
The PES of Krypton isotopes are presented in Fig.\ref{kr}.

As can be seen, inclusion of higher terms of expansion leads to
considerable complication of the PES geometry: for all considered
Krypton isotopes we run into potentials with PES of complicated
topology with many local minima. It's clear that, for all potential
surfaces, $C_{3v}$-symmetry is preserved.

\begin{figure}
\includegraphics[width=0.23\textwidth,draft=false]{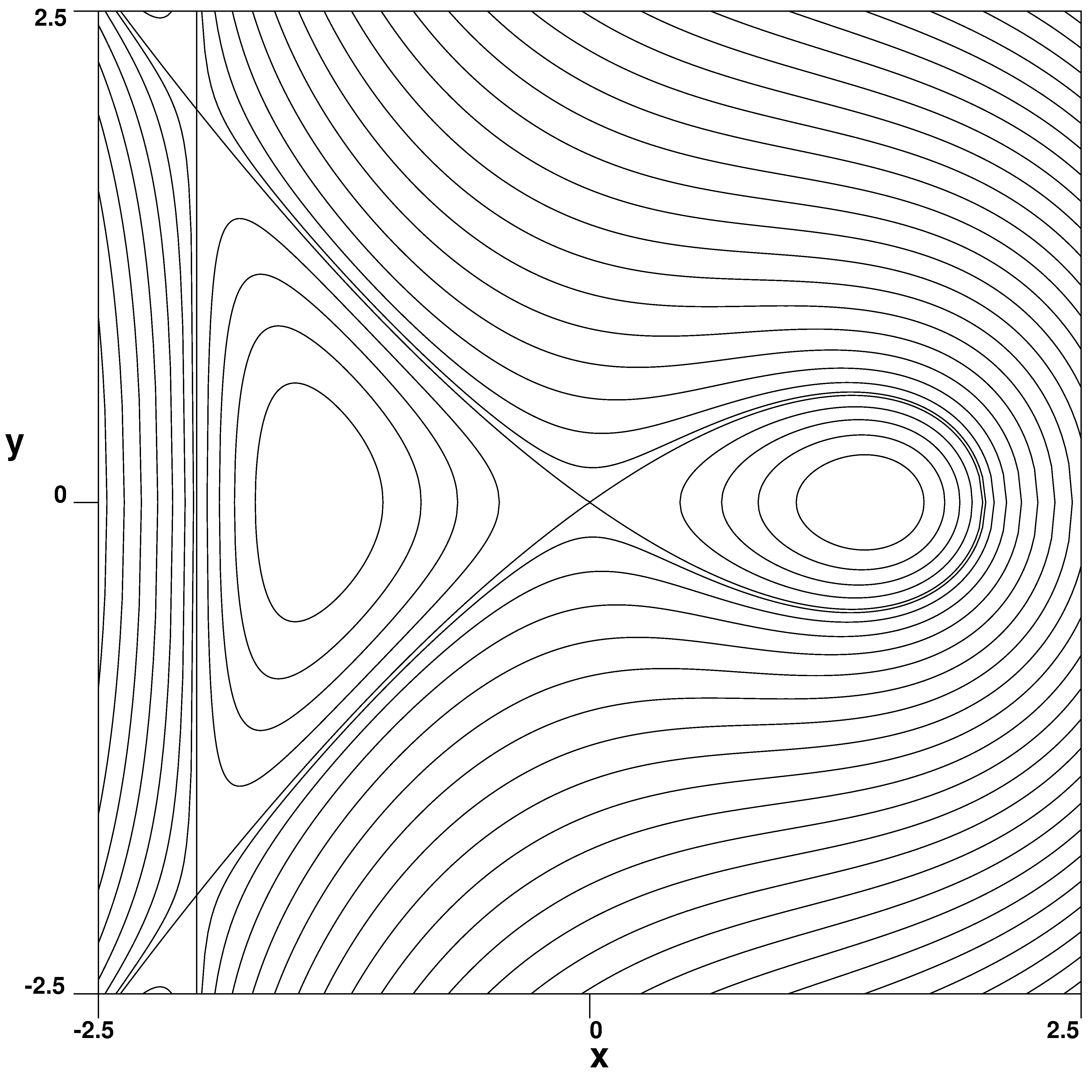}
\includegraphics[width=0.23\textwidth,draft=false]{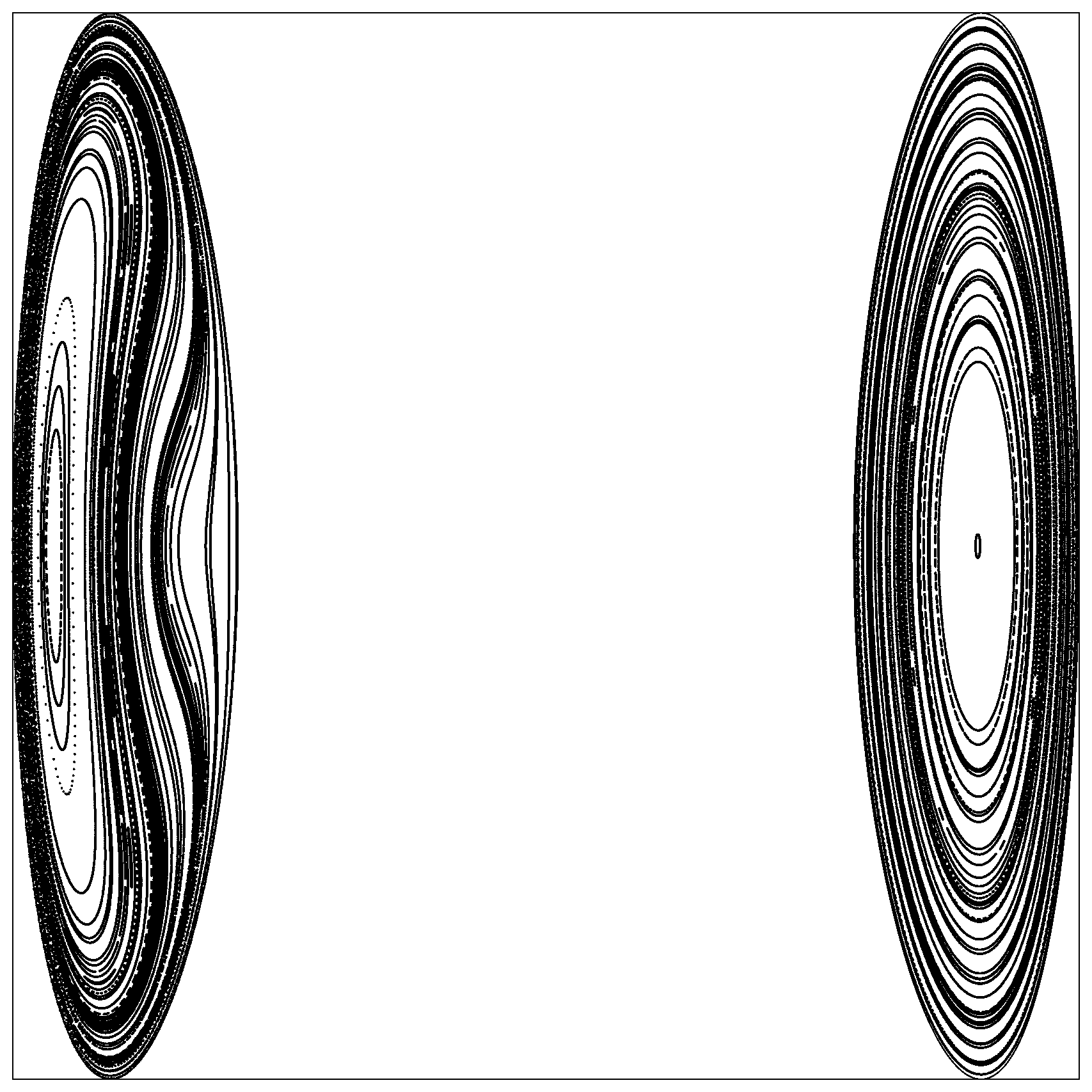}
\includegraphics[width=0.23\textwidth,draft=false]{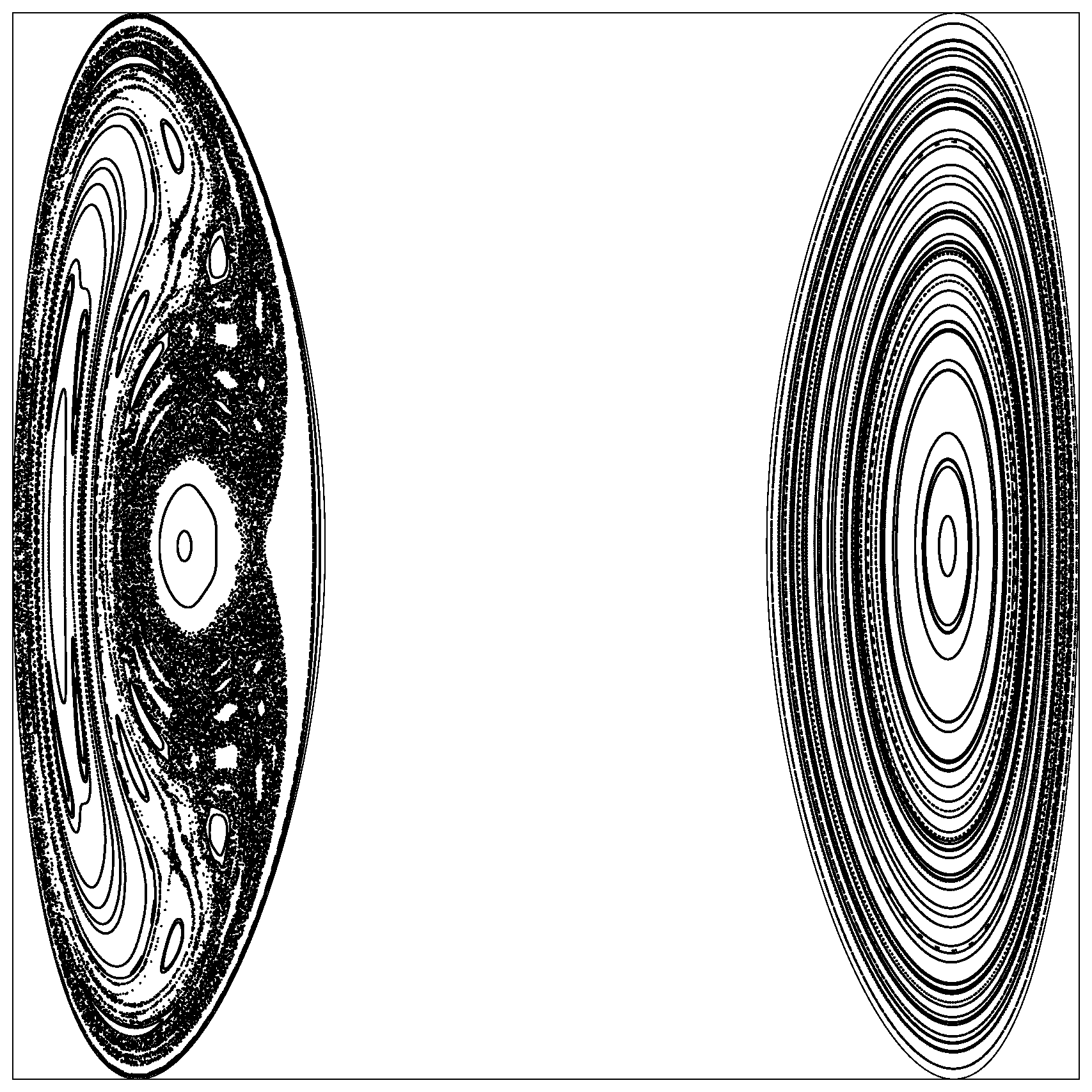}
\includegraphics[width=0.23\textwidth,draft=false]{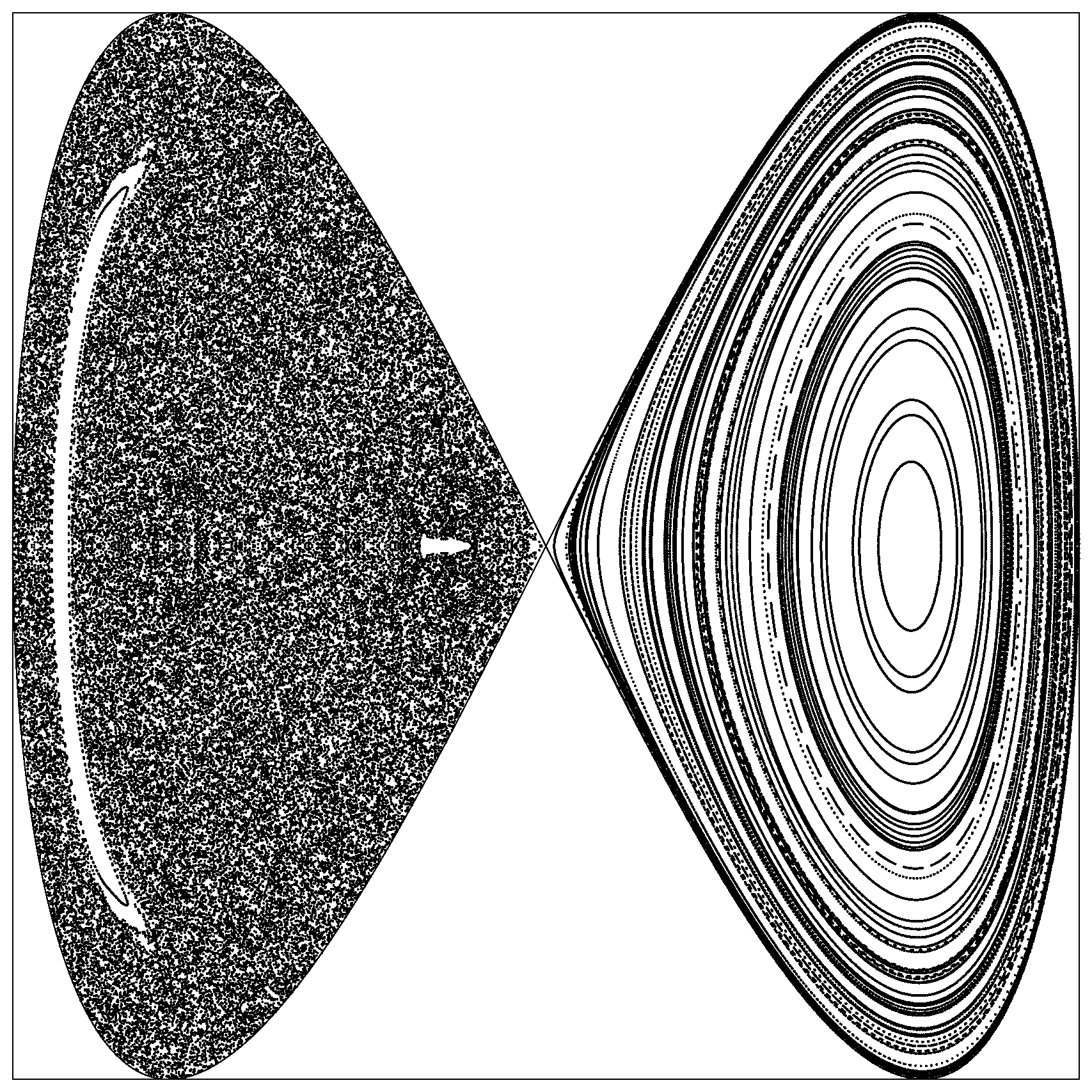}\\
\includegraphics[width=0.23\textwidth,draft=false]{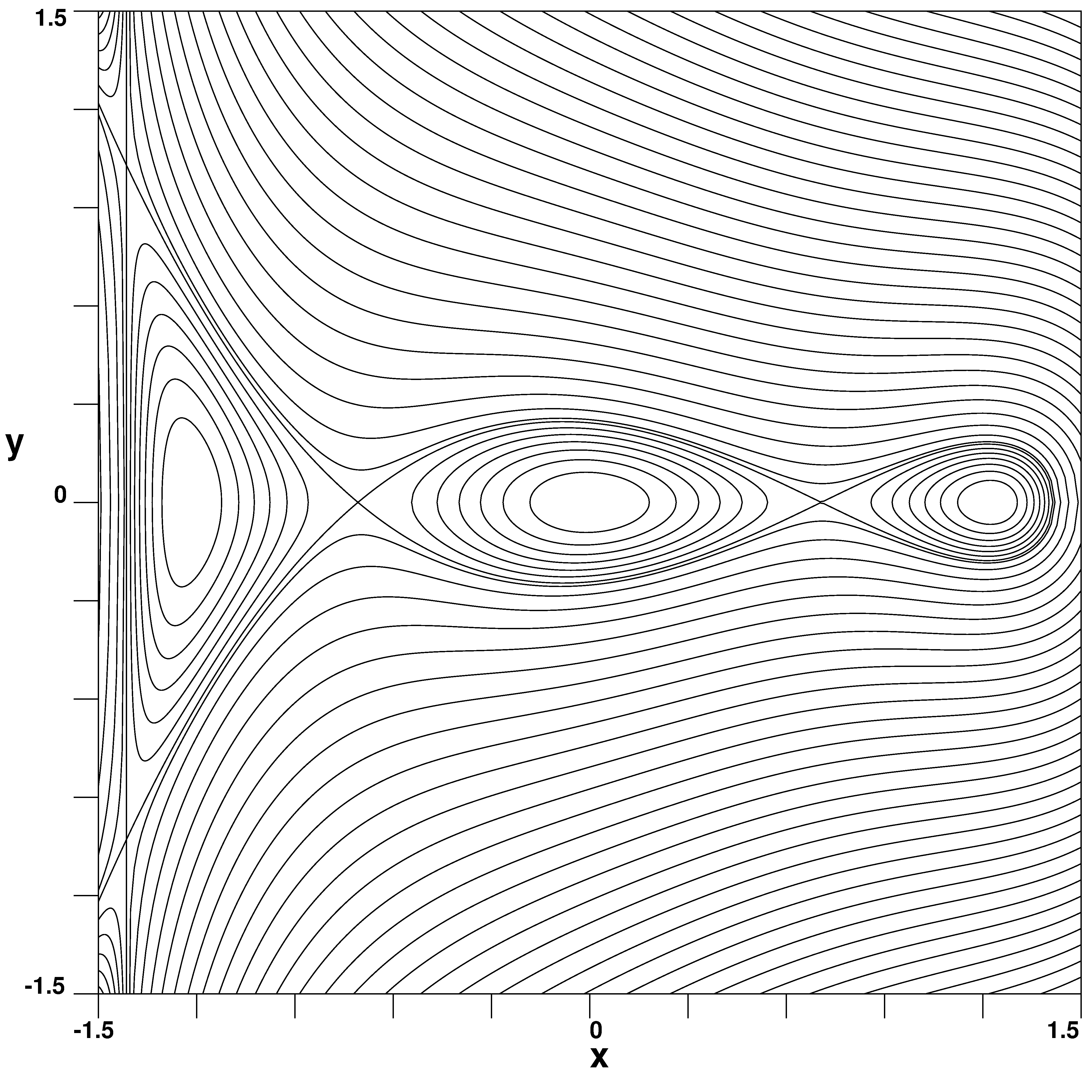}
\includegraphics[width=0.23\textwidth,draft=false]{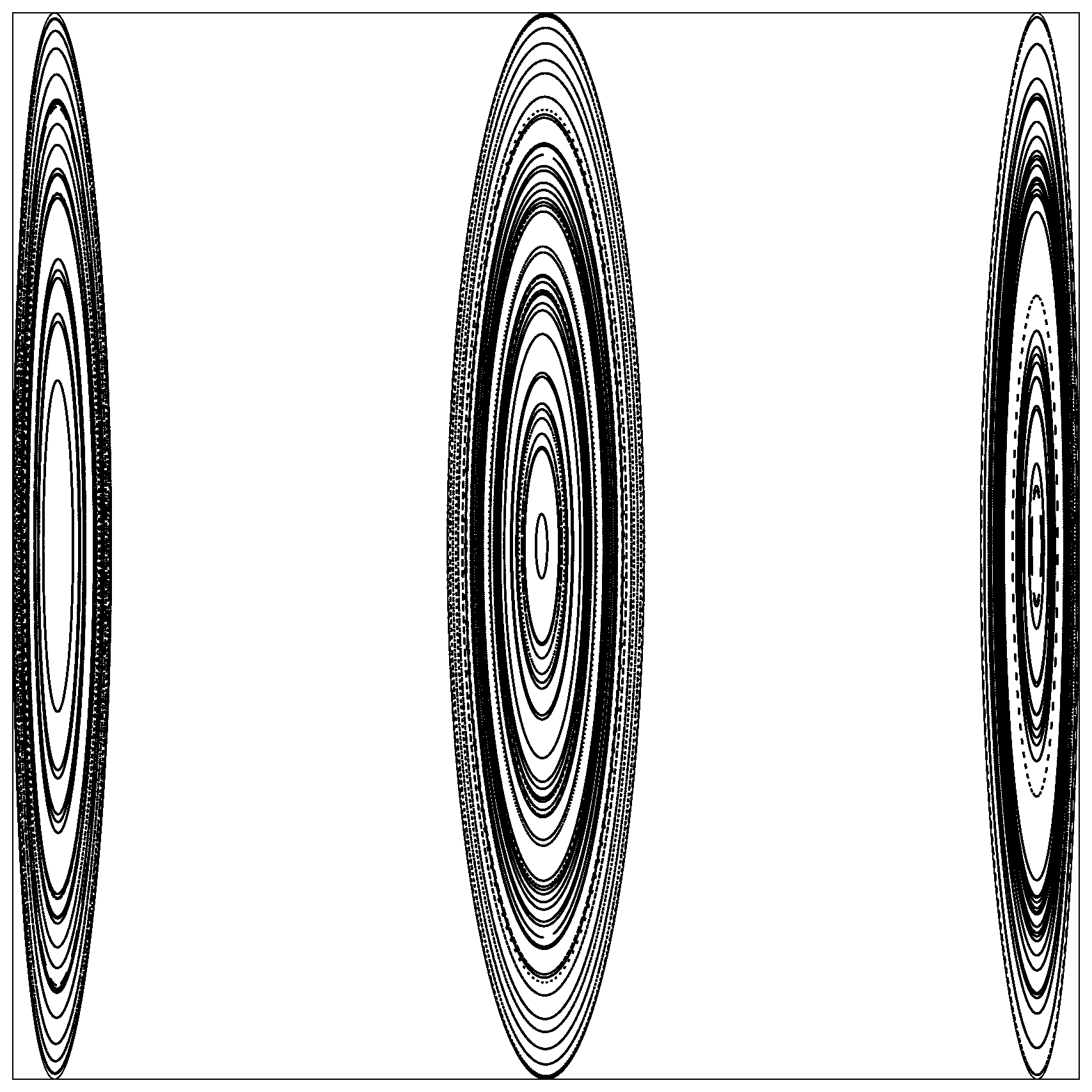}
\includegraphics[width=0.23\textwidth,draft=false]{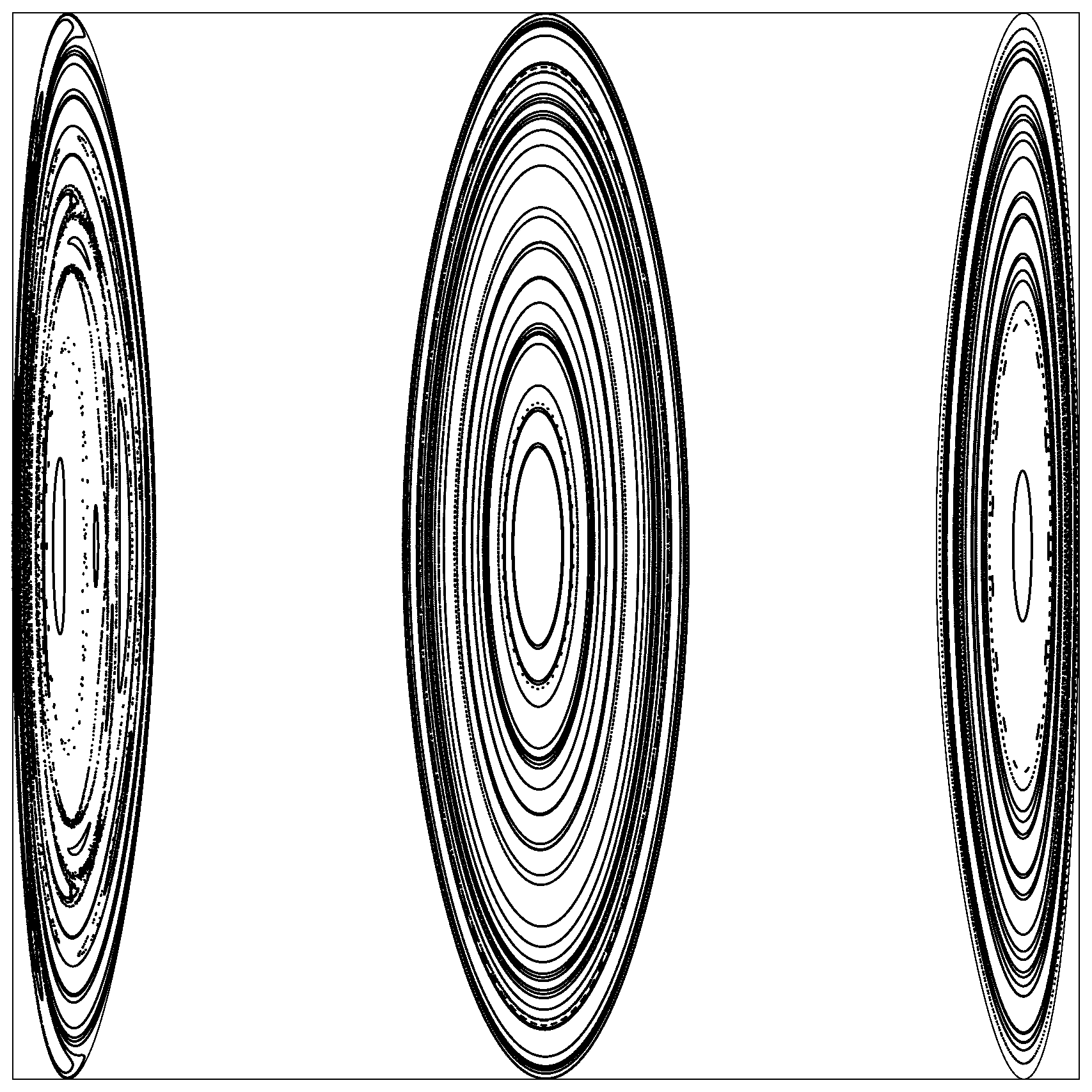}
\includegraphics[width=0.23\textwidth,draft=false]{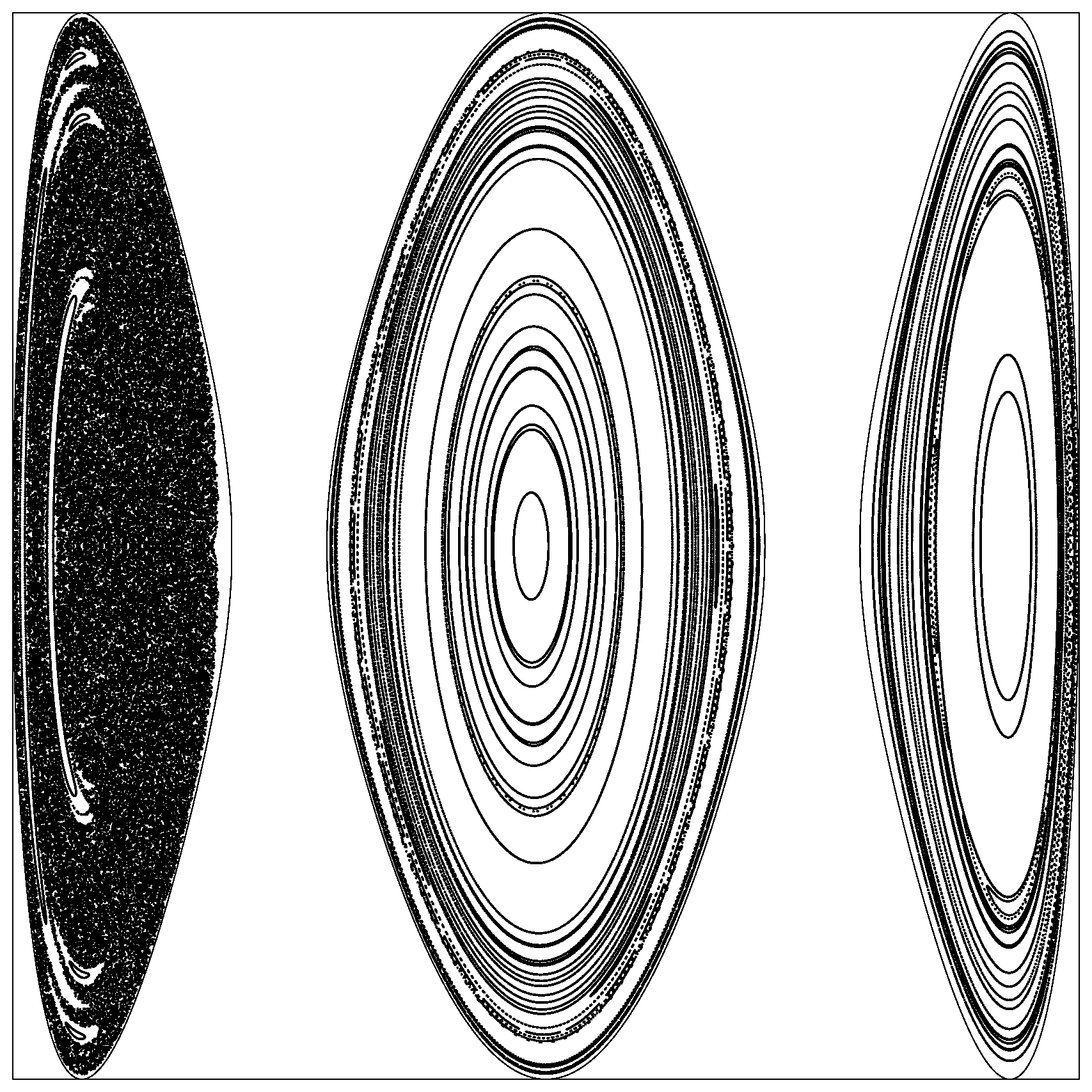}
\caption{Level lines and Poincar\'e sections for $D_5$ and $D_7$
potentials .\label{d5d7}}
\end{figure}

The mixed state, which was shown above for the potential of
quadrupole oscillations, is the representative state for a wide
class of two-dimensional potentials with few local minima. According
to the catastrophe theory \cite{cat_th}, a rather wide class of
polynomial potentials with several local minima is covered by the
germs of the lowest umbilical catastrophes, of type $D_5$, $D_4^-$,
$D_7$, subjected to certain perturbations. Let us note that the
H\'enon-Heiles potential coincides with the elliptic umbilic
$D_4^-$. Fig.\ref{d5d7} represents level lines and Poincar\'e
sections at different energies for multi-well potentials from a
family of umbilical catastrophes $D_5$ and $D_7$:
\begin{equation}\label{u_d5} U_{D_5}=2y^2-x^2+xy^2+\frac14 x^4 + 1\end{equation}
\[U_{D_7}=\sqrt2 y^2+\frac38 x^2+xy^2-\frac12 x^4+\frac16 x^6\]

The mixed state is observed for all considered potentials of
umbilical catastrophes in the interval of energies $E_{cr}<E<E_S$
(here $E_{cr}$ is the critical energy of the transition to chaos in
that local minimum, where chaos is observed at energies smaller than
the saddle energy).
\chapter{Chaoticity Criteria for Multi-Well Potential}\sat
\section{General formulation of the stochasticity criteria}
As noted above, stochasticity is understood as a rise of statistical
properties in a purely deterministic system due to local
instability. According to this definition values of parameters of a
dynamical system, under which local instability arises, are
identified as regularity-chaos transition values. However,
stochasticity criteria of such a type are not sufficient (their
necessity poses a separate and complicated question), since loss of
stability could lead to the transformation of one kind of regular
motion to another. Despite this serious limitation, stochastic
criteria in combination with numerical experiments facilitate an
analysis of motion and essentially extend efficacy of numerical
calculations.
\section{Non-linear resonance overlap criterion}
First among the widely used stochasticity criteria is the nonlinear
resonances overlap criterion presented by Chirikov \cite{chirikov}.
The essence of this criterion is easier to explain by the example of
a one-dimensional Hamiltonian system, which is subjected to periodic
perturbation
\[H=H_0(p,x)+Fx\cos\Omega t.\]

For the unperturbed system we can always introduce the action-angle
variables $(I,\theta)$ in which
\begin{equation}\label{h_aa}H=H_0(I)+\sum\limits_{k=-\infty}^{\infty}x_k(I)\cos(k\theta-\Omega
t),\end{equation} where
\[x_k(I)=\frac{1}{2\pi}\int\limits_0^{2\pi}d\theta e^{ik\theta}x(I,\theta).\]

In the new variables the scenario of stochasticity, on which
resonances overlap criterion is based, is the following. An external
periodic in time field induces a dense set of resonances in the
phase space of a nonlinear conservative Hamiltonian system. The
positions of these resonances $I_k$ are determined by the resonance
condition between the eigenfrequency \[\omega(I)=\frac{\partial
H_0}{\partial I}\] and the frequency of the external perturbation
$\Omega$. For very weak the external fields the principle resonance
zones remain isolated. As the amplitude $F$ of external field is
raised, the widths $W_k$ of the resonance zones increase
\[W_k=\left.4\left(\frac{fx_k}{\omega'(I)}\right)^{\frac12}\right|_{I=I_k}\]
and at $F>F_{cr}$ resonances overlap. When this overlap occurs, i.e.
under the condition
\begin{equation}\label{roc}\frac12(W_k+W_{k+1})=|I_k-I_{k+1}|,\end{equation}
it is said that there is transition to global stochastic behavior in
the corresponding region of the phase space. The averaged motion of
the system in the neighborhood of the nonlinear isolated resonance
on the plane of the action-angle variables is similar to the
particle behavior in the potential well.  Isolated resonances
correspond to isolated potential wells. The overlap of the
resonances means, that the potential wells are close enough to make
possible the random walk of a particle between these wells.

The outlined scenario can easily be "corrected" for the description
of the transition to chaos in the conservative system with several
degrees of freedom. The condition of the resonance between the
eigenfrequency and the frequency of the external field must be
replaced by the condition of the resonance between the frequencies,
which correspond to different degrees of freedom
\[\sum\limits_i m_i\frac{\partial H_0}{\partial I_i}=0.\]
The intensity of the interaction between different degrees of
freedom, i.e. the measure of nonlinearity of the original
Hamiltonian, acts as the amplitude of the external field in this
case.
\section{Stochastic layer destruction criterion}
This method could be modified for systems with unique resonance
\cite{doviel}. In this case, the origin of the large-scale
stochasticity is connected with the destruction of the stochastic
layer near the separatrix of the isolated resonance. The main point
of modification consists of the approximate reduction of the initial
Hamiltonian in the neighborhood of the resonance to the Hamiltonian
of the periodically perturbed nonlinear oscillator
\[H(\nu,x,\tau)=\frac12\nu^2 M\cos x - P\cos k(x-\tau).\]
The width of the resonance stochastic layer defined by
\cite{rechester}
\[W\sim \frac{\rho e^{-\frac1\rho}}{M\rho^{2k+1}}\]
where
\[\rho=\frac{2\sqrt M}{\pi k}.\]
If $P/M$ has an order $\rho^s$, then
\[W\sim \rho^{-\lambda}e^{-\frac1\rho}\]
where $\lambda=2k+1-s$. And when
\[\rho_i=\frac1\lambda\left(1-\frac{1}{\sqrt{1+\lambda}}\right)\]
function $W(\rho)$ has a bending flex point. Fast growth of $W$
allows us to define the threshold of the stochastic layer
destruction as
\[\rho_{cr}=\frac{1}{\lambda^2}\left(\sqrt{1+\lambda}-1\right),\]
when the tangent line in point $\rho_i$ to the function $W(\rho)$
transects the $\rho$-axis.

Application of these criteria in the presence of strong nonlinearity
(which is inevitable when considering multi-well potentials)
encounters an obstacle: action-angle variables effectively work only
in the neighborhood of the local minimum. Because of this, interest
in methods based on direct estimation of trajectories divergence
speed, arises. One of the criteria of such a type is so-called
negative curvature criterion \cite{toda}. This criterion is based on
the existence of the following scenario of the transition from
regular to chaotic motion.
\section{Negative curvature criterion}
At low energies, the character of motion near the minimum of the
potential energy, where the curvature is obviously positive, is
periodic or quasiperiodic and is separated from the instability
region by the zero curvature line. As the energy grows, the
"particle" will stay for some time in the negative curvature region
of the PES where initially close trajectories exponentially diverge.
After a long time these result in a motion which imitates a random
one and is usually called stochastic. According to this
stochastization scenario, the critical energy of the transition to
chaos $E_{cr}$ coincides with the lowest energy on the zero
curvature line
\[E_{cr}=U_{min}(K=0)\]
\subsection{Results for one-well case}
Now let's demonstrate the efficiency of negative curvature criterion
on the example of  one-parametric family of potentials
\begin{equation}\label{u_mu}U(x,y;\mu)=\frac12(x^2+y^2)+xy^2+\mu x^3\end{equation}
With $\mu=-1$ potential (\ref{u_mu}) reduces to H\'enon–-Heiles
potential and with $\mu=+1$ --- to a separable one that is called
anti-–H\'enon–-Heiles potential. Interaction in any three-body
system can be reduced to such a type of potential in the cubic
approximation if its potential has the form \cite{ford}
\[U(x_1,x_2,x_3)=U(x_1-x_2)+U(x_2-x_3)+U(x_3-x_1).\]

\begin{figure}
\center{\includegraphics[width=0.5\textwidth,draft=false]{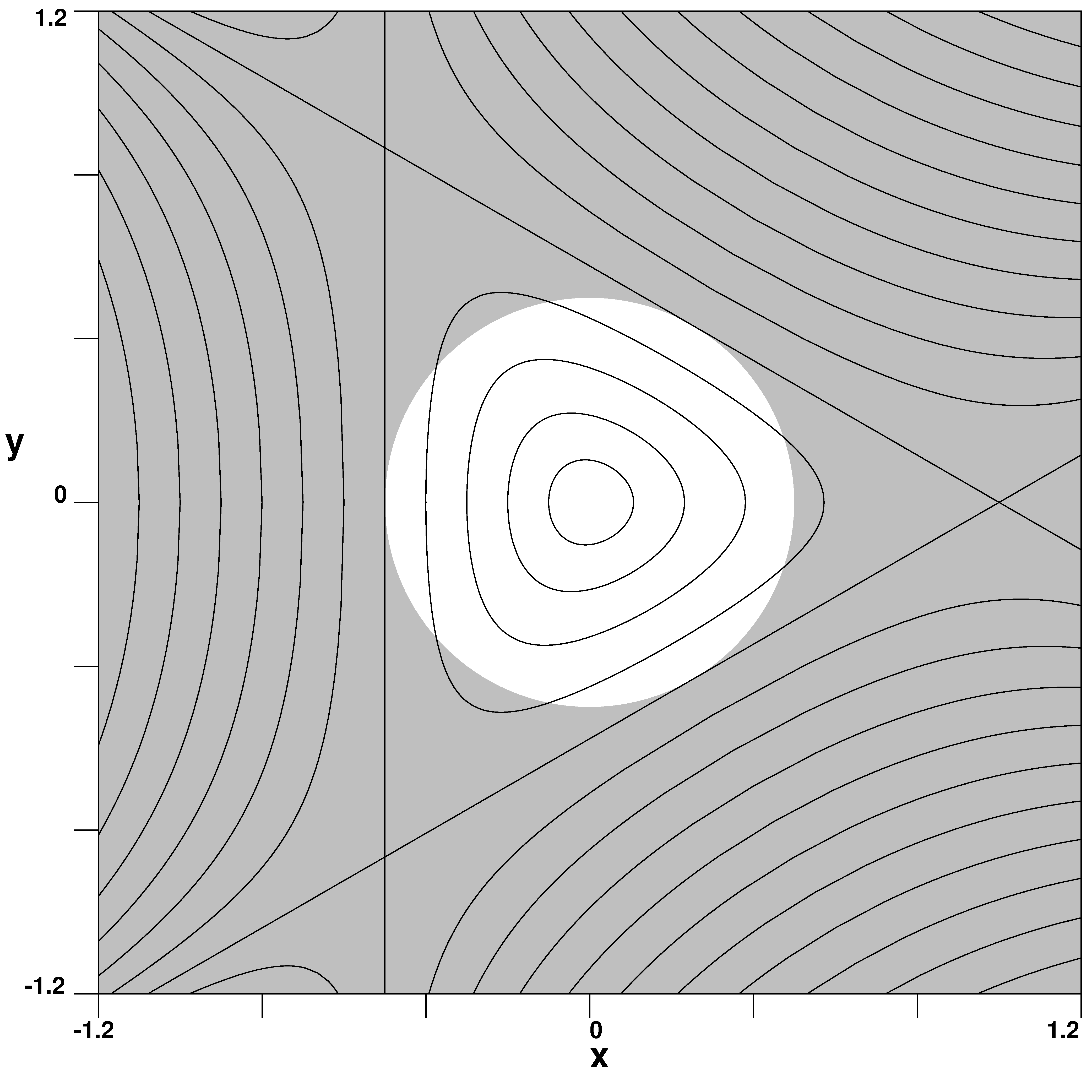}}
\caption{Level lines and negative curvature region (in grey) for
H\'enon-–Heiles potential \label{hh_pes}}
\end{figure}

\begin{figure}
\includegraphics[width=0.5\textwidth,draft=false]{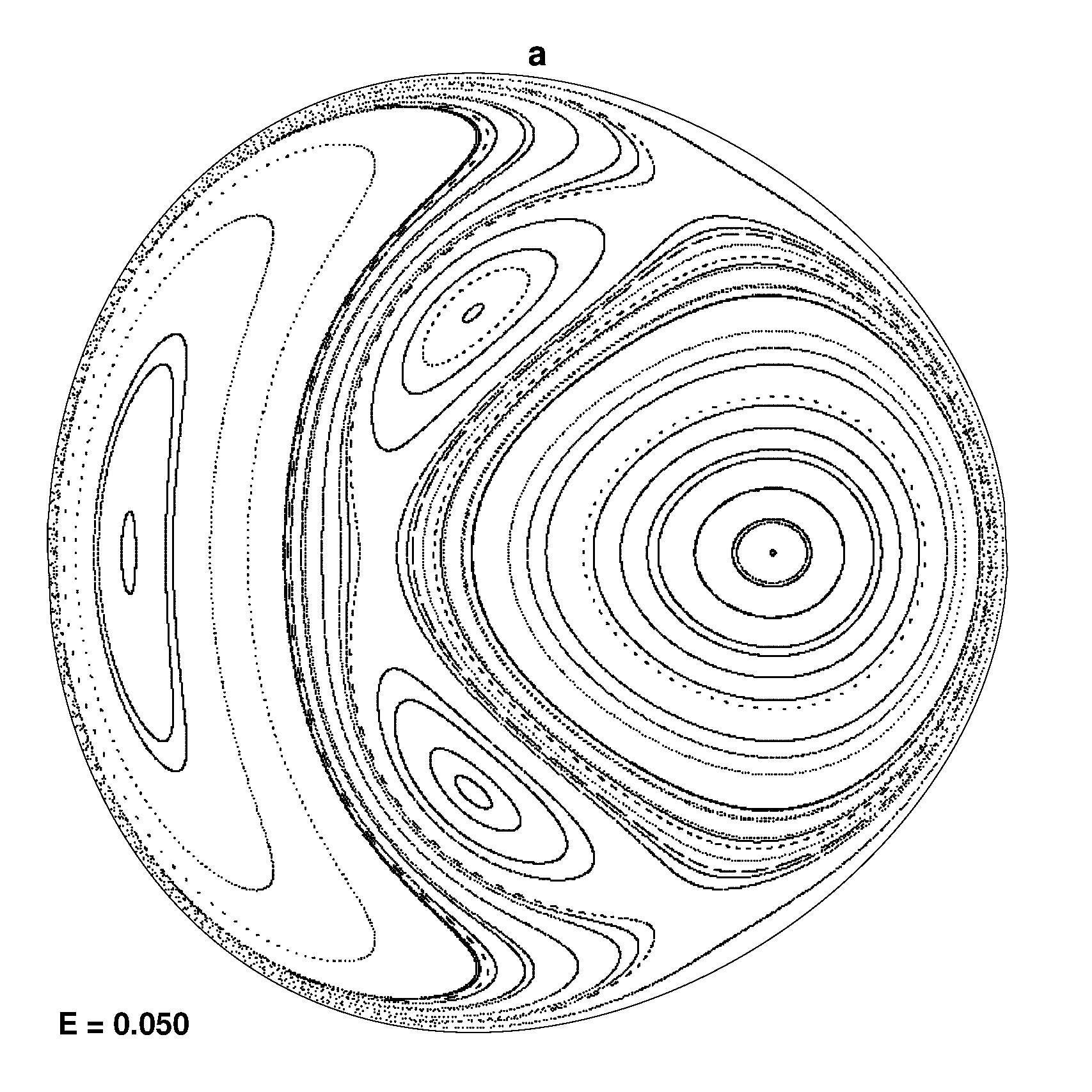}
\includegraphics[width=0.5\textwidth,draft=false]{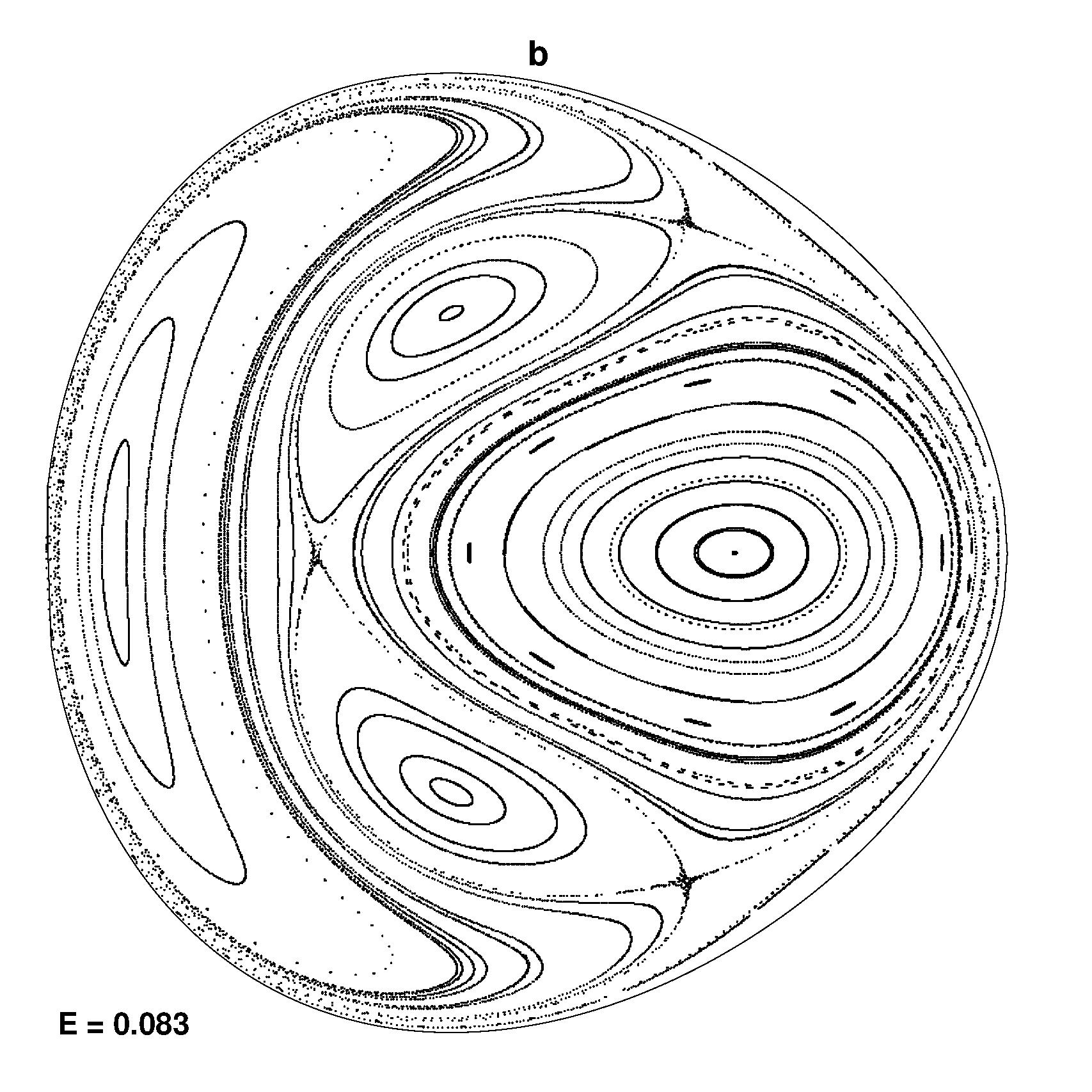}

\vspace{1cm}
\includegraphics[width=0.5\textwidth,draft=false]{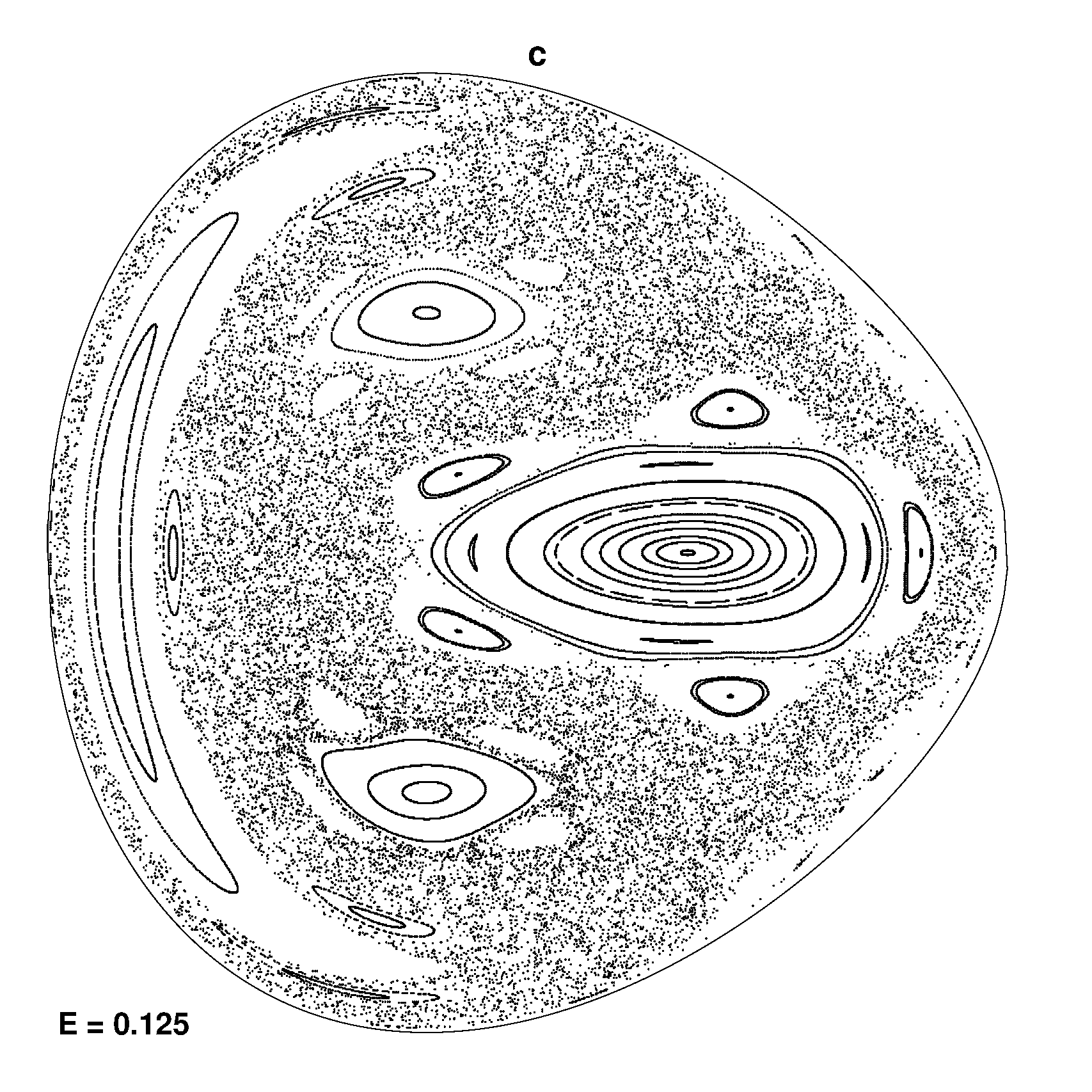}
\includegraphics[width=0.5\textwidth,draft=false]{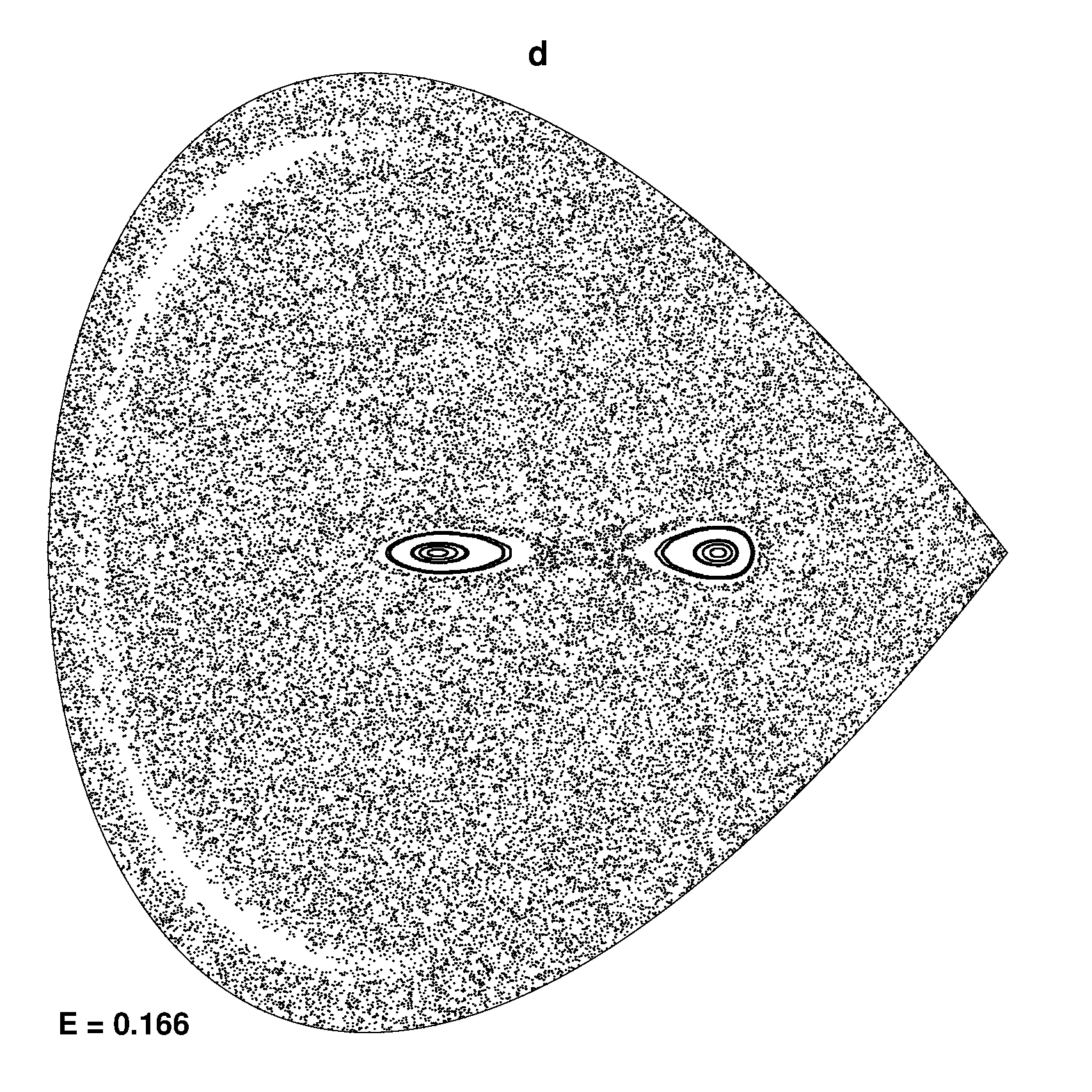}
\caption{PSS for the H\'enon–-Heiles potential at $E<E_S/2$ (a),
$E=E_S/2$ (b), $E_S/2<E<E_S$ (c), $E=E_S$ (d).\label{hh_pss}}
\end{figure}

Gaussian curvature of the considered PES turns to zero at the points
that satisfy the condition
\[-\frac1\mu y^2+(x-x_0)^2=R_0^2,\]
where
\[x_0=-\frac{1+\mu}{4\mu};\ R_0=\frac{1-\mu}{4\mu}.\]
When $\mu<0$, the zero-curvature line of (\ref{u_mu}) is an ellipse,
which reduces to a circle for the H\'enon–-Heiles potential
$(\mu=-1)$ (Fig.\ref{hh_pes})
\[x^2+y^2=\frac14\]
On the zero-curvature line, energy is defined by the expression
\[U(K=0)=\frac43\mu x^3+(1+\mu)x^2+\frac14(2+\mu)x +\frac18,\]
and the minimal value of energy on the zero-curvature line is
\[U_{min}(K=0)=\frac{3-\mu}{48}.\]
According to the considered stochastization scenario, this value
offers the critical energy when the transition from regular to
chaotic motion occurs. For the H\'enon–-Heiles potential
\[U_{min}(K=0)=\frac{1}{12}.\]
This result is in good agreement with numerical integration of the
equations of motion, which is presented on Fig.\ref{hh_pss} in the
form of the Poincar\'e sections.

It is necessary to make one important remark. Analysis of numerical
integration of the equations of motion (e.g. in the case of PSS)
allows us to introduce the critical energy of the transition to
chaos, determined as the energy such that part of phase space with
chaotic motion exceeds a certain arbitrary chosen value. Similar
uncertainty is connected with the absence of the sharp transition to
chaos when energy increases. Therefore a certain caution is required
for comparison of the "approximate" critical energy obtained by
numerical simulation, with the "exact" value obtained with the help
of analytical estimations, i.e. on the base of different criteria of
stochasticity.
\subsection{Failure for multi-well case}
Negative curvature criterion allows us to obtain a number of
interesting results in potentials with simple geometry (a single
minimum) \cite{bolotin-89}. However, when passing to multi-well
potentials, this criterion fails to work correctly. In particular,
for the above mentioned potentials ($D_5$ and $D_7$), the structure
of Gaussian curvature is similar in different wells. For example,
for $D_5$ potential according to negative curvature criterion we get
the same value of critical energy for both minima $E_{cr}\sim 5/9$,
but chaotic motion is only observed in the left well (see
Fig.\ref{d5d7}).

A natural question immediately arises: is it possible, using only
geometrical properties of PES but not numerically solving equations
of motion, to formulate an algorithm for finding the critical energy
for single local minima in multi-well potential? We will try to
answer this question below in the framework of the so-called
geometrical approach \cite{pettini_cs,pettini}.
\section{Geometrical approaches}\sat
The geometrical approach is based on application of differential
geometry to study the chaotic dynamics of Hamiltonian systems. It
turns out that as long as we consider the Hamiltonian of the form
(\ref{qo_ham}) we can restrict ourselves to the study of the
configuration space without loosing information. Thus application of
this method for the analysis of the features of Hamiltonian dynamics
in multi-well potentials seems natural because the characteristic of
the multi-well potential is formulated in terms of configurational
space.

As is known, geodesics are among the main objects in Riemannian
geometry. They are defined as the shortest curves that connect two
points on a manifold. The manifold in its turn is defined by
metrics. Having once fixed the metrics we thus define the distance
on the manifold:
\[ds^2=g_{ik}dx^i dx^k.\]
We have the following condition for geodesics in this case:
\begin{equation}\label{gc}\delta\int ds=0.\end{equation}
After variation we could obtain the differential equation for
geodesics:
\begin{equation}\label{ge}\frac{d^2
q^i}{ds^2}+\Gamma^{i}_{\,jk}\frac{dq_j}{ds}\frac{dq_k}{ds}=0,\end{equation}
where $\Gamma^{i}_{\,jk}$ are Christoffel symbols.

Using variational principles it is possible to formulate Hamiltonian
mechanics in a geometrical way. Let us consider this question more
closely. A trajectory of dynamical system is defined according to
Maupertuis \cite{landau} principle:
\begin{equation}\label{mp}\delta\int\limits_\gamma
2Tdt=0,\end{equation} where $\gamma$ are all isoenergetic paths
connecting end points, or according to Hamilton's principle:
\[\delta\int\limits_{t_1}^{t^2} Ldt=0.\]

To connect mechanics with Riemannian geometry we must choose the
metrics that convert the expression under the integral into the
length element. By such a procedure we will specify the manifold.
Then trajectories will be geodesics on this manifold. We will call
this CM –-- configurational manifold. This approach has an evident
advantage: potential energy function includes all information about
the system, so one needs to consider only configurational space but
not phase space. Let's emphasize that Christoffel symbols in this
approach act as counterparts of forces in ordinary mechanics and
metrics --- as a potential.

The simplest metric is the Jacobi metric. It has the form:
\begin{equation}\label{jm} g_{ik}=2\left(E-V(q)\right)\delta_{ik}\end{equation} By means of this
metric Maupertuis principle (\ref{mp}) could be rewritten in the
form equivalent to the condition for geodesics (\ref{gc}) so that
trajectories are defined by equation (\ref{ge}). It could be shown
that the geodesic equation (\ref{ge}) with Jacobi metric (\ref{jm})
leads to Newton's equations.

Having equations of motion we now could consider local instability
in geometrical form. Let $\mathbf{q}$ and $\mathbf{q'}$ be two
nearby trajectories at $t=0$. Then let us define the deviation
\[J_i=q'_i-q_i.\]
The dynamics of deviation are governed by the well known
Jacobi-Levi-Civita (JLC) equation:
\begin{equation}\label{jlc}\frac{d^2
J^i}{ds^2}+R^{i}_{\,jkl}\frac{dq^j}{ds}J^k\frac{dq^l}{ds}=0,\end{equation}
where $R^{i}_{\,jkl}$ is the curvature tensor. The two-dimensional
case that we are interested in is very simple to consider because
the only nonvanishing curvature tensor component is $R_{1212}$. To
write JLC equation (\ref{jlc}) explicitly we need to present local
orthonormal basis. The simple choice is the following:
\[\mathbf{e}_1=\left(-\frac{dq^2}{ds},\frac{dq^1}{ds}\right),\
\mathbf{e}_2=\left(\frac{dq^1}{ds},\frac{dq^2}{ds}\right).\] On this
basis the deviation takes the form:
\[\mathbf{J}=\sum\limits_i \xi_i(s)\mathbf{e}_i(s)\]
It can be shown that in the two-dimensional case the JLC equation
leads to two equations for deviation components on the chosen basis:
\begin{equation}\label{deviation}\begin{array}{c}
\frac{d^2\xi_1}{ds^2}+\frac12R\xi_1=0\\
\\
\frac{d^2\xi_2}{ds^2}=0
\end{array}\end{equation} where $R$ is scalar curvature. We can see that stability is
determined by scalar curvature. For two-degrees-of-freedom systems
Riemannian curvature has the form
\[R=\frac{(E-V)\Delta V+|\Delta V|^2}{2(E-V)^2}\]
where $\Delta V$ is positive for the considered potentials; thus
Riemannian curvature is positive too. Due to this we cannot connect
divergence of trajectories with negative Riemannian curvature.

Pettini et al. \cite{pettini} point out that instability is caused
by oscillations of positive curvature and has parametric nature.
Calculations of deviation dynamics of regular and chaotic
trajectories in H\'enon–-Heiles are presented in \cite{pettini_cs}.
It is shown that, for regular trajectories, the normal component of
deviation is bounded or grows linear in time, while for chaotic it
grows exponentially. The initial conditions were chosen in specific
areas in the Poincar\'e section. It is interesting to mention that
initial conditions lying on the border of a regular island in
section exhibit very slow exponential growth. This illustrates the
so-called effect of sticky orbits.

Let us briefly discuss the behavior of deviation in equation
(\ref{deviation} second order for $\xi_1$). In order to do that we
need firstly to transform (\ref{deviation}) to physical values, i.e.
to time instead of interval \cite{pettini}. This procedure leads to
the following equation:
\begin{equation}\label{xi_1_pettini}\frac{d^2\xi_1}{dt^2}-\frac{\dot{W}}{W}\frac{d\xi_1}{dt}+\left[\Delta
V+\frac{(\nabla V)^2}{W}\right]\xi_1=0.\end{equation} To simplify
the equation and clarify its physical meaning, authors of paper
\cite{pettini} make the substitution:
\begin{equation}\label{subs}\xi_1(t)=Y(t)\sqrt{W}\end{equation}
Obviously, $\xi_1(t)$ and $Y(t)$ have the same behavior in the
meaning of stability hence $W(t)$ is bounded. Thus, using
(\ref{subs}):
\[\frac{d^2Y}{dt^2}+\Omega(t)Y=0\]
with
\[\Omega(t)=\left[\Delta V+\frac{3(\nabla V)^2}{2W}\right] -
\frac{3}{4W^2}\left[\dot{q}_1\frac{\partial V}{q_1} +
\dot{q}_2\frac{\partial V}{q_2}\right]^2 -
\frac{1}{2W}\sum\limits_{i,k}\frac{\partial^2V}{\partial q_1\partial
q_2}\dot{q}_i \dot{q}_k\]

Another geometrical approach, based on redefinition of covariant
derivative was proposed by Kocharyan in \cite{kocharyan}. In the
two-dimensional case these two approaches lead to the same equations
for deviation.

There are two possibilities for the solution to be unstable. The
first case appears when $\Omega(t)$ is negative. The second
possibility is parametric resonance. As was mentioned before, the
second case is more actual since the Riemannian curvature is
positive.

\begin{figure}
\includegraphics[width=\textwidth]{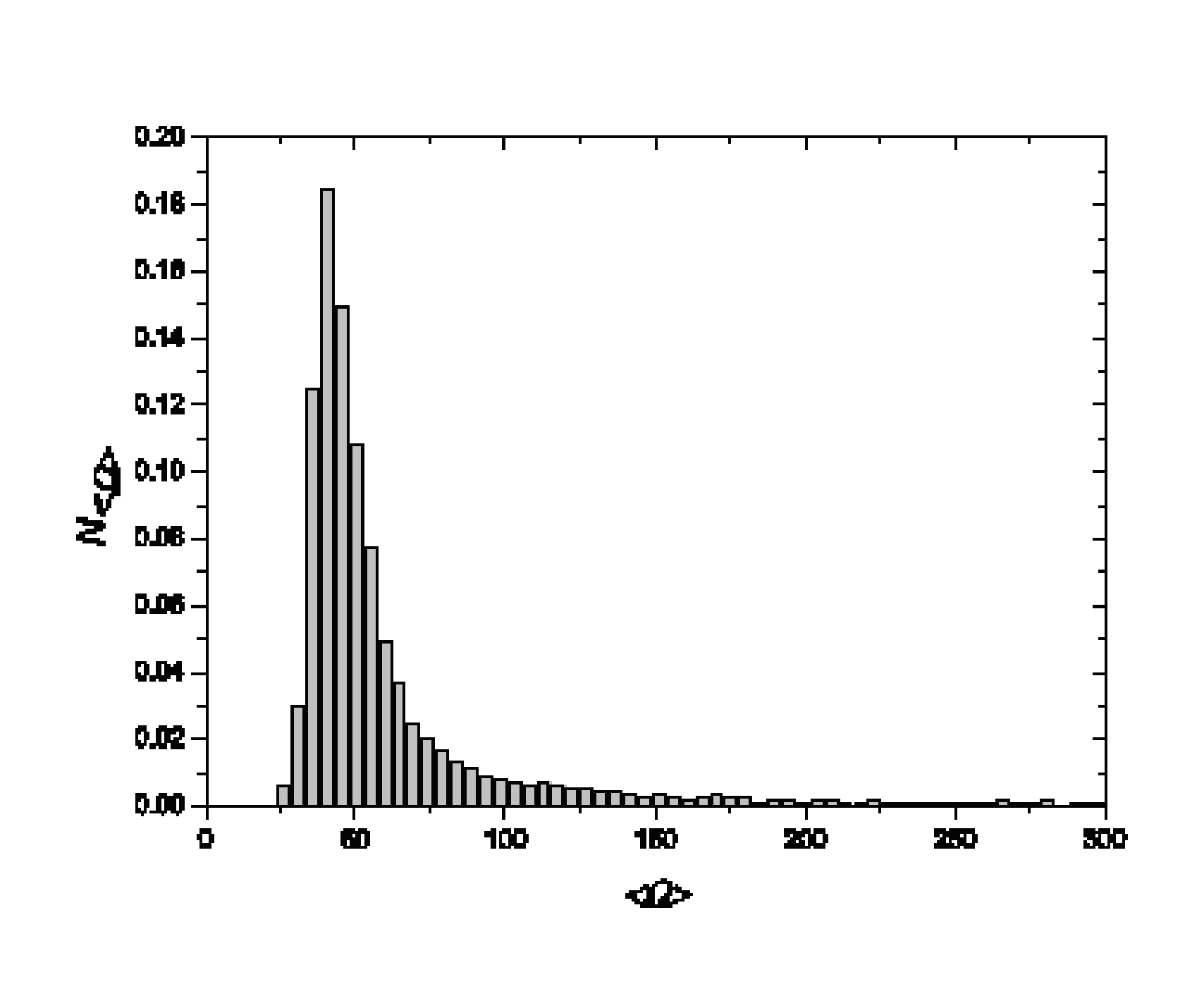}
\caption{\label{omega_t_distr}The distribution of
$\langle\Omega(t)\rangle$ for an ensemble of $10^4$ trajectories in
the chaotic well of $D_5$ potential for $E=E_S-0.1$}
\end{figure}

Fig.\ref{omega_t_distr} presents the distribution of
$\langle\Omega(t)\rangle$ ($\langle\ldots\rangle$ means time
averaged value) for an ensemble of $10^4$ trajectories in the
chaotic well of potential $D_5$ with energy $E=E_S-0.1$.

As we can see, there is no trajectory with negative
$\langle\Omega(t)\rangle$. Further, only $10\%$ of trajectories have
somewhere $\Omega(t)<0$, but even for them $\langle\Omega(t)\rangle$
is positive. Although this is only a brief survey of the situation,
we can expect that instability of the solution of
(\ref{xi_1_pettini}) to have parametric nature.

As was mentioned earlier, stochasticity criterion must derive
critical energy. However it was pointed out by Pettini et al. that
we must know all information about curvature oscillations to
consider parametric instability. So, dynamical calculations are
unavoidable in this investigation. Nevertheless, we could derive
purely geometrical criterion by introduction of additional
coordinates. Let us rewrite the JLC equation in the form which does
not depend on dimensionality of the manifold:
\[\frac12\frac{d^2 \left\|\mathbf{J}\right\|^2}{ds^2} +
 K^{(2)}(\mathbf{J},\mathbf{v})\left\|\mathbf{J}\right\|^2 -
 \left\|\frac{d}{ds}\mathbf{J}\right\|^2=0\]
where $K^{(2)}(\mathbf{J},\mathbf{v})$ is the so-called sectional
curvature:
\[K^{(2)}(\mathbf{J},\mathbf{v})=R_{iklm}\frac{J^i}{\left\|\mathbf{J}\right\|}\frac{dq^k}{ds}
\frac{J^l}{\left\|\mathbf{J}\right\|}\frac{dq^m}{ds}\] and
$\langle\mathbf{J},\mathbf{v}\rangle=0$. Note that the point where
$K^{(2)}(\mathbf{J},\mathbf{v})<0$ is unstable. Since there are more
than one sectional curvatures for the case $N>2$, we could connect
instability with the negative sign of some of them. It is assumed
that negativity of some of the sectional curvatures is sufficient
condition for the rise of instability. One of the enlarged metrics
is the Eisenhart metric:
\[ds^2_E=(g_E)_{\mu\nu}dq^\mu dq^\nu = a_{ij}dq^i dq^j - 2V(\mathbf{q})(dq^0)^2 + 2dq^0dq^{N+1}\]
where $a_{ij}$ is the kinetic energy matrix. The additional
coordinates are $q^0=t$ and $q^{N+1}$ (the latter is connected with
action). The nonvanishing components of curvature tensor are
\[R_{0i0j}=\frac{\partial^2 V}{dq^i dq^j}.\]

Pettini et al. \cite{pettini_cs} considered
$K^{(2)}(\mathbf{J},\mathbf{v})$ on the constant energy surface for
vectors
\[\mathbf{J}=(0,\dot{q}_2,-\dot{q}_1,0),\ \mathbf{v}=(1,\dot{q}_1,\dot{q}_2,\dot{q}_3).\]
Thus $K^{(2)}(\mathbf{J},\mathbf{v})$ takes on the form:
\[K^{(2)}(\mathbf{q},\mathbf{\dot{q}})=\frac{1}{2(E-V(\mathbf{q}))}
\left(\frac{\partial^2 V}{\partial q_1^2} \dot{q}_2^2 +
\frac{\partial^2 V}{\partial q_2^2} \dot{q}_1^2 - \frac{\partial^2
V}{\partial q_1\partial q_2} \dot{q}_1\dot{q}_2 \right)\] This value
is easy to calculate at any point of phase space. In
\cite{pettini_cs} the averaged value of
$K^{(2)}(\mathbf{J},\mathbf{v})$ is introduced. It is shown that for
H\'enon–-Heiles potential there exists a correlation between chaotic
trajectories relative measure and averaged value of sectional
curvature. This approach correctly predicts the value of critical
energy.

\begin{figure}
\includegraphics[width=\textwidth,draft=false]{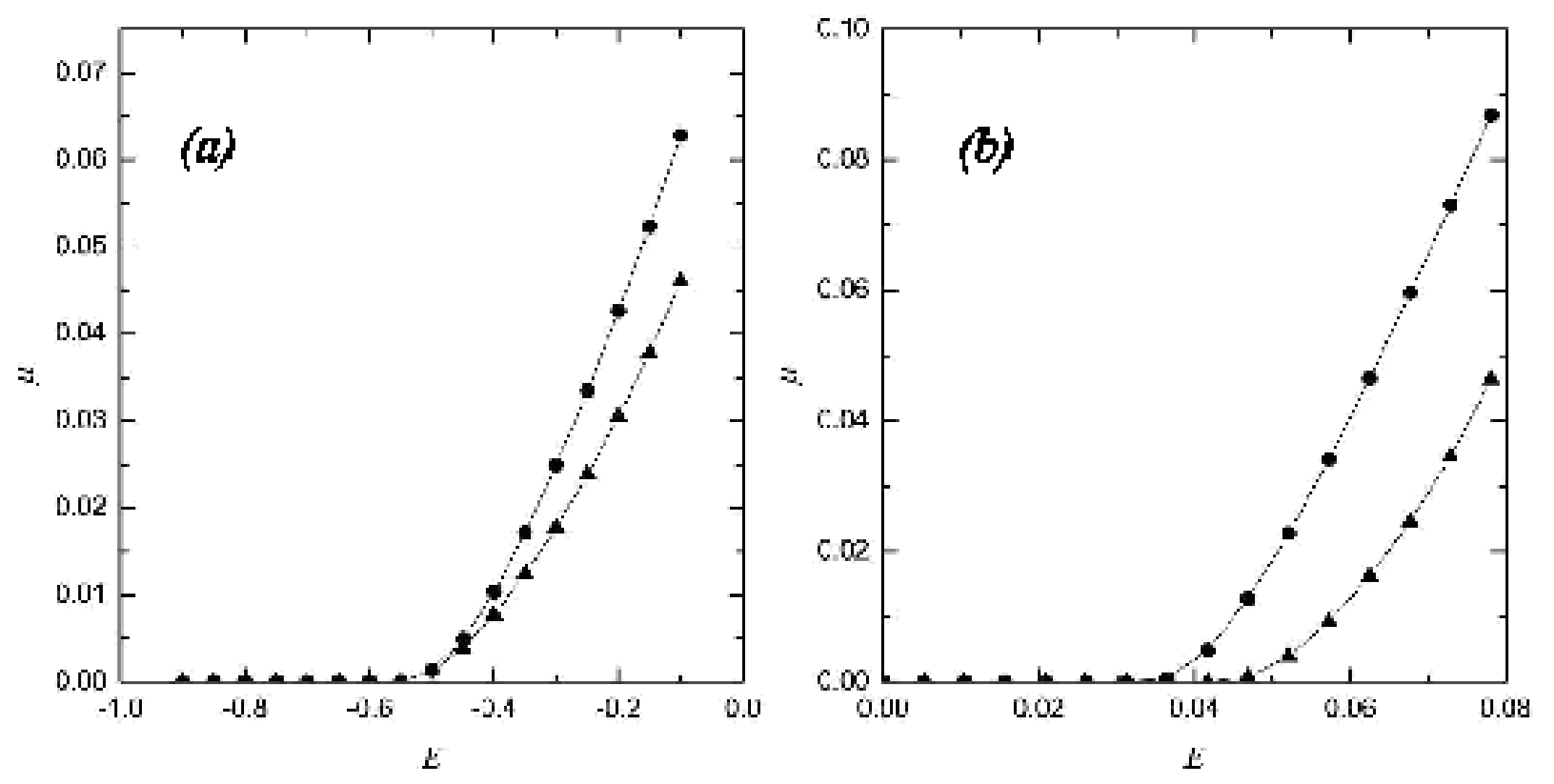} \caption{Function $\mu(E)$ for $D_5$ (a) and $D_7$ (b) potentials.
$\mu(E)$ for chaotic wells are represented by dotted lines, for
regular --- by triangles. \label{mu}}
\end{figure}

The case of multi-well potential is more complex. It is necessary to
clarify whether this condition is sufficient for the development of
chaoticity or not; clearly speaking we needs to answer the question,
does the presence of negative curvature parts on CM always lead to
chaos? Potentials with mixed state are a very convenient model for
investigation of this question, since there exist both regimes of
motion. So, we need to study, how the structure of
$K^{(2)}(\mathbf{J},\mathbf{v})$ differs in different wells. For
that we calculate in \cite{gleb} a part of phase space with negative
curvature as a function of energy, i.e. a volume of phase space
where $K^{(2)}(\mathbf{J},\mathbf{v})<0$ referred to the total
volume:
\[\mu(E)=\frac{\int d\mathbf{q} d\mathbf{p}\Theta(-K^{(2)})\delta(H(\mathbf{q},\mathbf{p})-E)}{\int d\mathbf{q} d\mathbf{p}\delta(H(\mathbf{q},\mathbf{p})-E)}.\]
We carried out calculations for two potentials: $D_5$ and $D_7$.
Calculations of $\mu(E)$ (Fig.\ref{mu}) show that there are parts,
where $K^{(2)}(\mathbf{J},\mathbf{v})<0$ in all wells, but
nevertheless chaos exists only in one well. Moreover, for the well
with chaotic motion, function $\mu(E)$ gives the correct value of
critical energy --- at this energy $\mu(E)$ becomes positive.

\begin{figure}
\includegraphics[width=\textwidth,draft=false]{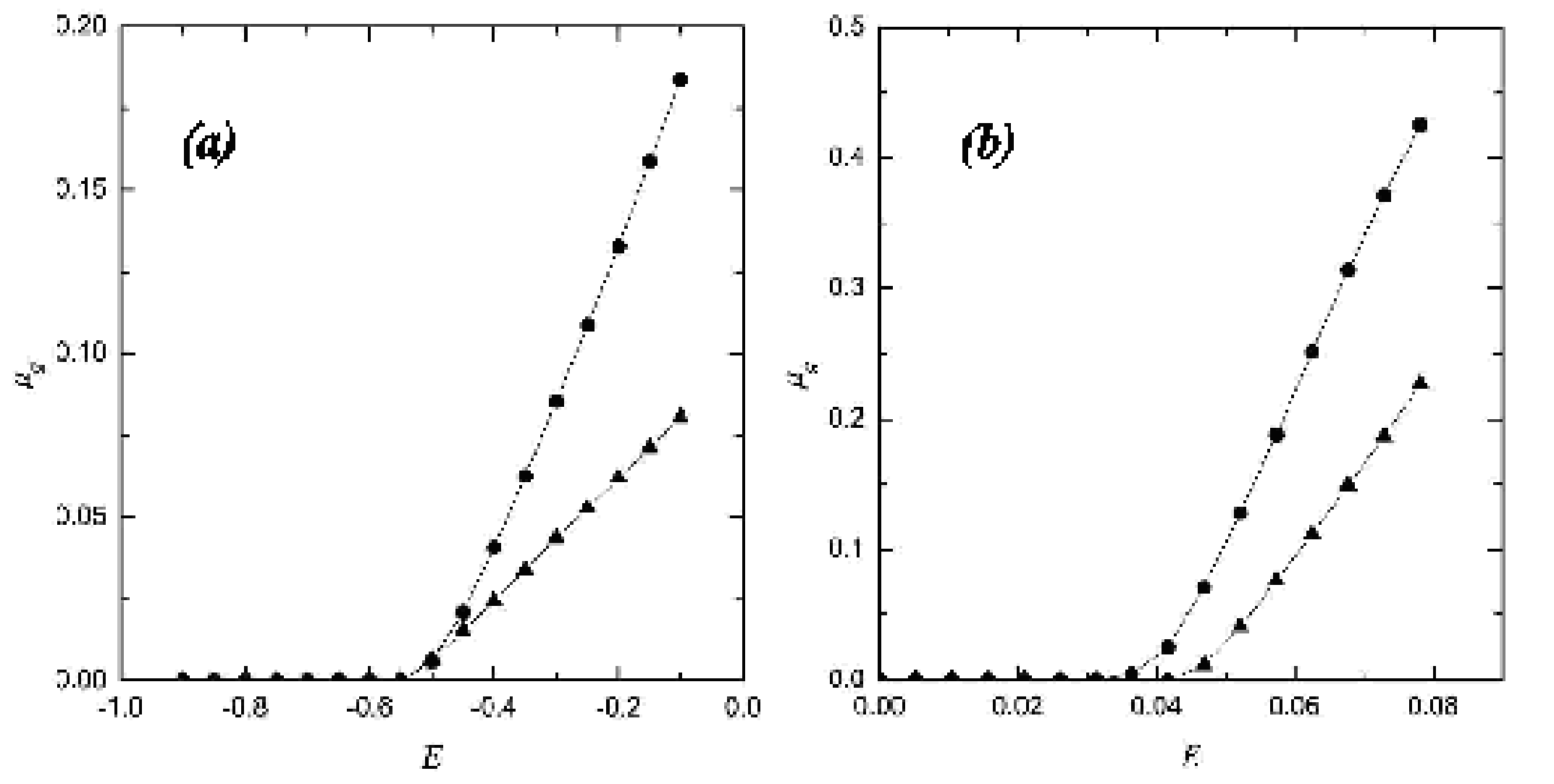}
\caption{Function $\mu_G(E)$ for $D_5$ (a) and $D_7$ (b) potentials.
\label{mug}}
\end{figure}

The situation with regular wells is more complicated. Although part
of phase space, where $K^{(2)}(\mathbf{J},\mathbf{v})<0$, is
nonzero, chaos in the well does not exist. This can be viewed on the
Poincar\'e sections. For comparison in Fig.\ref{mug} part of CS with
negative Gaussian curvature ($\mu_G$) is shown. We can see that
structure of negative Gaussian curvature is similar to the
$K^{(2)}(\mathbf{J},\mathbf{v})$-structure.

Investigation of the curvature of the manifold, as we can see from
the above cited data, does not give a plain method for
identification of chaos in any minimum, especially if there exist
both regular and chaotic regimes of motion. It is impossible to
determine a priori whether chaos existed in the system without using
dynamical description (in our case which are Poincar\'e sections).
Nevertheless, we can efficiently use geometrical methods for
investigation of chaos in multi-well potentials. In thee potentials
considered above chaos exists only in wells, which have two details:
a non-zero part of negative curvature on the manifold and at least
one hyperbolic point in the Poincar\'e section. According to this,
we can use the following method for identification of chaos and
calculation of critical energy. At the first step the Poincar\'e
section at low energy is drawn for the well and the presence of the
hyperbolic point is detected. The quantity $\mu(E)$ must then be
calculated (or the part of CS with negative Gaussian curvature). The
value of energy at which $\mu(E)$ becomes positive could be
associated with critical energy. If there are no hyperbolic points
in the section than chaos does not exist in the well. Consequently,
geometrical methods could be efficiently used for determination of
critical energy in complex potentials and identification of chaos in
general. However, we must carefully use these methods and combine
them with qualitative methods, such as Poincar\'e sectioning method.
\sat\section{Normal forms}\sat
The structure of the Poincar\'e
surfaces of section can be reproduced without resorting to the
numerical solution of the equations of motion. For this, let us use
the method of treating non separable classical systems that was
originally developed by Birkhoff \cite{birkhoff} and later was
extended by Gustavson \cite{gustavson}.

Every two-dimensional Hamiltonian near equilibrium point can be
represented in polynomial form as follows
\[\begin{array}{c}
H(\mathbf{p},\mathbf{q})= H^{(2)}(\mathbf{p},\mathbf{q}) +
V(\mathbf{q}),\\
H^{(2)}(\mathbf{p},\mathbf{q}) = \sum\limits_{\nu=1}^2 \frac12
\omega_\nu^2(p_\nu^2+q_\nu^2),\\
V(\mathbf{q})=\sum\limits_{j\le3}V_{j_1 j_2} q_1^{j_1} q_2^{j_2},\\
\mathbf{q}=(q_1,q_2),\ \mathbf{p}=(p_1,p_2).
\end{array}\]

The procedure of reducing to normal form depends on whether the
frequencies $\omega_\nu$ are commensurable or not. If they are
incommensurable then there exists a canonical transformation
$(\mathbf{q},\mathbf{p})\rightarrow(\mathbf{\xi},\mathbf{\eta})$
such that in variables $(\mathbf{\xi},\mathbf{\eta})$ Hamiltonian
$\Gamma(\mathbf{\xi},\mathbf{\eta})$ will be a function of only two
combinations
\[I_\nu=\frac12\left(\xi_\nu^2+\eta_\nu^2\right),\ \nu=1,2.\]
In other words, the Birkhoff normal form is an expansion of the
original Hamiltonian over two one-dimensional harmonic oscillators
\[H(\mathbf{p},\mathbf{q})\rightarrow\Gamma(\mathbf{\xi},\mathbf{\eta}) =
\omega_1 I_1 + \omega_2 I_2 +\sum\limits_{\mu,\nu}a_{\mu\nu}I_\mu
I_\nu + \ldots\] If the frequencies $\omega_\nu$ are commensurable,
i.e. if there exist resonance relations of the type
$m\omega_1+n\omega_2$, the normal form becomes more complicated and
will contain apart from $I_\nu$ other combinations of variables
$\xi_\nu$ and $\eta_\nu$ as well. Such extended normal forms are
called the Birkhoff-Gustavson normal forms.

We cite as an example the Birkhoff normal form (up to the terms of
the sixth degree with respect to $(\mathbf{\xi},\mathbf{\eta})$) for
the umbilical catastrophe $D_5$ in the neighborhood of right
minimum:
\[\begin{array}{c}H(I_1,I_2)=2.613I_1+2I_2-0.219I_1 I_2-0.017I_1^2-0.375I_2^2-\\
-0.005I_1^3 - 0.028I_1^2 I_2 -0.122I_1 I_2^2-0.133I_2^3\end{array}\]

The reduction of the Hamiltonian to the normal form solves the
problem of the construction of a full set of approximate integrals
of motion. The solution of the equations
\[\begin{array}{c}
H(p_x,p_y,x,y)=E\\
H(p_x,p_y,x,y)=I_0\\ x=const
\end{array}\]
allows us to find the set of intersections of the phase trajectory
with the selected plane ($x=const$) and to reconstruct the structure
of PSS.

\begin{figure}
\center{\includegraphics[width=0.5\textwidth,draft=false]{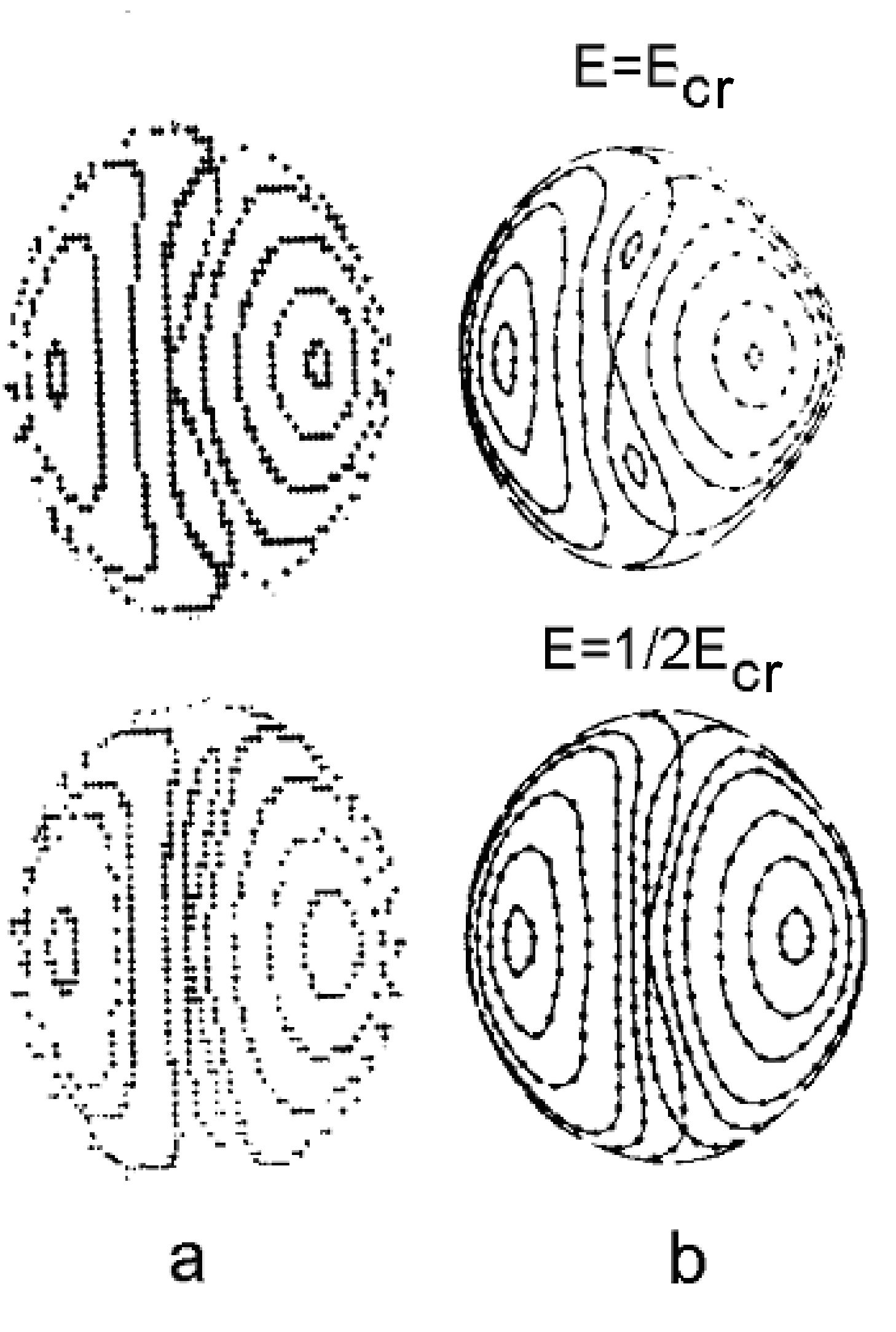}}
\caption{The PSS for different values of energy for central minimum
of ${}^{74}Kr$ obtained a) by numerical integration of equation of
motion, b) by normal forms. \label{kr_pss}}
\end{figure}

The PSS for the quadrupole oscillations of nuclei ${}^{74}Kr$, which
are constructed in such a way, are shown in Fig.\ref{kr_pss}. The
qualitative coincidence of topology of PSS calculated with the help
of normal forms and the numerical integration of the equation of
motion is noteworthy.

Having failed in an attempt to formulate adequate criterion of
stochasticity that is based on the estimation of the rate of
divergence of the two initially close trajectories (nevertheless, we
do not doubt the existence of such criterion), let us now return to
the resonance overlap criterion. By means of it we will try to
understand the differences in the phase space structure of the
different local minima that realize different dynamical regimes.

As an example let us consider the application of this criterion to
the dynamics that are generated by the potential of umbilical
catastrophe $D_5$. Accordingly the Hamiltonian in the reference
frame connected with left (upper sign) and right (lower sign) well
has the form
\begin{equation}\label{h_d5}H=\frac12 (\dot{x}^2 + \omega_1^2 x^2) +\frac12 (\dot{y}^2 +
\omega_2^2 y^2) + xy^2 \mp \sqrt2 x^3 + \frac14 y^4 \end{equation}
where
\[\omega_1=\sqrt{2(2\mp\sqrt2)};\ \omega_2=2.\]
Now we make a canonical transform to the action-angle variables
\begin{equation}\label{nfc}\begin{array}{cc}
x=\sqrt{\frac{2I_1}{\omega_1}}\cos\varphi_1; &
y=\sqrt{\frac{2I_2}{\omega_2}}\cos\varphi_2;\\
\dot{x}=\sqrt{2I_1\omega_1}\sin\varphi_1; &
\dot{x}=\sqrt{2I_2\omega_2}\sin\varphi_2.
\end{array}\end{equation}
Thus Hamiltonian (\ref{h_d5}) takes the form
\begin{equation}\label{h_nf}\begin{array}{c}
H(I_1,I_2,\varphi_1,\varphi_2)=H_0(I_1,I_2)+
\sum\limits_{n_1,m_2\in\Lambda} f_{m_1
m_2}\cos(m_1\varphi_1+m_2\varphi_2)\\
\Lambda:[0,1],[0,2],[0,3],[0,4],[2,1],[2,-1]
\end{array}\end{equation}
where
\[\begin{array}{c}
H_0(I_1,I_2)=I_1\omega_1 + I_2\omega_2 + \frac38
\frac{I_2^2}{\omega_2^2} -1\\
f_{01}=\frac{I_1}{\omega_1}\left(\frac{2I_2}{\omega_2}\right)^{1/2}
\pm 3\left(\frac{I_2}{\omega_2}\right)^{3/2}\\
\begin{array}{cc}
f_{02}=\frac12 \frac{I_2^2}{\omega_2^2}; &
f_{03}=\pm\left(\frac{I_2}{\omega_2}\right)^{3/2}\\
f_{04}=\frac18 \frac{I_2^2}{\omega_2^2}; & f_{21}=f_{2,-1}=\frac12
\frac{I_1}{\omega_1} \left(\frac{2I_2}{\omega_2}\right)^{1/2}
\end{array}\end{array}\]

An item with indexes $r=(r_1,r_2)$ is called the resonance for a
given value of energy $E$ if there exist action variables
$(I_1^r,I_2^r)$ such, that $E=H_0(I_1^r,I_2^r)$ and
\[r_1\bar{\omega}_1(I_1^r,I_2^r)+r_2\bar{\omega}_2(I_1^r,I_2^r)=0\]
where
\[\bar{\omega}_i=\frac{\partial H_0}{\partial I_i},\ i=1,2.\]

If the system is far enough from resonance, i.e. for all $(m_1,m_2)$
\[|m_1\bar{\omega}_1+m_1\bar{\omega}_1|\gg f_{m_1 m_2},\]
then avoiding the small denominator problem we could make a
canonical transform to new action-angle variables thus eliminating
angle dependence in lower orders under some small parameter. This
procedure results in the redefinition of the integrable part $H_0$
of the initial Hamiltonian and enlargement of the set $\Lambda$ of
angle-dependent terms. After that we could encounter one of
following three possibilities:
\begin{enumerate}
\item resonance terms are still absent
in the considered region;
\item the single resonance term arises;
\item
multiple resonances arise.
\end{enumerate}
In the first case we should execute a new canonical transformation
and keep carrying out the procedure until case 2 or 3 appears. In
the second case critical energy of the transition to large-scale
stochasticity could be defined with stochastic layer destruction
criterion \cite{doviel}. And finally in the third case we could use
the Chirikov's resonance overlap criterion \cite{chirikov} to
determine the critical energy.

Small unbalancing near resonance
\[|m_1\bar{\omega}_1+m_1\bar{\omega}_1|\le f_{m_1 m_2}\]
could be compensated by higher order terms that have arisen from the
repeated canonical transformation of the non-resonance terms. In the
Hamiltonian (\ref{h_nf}) the term $[2,-1]$ is subjected to condition
(\ref{nfc}), and the considered procedure leads to
\[\begin{array}{c}
H_0(I_1,I_2)=I_1\omega_1 -1 + \frac38
\frac{I_2^2}{\omega_2^2} - \frac{4\omega_1+5}{32\omega_1^2(\omega_1+1)}I_1^2+\\
+\left[\pm\frac{3\sqrt2}{2}-\frac{1}{\omega_1(\omega_1+1)}\right]\frac{I_1
I_2}{8\omega_1}.
\end{array}\]

In the left well let us take into account only two terms: resonance
\[\frac12 \frac{I_1}{\omega_1}\left(\frac{2I_2}{\omega_2}\right)^{1/2}\cos(2\varphi_1-\varphi_2)\]
and "shaking" ones \cite{doviel}:
\[\frac{\sqrt2}{64\omega_1} \left[\frac{3}{\omega_1+1}+\frac{1}{\omega_1}\right]I_1 I_2\cos(2\varphi_1-2\varphi_2).\]
After this direct application of the stochastic layer destruction
criterion leads to the value of the critical energy in the left well
\[E_{cr}\simeq-0.51\]
This value is in a good agreement with numerical integration results
and in qualitative agreement with the estimation obtained by
negative curvature criterion $E_{cr}=-5/9$. Straightforward analysis
of the integrable part of the Hamiltonian (\ref{h_nf}) shows that
there are no resonance terms in the right well. Thus transition to
large-scale stochasticity in the right well could be attained only
when passing the saddle energy. This fact is in full agreement with
numerical results.

Resonance overlap criterion allows better understanding of the
mechanism of the above mentioned regularity-chaos-regularity
transition that exists in any (multi- or single-well) potential with
a localized region of instability. Let us use similarity in
structure of phase space of the considered two-dimensional
autonomous Hamiltonian system with the compact region of negative
Gaussian curvature and one-dimensional system with periodic
perturbation (\ref{h_aa}) \cite{bolotin95}.

\begin{figure}
\includegraphics[width=\textwidth,draft=false]{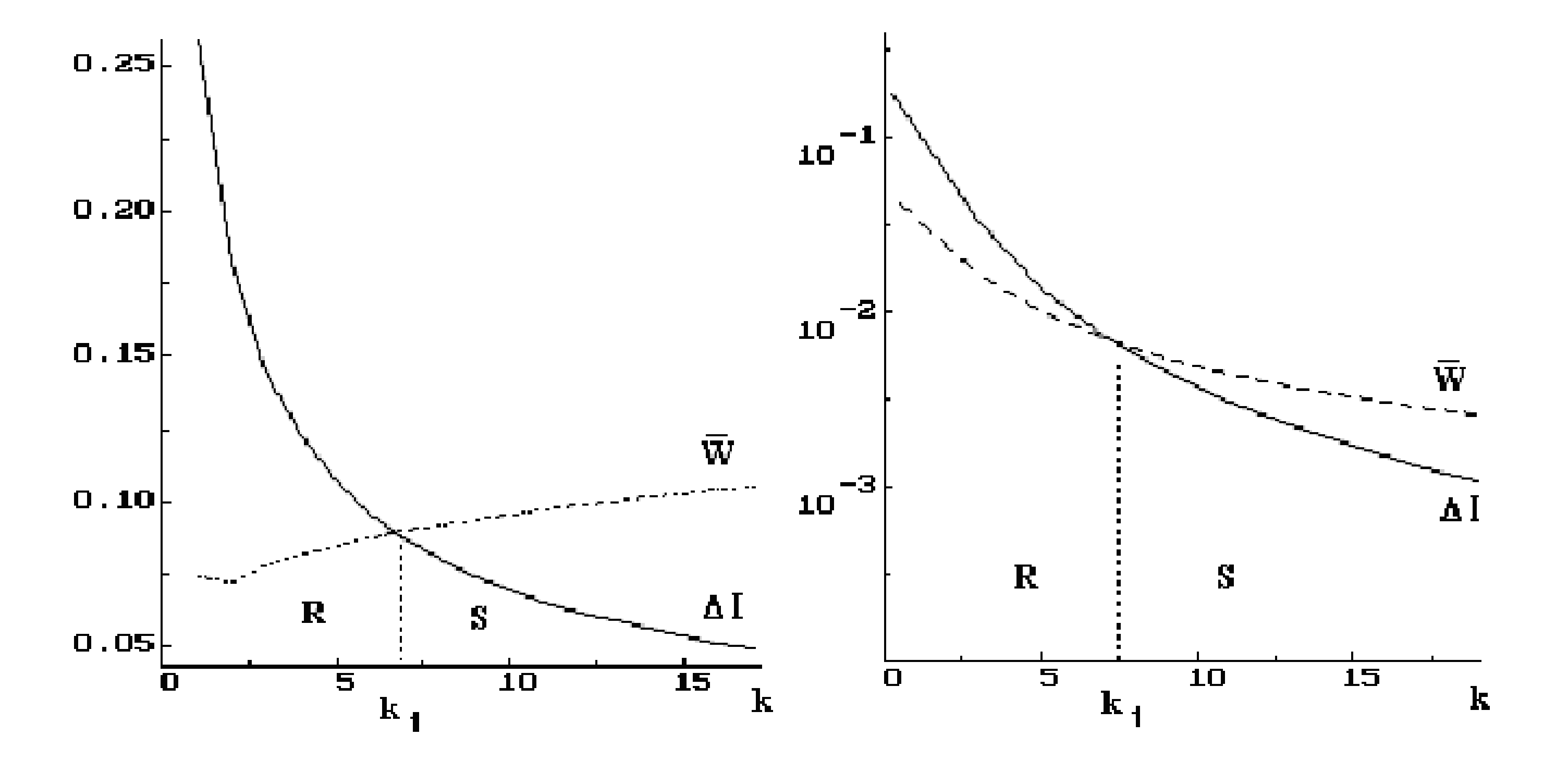}
\caption{The resonant spacing $\Delta I_k$ (solid lines) and the
mean widths $\bar{W}_k$ (dashed lines) as functions of the resonance
numbers $k$. On the left: for one-dimensional Coulomb, on the right:
for square-well potential. The critical point $k_1$ separates the
regular range ($R$) from the chaotic one ($S$). \label{iw}}
\end{figure}

The behavior of the widths of the resonances
\[W_k\equiv\frac12(W_{k+1}+W_k),\]
and the distances between them
\[\Delta I_k\equiv|I_{k+1}-I_k|\]
as a function of the resonance number is simplest when the
satisfaction of resonance overlap condition (\ref{roc}) for number
$k_1$ (at a fixed level of the external perturbation) guarantees
that this condition holds for arbitrary $k>k_1$. This is precisely
the situation, that prevails in the extensively studied systems of a
Coulomb potential \cite{jensen} and a square well \cite{lin}
subjected in each case to a monochromatic perturbation. In the
former case we have $\bar{W}_k\approx k^{1/6}$ and $\Delta
I_k\approx k^{-2/3}$, while in the latter we have $\bar{W}_k\approx
k^{-1}$ and $\Delta I_k\approx [k(k+1)]^{-1}$. As can be seen from
Fig.\ref{iw} there is R-C transition (we will call this transition a
"normal" transition) for both the Coulomb problem and a square well,
since there exists a unique point $k_1$ such that at $k>k_1$ the
condition $\bar{W}_k>\Delta I_k$ always holds. The motion is
therefore chaotic. However, as the behavior of the widths of the
resonances and of the distances between them as a function of the
resonance number becomes more complicated, we can allow the
appearance of an additional intersection point and thus a new
transition: C-R transition, which we will call "anomalous". There is
also the exotic possibility of the intermitting occurrence of the
regular and chaotic regions in the phase space.

We demonstrate that an anomalous C-R transition occurs in a simple
Hamiltonian system: an anharmonic oscillator, subjected to a
monochromatic perturbation \cite{bolotin94,bolotin95}. The dynamics
of such a system is generated by the Hamiltonian
\begin{equation}\label{h_pxt}H(p,x,t)=H_0(p,x)+Fx\cos\Omega t\end{equation} where the unperturbed
Hamiltonian is
\begin{equation}\label{h_pxt_0}H_0(p,x)=\frac{p^2}{2m}+Ax^n=E,\ (n=2l,\ l>1)\end{equation}
The considered system fills a gap between two extremely important
physical models: the harmonic oscillator ($n=2$) and the square well
($n=\infty$).

In terms of action-angle variables $(I,q)$, the Hamiltonian
$H_0(p,x)$ (\ref{h_pxt_0}) becomes \cite{bolotin95}
\[H_0(I)=\left(\frac{2\pi}{\alpha G(n)}I\right)^\alpha ,\
G(n)=\frac{2\sqrt{2\pi
m}\Gamma\left(1+\frac1n\right)}{A^{\frac1n}\Gamma\left(\frac12+\frac1n\right)},\
\alpha=\frac{2n}{n+2}\] The resonant values of the action $I_k$ that
can be found from the conditions \[k\omega(I_k)=\Omega,\
\omega(I)=\frac{dH_0}{dI}\] are
\[I_k=\alpha\left(\frac{G(n)}{2\pi}\right)^{2n\beta}\left(\frac\Omega k\right)^{\frac{2n\alpha}{\beta}},\ \beta=\frac{1}{n-2}.\]
A classical analysis, based on the resonance overlap condition,
leads to the following expression for the critical amplitude of the
external perturbation
\begin{equation}\label{fcr}F_k^{cr}=2^{(2-3n)\beta}\frac{1}{4n}\frac{\alpha^2}{\beta}\frac{1}{x_k}
\left(\frac{G(n)}{2\pi}\right)^{2n\beta} \Omega^{2n\beta} k^{4\beta}
\left[k^{(n+2)\beta}-(k+1)^{(n+2)\beta}\right],\end{equation} where
$x_k$ is the Fourier component of the coordinate $x(I,\theta)$.
Expression (\ref{fcr}) solves the problem of reconstructing the
structure of the phase space for arbitrary values of the parameters.
\begin{figure}
\includegraphics[width=\textwidth,draft=false]{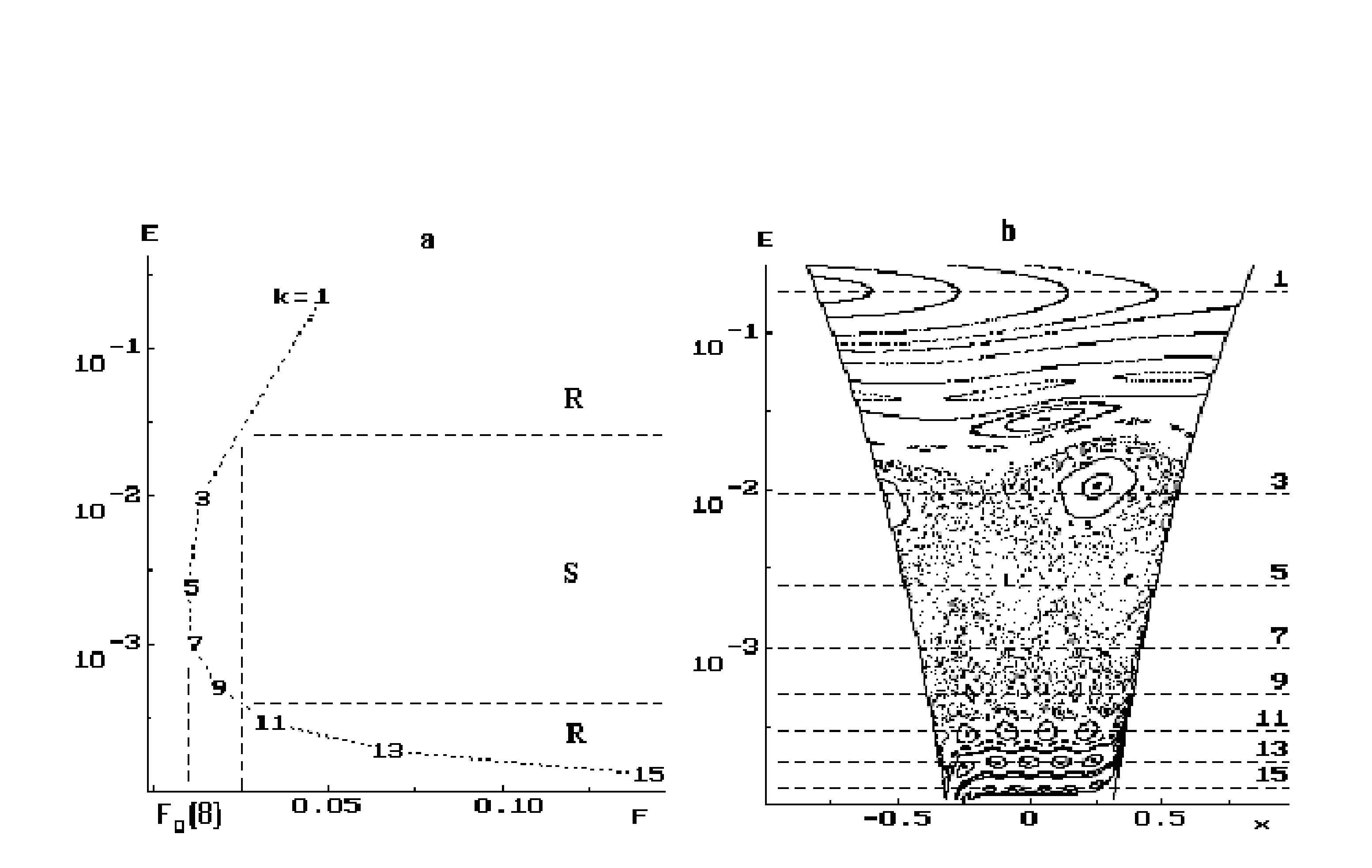}
\caption{a) The phase diagram of the R-C-R transition for
Hamiltonian (\ref{h_pxt}): resonance energy $E_k$ versus the
critical values of the external perturbation $F(n=8)$. b) The
snap-shot $E(x)$ confirms that the anomalous C-R transition occurs.
\label{rcr_1d}}
\end{figure}

The phase diagram in Fig.\ref{rcr_1d}a can be used to determine, at
the fixed level of the external perturbation, the energy intervals
of regular and chaotic motion. The snap-shot of $E(x)$ at the right
in Fig.\ref{rcr_1d}b confirms that an anomalous C-R transition
occurs. We can clearly see isolated nonlinear resonances which
persist at large values of $k$, and near which the motion remains
regular. The reason for this anomaly is explained by Fig.3.8. The
plots of the resonance widths and  the distances between resonances
in this figure demonstrate that there are two rather than one
intersection points: $k=k_1$ and $k=k_2$. Consequently, there is an
anomalous C-R transition.

Thus, for $1D$ system with periodic perturbation R-C-R transition
can be observed just as in the case of a $2D$ autonomous Hamiltonian
system. The reason for the additional transition in both cases is
common: the localized region of instability. In the first case this
reason is the localized domain of overlap resonances, while in the
second one this reason is the localized domain of negative Gaussian
curvature.
\begin{figure}
\includegraphics[width=\textwidth,draft=false]{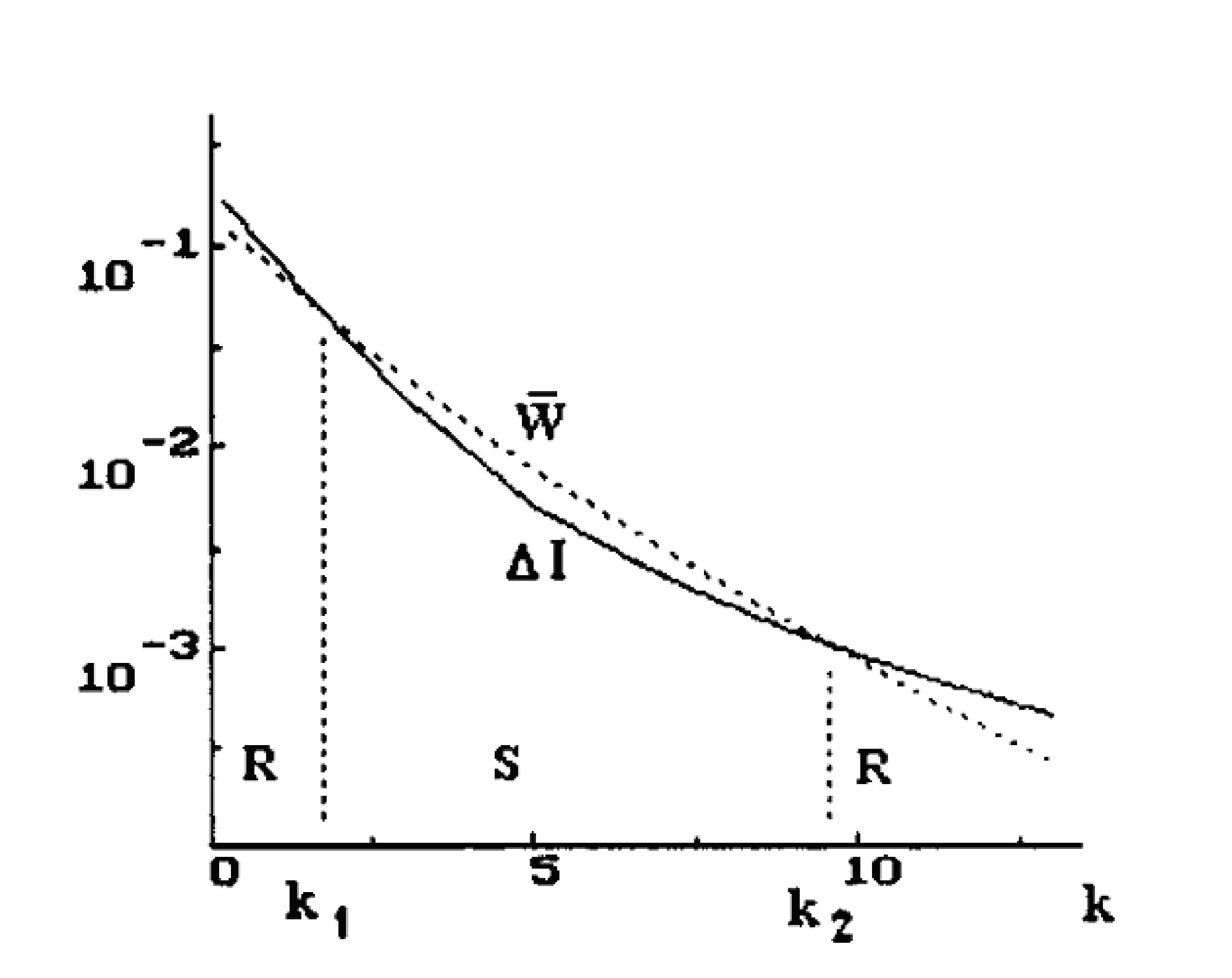}
\caption{The resonant spacings $\Delta I_k$ (solid line) and mean
widths $\bar{W}_k$ (dashed line) as functions of the resonance
numbers $k(n=8)$. There are two critical points $k_1,k_2$ . The
second critical point corresponds to the anomalous C-R transition.
\label{iw_an}}
\end{figure}
\chapter{Quantum Manifestations of Classical Stochasticity --- Formulation of the Problem}\sat
After almost a hundred years of development, quantum mechanics
became a universal picture of the world. On any observable scales of
energy we could not find any violations of quantum mechanics. But
this does not mean that from time to time quantum mechanics does not
confront another challenge. The problem that arose at the face of
quantum mechanics in the second part of the last century is called
quantum chaos. The essence of the problem is the fact that, on the
one hand, the energy spectrum of any quantum system with finite
motion is discrete and thus its evolution is quasi-periodic, but, on
the other hand, the correspondence principle demands transition to
classical mechanics which demonstrates not only regular solutions
but chaotic too. This deep and serious problem requires an answer
firstly to the question: what does it mean when one theory is a
limiting case of another? \cite{berry91}

Usually a more general theory $G$ is connected with a special theory
$S$ with a dimensionless parameter $\delta$ such as
\[G\rightarrow S\ \mathrm{if}\ \delta\rightarrow0\]
For example, if under $G$ we understand special relativity theory,
and under $S$ classical mechanics, then $\delta=v^2/c^2$. In the
simplest case we can represent the general theory as a Tailor series
over parameter $\delta$.

However such a simple situation is a very rare exception. In the
most general (and the most interesting) case the limit
$\lim\limits_{\delta\rightarrow0}G$ is singular and the transition
$G\rightarrow S$ is far from being trivial. So, for example, the
transition from Navier-Stocks equations (viscous fluid) to Euler
equations (ideal fluid) is singular: dissipation does not turn to
zero smoothly at zero viscosity. So difficult for investigation, the
no man's land between the two theories contains new physics, like
turbulence or critical behavior at phase transitions. It is a
similar region where we have to study the influence of classical
stochasticity on quasi-classical behavior. In our case $G$ stands
for quantum mechanics, $S$ --- classical mechanics and $\delta$ is
some dimensionless combination of physical quantities with $\hbar$
in the numerator. According to Berry \cite{berry91}, the
$\hbar\rightarrow0$ limit swarms with non-analyticities.

As we are interested in absolutely concrete aspects of the
semiclassical limit, namely: how the presence of classical chaos is
reflected in quantum quantities, we shall discuss one more principal
difficulty. As in classical mechanics chaos is realized only on
large time scales (required for complete mixing, i.e. for
realization of the limiting tendency to zero of the correlation
function), any useful discussion of semiclassical limit must
simultaneously account for both the limit $t\rightarrow\infty$ and
the limit $\hbar\rightarrow0$ \cite{chirikov_budker}. A natural
question arises whether the two individually non-trivial limits
$t\rightarrow\infty$ and $\hbar\rightarrow0$ commute? The answer is
negative: long-time semiclassical evolution fundamentally differs
from long-time classical evolution --- in the common case situation
the classical long-time limit is chaotic, while in semiclassics the
temporal asymptote is not, and any chaos represents just a
transition process. Therefore in an attempt to construct the quantum
theory of dynamical chaos we immediately confront a row of evident
and very deep contradictions between well-established principles of
classical chaos and fundamental principles of quantum mechanics.
What is the reason for those contradictions?

As is well known, the energy spectrum of any quantum system that
undergoes finite motion, is always discrete. And it is not a
property of a concrete equation, but a consequence of the
fundamental principles of quantum mechanics: the discrete nature of
the phase space or, more formally, the non-commutativity of quantum
phase space. Indeed, according to the uncertainty principle, an
individual quantum state cannot occupy the phase volume $V_1\le
h^N$, where $N$ is the dimensionality of the configuration space.
Therefore a motion limited by a region $V$ will contain $V/V_1$
eigenstates. According to existing ergodic theory such motion is
considered as regular, in contrast to chaotic motion with continuous
spectrum and exponential instability. The latter statement can be
verified using the notion of algorithmic complexity, which can be
defined as the relation:
\begin{equation}\label{ac}C=\frac{N_{in}}{N_{out}},\end{equation} where $N_{in}$ and
$N_{out}$ are expressed in bits input and output routine lengths
respectively. This quantity can be determined for any moment of
time; however the distinction between regular and chaotic motion
manifests only in the limit $t\rightarrow\infty$. If the motion is
chaotic, then $C\rightarrow const>0$, if it is regular
--- $C\rightarrow0$. In order to understand the reason for that we should
note that the output length of routine --- information about the
orbit in an arbitrary moment of time --- grows proportional to $t$.
The input data sequence consists of two main parts. The first is the
algorithm for solution of equations of motion, its length does not
depend on $t$. The second is the definition of initial conditions
with the precision required to reproduce the required final result.
For chaotic systems, where errors grow exponentially, this part is
proportional to $t$ and therefore dominates in the input routine
length. Therefore in that case the algorithmic complexity (\ref{ac})
will tend to constant. For non-chaotic systems part of the input
routine, connected with initial conditions, grows slower (for
example as $\ln t$ when the errors grow linearly), and the limiting
value of the algorithmic complexity is $C=0$.

All experiments performed up to the present time showed strict
fulfillment of that rule for classical chaotic systems. For quantum
systems, as for those that are chaotic in the classical limit, and
for those that are regular, only zero algorithmic complexity was
observed. This result can be briefly formulated in the spirit of
Bohr complementarity: classical evolution is deterministic, but
random, quantum evolution is not deterministic and it is not random.
In other words, the problem consists in the fact that the discrete
nature of the spectrum never implies chaos, or more exactly any
resemblance to chaos in the sense of the ergodicity theory, in any
quantum system with finite motion. Meanwhile the correspondence
principle in the semiclassical limit requires the presence of chaos,
connected with the nature of motion in the classical case.

If to state the point of view that chaos never appears in quantum
mechanics, then a possible reaction to that is just to give up the
study of the question. But it will mean that we avoid the challenge
that Nature gives us in the limit of small $\hbar$ and large $t$,
which is equivalent to ignoring other singular phenomena, such as
turbulence or phase transitions. An alternative point of view
consists in the fact that not waiting for the complete solution of
the problem (or rather for its correct formulation) we can study its
limited variant: investigation of special features of quantum
systems behavior whose classical analogues are chaotic , or, in
other words, search for quantum manifestations of classical
stochasticity (QMCS). It is the problem that will be considered in
the following chapters on quantum systems with potential energy
surfaces of non-trivial topology.

Deterministic chaos is a general feature of Hamiltonian systems with
the number of simple integrals of motion less than the number of
degrees of freedom. Lack of the full set of integrals of motion
(full set consists of such number of integrals that is equal to the
number of degrees of freedom of the quantized system) makes the
traditional procedure of multi-dimensional systems quantization
unrealizable. Let us consider this statement in details.

As is well known \cite{landau}, in the one-dimensional case it is
always possible to introduce such canonically conjugate
"action-angle" variables that the Hamiltonian becomes a function of
action variable only. The standard definition of the action variable
concerns integral along the periodic orbit
\[I=\frac{1}{2\pi}\oint p(x)dx\]
where $p(x)$ is the particle's momentum. In the context of the
semi-classical approach we could construct an approximate solution
of Schr\"odinger equation in the terms of the integral along
classical trajectory \cite{landau_qm}
\begin{equation}\label{psi_sc}\psi(x)=\frac{1}{\sqrt p}e^{\frac i
\hbar\int\limits_{x_0}^xp(x')dx'}.\end{equation} This construction
makes sense only in the case when phase grows on multiples of $2\pi$
along the periodic orbit. This limitation immediately leads to the
semi-classical quantization condition
\[I=\frac{1}{2\pi}\oint p(x)dx=\left(n+\frac\mu4\right)\hbar\]
where $n$ is nonnegative integer number and $\mu$ is the so-called
Maslov index which is equal to the number of points along periodic
orbit where the semi-classical approximation is violated (in the
one-dimensional case this occurs in turning points and  $\mu=2$).
Semi-classical energy eigenvalues $E_n$ are obtained by the
computation of the Hamiltonian $H(I)$ for quantized actions
\[E_n=H\left(I=\left(n+\frac\mu4\right)\hbar\right).\]
For multi-dimensional systems this procedure could be executed only
in the case when the number of integrals of motion is equal to
number of degrees of freedom, i.e. for integrable systems. In this
case the procedure is called the Einstein--Brillouin--Keller
quantization. Let us restrict with the two-dimensional case for
simplicity. If the system is integrable, then there exist two
couples of canonically conjugate action-angle variables
$(I_1,\theta_1)$ and $(I_2,\theta_2)$ such that classical the
Hamiltonian depends only on action variables
\begin{equation}\label{h_i}H=H(I_1,I_2)\end{equation} Finite classical motion is periodic in every
angle variable with frequencies
\[\Omega_i=\frac{\partial H}{\partial I_i},\ (i=1,2).\]
In the general case frequencies $\Omega_i$ are not close to each
other, and motion in four-dimensional phase space is quasi-periodic.
Phase trajectories lie on invariant tori that are defined by
integrals of motion $I_i$. Semi-classical wave functions could be
constructed in the form that is analogous to (\ref{psi_sc}), but
turning points must be replaced by caustic surfaces. Uniqueness of
the wave function demands the quantization condition
\begin{equation}\label{i2}I_i=\left(n_i+\frac{\mu_i}{4}\right)\hbar\end{equation}

As in the one-dimensional case energy eigenvalues could be obtained
by the substitution of (\ref{i2}) into Hamiltonian (\ref{h_i}). It
was understood by Einstein in 1917 that this method could be applied
only to quantum integrable systems with trajectories lying on tori.
For non-integrable (i.e. chaotic) systems consistent quantization
method did not exist for half of the century. But how to implement
quantization in the non-integrable situation?

Progress in the problem of chaotic systems quantization was obtained
with the help of Feynman's formulation of quantum mechanics. The
first mention of the applicability of path integrals to chaotic
systems was given by Selberg (1956), who built the dynamics of the
particle on the Riemannian surface with negative curvature in the
terms of path integrals. This is certainly a chaotic system although
this term did not exist at that time.

Gutzwiller was the first who successfully applied an analogous
approach to quantization of chaotic systems. In 1982, he had shown
that semi-classical approximation in the form of path integrals
allows us to obtain the spectrum of a chaotic system
\cite{gutzwiller_sc}. This study was the culmination of the large
series of his works
\cite{gutzwiller1,gutzwiller2,gutzwiller3,gutzwiller4,gutzwiller5}.
Works of Balian, Block \cite{bb1,bb2} concern the same approach
--- they connect classical periodic orbits with the quantum spectrum
of the underlying system.

Periodic orbits play the main role in Gutzwiller's non-integrable
systems quantization method. The final aim of the method consists in
the evaluation of the density of levels
\[\rho(E)=\sum\limits_n\delta(E-E_n)\]
in the terms of solutions of classical equations of motion.

Using the expression
\[\delta(E-E_n)=\frac1\pi\Im\lim_{\varepsilon\rightarrow0}\frac{1}{E_n-E-i\varepsilon},\]
we could obtain
\begin{equation}\label{rho_green}\rho(E)=-\frac1\pi\Im
Sp\left(\frac{1}{E-\hat{H}}\right).\end{equation} Operator under the
trace is the Green function:
\[G(q_A,q_B,E)=\sum\limits_n \frac{\Psi_n^*(q_A)\Psi_n(q_B)}{E-E_n}=
\sum\limits_n \Psi_n^*(q_A)\frac{1}{E-\hat{H}}\Psi_n(q_B)\]
\[\sum\limits_n \Psi_n^*(q_A)\Psi_n(q_B)=\int\delta(q_A-q)\delta(q_B-q)dq\]
And therefore
\[G(q_A,q_B,E)=\langle q_A|\frac{1}{E-\hat{H}}|q_B\rangle\]
Thus (\ref{rho_green}) could be rewritten in the form
\[\rho(E)=-\frac1\pi\Im SpG\]

The further procedure implies the construction of Green's function
semi-classical approximation (and then its Fourier transform) and
calculation of the trace. Gutzwiller had shown that this procedure
results in the following expression for the levels density:
\begin{equation}\label{gutzwiller_formula}\rho(E)\simeq\bar{\rho}(E)+\sum\limits_p\sum\limits_{k=1}^\infty
A_{p,k}(E)\cos\left[k\left(I_p(E)-\frac\pi2\mu_p\right)\right]\end{equation}
Here $\bar{\rho}(E)$ is the smoothed density of levels that could be
obtained via Thomas--Fermi approximation or Weyl's formula for
billiards. The sum marked by index $p$ is evaluated over all
"primitive" periodic orbits and sum in $k$ --- over  $k$-reiteration
of these orbits. The phase for every periodic orbit consists of the
action along this orbit $I_p(E)$ and Maslov index $\mu_p$. Amplitude
$A_{p,k}$ is determined by the expression
\[A_{p,k}(E)=\frac{T_p(E)}{\pi k\sqrt{\det(\hat{M}-\hat{I})}}\]
where $T_p(E)=\partial I/\partial E$ is the orbit's period and
$\hat{M}$ is the monodromy matrix, that is well known from the
classical analysis of motion stability. The formula
(\ref{gutzwiller_formula}) is called the Gutzviller trace
formula\footnote{Feynman called it one of the main achievements of
theoretical physics of the twentieth century} and expresses density
of quantum spectrum through values that are calculable in the
context of classical mechanics. At the same time this expression
could be understood as universal semi-classical quantization
condition that is correct both for integrable and non-integrable
systems: highly excited (semi-classical) energy levels are the
points where right hand side of the trace formula has poles.

Although many important results were obtained with the trace
formula, not all its analytical features are clear enough for now.
Mainly this is due to the difficulties in the corresponding
classical calculations. First of all we encounter the problem of
periodic orbit evaluation: their number grows exponentially with
growth of period and all of them are by definition unstable. At the
same time there is a problem of adequate description of their
contributions
--- the summation problem. And finally the generalization of
Gutzwiller formula on the considered multi-well case is nontrivial.
Thus the problem of numerical integration of the Schr\"odinger
equation becomes the basic one for calculation of the semi-classical
part of the spectrum of quantum systems that are chaotic in
classical limit. In the following we will briefly review the main
types of calculation problem arising in research of the
manifestations of quantum chaos in specific physical models:
evaluation of the energy spectrum and investigation of its features,
obtaining stationary wave functions and their analysis in different
representations, modeling of time evolution of time-dependent
states. We will analyze in detail the main numerical methods for
these problems --- the matrix diagonalization technique and the
spectral method.
\chapter{Numerical Methods In Multi-Well Potentials}\sat
Studies of deterministic chaos, both classical and quantum, more
than other domains of modern physics derive their development from
computational power growth. The number of scientific papers on that
topic published per year grows as $N\approx e^{\lambda t}$, where
$\lambda\approx0.23\ \rm{year}^{-1}$, That value of $\lambda$ is
significantly greater than the growth rate for the total volume of
scientific publications ($\lambda\approx0.0 46\ \rm{year}^{-1}$),
but is very close to the growth rate of global computational power
$\lambda\approx0.25\ \rm{year}^{-1}$. And it is not surprising at
all because the main body of those papers is devoted to quantum
chaos researches in numerical experiments.

In the present chapter we develop numerical methods for analysis of
quantum chaos problems and demonstrate the advantages of the
spectral method (SM) in comparison with the matrix diagonalization
technique (MD) in application to the solution of Schr\"odinger
equation in smooth potential systems, in particular with multiple
well PPS.
\section{The Matrix Diagonalization Technique}\sat
\subsection{General theory}\sat Let us consider
the Schr\"odinger equation for a system with discrete energy
spectrum
\begin{equation}H\Psi_n=E_n\Psi_n,\ n\in
\mathbb{N}\label{schr_eq}\end{equation}
 and let there be full orthonormal
basis of functions
\[\varphi_k, k\in\mathbb{K},\ \langle\varphi_{k'}|\varphi_{k''}\rangle=\delta_{k' k''}\]
where $\mathbb{N}$ and $\mathbb{K}$ are countable sets.

The basis functions $\varphi_k$ are solutions of another
Schr\"odinger equation
\[h\varphi_n=e_n\varphi_n,\ n\in \mathbb{K},\]
they are given analytically or are obtained numerically in an
independent way.

Obviously there exists a decomposition
\[\Psi_n=\sum\limits_{k\in\mathbb{K}}a_k^{(n)}\varphi_k,\ a_k^{(n)}=\langle\varphi_k|\Psi_n\rangle.\]

The solution of the Schr\"odinger equation (\ref{schr_eq}) by the
matrix diagonalization technique implies the following:
\begin{enumerate}
\item the set $\mathbb{K}$ is presented as a direct sum of the subsets
\[\mathbb{K}=\mathbb{\bar{K}}\oplus\mathbb{K'},\] such as
$\mathbb{\bar{K}}$ is finite and $\mathbb{K'}$ is a countable set.
\item original Hamiltonian of the problem (\ref{schr_eq}) is
presented in the form
\[H=\bar{H}+H',\]
where by definition
\begin{equation}\label{defin}\langle\varphi_{k'}|\bar{H}|\varphi_{k''}\rangle=\langle\varphi_{k'}|H|\varphi_{k''}\rangle,\
k',k''\in\mathbb{\bar{K}}\end{equation} and all other matrix
elements of $\bar{H}$ are zeros.
\item the eigenvalue problem
\begin{equation}\bar{H}\bar{\Psi}_n=\bar{E}_n\bar{\Psi}_n,\
n\in\mathbb{\bar{K}}\label{eigen_pr}\end{equation}
 is solved numerically, where
\begin{equation}\label{decomp}\bar{\Psi}_n=\sum\limits_{k\in\mathbb{\bar{K}}}\bar{a}_k^{(n)}\varphi_k,\
\bar{a}_k^{(n)}=\langle\varphi_k|\bar{\Psi}_n\rangle,\
\langle\bar{\Psi}_{n'}|\bar{\Psi}_{n''}\rangle=\delta_{n'n''}.\end{equation}
\end{enumerate}

Let us find the conditions under which the numerically obtained
$\bar{E}_n$ and $\bar{\Psi}_n$ are good approximations to the
original $E_n=\bar{E}_n + {E}_n'$ and $\Psi_n=\bar{\Psi}_n +
{\Psi}_n'$, or in other words the conditions for smallness of
${E}_n'$ and ${\Psi}_n'$.

Let us rewrite (\ref{schr_eq}) in the form
\[\left(\bar{H}+H'\right)\left(\bar{\Psi}_n+\Psi_n'\right)=\left(\bar{E}_n+E_n'\right)\left(\bar{\Psi}_n+\Psi_n'\right)\]
and simplify it using (\ref{eigen_pr}) to obtain
\begin{equation}\label{simpl}
H'\bar{\Psi}_n - E_n'\bar{\Psi}_n + \bar{H}\Psi_n' -
\bar{E}_n\Psi_n' + H'\Psi_n' - E_n'\Psi_n'=0.
\end{equation}
Taking scalar product of $\bar{\Psi}_n$ in (\ref{simpl}), and taking
into account that, according to (\ref{defin}) and (\ref{decomp}),
\[\langle\bar{\Psi}_n|H'|\bar{\Psi}_n\rangle=0,\ \left|\bar{\Psi}_n\right|^2=1,\]
we obtain
\[E_n'=\frac{\langle\bar{\Psi}_n|H'|\Psi_n'\rangle}{\left|\bar{\Psi}_n\right|^2+\langle\bar{\Psi}_n|\Psi_n'\rangle}=
\frac{\sum\limits_{k'\in\mathbb{\bar{K}},\
k''\in\mathbb{K'}}\bar{a}_{k'}^{(n)}a_{k''}^{(n)}H_{k'k''}'}{\sum\limits_{k'\in\mathbb{\bar{K}}}\bar{a}_{k'}^{(n)}a_{k'}^{(n)}}=
\frac{\langle\bar{\Psi}_n|H'|\Psi_n\rangle}{1-\left|\Psi_n'\right|^2/2}.\]
It is easy now to see, that smallness of
\begin{equation}\label{small}
a_k^{(n)}=\langle\Psi_n|\varphi_k\rangle,\ k\in\mathbb{K'}
\end{equation}
is a sufficient but not necessary condition for smallness of $E_n'$.
In order to find the conditions for smallness of (\ref{small}) it is
convenient to make use of the Wigner representation
\begin{equation}\label{wigner_repr}
\begin{array}{c}
\Psi_n^{(W)}(p,q)=\int dx e^{-\frac i \hbar px}\Psi_n\left(q+\frac x
2\right)\Psi_n^{*}\left(q-\frac x 2\right)\\
\varphi_k^{(W)}(p,q)=\int dx e^{-\frac i \hbar
px}\varphi_k\left(q+\frac x 2\right)\varphi_k^{*}\left(q-\frac x
2\right),
\end{array}
\end{equation}
and then we have
\[\langle\Psi_n|\varphi_k\rangle^2=\frac{1}{\left(2\pi\hbar\right)^D}\int dp dq \Psi_n^{(W)}(p,q)\varphi_k^{(W)}(p,q),\]
where $D$ is the dimension of the configuration space.

According to the principle of uniform semiclassical condensation of
quantum states \cite{robnik_uscp} in semiclassical limit, the Wigner
functions $\Psi_n^{(W)}(p,q)$ and $\varphi_n^{(W)}(p,q)$ are
localized on energy surfaces $H(p,q)=E_n$ and $h(p,q)=e_k$
respectively
\[\Psi_n^{(W)}(p,q)\sim\delta\left(H(p,q)-E_n\right),\ \varphi_k^{(W)}(p,q)\sim\delta\left(h(p,q)-e_k\right).\]
The coefficients (\ref{small}) are negligibly small when the
corresponding energy surfaces do not have common points, and
therefore analysis of applicability of the matrix diagonalization
method is reduced to analysis of the classical phase space of the
system under consideration \cite{bohigas}.
\subsection{1D harmonic oscillator}\sat
As the simplest example let us consider calculation of the energy
spectrum of a one-dimensional harmonic oscillator
\begin{equation}\label{ho_1d}H(p,q)=\frac{p^2}{2}+\frac{q^2}{2}\end{equation} on the basis of
eigenfunctions of another harmonic oscillator
\[h(p,q;\omega)=\frac{p^2}{2}+\omega^2\frac{q^2}{2}.\]
The basis functions, corresponding to energy levels
\begin{equation}\label{ho_spectr}e_k=\hbar\omega\left(k+\frac 1 2\right)\end{equation}
 have the well-known form
\begin{equation}\label{ho_basis}\varphi_k(x;\omega)=\left(\frac{\omega}{\pi\hbar}\right)^{\frac
1 4} \frac{H_k(\xi)}{\sqrt{2^k k!}}e^{-\frac{\xi^2}{2}},\
\xi=\sqrt{\frac \omega \hbar}x,\end{equation} where
\[H_n(\xi)=(-1)^n e^{\xi^2}\frac{d^n}{d\xi^n}e^{-\xi^2}=
\sum\limits_{m=0}^{[n/2]}\frac{(-1)^m n!}{m!(n-2m)!}(2\xi)^{n-2m}\]
are Hermit polynomials and $[n/2]$ means integer part.

The Wigner form for the basis functions (\ref{ho_basis}) reads
\[\varphi_k(p,q;\omega)=\int\limits_{-\infty}^{+\infty}dx e^{-\frac i \hbar px}
\varphi_k\left(q+\frac x 2;\omega\right)\varphi_k^*\left(q-\frac x
2;\omega\right)= (-1)^k 2e^{-2\rho^2}L_k(4\rho^2),\] where
\[\rho^2=\frac{h(p,q;\omega)}{\hbar\omega}\]
and $L_k$ are Laguerre polynomials.

\begin{figure}
\vspace{0.02\textheight}
\includegraphics[width=0.5\textwidth,draft=false]{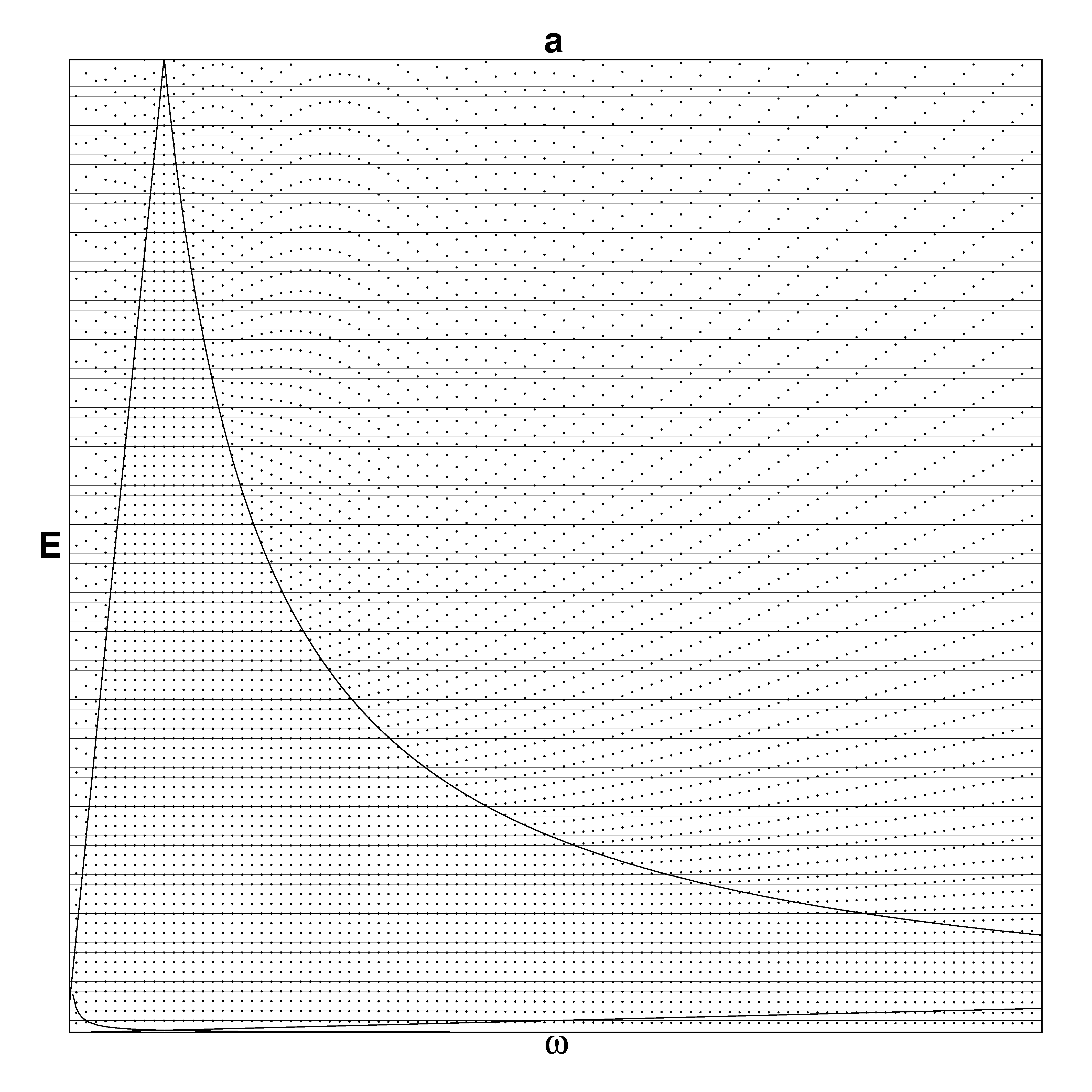}
\includegraphics[width=0.5\textwidth,draft=false]{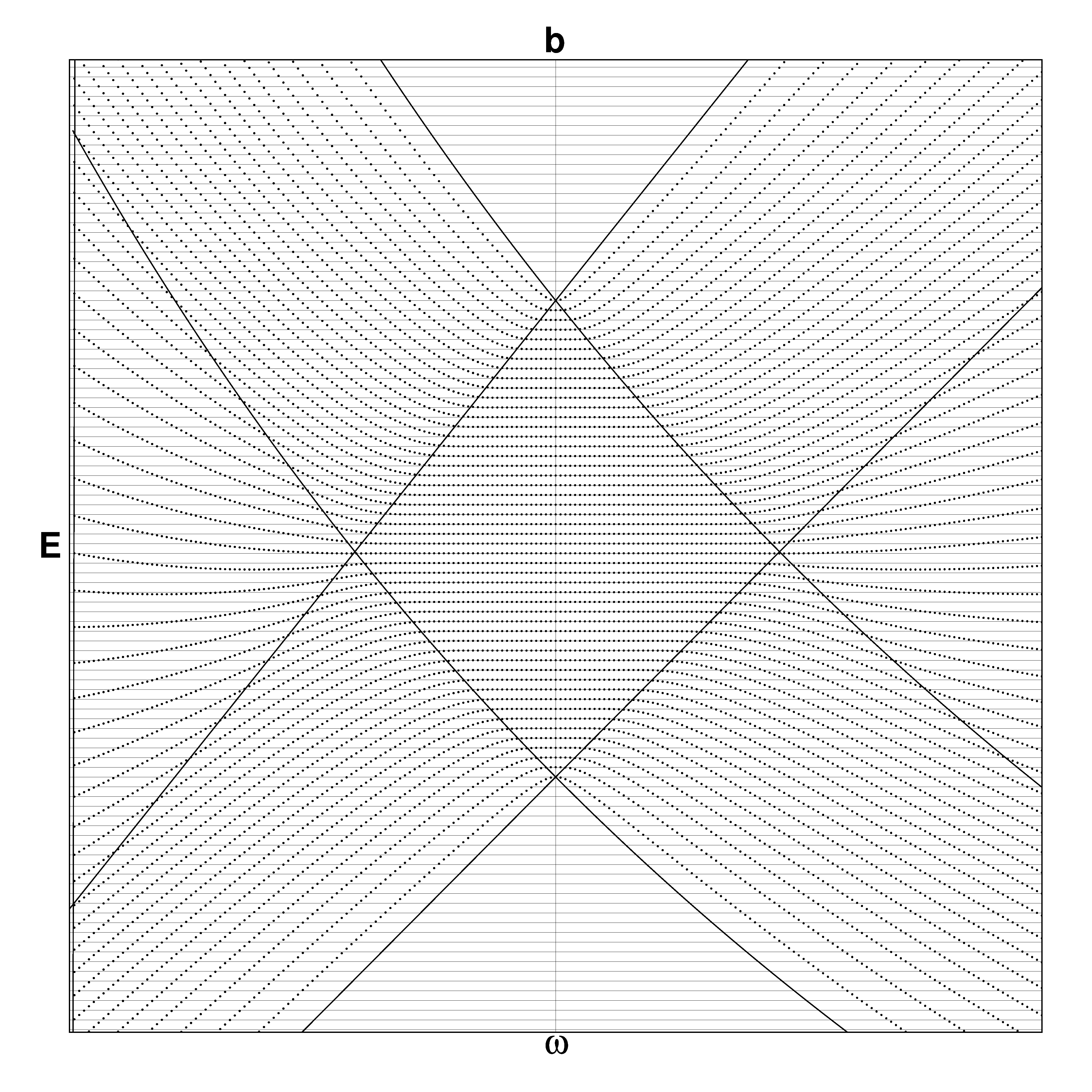}
\caption{Dependence of numerically found energy levels of harmonic
oscillator (\ref{ho_1d}) on the auxiliary basis frequency $\omega$
for $0.03<\omega<10.0$, $k_1=2$, $k_2=100$ (a) and
$0.75<\omega<1.25$, $k_1=400$, $k_2=500$ (b): points represent the
numerical data, thin solid lines are exact energy levels and thick
solid lines limit the computational stability region determined from
semiclassical analysis (\ref{ho_ho_sr})\label{ho_ho_e_w}}
\end{figure}

Taking $\omega=1$ in the expressions for energy levels
(\ref{ho_spectr}) and basis functions (\ref{ho_basis}) we
immediately obtain the exact result for the original problem
\[E_n=\hbar\left(n+\frac 1 2\right)\]
but for arbitrary $\omega$ diagonalization of the original
Hamiltonian (\ref{ho_1d}) in the chosen basis represents already
non-trivial procedure, because the corresponding matrix is
non-diagonal
\[\begin{array}{c}
H_{mn}=\frac \hbar 2 \left(\omega+\frac 1 \omega\right)\left(n+\frac
1 2\right)\delta_{mn}\\
-\frac \hbar 4
\left(\sqrt{(n+1)(n+2)}\delta_{m,n+2}+\sqrt{(m+1)(m+2)}\delta_{n,m+2}\right)
\end{array}.\]
As the original and the basis functions are semiclassically
localized in the phase space on the ellipses
$\frac{p^2}{2}+\frac{q^2}{2}=E_n$ and
$\frac{p^2}{2}+\omega^2\frac{q^2}{2}=e_k$ respectively, simple
geometrical analysis of the intersection conditions for the
corresponding manifolds defines the basis set, which is necessary
and, as the numerical experiment confirms (fig.\ref{ho_ho_e_w}),
sufficient, to correctly determine the required state $E_n$
\[\min\left(\omega,\frac 1 \omega\right)E_n<\hbar\left(k+\frac 1
2\right)<\max\left(\omega,\frac 1 \omega\right)E_n.\]

\begin{figure}
\includegraphics[width=0.5\textwidth,draft=false]{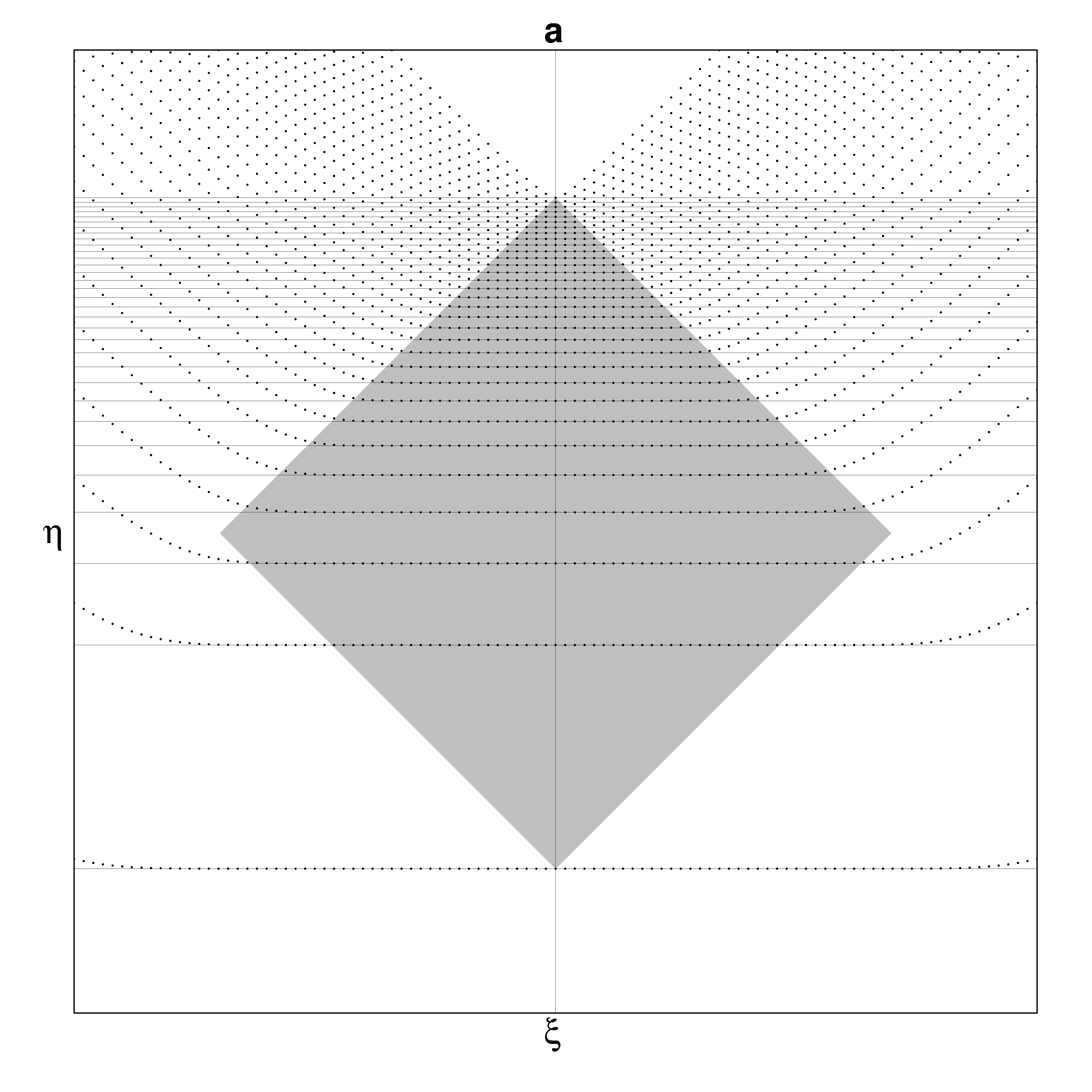}
\includegraphics[width=0.5\textwidth,draft=false]{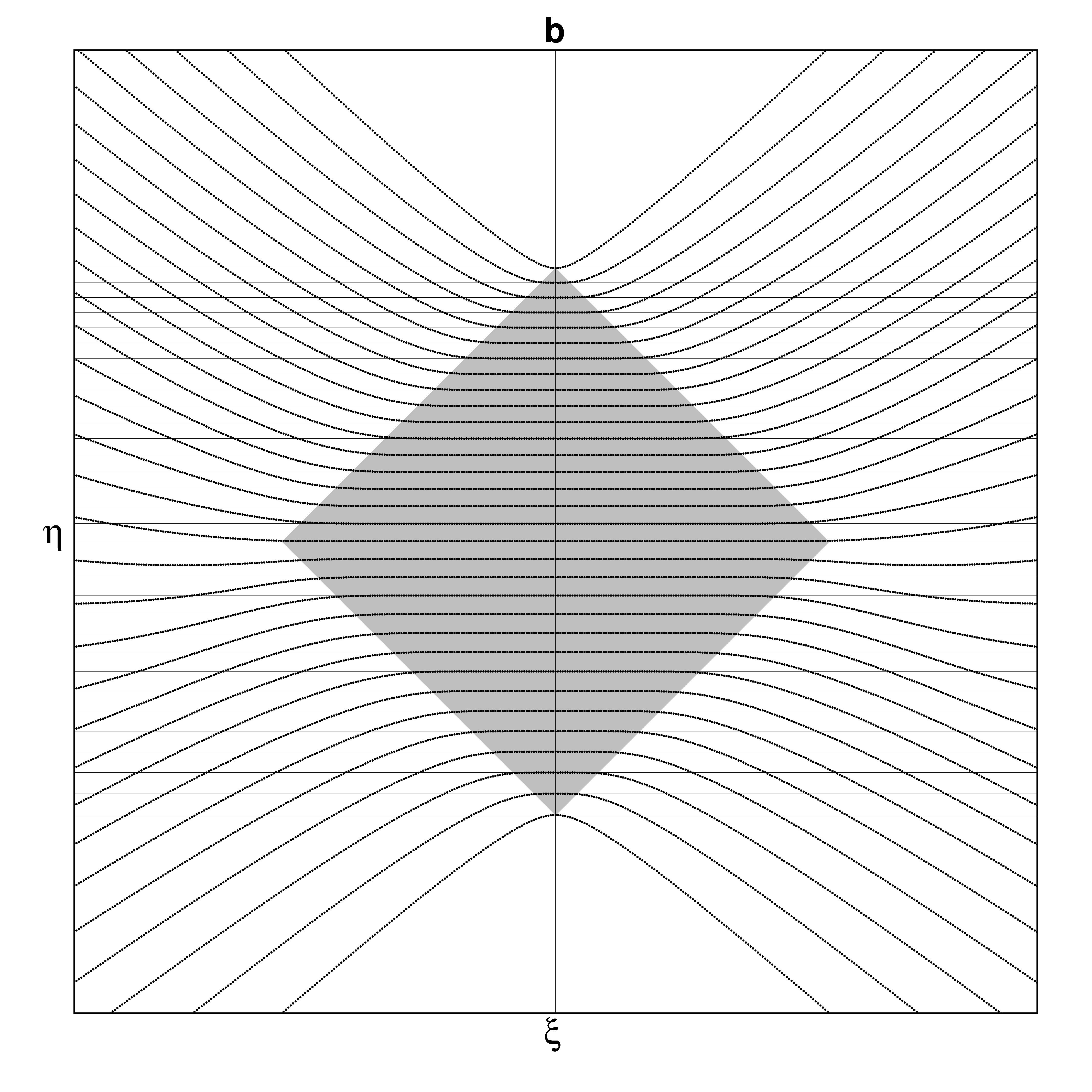}
\caption{Dependence of numerically found energy levels of harmonic
oscillator (\ref{ho_1d}) on the auxiliary basis frequency $\omega$
for $k_1=2$, $k_2=60$ (a) and $k_1=130$, $k_2=190$ (b): points
represent the numerical data, solid lines are exact energy levels
and gray color fills the stability region determined from
semiclassical analysis (\ref{ho_ho_sr_l})\label{ho_ho_e_w_l}}
\end{figure}

As can be seen on Fig.\ref{ho_ho_e_w}, at fixed set of basis
functions $k_1<k<k_2$ the region of computationally stable results
for energy levels represents the interior of a curvilinear tetragon
formed by a pair of hyperbolas and a pair of straight lines:
\begin{equation}\label{ho_ho_sr}\left\{\begin{array}{c} \hbar\omega\left(k_1+\frac 1
2\right)<E_n<\hbar\omega\left(k_2+\frac
1 2\right)\\
\frac \hbar \omega\left(k_1+\frac 1 2\right)<E_n<\frac \hbar
\omega\left(k_2+\frac 1 2\right)
\end{array}\right. .\end{equation}
It is convenient to introduce new variables
\[\eta=\ln E,\ \xi=\ln\omega,\ \nu_{1,2}=\ln\left(k_{1,2}+\frac 1 2\right),\]
in which the stability region (\ref{ho_ho_sr}) transforms into the
interior of the square (Fig.\ref{ho_ho_e_w_l})
\begin{equation}\label{ho_ho_sr_l}\nu_1<\eta\pm\xi<\nu_2.\end{equation}
Obviously the optimal basis will be that which corresponds to
maximum vertical dimensions of the stability region. In the
considered example it evidently corresponds to frequency $\omega=1$.

\begin{figure}
\includegraphics[width=0.5\textwidth,draft=false]{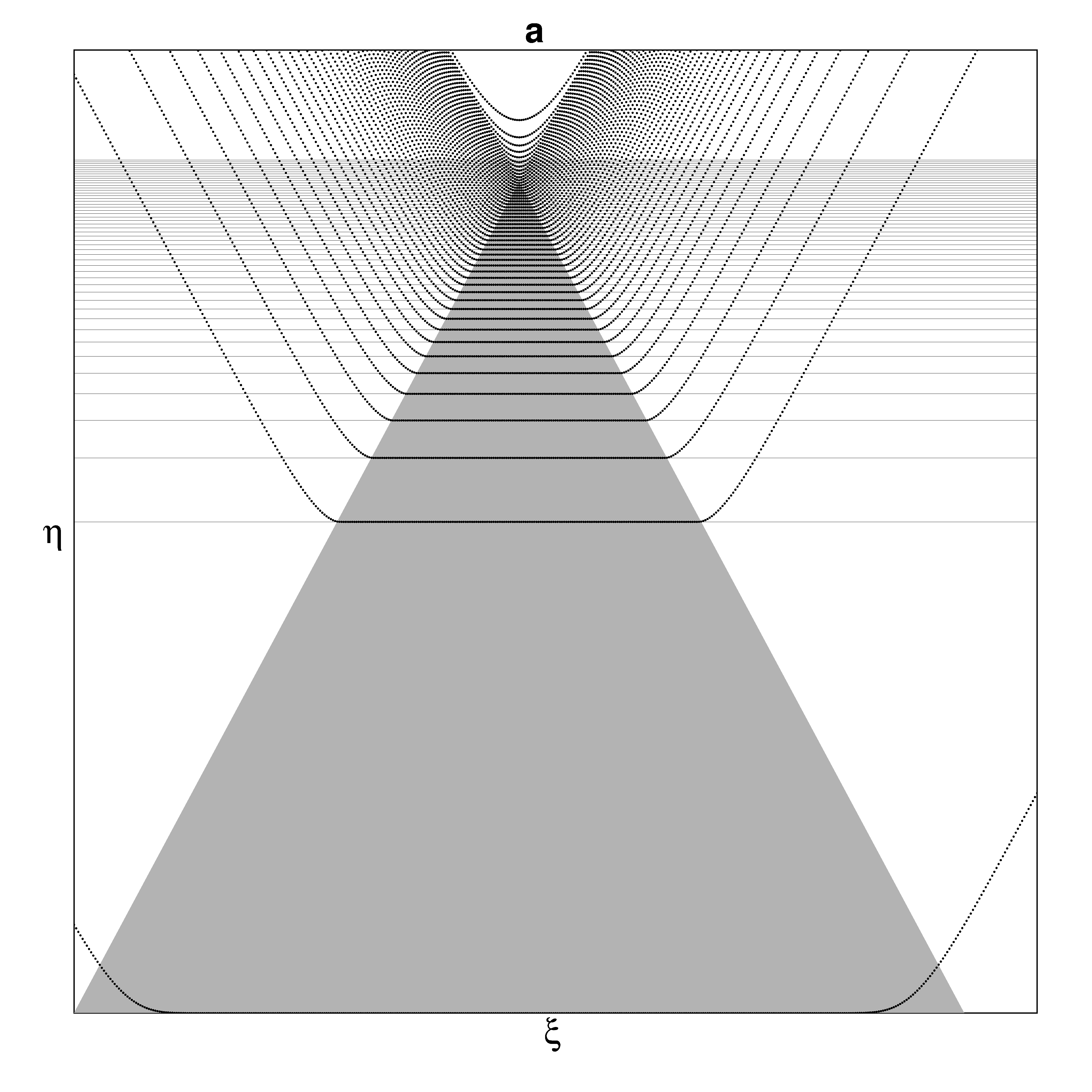}
\includegraphics[width=0.5\textwidth,draft=false]{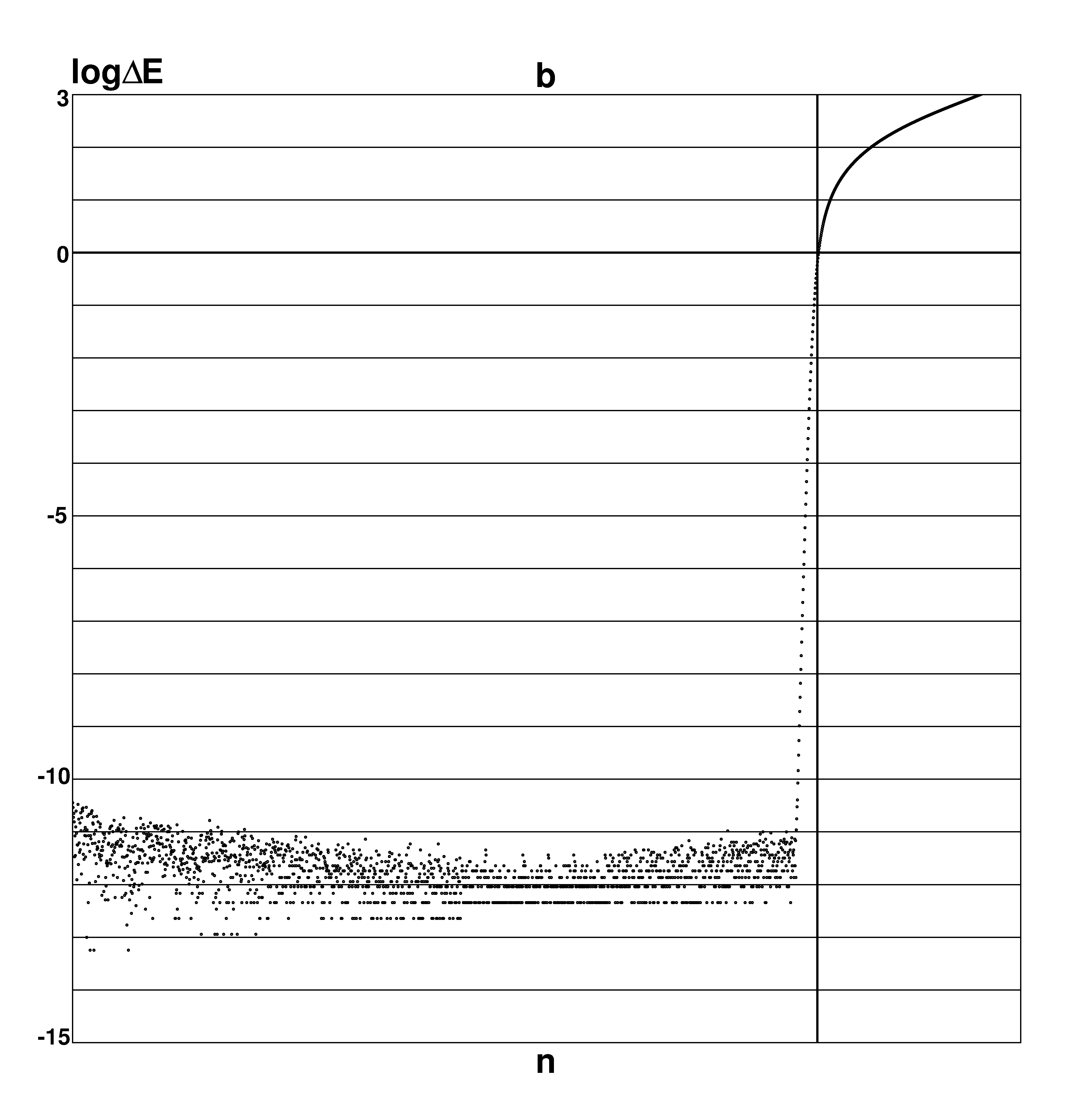}
\caption{Calculation of the harmonic oscillator energy levels
(\ref{ho_spectr}) in the plane waves basis (\ref{pw_basis}) for
$k_1=1$, $k_2=2 500$ . a) Dependence of numerically found energy
levels of harmonic oscillator (\ref{ho_1d}) on the auxiliary basis
parameter $a$: points represent the numerical data, solid lines are
exact energy levels and gray color fills the stability region
determined from the semiclassical analysis (\ref{ho_pw_sr_l}) b)
Dependence of absolute error on main quantum number $n$ for optimal
basis parameter (\ref{ho_pw_opt}). The vertical line corresponds to
the applicability limit (\ref{ho_pw_lim}) of the matrix
diagonalization method. \label{ho_pw}}
\end{figure}

An even simpler basis set can be built on plane waves ---
eigenfunctions of one-dimensional billiard, i.e. infinitely deep
potential well with walls in the points $x_{1,2}=\pm a$:
\begin{equation}\label{pw_basis}\varphi_k(x;a)=\left\{\begin{array}{cc}
\frac{1}{\sqrt a}\sin\frac{\pi k}{2a}(x+a), & |x|\le a\\
0, & |x|>a
\end{array}\right. .\end{equation}
The semiclassical phase space localization analysis can be applied
to the basis functions (\ref{pw_basis}) as well, if we take into
account that they are localized on the straight lines
\[\left\{\begin{array}{l}
x=\pm a\\ p=\pm\frac{\pi\hbar}{2a}k\end{array}\right. .\] We should
note that, unlike the harmonic oscillator, the plane waves basis
(\ref{pw_basis}) cannot be cut from below: $k_1=1$. As result the
stability region on the $(E,a)$-diagram for computation of the
harmonic oscillator spectrum (\ref{ho_spectr}) in the plain waves
basis (\ref{pw_basis}) with indexes $k=1,\ldots,k_2$ is represented
by the interior of a curvilinear triangle with base $E=0$ and
bordered by a parabola and a quadratic hyperbola, depending on
$k_2$:
\[\left\{\begin{array}{l} E_n<\frac{a^2}{2}\\
E_n<\frac{\pi^2\hbar^2}{8a^2}k_2^2\end{array}\right. .\] In the
variables
\[\left(\eta=\log_2 E,\ \xi=\log_2 a, \ \nu=\log_2 \pi\hbar k_2\right)\]
the computational stability region takes the form of a normal
triangle (see fig.\ref{ho_pw}a)
\begin{equation}\label{ho_pw_sr_l}\left\{\begin{array}{l} \eta<2\xi-1\\
\eta<2(\nu-\xi)-3\end{array}\right. .\end{equation} The vertex of
the triangle (\ref{ho_pw_sr_l})
\[\left(\xi=\frac{\nu-1}{2},\eta=\nu-2\right)\] corresponds to the optimal
basis with parameter \begin{equation} \label{ho_pw_opt}
a=\sqrt{\frac{\pi\hbar k_2}{2}},\end{equation} which allows correct
calculation of energy levels up to
\[E_{max}=\frac{\pi\hbar}{4}k_2.\]
Therefore the plane waves basis (\ref{pw_basis}) with indexes
$k=1,\ldots,k_2$ allows the correct calculation of the harmonic
oscillator energy levels (\ref{ho_spectr}) with indices
$n=0,\ldots,k_2=N$ (see Fig.\ref{ho_pw}b), where
\[N=\frac \pi 4 k_2 - \frac 1 2.\]
For large basis dimensions the fraction of correctly calculated
energy levels tends to the limit
\begin{equation}\label{ho_pw_lim}\lim_{k_2\rightarrow\infty}\frac N
k = \frac \pi 4 = 75\%
\end{equation}
\subsection{Quartic oscillator in 1D and in 2D}\sat
As a more complicated example let us consider  calculation of energy
levels for the quartic oscillator
\begin{equation}\label{qo}H(p,x)=x^2+p^4.\end{equation}

\begin{figure}
\includegraphics[width=0.5\textwidth,draft=false]{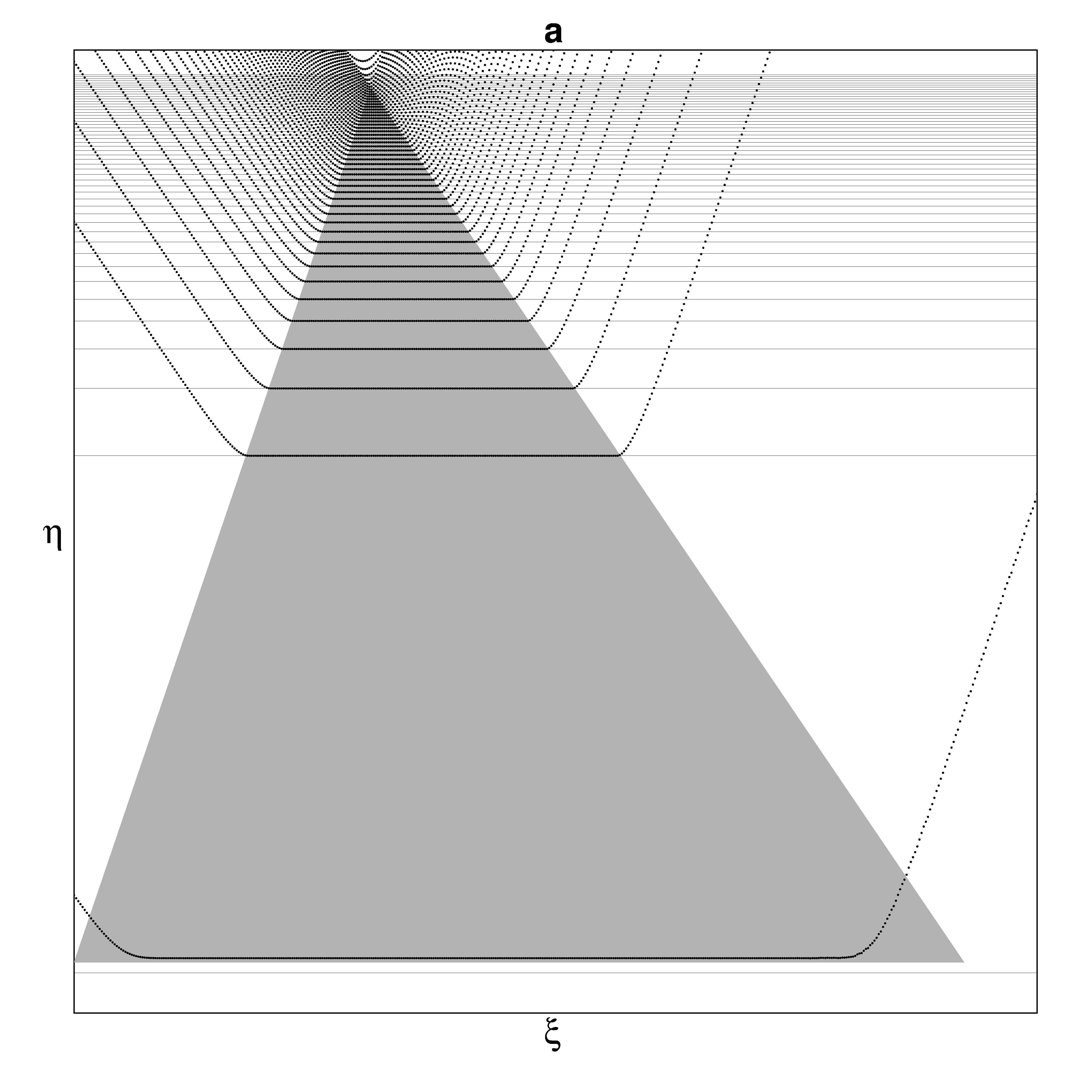}
\includegraphics[width=0.5\textwidth,draft=false]{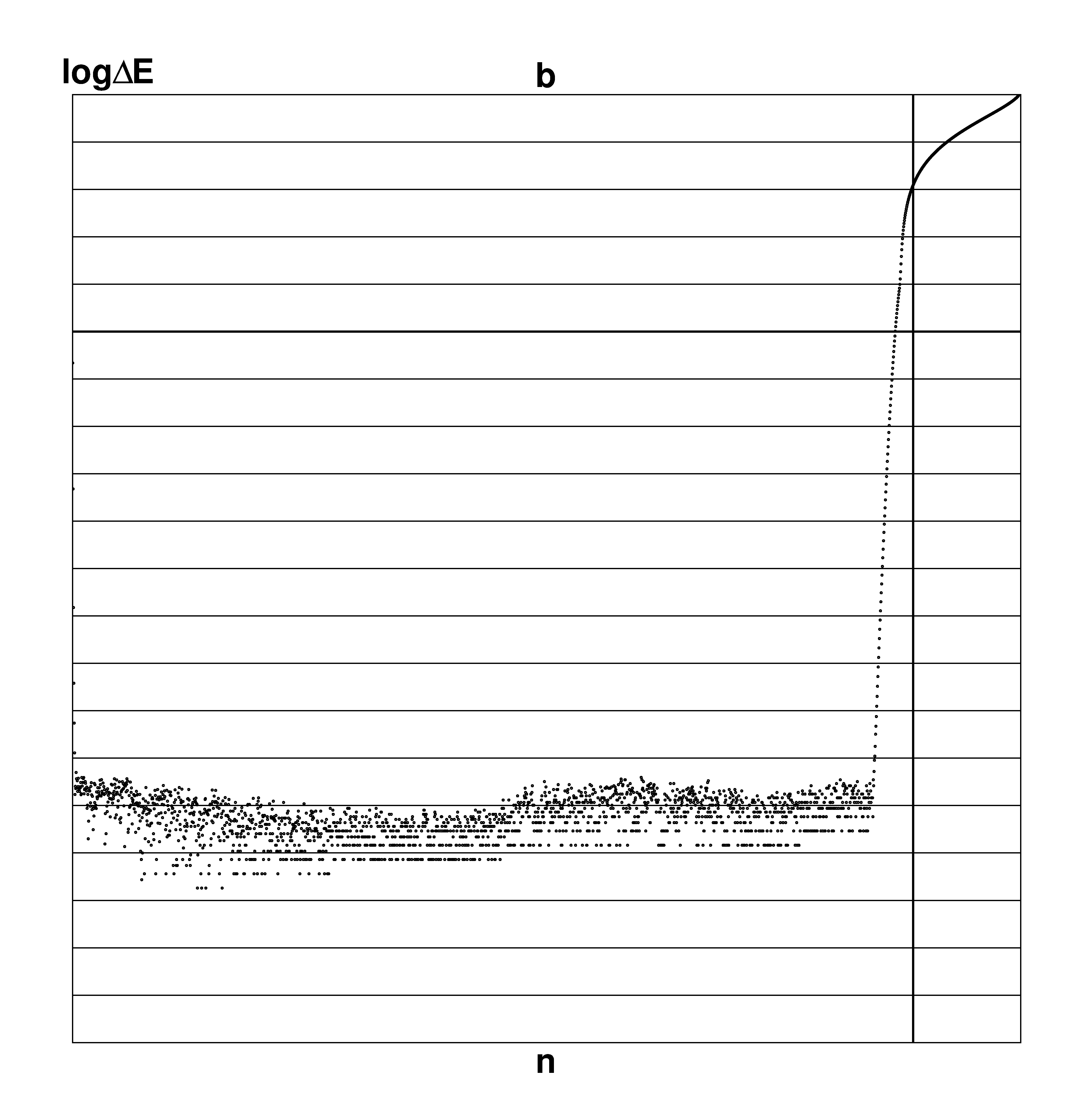}
\caption{Calculation of energy levels of quartic oscillator
(\ref{qo}) in the plane waves basis (\ref{pw_basis}) for $k_1=2$,
$k_2=2 500$. a) Dependence of numerically found energy levels on the
auxiliary basis parameter $a$: points represent the numerical data,
solid lines are exact energy levels and gray color fills the
stability region determined from the semiclassical analysis b)
Dependence of absolute errors on main quantum number $n$ for optimal
basis parameter. The vertical line corresponds to the applicability
limit of the matrix diagonalization method.\label{qo_pw}}
\end{figure}

\begin{figure}
\includegraphics[width=0.5\textwidth,draft=false]{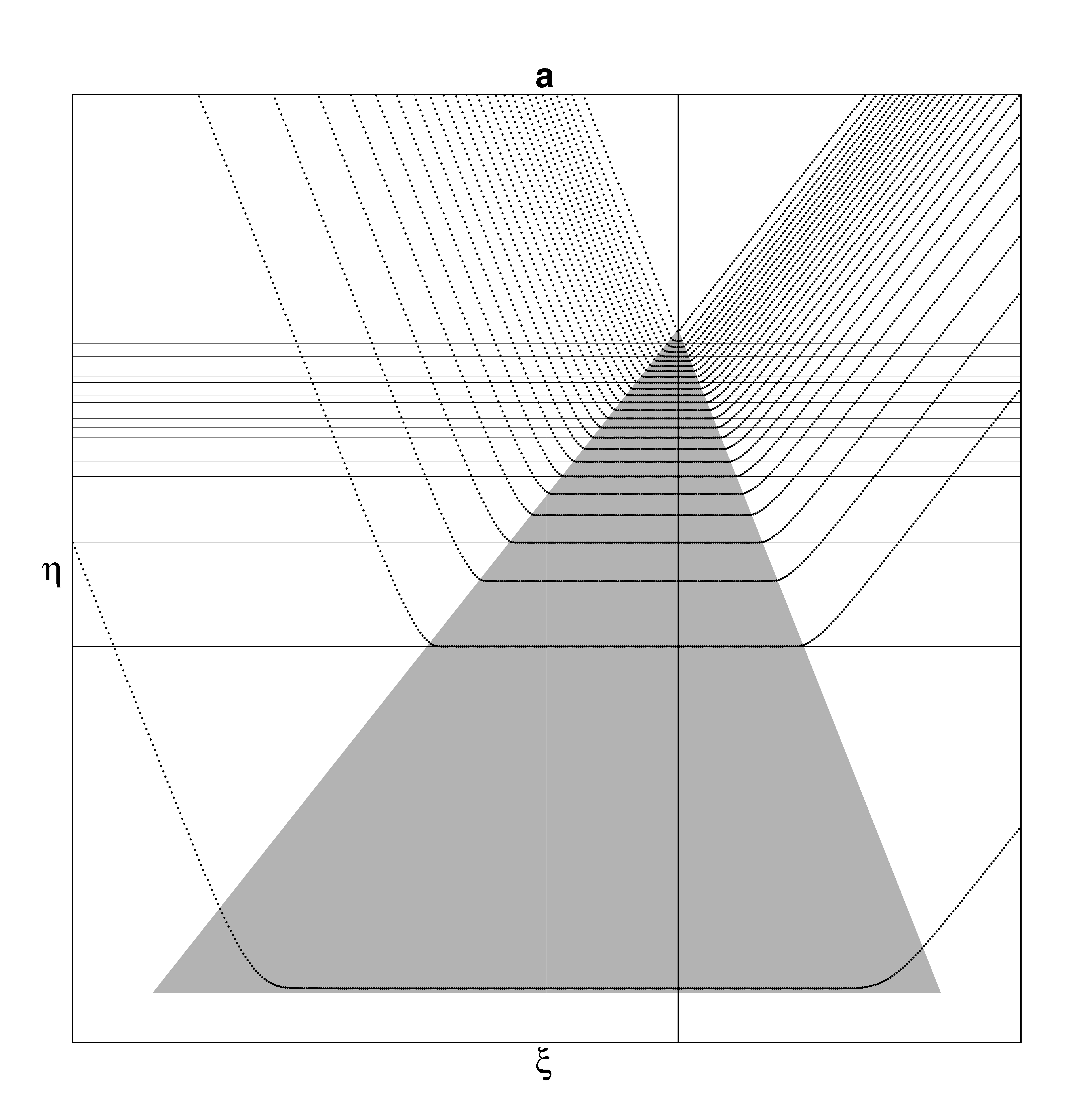}
\includegraphics[width=0.5\textwidth,draft=false]{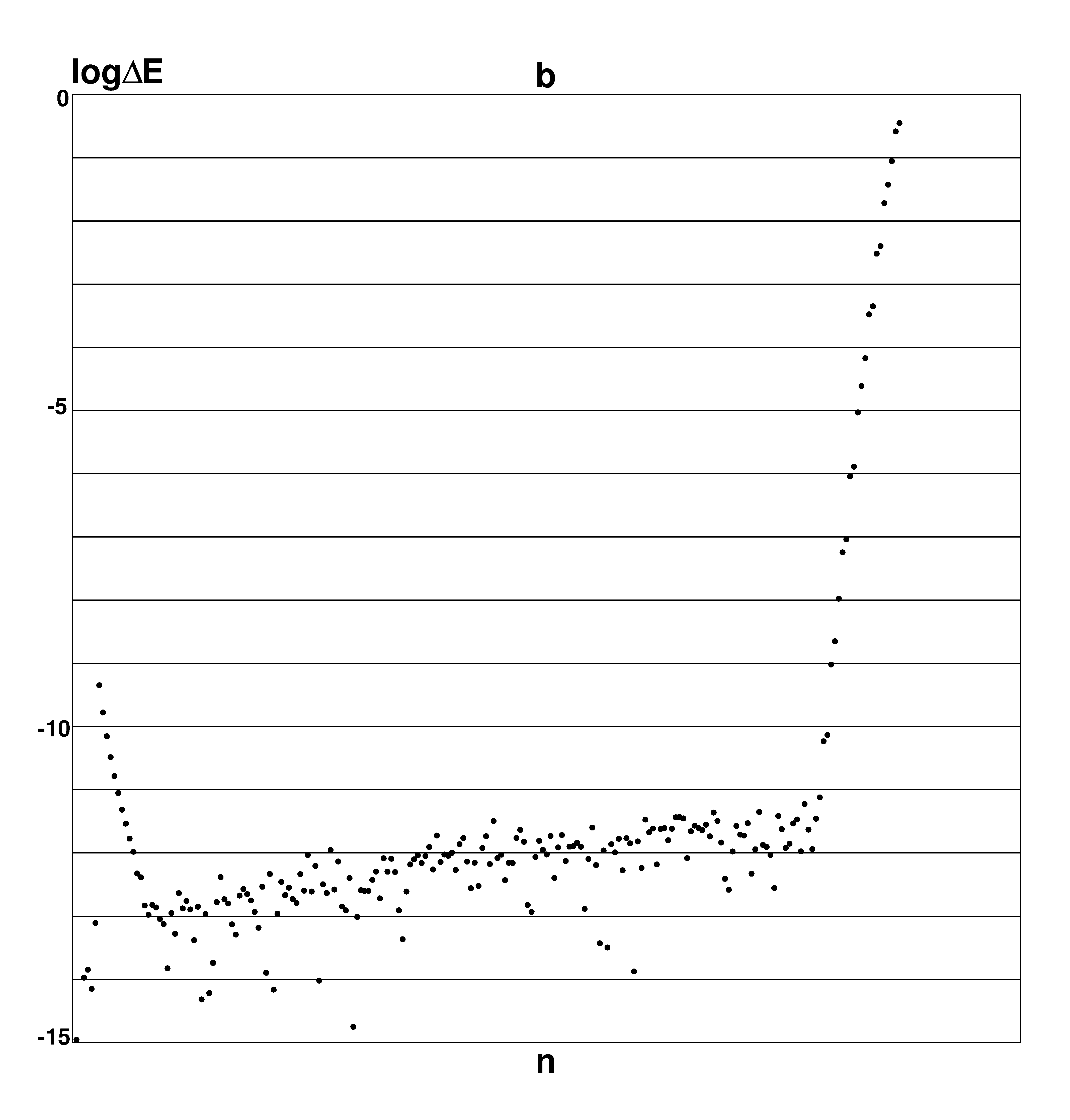}
\caption{Calculation of energy levels of quartic oscillator
(\ref{qo}) in the harmonic oscillator basis (\ref{ho_basis}) for
$k_1=2$, $k_2=250$. (a) Dependence of numerically found energy
levels on the auxiliary basis frequency $\omega$: points represent
the numerical data, solid lines are exact energy levels and gray
color fills the stability region determined from the semiclassical
analysis b) Dependence of absolute errors on main quantum number $n$
for optimal basis parameter.\label{qo_ho}}
\end{figure}

As distinct from the harmonic oscillator, this quantum mechanical
problem does not allow analytical solution, but there is a very
accurate semiclassical approximation for the quartic oscillator
spectrum \cite{bohigas93}
\begin{equation}\label{qo_semi}\begin{array}{c}E_n\approx\pi^2\left(\frac{3\sqrt2\left(n+\frac12\right)}{\Gamma\left(\frac14\right)}\right)^{\frac43}
\left\{1+\frac{1}{9\pi\left(n+\frac12\right)^2}\right.\\
\left.-\frac{\frac58+\frac{11}{6}\left(\frac{\Gamma\left(\frac14\right)^2}{4\pi}\right)^4}{\left[9\pi\left(n+\frac12\right)^2\right]^2}
+\frac{\frac{11}{12}+\frac{341}{10}\left(\frac{\Gamma\left(\frac14\right)^2}{4\pi}\right)^4}{\left[9\pi\left(n+\frac12\right)^2\right]^3}+\ldots\right\}
\end{array},\end{equation}
Formula (\ref{qo_semi}) gives the energy levels with $15$ correct
digits, which is quite sufficient to check the numerical results in
the given case. The results of such a check are presented on
Fig.\ref{qo_pw} for the plane waves basis (\ref{pw_basis}) and on
Fig.\ref{qo_ho} for the harmonic oscillator basis (\ref{ho_basis}).

Semiclassical phase space analysis gives the following results for
the quartic oscillator (\ref{qo}): the stability region is
determined by the conditions
\[\left\{\begin{array}{c}E_n<a^4\\
E_n<\frac{\pi^2\hbar^2}{4a^2}k_2^2\end{array}\right.\] for the plane
waves basis (\ref{pw_basis}) and
\[\left\{\begin{array}{c}E_n<2\hbar\omega\left(k+\frac12\right)-\frac{\omega^4}{4}\\
E_n<\frac{4\hbar^2}{\omega^2}\left(k+\frac12\right)^2\end{array}\right.\]
in the harmonic oscillator basis (\ref{ho_basis}). Now it is easy to
obtain expressions for optimal basis parameters and maximum energies
of correctly calculable states:
\[\begin{array}{cc}
a=\left(\frac{\pi\hbar k}{2}\right)^\frac13, &
E_{max}^{PW}=\left(\frac{\pi\hbar k}{2}\right)^\frac43\\
\omega=\left[2\hbar\left(k+\frac12\right)\right]^\frac13, &
E_{max}^{HO}=\frac34\left[2\hbar\left(k+\frac12\right)\right]^\frac43.
\end{array}\]
Taking into account (\ref{qo_semi}), we obtain for the relative
number of correctly calculated energy levels
\[\frac{N}{k_2}\approx\frac{\Gamma^2\left(\frac14\right)}{6\sqrt{2\pi}}\approx0.874\] for the
plane waves basis (\ref{pw_basis}) and
\[\frac{N}{k_2}\approx\frac{\Gamma^2\left(\frac14\right)}{2\pi^\frac32 3^\frac14}\approx0.897\]
in the harmonic oscillator basis (\ref{ho_basis}). Hence it follows
that the latter is slightly more efficient.

Harmonic oscillator (\ref{ho_basis}) and infinitely deep potential
well (\ref{pw_basis}) in fact exhaust the set of exactly solvable
one-dimensional quantum systems whose eigenfunctions can be used as
a basis for matrix diagonalization. There are many more
possibilities in the models with dimensionality of more then one.
Further, for simplicity we consider only two-dimensional systems,
but the results can be trivially generalized for higher dimensions.

\begin{figure}
\includegraphics[width=0.5\textwidth,draft=false]{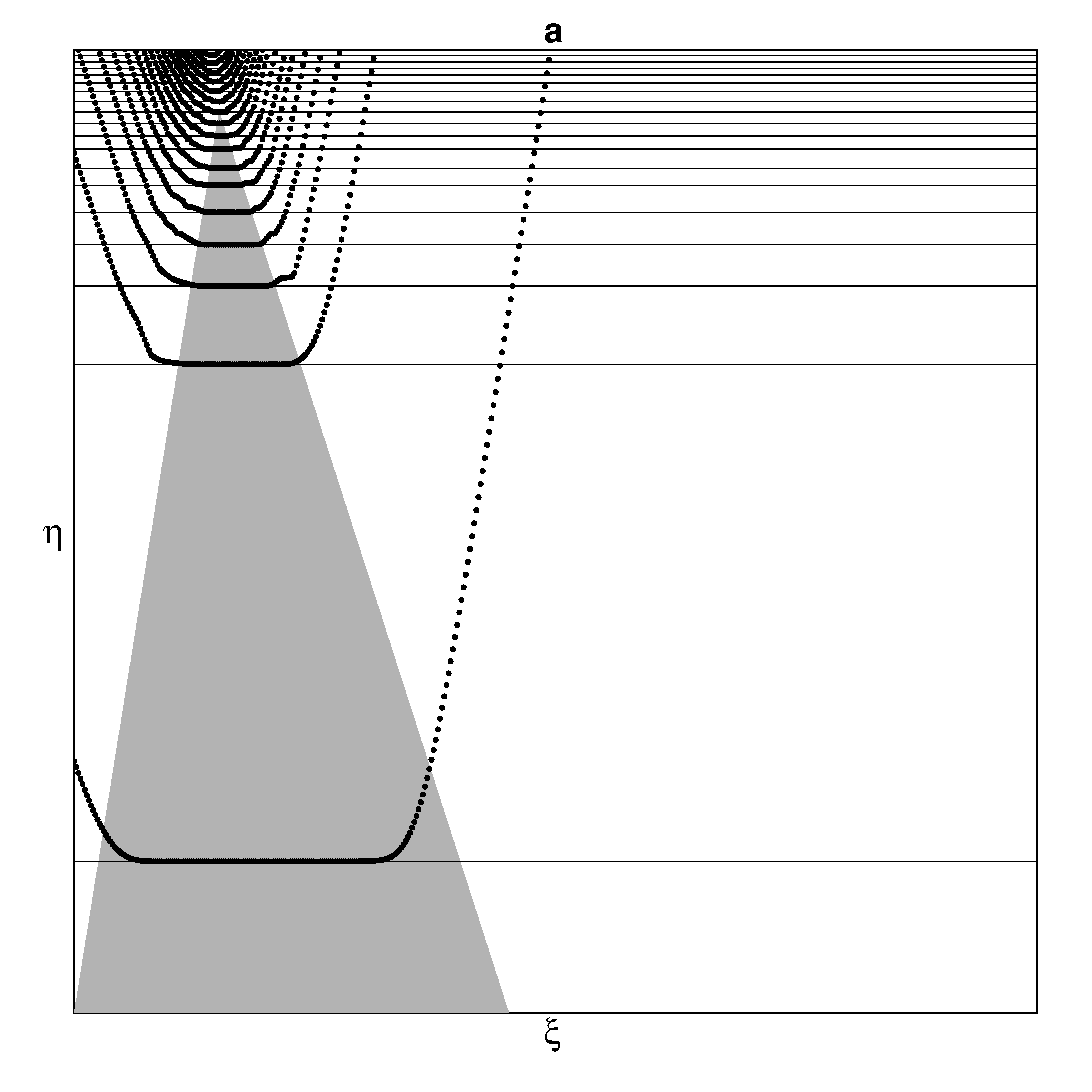}
\includegraphics[width=0.5\textwidth,draft=false]{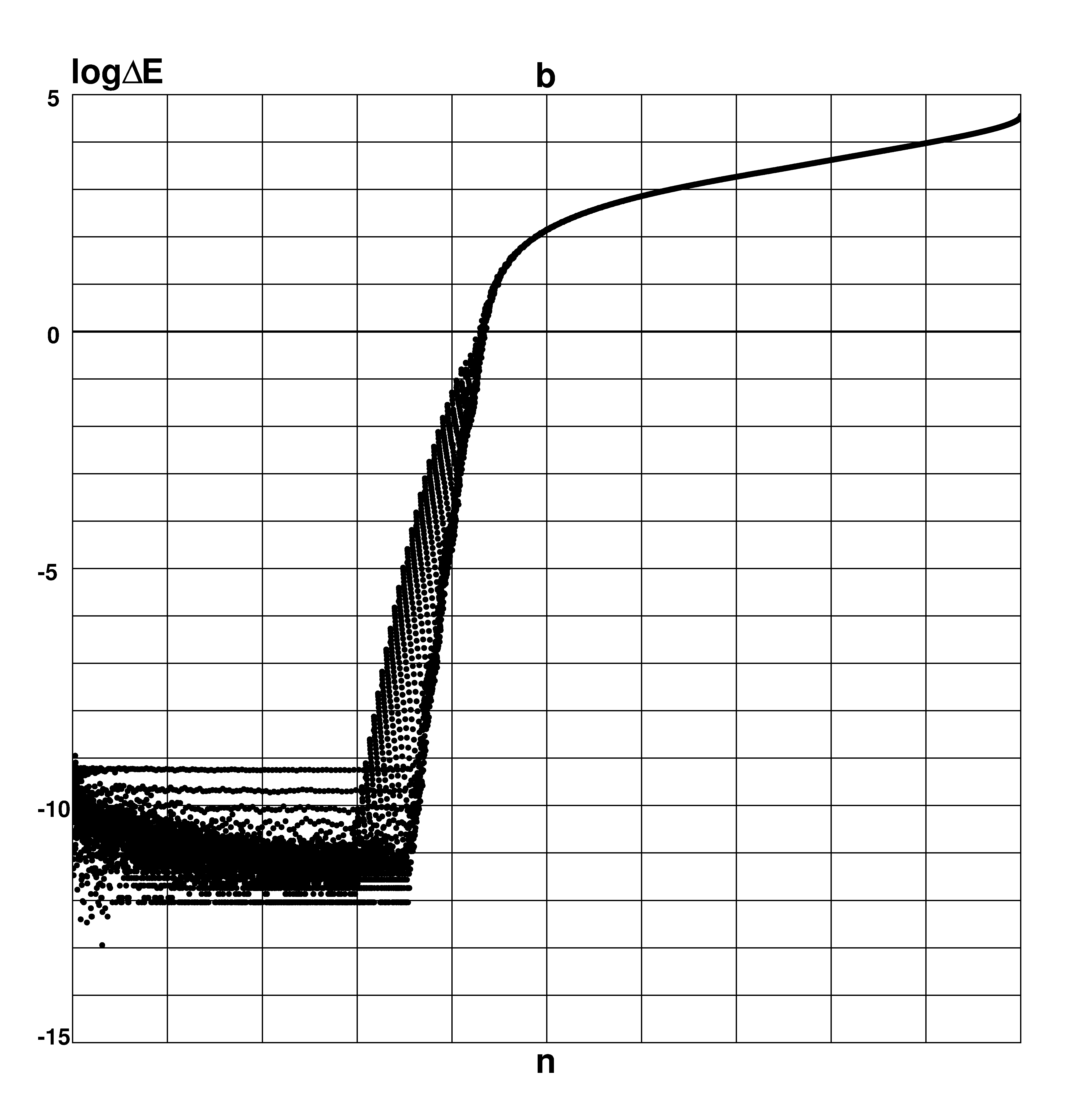}
\caption{Calculation of energy levels of CQO potential (\ref{cqo})
with $\alpha=6$ in the harmonic oscillator basis (\ref{hoho_basis}).
a) Dependence of numerically found energy levels on the auxiliary
basis frequency $\omega=\omega_x=\omega_y$: points represent the
numerical data, solid lines are exact energy levels and gray color
fills the stability region determined from the semiclassical
analysis b) Dependence of absolute errors on main quantum number $n$
for optimal basis parameter.\label{cqo_ho}}
\end{figure}

\begin{figure}
\includegraphics[width=0.5\textwidth,draft=false]{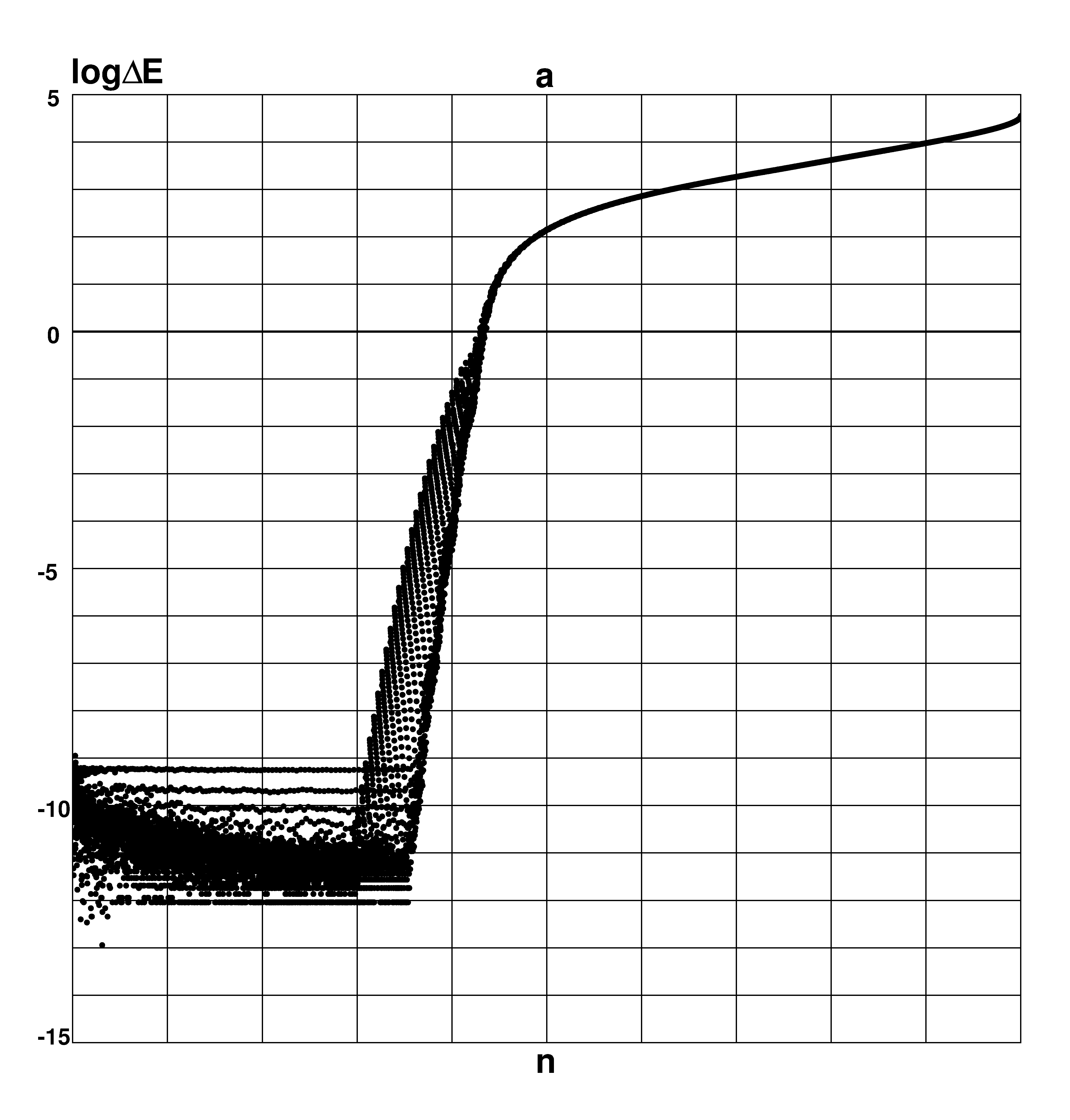}
\includegraphics[width=0.5\textwidth,draft=false]{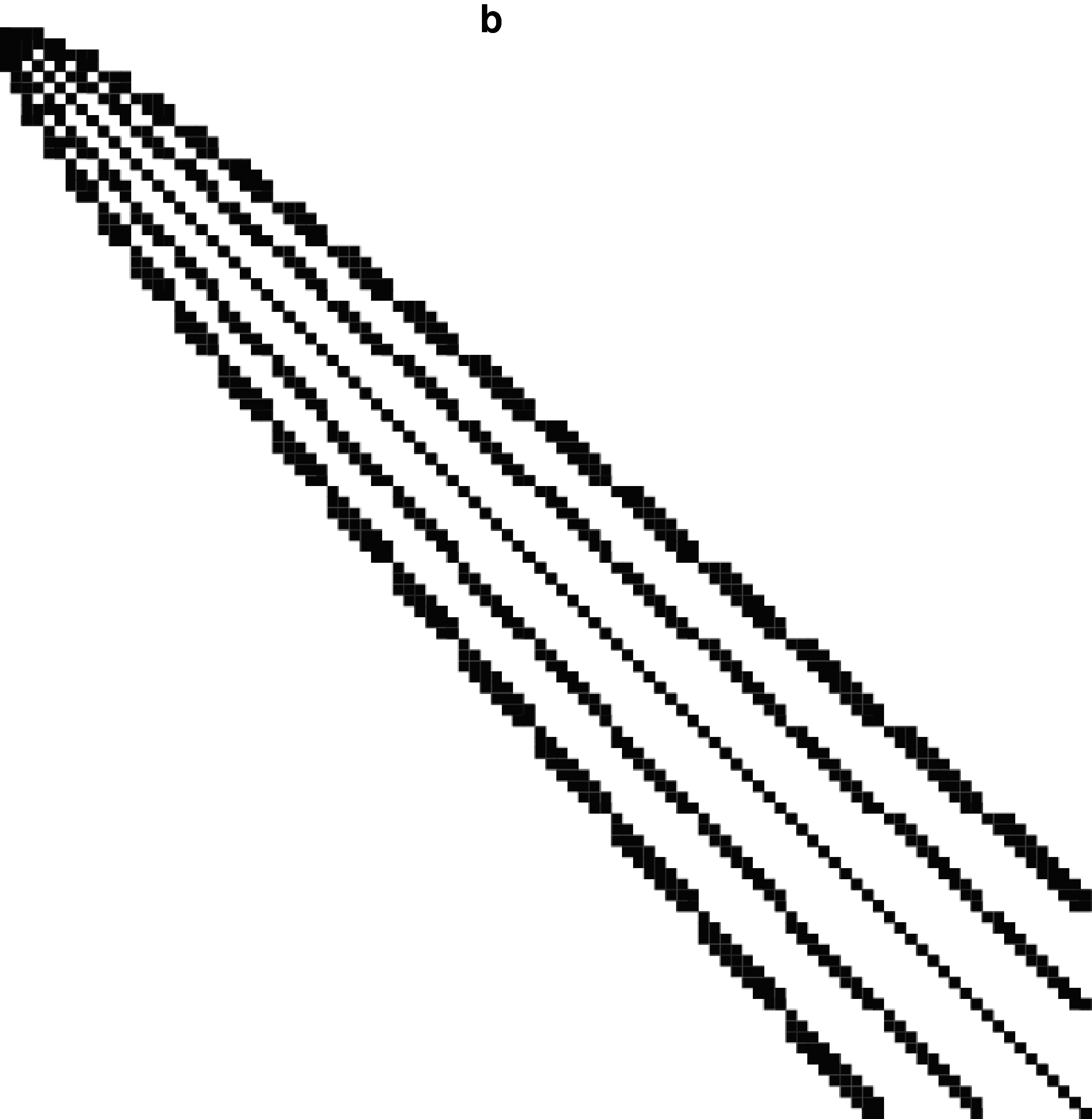}
\caption{a) Comparison of numerically found energy levels of CQO
potential (\ref{cqo}) with $\alpha=6$ in the plane waves basis
(\ref{pwpw_basis}) and in the harmonic oscillator basis
(\ref{hoho_basis}) for optimal parameter of both bases. b)
Distribution of non-zero Hamiltonian matrix elements in the harmonic
oscillator basis (\ref{hoho_basis}) for $A_1$-type states in the CQO
potential (\ref{cqo}). \label{cqo_de}}
\end{figure}

The simplest type of two-dimensional basis can be constructed from
products of eigenfunctions of exactly solvable one-dimensional
problems, for example, from plane waves
\begin{equation}\label{pwpw_basis}\varphi_{k_x,k_y}(x,y;a_x,a_y)=\frac{1}{\sqrt{a_x
a_y}}\sin{\frac{\pi k_x}{2a_x}(x+a_x)}\sin{\frac{\pi
k_y}{2a_y}(x+a_y)},\end{equation} harmonic oscillator eigenfunctions
\begin{equation}\label{hoho_basis}\varphi_{k_x,k_y}(x,y;\omega_x,\omega_y)=\frac{\left(\omega_x\omega_y\right)^{\frac14}}{\sqrt{\pi\hbar}}
\frac{H_{k_x}(\sqrt{\frac{\omega_x}{\hbar}}x)H_{k_y}(\sqrt{\frac{\omega_y}{\hbar}}y)}{\sqrt{2^{k_x+k_y}k_x!k_y!}}e^{-(\omega_x
x^2+\omega_y y^2)/(2\hbar)},\end{equation} or as a combination of
both of them
\begin{equation}\label{pwho_basis}\varphi_{k_x,k_y}(x,y;a,\omega)=\frac{1}{\sqrt{a}}\left(\frac{\omega}{\pi\hbar}\right)^{\frac14}
\sin{\frac{\pi k_x}{2a}(x+a)}
\frac{H_{k_y}(\sqrt{\frac{\omega}{\hbar}}y)}{\sqrt{2^{k_y}k_y!}}e^{-\frac{\omega
y^2}{2\hbar}}\end{equation} Efficiency of the basis types
(\ref{pwpw_basis},\ref{hoho_basis},\ref{pwho_basis}) depends on the
problem under consideration, but in any case simple semiclassical
phase space analysis allows us to choose the most preferable among
them. Skipping cumbersome exact expressions, let us give only the
ultimate results (Fig.\ref{cqo_ho},\ref{cqo_de}a) of
the semiclassical optimization of the spectrum calculation in the
two-dimensional potential of coupled quartic oscillators (CQO)
\begin{equation}\label{cqo} U_4(x,y;\alpha)=x^4+y^4+\alpha x^2
y^2.\end{equation}

Quite often the Hamiltonian of the system under consideration has a
discrete symmetry. For example, the CQO potential (\ref{cqo}) is
invariant under transformations of square symmetry group $C_{4v}$.
In such cases it is convenient for many reasons to calculate the
states of different symmetry types independently. Firstly, for
investigation of statistical properties of energy spectra in quantum
chaology we have anyway to exclude pure spectral series ---
sequences of states with one and the same symmetry type. Secondly,
as a rule, states of different symmetry, even and odd for example,
are usually very close in energy if not degenerate. Therefore
numerical computation of all symmetry types of states together leads
to very ill-conditioned matrices, while exclusion of certain
symmetry type improves the conditionality. And lastly, computation
of different symmetry types one-by-one runs evidently faster than
determination of all the states altogether. For example, the basis
for determination of $A_1$-type states in the CQO potential
(\ref{cqo}) is constructed from symmetrized combinations of the form
$\varphi_{mn}+\varphi_{nm}$ of basis vectors (\ref{pwpw_basis}) with
 $a_x=a_y$, ($m,n$ odd) or (\ref{hoho_basis}) with $\omega_x=\omega_y$ ($m,n$
 even).

The characteristic feature of polynomial potential is the sparse
band structure of the Hamiltonian matrix (Fig.\ref{cqo_de}b) in the
basis of harmonic oscillator (\ref{hoho_basis}). For example, for
large basis dimensions $n$ the bandwidth for $A_1$-type states in
the CQO potential (\ref{cqo}) is equal $m\approx2\sqrt{2n}$. Clearly
for polynomial potentials it makes sense to use special routines for
band matrix diagonalization which allows us to economize both the
CPU time and memory usage for computations.

Simple analysis shows that the number of non-zero matrix elements
for $A_1$-type states in the CQO potential (\ref{cqo}) never exceeds
$11n$. It means that the used basis vectors ordering is not optimal
and that there possibly exists another ordering which leads to band
matrices with constant bandwidth $m=5$, or at least with slower
growing bandwidth. Such ideal ordering would considerably speed up
the computations but its search represents a non-trivial task.
\sat\section{The Spectral Method}\sat The spectral method (SM) for
the solution of the Schr\"odinger equation was proposed in the paper
\cite{sm} in application to 1D and 2D potential systems, but it can
be easily generalized for the Schr\"odinger equations of arbitrary
dimensions:
\[\left[-\frac{\hbar^2}{2}\sum\limits_{i=1}^{D}\partial_i^2 + U(x_1,\dots,x_D)\right] \psi_n (x_1,\dots,x_D) = E_n \psi_n (x_1,\dots,x_D),\]
where $D$ is the dimensionality of the system configuration space.
Let us assume that the potential $U(x_1,\dots,x_D)$ allows only
finite motion for all energies, therefore our task is to find
discrete energy spectrum $E_n$ and stationary wave functions $\psi_n
(x_1,\dots,x_D)$.

Let us consider time-dependent solution $\psi (x_1,\dots,x_D;t)$ for
the corresponding non-stationary Schr\"odinger equation
\[\left[-\frac{\hbar^2}{2}\sum\limits_{i=1}^{D}\partial_i^2 + U(x_1,\dots,x_D)\right] \psi(x_1,\dots,x_D;t) = i\hbar\partial_t \psi(x_1,\dots,x_D;t)\]
with some in principle arbitrary initial condition \[\psi_0
(x_1,\dots,x_D)=\psi_n (x_1,\dots,x_D;t=0)\]. Applying the
decomposition
\[\psi_0 (x_1,\dots,x_D)=\sum\limits_{n=1}^{\infty}a_n \psi_n (x_1,\dots,x_D),\]
we obtain
\[\psi (x_1,\dots,x_D;t)=\sum\limits_{n=1}^{\infty}a_n \psi_n (x_1,\dots,x_D)e^{-i\frac{E_n t}{\hbar}}.\]
Here and further we imply that the wave functions $\psi_n
(x_1,\dots,x_D)$ are orthonormal
\[\int dx_1 \ldots dx_D \bar{\psi}_i
(x_1,\dots,x_D) \psi_k (x_1,\dots,x_D)=\delta_{ik}.\]

Let us consider autocorrelation function of the form
\begin{equation}\label{autocor}P(t)=\int dx_1 \ldots dx_D \bar{\psi}_0
(x_1,\dots,x_D) \psi
(x_1,\dots,x_D;t)=\sum\limits_{n=1}^{\infty}\left|a_n\right|^2
e^{-i\frac{E_n t}{\hbar}}.\end{equation} We assume the initial wave
function $\psi_0 (x_1,\dots,x_D)$ to be normalized
\[\int dx_1 \ldots dx_D \bar{\psi}_0
(x_1,\dots,x_D) \psi_0
(x_1,\dots,x_D)=\sum\limits_{n=1}^{\infty}\left|a_n\right|^2=1,\] so
that $P(0)=1$. The Fourier transform of the autocorrelator
(\ref{autocor}) contains information about the energy spectrum $E_n$
of the system
\begin{equation}\label{p_e}P(E)=\int\limits_{-\infty}^{\infty}dt
e^{i\frac{Et}{\hbar}}P(t)=\hbar\sum\limits_{n=1}^{\infty}\left|a_n\right|^2
\delta\left(E-E_n\right).\end{equation} For determination of
stationary wave functions $\psi_n (x_1,\dots,x_D)$ we will need the
Fourier transform of $\psi (x_1,\dots,x_D;t)$ itself
\begin{equation}\label{psi_e}\psi (x_1,\dots,x_D;E)=\int\limits_{-\infty}^{\infty}dt
e^{i\frac{Et}{\hbar}}\psi
(x_1,\dots,x_D;t)=\hbar\sum\limits_{n=1}^{\infty}a_n \psi_n
(x_1,\dots,x_D) \delta\left(E-E_n\right).\end{equation}

Naturally in practice we never try to find all $E_n$ and $\psi_n
(x_1,\dots,x_D)$. Usually the task is to determine all the energy
levels $E_n$ inside a given interval $E_1<E<E_2$ with certain
accuracy $\delta_E$ and to calculate the corresponding stationary
wave functions $\psi_n (x_1,\dots,x_D)$ on some finite set of points
$\left\{x_i^{(k_i)}=x_{i}^{(0)}+k_i\Delta x_i,\
k_i=0,\ldots,N_i\right\}$ also with finite accuracy. Further we will
assume for simplicity that all $N_i$ are equal.

\begin{figure}
\includegraphics[width=0.5\textwidth,draft=false]{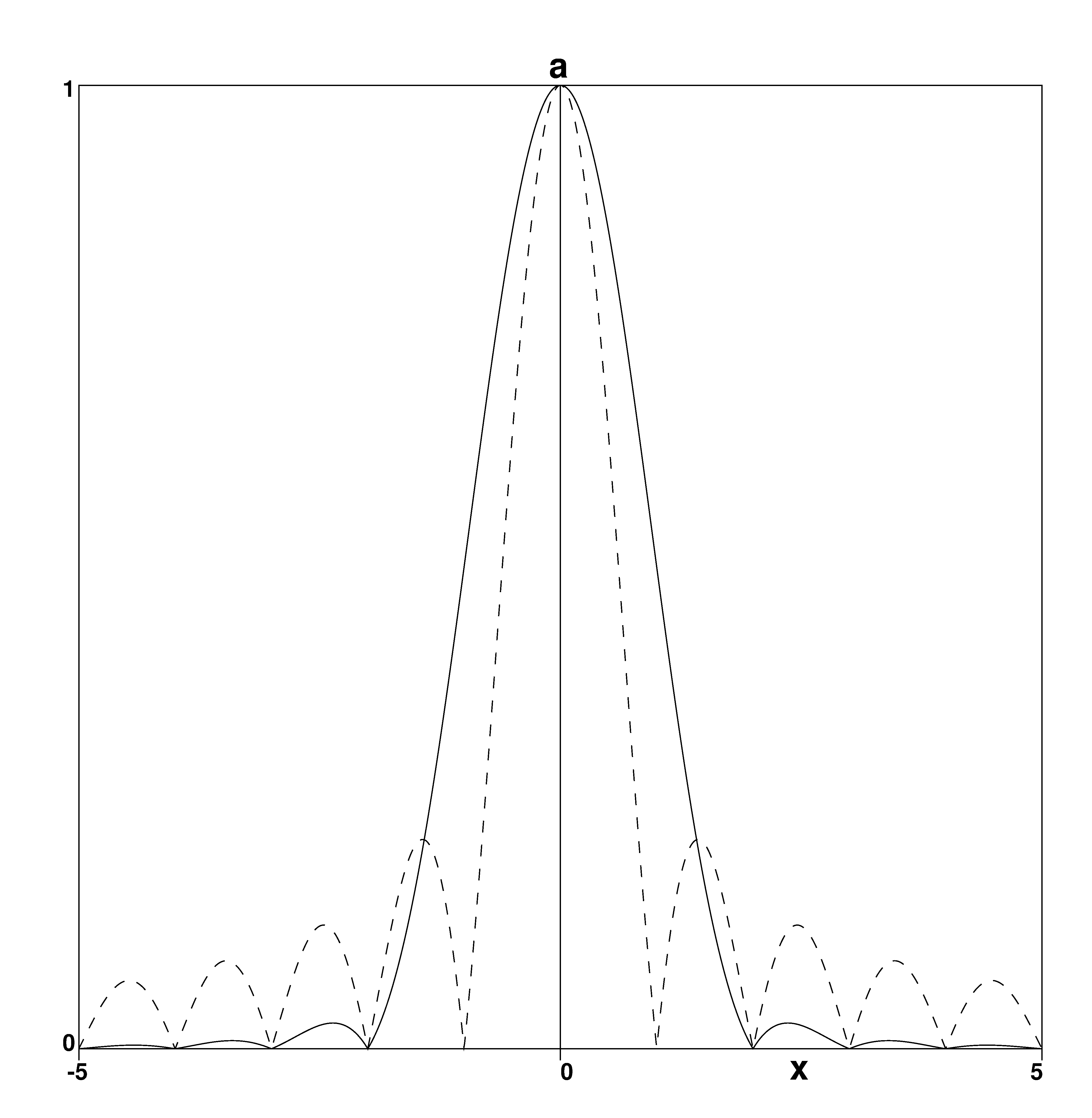}
\includegraphics[width=0.5\textwidth,draft=false]{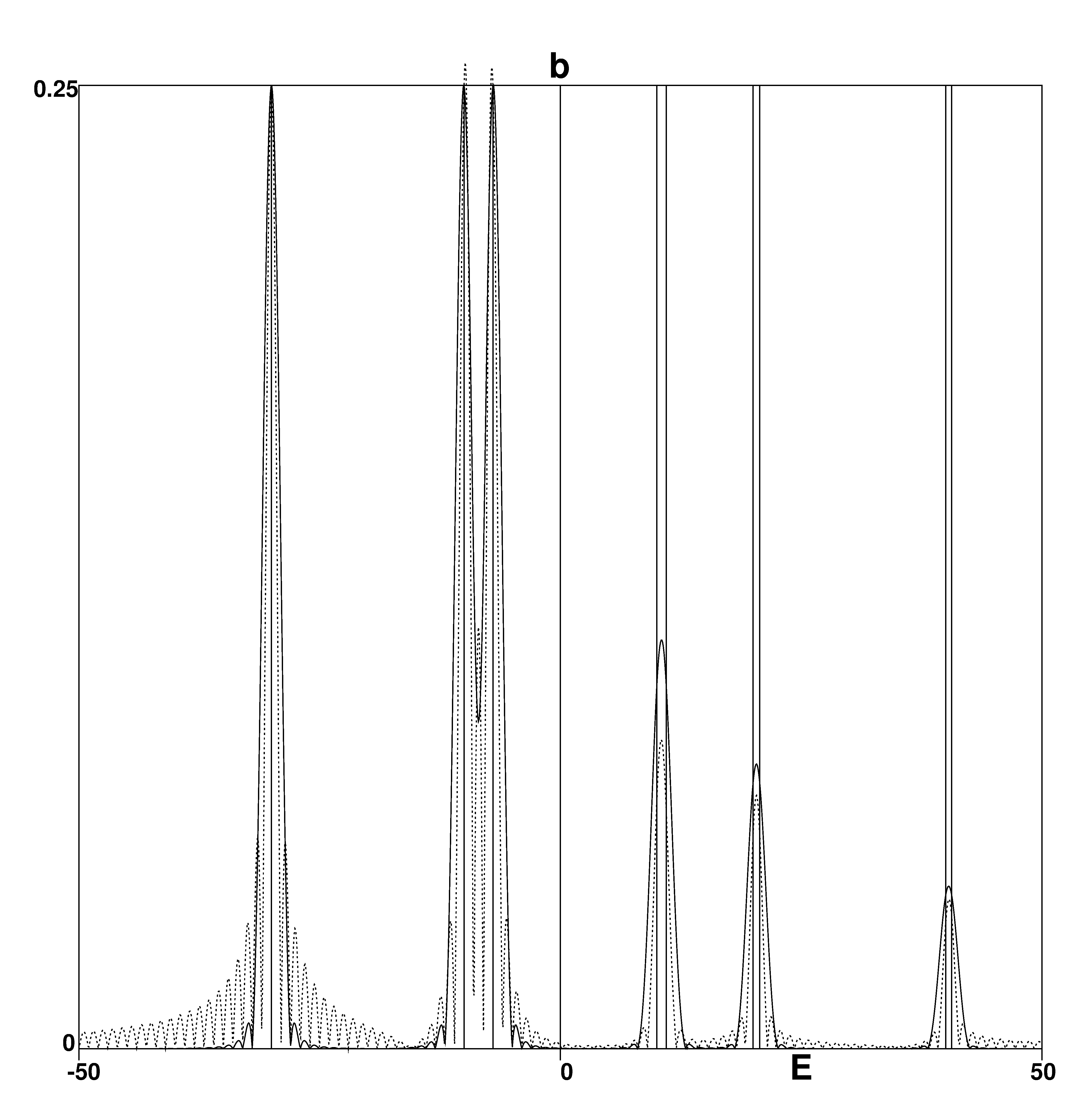}
\caption{a) $\left|f_T(x)\right|$ profile (dashed line) and
$\left|f_H(x)\right|$ profile (solid line). b) $\left|P_T(E)\right|$
 (dashed line) and $\left|P_H(E)\right|$ (solid line) for the model
 system with energy spectrum
 $E_{1,2,\ldots,9}=\{-30,-10,-7,10,11,20,20.7,40,40.63\}$ (marked by solid vertical
 lines) and
 $a_{1,2,\ldots,9}=\{0.5,0.5,0.5,0.25,0.25,0.2,0.2,0.15,0.15\}$ with
 $h=T=1$.
\label{d_t}}
\end{figure}

\begin{figure}
\includegraphics[width=\textwidth,draft=false]{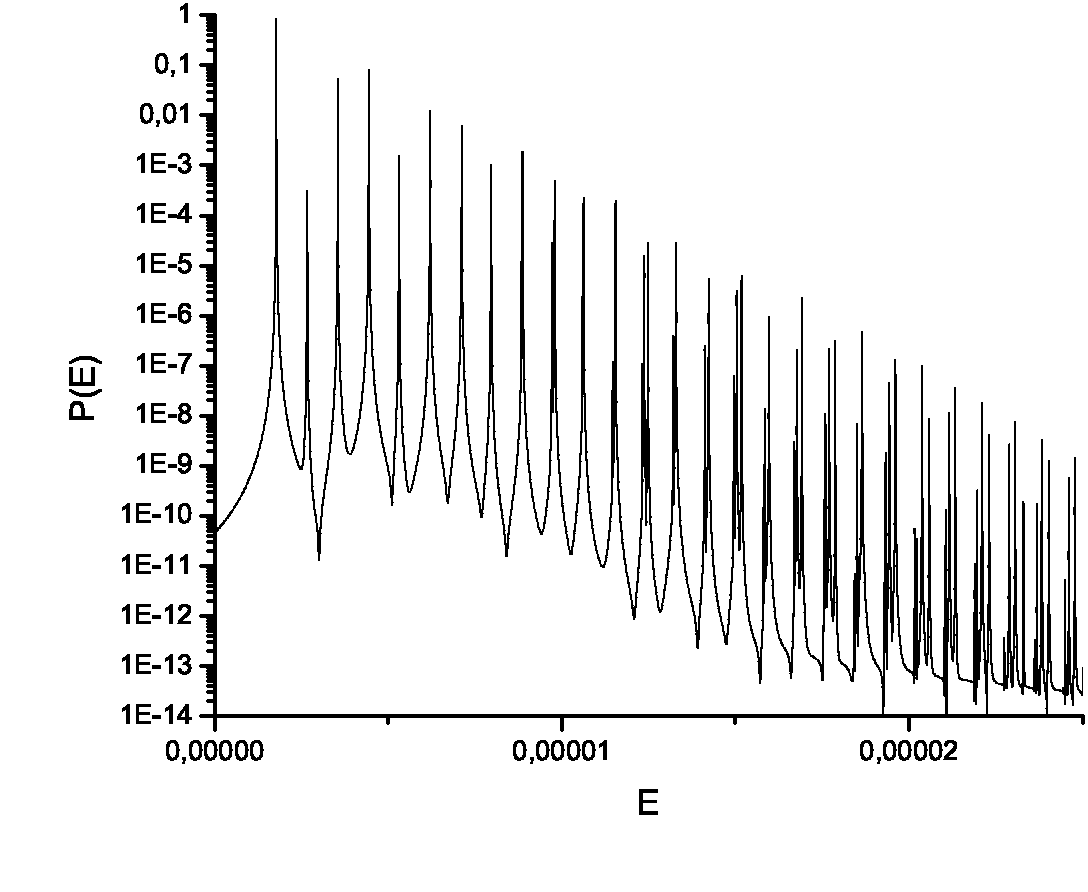}
\caption{$\left|P_T(E)\right|$ for calculation of $E$-type energy
levels in the quadrupole oscillation potential. \label{p_e_typ}}
\end{figure}

Let us show how to apply expressions (\ref{p_e},\ref{psi_e}) for
computation procedure construction. It is evident that in practical
calculations we always deal with only a finite number of data known
also with finite accuracy. In our case it means that the principal
function --- $\psi (x_1,\dots,x_D;t)$ --- will be known on a finite
set of points both on temporal and spacial coordinates. Accordingly
the autocorrelation function (\ref{autocor}) can be calculated also
only on a finite set of points $t_k=k\Delta t,\ k=1,\ldots,M$, and
therefore we have to apply the finite analogues of the Fourier
transforms (\ref{p_e}) and (\ref{psi_e})
\begin{equation}\label{p_e_f}P_T(E)=\frac1T\int\limits_{0}^{T}dt
e^{i\frac{Et}{\hbar}}P(t)=\hbar\sum\limits_{n=1}^{\infty}\left|a_n\right|^2
\delta_T \left(E-E_n\right)\end{equation}
\begin{equation}\label{psi_e_f}\begin{array}{c}\psi_T (x_1,\dots,x_D;E)=\frac1T\int\limits_{0}^{T}dt
e^{i\frac{Et}{\hbar}}\psi
(x_1,\dots,x_D;t)\\
=\hbar\sum\limits_{n=1}^{\infty}a_n \psi_n (x_1,\dots,x_D) \delta_T
\left(E-E_n\right),\end{array}\end{equation} where $T=M\Delta t$ and
the finite analogue of $\delta$-function takes the form
\[\begin{array}{c}
\delta_T(E)=\frac1T\int\limits_0^T dt e^{i\frac{Et}{\hbar}}=\frac{e^{i\frac{ET}{\hbar}}-1}{i\frac{ET}{\hbar}}=f_T\left(i\frac{ET}{h}\right)\\
f_T(x)=\frac{\sin\pi x}{\pi x}e^{i\pi x}.
\end{array}\]
As distinct from the usual $\delta$-function, $\delta_T(0)=1$ (see
Fig.\ref{d_t}a).
Therefore each energy level $E_n$ corresponds to a sufficiently
sharp peak --- local maximum of the function $|P_T(E)|$, situated at
the point $E=E_n$ (Fig. \ref{d_t}b). The typical profile of the
function $|P_T(E)|$ has many local maxima (Fig.\ref{p_e_typ}), but a
major part of them has absolutely nothing to do with the energy
levels of the system under consideration --- they appear due to the
oscillating character of $\delta_T(E)$. So the full number of maxima
on Fig.\ref{d_t}b equals $94$, while the corresponding system has
only $9$ physical energy levels in reality. Therefore, formally
analyzing $|P_T(E)|$, we will obtain plenty of extra "parasite
levels". The unique characteristic that allows us to distinguish
such phantom levels from the real ones is the relative smallness of
$|P_T(E)|$ in corresponding local maxima. However if we were just to
ignore all the levels that have $|P_T(E)|$ peak amplitude less than
a certain fixed threshold value, we risk loosing some real physical
levels, which correspond to small values of $|a_n|$. In this
connection it is very useful to apply the weighted Fourier
transform:
\begin{equation}\label{wft}\begin{array}{c}P_w(E)=\frac1T\int\limits_0^T dt
e^{i\frac{Et}{\hbar}}P(t)w(t) =
\sum\limits_{n=1}^{\infty}\left|a_n\right|^2\delta_w\left(E-E_n\right)\\
\psi_w(x_1,\ldots,x_D;E)=\frac1T\int\limits_0^T dt
e^{i\frac{Et}{\hbar}}\psi(x_1,\ldots,x_D;t)w(t)=\\
=\sum\limits_{n=1}^{\infty}a_n\psi_n(x_1,\ldots,x_D)\delta_w\left(E-E_n\right)\\
\delta_w(E)=\frac1T\int\limits_0^T dt e^{i\frac{Et}{\hbar}}w(t)
\end{array}\end{equation}
where the weight function $w(t)$ satisfies the conditions
\begin{equation}\label{w_cond}\begin{array}{c}w(0)=w(T)\\\int\limits_0^T w(t) dt=1.\end{array}\end{equation}
Therefore $\delta_w(0)=1$ for any such $w(t)$. The simplest
function, satisfying the conditions (\ref{w_cond}), is Hann function
\[w_H(t)=1-\cos\frac{2\pi t}{T},\]
for which
\[\begin{array}{c}\delta_H (E)=\frac{e^{i\frac{Et}{\hbar}}-1}{i\frac{Et}{\hbar}}
-
\frac12\left[\frac{e^{i\left(\frac{Et}{\hbar}+2\pi\right)}-1}{i\frac{Et}{\hbar}+2\pi}
+
\frac{e^{i\left(\frac{Et}{\hbar}-2\pi\right)}-1}{i\frac{Et}{\hbar}-2\pi}\right]=f_H\left(\frac{Et}{h}\right)\\
f_H(x)=\frac{\sin(\pi x)e^{i\pi x}}{\pi
x(1-x^2)}=\frac{f_T(x)}{1-x^2}.\end{array}\]

Inclusion of the weight function in the modified Fourier transform
allows us to diminish the relative amplitude of the phantom peaks in
$|P_H(E)|$ (Fig.\ref{d_t}a) and even slightly decrease their number
--- to $76$ for $|P_H(E)|$ instead of $86$ for $|P_T(E)|$
(Fig.\ref{d_t}b). Numerical analysis of the $|f_T(x)|$ shape shows
that the second maximum is situated in points
$x_{Tmax}\approx\pm1.43$ and has amplitude $f_{Tmax}\approx0.217$,
while for the $|f_H(x)|$ the analogous analysis maximum is situated
almost twice farther from the principal one ---
$x_{Hmax}\approx\pm2.36$, and its amplitude appears less for whole
order of magnitude --- $f_{Tmax}\approx0.0267$. Amplitude of more
distant maxima decrease even faster (Fig.\ref{d_t}a).

\begin{figure}
\includegraphics[width=\textwidth,draft=false]{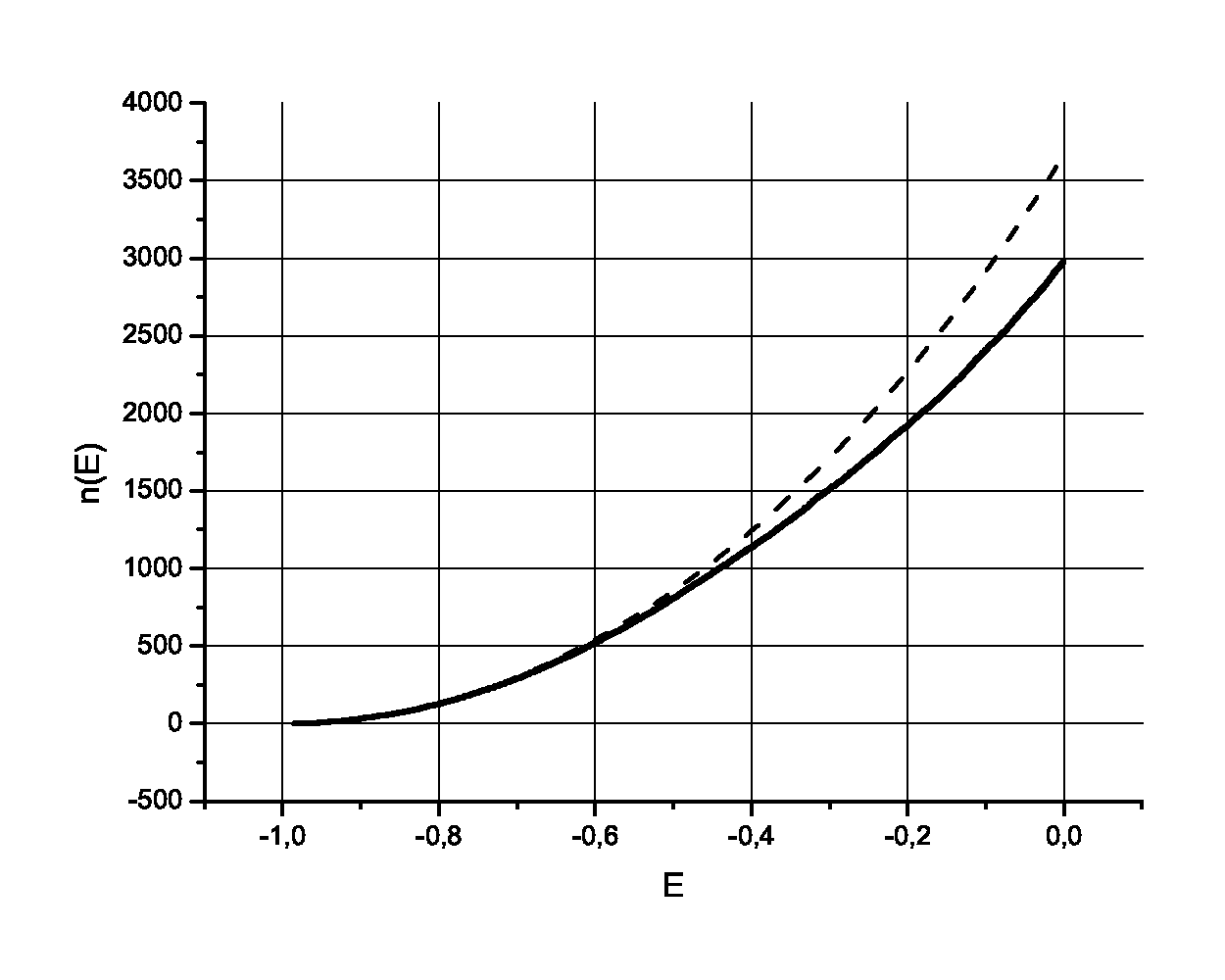}
\caption{State number $n(E)$ (dashed line) in Thomas-Fermi
approximation versus the numerically obtained $n(E)$ (solid line)
for lower umbilic catastrophe $D_5$ potential. \label{n_e}}
\end{figure}

Besides that, an important independent criterion to estimate the
accuracy of numerical results can be realized using the
semiclassical approximation for the stair-case states number
function $n(E)$: comparison of numerical $n(E)$ with the
semiclassical one allows us to roughly estimate the number of lost
or acquired extra levels (fig.\ref{n_e}).

Analyzing the positions and amplitudes of the $\left|P_H(E)\right|$
local maxima one can determine the energy levels $E_n$ and absolute
magnitudes of the coefficients $a_n$
\[\left|a_n\right|=\sqrt{\left|P_H(E_n)\right|}.\]
If in the expression
\[\psi_H(x_1,\ldots,x_D;E_n)=a_n \psi_n(x_1,\ldots,x_D) + \sum\limits_{k\ne n}a_k \psi_k(x_1,\ldots,x_D)\delta_H(E_n-E_k)\]
we now neglect the terms containing $\delta_H(E_n-E_k)$ (compare
with \ref{wft}), we can determine the eigenfunctions up to a phase
factor:
\begin{equation}\label{psi_n}\psi_n(x_1,\ldots,x_D)\approx\frac{1}{\left|a_n\right|}\psi_H(x_1,\ldots,x_D;E_n)\end{equation}
Here and forth we consider that there are no degenerate levels among
$E_n$. In practice in any system with a degenerate spectrum we can
(and should) remove the degeneracy by a certain choice of the
initial state $\psi_0(x_1,\ldots,x_D)$.

However such an approach is applicable only while the given energy
level $E_n$ is sufficiently separated from the neighbors. Indeed, if
one or more levels are situated at too short a distance, the
position of corresponding maxima $\left|P_H(E_n)\right|$
considerably differs from the actual values $E_n$. Moreover for
sufficiently close levels the $\left|P_H(E_n)\right|$ profile will
have only a single common peak (Fig.\ref{d_t}b).

In general, it looks impossible to determine exactly the minimum
separation between close levels at which they still can be resolved,
but we can state with confidence that this value is of the order of
$\Delta E_{min}=h/T$ --- the natural width of level. Therefore all
duplets and multiplets with separation less than $\Delta E_{min}$
will give only single local maxima, and some real levels will be
lost in the computed spectrum (see Table \ref{table}).
\begin{table}
\begin{tabular}{lrrrrrr}
& $E_1$ & $|a_1|$ & $E_2$ & $|a_2|$ & $E_3$ & $|a_3|$ \\
Exact                      & -30.00 & 0.50 & -10.00 & 0.50 & -7.00 & 0.50 \\
From $\left|P_T(E)\right|$ & -30.03 & 0.50 & -10.09 & 0.50 & -6.91 & 0.50 \\
From $\left|P_H(E)\right|$ & -30.00 & 0.50 &  -9.97 & 0.50 & -7.03 &
0.50
\\\\
& $E_4$ & $|a_4|$ & $E_5$ & $|a_5|$ & $E_6$ & $|a_6|$ \\
Exact                      &  10.00 & 0.25 &  11.00 & 0.25 & 20.00 & 0.20 \\
From $\left|P_T(E)\right|$ &   9.89 & 0.25 &  11.24 & 0.27 & 20.22 & 0.19 \\
From $\left|P_H(E)\right|$ &   9.53 & 0.21 &  11.47 & 0.21 & 20.35 &
0.18
\\\\
& $E_7$ & $|a_7|$ & $E_8$ & $|a_8|$ & $E_9$ & $|a_9|$ \\
Exact                      &  20.70 & 0.20 &  40.00 & 0.15 & 40.63 & 0.15 \\
From $\left|P_T(E)\right|$ &  21.10 & 0.19 &  40.32 & 0.16 &  -    &  -   \\
From $\left|P_H(E)\right|$ &   -    & -    &  40.31 & 0.15 & - & -
\end{tabular}
\caption{Energy spectrum $E_n$ and eigenstates amplitudes $a_n$ for
the model system from Fig. \ref{d_t}b\label{table}}
\end{table}

From Table \ref{table} we can see that analysis of
$\left|P_H(E)\right|$ gives considerably more accurate results for
isolated levels. On the other hand, $\left|P_T(E)\right|$ has a
smaller natural level width and gives more adequate results for
close levels. However the preference of $\left|P_T(E)\right|$ in
resolution of close levels is so insignificant that
$\left|P_H(E)\right|$ appears more preferable in the majority of
cases.

Taking into account that $P(t)$ as well as $\psi(x_1,\ldots,x_D;t)$
is calculated only on finite set of points $t_k=k\Delta t,\
k=0,\dots,M$, the most natural way to obtain $P_H(E)$ is the
discrete Fourier transform
\begin{equation}\label{dft}P_H(E_k)=\frac1M\sum\limits_{j=0}^{M}P(t_j)w(t_j)e^{2\pi
i\frac{jk}{M}} = \frac{\Delta
t}{T}\sum\limits_{j=0}^{M}P(t_j)w(t_j)e^{i\frac{t_j
E_k}{\hbar}},\end{equation} where $E_k=\frac h T k,\ k
=-M/2,\ldots,M/2$. Therefore $P_H(E)$ appears to be calculated in
energy interval $-\frac{h}{2\Delta t}<E<\frac{h}{2\Delta t}$ with
step $\Delta E=h/T$ exactly equal to natural level width.

As $w(T)=w(0)=0$, the discrete Fourier transform (\ref{dft}) in fact
coincides with the formula for numerical integration by the
trapezium rule\cite{antia}, applied to (\ref{wft}), and it is easy
to estimate its error
\[R_H(E)=\frac{T{\Delta t}^2}{12}\left|\frac{d^2}{{dt}^2}\left.\left(P(t)w(t)e^{i\frac{Et}{\hbar}}\right)\right|_{t=\tau}\right|\]
where $0<\tau<T$. But
\[\begin{array}{c}
\left|\frac{d^2}{{dt}^2}\left(P(t)w(t)e^{i\frac{Et}{\hbar}}\right)\right|=\\
\left|\frac{d^2}{{dt}^2}\left[\sum\limits_{n=1}^{\infty}|a_n|^2\left(e^{2\pi
i \frac t h(E-E_n)} - \frac12 e^{2\pi i \frac t h(E-E_n+\frac h T)}
- \frac12 e^{2\pi i \frac t h(E-E_n-\frac h
T)}\right)\right]\right|\le\\
\le\sum\limits_{n=1}^{\infty}\frac{|a_n|^2}{\hbar^2} \left|(E-E_n)^2
- \frac12 \left(E-E_n+\frac h T\right)^2 e^{2\pi i \frac t T} -
\frac12 \left(E-E_n-\frac h
T\right)^2 e^{-2\pi i \frac t T}\right|\le\\
\le\frac{1}{\hbar^2}\sum\limits_{n=1}^{\infty}|a_n|^2
\left[(E-E_n)^2 + \frac12 \left(E-E_n+\frac h T\right)^2 + \frac12
\left(E-E_n-\frac h T\right)^2 \right] =\\
=\sum\limits_{n=1}^{\infty}|a_n|^2 \left[\frac{(E-E_n)^2}{\hbar^2} +
\left(\frac{2\pi}{T}\right)^2\right]=\frac{(E-\bar{E})^2}{\hbar^2} +
2\left(\frac{D_E}{\hbar}\right)^2 + \left(\frac{2\pi}{T}\right)^2,
\end{array}\]
where $\bar{E}$ and $D_E$ are respectively mean energy and
dispersion in the initial state $\psi_0(x_1,\ldots,x_D)$:
\[\begin{array}{c}
\bar{E}=\int {dx}_1\ldots {dx}_D
\bar{\psi}_0(x_1,\ldots,x_D)\hat{H}\psi_0(x_1,\ldots,x_D)=\sum\limits_{n=1}^{\infty}|a_n|^2
E_n\\
D_E^2=\int {dx}_1\ldots {dx}_D
\bar{\psi}_0(x_1,\ldots,x_D)(\hat{H}-\bar{E})^2\psi_0(x_1,\ldots,x_D)=\sum\limits_{n=1}^{\infty}|a_n|^2
(E_n^2-\bar{E}^2)\\
\hat{H}=-\frac{\hbar^2}{2}\sum\limits_{i=1}^D \partial_i^2 +
U(x_1,\ldots,x_D),
\end{array}\]
and we finally get
\[R_H(E)\le\frac{T\Delta t^2}{6\hbar^2}\left[(E-\bar{E})^2+D_E^2\right]+\frac{\pi^2\Delta t^2}{3T}.\]

Therefore formally integration in (\ref{wft}) allows us to calculate
$P_H(E)$ for any energy values, but in fact applicability of such an
approach is limited by the energy region
\[|E-\bar{E}|\ll\frac{\hbar^2}{\Delta t^2},\]
which is definitely not better than for (\ref{dft}). Application of
the approximated integration formulae of higher orders also does not
give any improvement to (\ref{dft}), because the error estimate for
numerical integration of (\ref{wft}) by $n$-th order method reads
\[R^{(n)}_H(E)\approx\frac{T\Delta t^n}{n!\hbar^n}(E-\bar{E})^n\]

For calculation of $\psi(x_1,\ldots,x_D;t)$ at $t_k=k\Delta t,\
k=1,\ldots,M$ we can apply step-by-step the split operator method
\[\psi(x_1,\ldots,x_D;t+\Delta t) = e^{i\frac{\hbar^2\Delta
t}{4}\sum\limits_{i=1}^{D}\partial_i^2} e^{-i\Delta t
U(x_1,\ldots,x_D;t)} e^{i\frac{\hbar^2\Delta
t}{4}\sum\limits_{i=1}^{D}\partial_i^2} \psi(x_1,\ldots,x_D;t)
+O(\Delta t^3).\]

Action of the differential operator $e^{i\frac{\hbar^2\Delta
t}{4}\sum\limits_{i=1}^{D}\partial_i^2}$ is also calculated with the
help of the discrete Fourier transform
\[\begin{array}{c}
e^{i\frac{\hbar^2\Delta t}{4}\sum\limits_{i=1}^{D}\partial_i^2}
\psi(x_1,\ldots,x_D;t) = \\ = \sum\limits_{n_1=-N_1/2}^{N_1/2}
\sum\limits_{n_2=-N_2/2}^{N_2/2} \ldots
\sum\limits_{n_D=-N_D/2}^{N_D/2} \psi_{n_1 n_2 \ldots
n_D}(t)e^{\sum\limits_{k=1}^{D}\left(2\pi i\frac{n_k x_k}{N_k \Delta
x_k} - i \Delta t \left(\frac{hn_k}{2N_k \Delta
x_k}\right)\right)}\\
\psi_{n_1 n_2 \ldots n_D}(t) = \frac{1}{\prod\limits_{k=1}^D
N_k}\sum\limits_{n_1=-N_1/2}^{N_1/2}
\sum\limits_{n_2=-N_2/2}^{N_2/2} \ldots
\sum\limits_{n_D=-N_D/2}^{N_D/2}\psi(x_1,\ldots,x_D;t) e^{-2\pi
i\sum\limits_{k=1}^D\frac{n_k x_k}{N_k \Delta x_k}}.
\end{array}\]

With sufficiently large numbers of time steps $M$ and coordinate
grid nodes $N$ the results of computations by the spectral method
really do not depend on the arbitrary initial wave function
$\psi_0(x_1,\ldots,x_D)$, but it is a reasonable choice of such a
function that is the most powerful means for the optimization of
computations by the spectral method in order to achieve sufficient
accuracy of the results with minimal expenses of computational
power.

It is convenient to choose the initial wave function
$\psi_0(x_1,\ldots,x_D)$ as a linear combination of Gaussian wave
packets of minimal uncertainty
\begin{equation}\label{gwp0}
\psi_G(x_1,\ldots,x_D)=e^{-\sum\limits_{i=1}^D \frac{(x_i-
x_{i0})^2}{2\sigma_i^2}+i\frac{p_i(x_i-x_{i0})}{\hbar}}.
\end{equation}
Combining functions of the form (\ref{gwp0}) with different
parameters $\sigma, p_i$ and $x_{i0}$ it is possible to selectively
excite eigenstates $\psi_n(x_1,\ldots,x_D)$ with the desired
symmetry properties and lying in the desired energy interval. Its
position
\[\langle E\rangle=\langle\psi_0(x_1,\ldots,x_D)|\bar{H}|\psi_0(x_1,\ldots,x_D)\rangle\]
and approximate width
\[\Delta E^2=\langle\psi_0(x_1,\ldots,x_D)|((\bar{H}-\langle E\rangle)^2|\psi_0(x_1,\ldots,x_D)\rangle\]
are determined by parameters of the Hamiltonian and Gaussian wave
packets that generate the initial state
\[\psi_0(x_1,\ldots,x_D)=\sum\limits_n a_n \psi_n(x_1,\ldots,x_D),\]
and can be varied in a wide range if desired. This in particular
allows us to calculate low-energy and high-energy states separately,
which leads to more efficient distribution of the computational
efforts.

In the spectral method computations the preponderant fraction of CPU
time is invested in the calculation of $\psi(x_1,\ldots,x_D;t)$, and
namely to multiple direct and inverse discrete Fourier transforms.
For two-dimensional problems the characteristic calculation
time-scales are $MN^2\ln N$ which is substantially better than $N^6$
for the matrix diagonalization method. Another important advantage
of the spectral method is the possibility to calculate multiple
eigenfunctions in parallel computations by (\ref{psi_n}), which
allows us significantly to economize the CPU time. For example, the
computation time for $100$ eigenfunctions is only twice longer that
for one eigenfunction.

\begin{figure}
\includegraphics[width=\textwidth,draft=false]{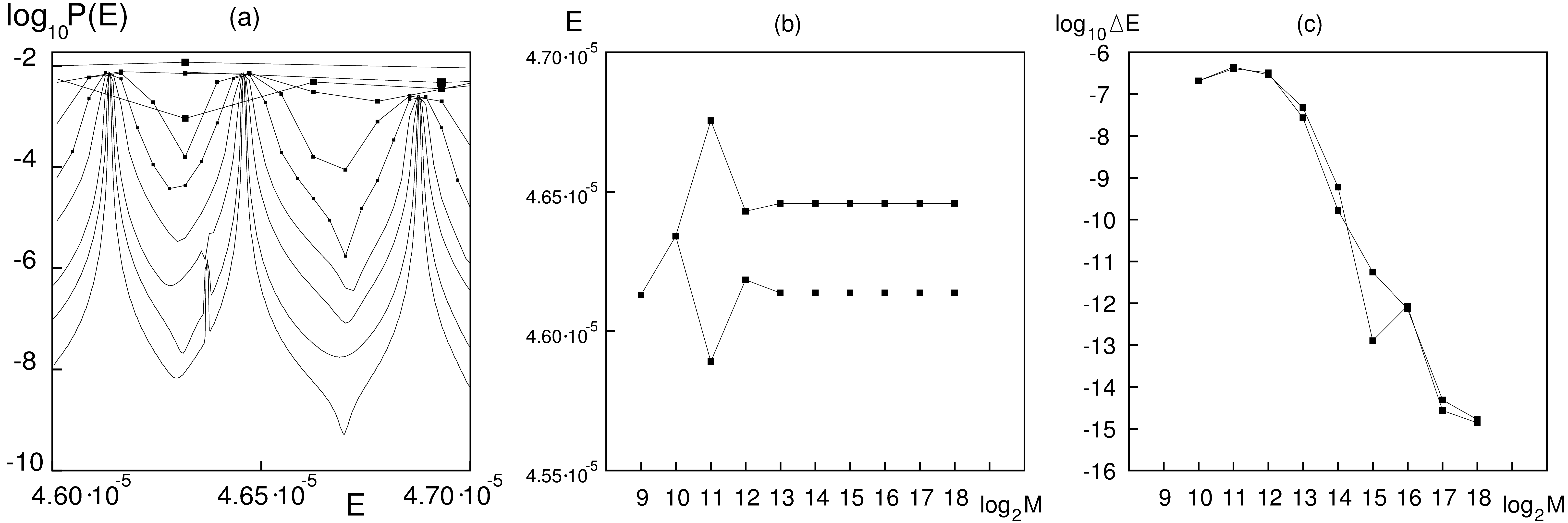}
\caption{Determination of two close energy levels $E_n$ and
$E_{n+1}$ in the quadrupole oscillations potential with $W=18$ and
different time step numbers $M=2^k,\ k=9,10,\ldots,18$: a)
$|P^{(k)}(E)|$ in logarithmic scale; b) values of $E_n$ and
$E_{n+1}$; c) consecutive corrections $E_n^{(k-1)}-E_n^{(k)}$ and
$E_{n+1}^{(k-1)}-E_{n+1}^{(k)}$. \label{levels}}
\end{figure}

The unlimited possibilities of the spectral method in computational
accuracy improvement are demonstrated in Fig.\ref{levels} for a
problem of determination of two close energy levels in the
quadrupole potential. With time step number $M$ increasing from
$2^9$ to $2^{18}$, peaks of $P(E)$ became more and more pronounced
(Fig.\ref{levels}a). For $M=2^9$ and $M=2^{10}$ energy resolution is
yet too small, and two neighboring levels look as one. At $M=2^{11}$
the duplet is already resolved but obtained values for energy levels
differ significantly from the real ones. Sufficient accuracy of
energy level determination is achieved at $M=2^{13}$ and further
increasing of $M$ does not lead to remarkable changes in calculated
energy levels values (Fig.\ref{levels}b). The accuracy of obtained
results grows very fast with increasing of $M$ and can be made
arbitrarily high (Fig.\ref{levels}c).
\sat\section{Comparative analysis of matrix diagonalization and
spectral methods}\sat The method of Hamiltonian diagonalization is
the most traditional way for numerical solution of the Schr\"odinger
equation. The spectral method for solution of the same problem
represents in its turn a newer one and for many reasons a more
preferable approach.

One of the most fundamental disadvantages of the matrix
diagonalization technique is the rather poor choice of exactly
solvable models whose eigenfunctions can be taken as a basis for
subsequent diagonalization of the Hamiltonian under consideration.
As a rule, properties of the Hamiltonian pose very rigid limitations
on the auxiliary basis parameters, therefore in most cases of matrix
diagonalization implementation, only one free parameter remains ---
it is the auxiliary basis dimension $N$. Further it is necessary
only to determine the minimal dimensionality $N$ sufficient for the
achievement of desired resulting accuracy. Such simplicity of the
matrix diagonalization method results in its insufficient
flexibility: in practice the application of matrix diagonalization
is justified only for those potentials that can be approximated at
least locally by some exactly solvable model. But such a limitation
cannot be satisfied for many important problems, especially in the
potentials with many local minima.

The spectral method uses a natural basis of free particle wave
functions --- such a basis is equally good, or better to say equally
bad, for potentials of any form. Such fundamental indifference of
the spectral method to shapes of potential energy surface is the
main reason of its universality. Compared to matrix diagonalization,
the spectral method has much more flexibility --- the researcher is
free to choose both the length and step of the computation grid in
time ($T$ and $\Delta t$) as well as in coordinate space ($L_i$ and
$\Delta x_i$). In the same time choice of the nodes number $N$ is
limited by the computational efficiency requirement: the
applicability condition for the fast Fourier transform algorithm
--- the main basis of the spectral method efficiency --- assumes that
all $N_i$ do not contain large simple factors; ideally all of them
should be integer powers of two $(N_i=2^{k_i})$. And, last but not
least, the main freedom lies in the choice of the initial state for
the spectral method computations. As it is very difficult to give
any general recommendation on that point, the spectral method
computations have become a real art rather than plain technique,
requiring great experience and constant practice. Because the
spectral method is not standardized up to the present time, the fast
Fourier transform represents the only one ready-to-use ingredient
for its realization, available in many well-known software
libraries. Other stages of computations require rather extended
although principally simple software development. On the other hand,
the spectral method algorithm itself can be easily generalized for
problems of any dimensions, which is not the case with the matrix
diagonalization technique --- reasonable construction of finite
multi-dimensional basis from one-dimensional eigenstates always
represents a non-trivial task because of the basis vectors ordering
problem.

A quantitative measure of the numerical method efficiency is the
growth of computational expense --- CPU time and RAM usage --- with
increase in the results both quantity and quality. It is useful to
compare the matrix diagonalization and the spectral method
efficiencies in calculation of $n$ energy levels of a quantum system
with fixed relative error $\varepsilon$. Taking into account the
fact that for studies of statistical properties the calculated
spectrum is inevitably unfolded, it is reasonable to define the
accuracy as the maximum ratio of absolute error of the computed
energy levels to mean level spacing
\[\varepsilon=\frac{\delta_E}{\Delta_E}=\rho(E)\delta_E.\]

For most potential systems the level density $\rho(E)$ grows quite
fast with the energy:
\[\rho(E)\approx E^{\lambda E-1},\]
where $D$ is the system dimensionality and $\lambda$ is close to
unity (it exactly equals unity for a harmonic oscillator). Therefore
the condition to achieve the desired accuracy will be the most
critical for levels with maximum energy, while the lowest levels
will be obtained with higher accuracy than needed --- this
inconvenient feature is due to the very nature of smooth potential
systems and is equally shared by both methods under consideration.

In the matrix diagonalization method the absolute computational
error for sufficiently low energies does not exceed the round-off
errors:
\[\delta_{E_k}\approx\delta_0 E_k N^2,\ k<\eta N, 0<\eta<1\]
where $N$ is the basis dimensionality, $\eta$ represents the
problem-dependent rate of correctly calculable states and $\delta_0$
is the machine round-off error ($\delta_0\sim10^{-15}$ for standard
double accuracy numbers). For such levels the relative error
$\varepsilon\approx E^{\lambda D}\delta_0 N^2$ appears to be
negligibly small and for required basis dimensions we get the
condition:
\[N_{MD}=\frac n \eta.\]

In the general case of matrix diagonalization the computation time
scales as $T_{MD}\approx N^3=n^3/\eta^3$ and memory usage is
$M_{MD}\approx N^2=n^2/\eta^2$. However in many important particular
cases, for example for polynomial potentials, the matrix
diagonalization implementation involves band matrices, and in such
cases $T_{BMD}\approx N^2$ and $M_{BMD}\approx N$.

In the spectral method the computational accuracy is determined by
the size of time computational grid: $\delta_E\approx1/T$, while the
time step determines the spectral bandwidth where the levels can be
determined: $\Delta t\approx 1/E_{max}$. Therefore the maximum
relative error in the spectral method scales as $\varepsilon\approx
E^{\lambda D-1}/T$, and the required time step number $N_T=T/\Delta
t\approx n/\varepsilon$. Because the spectral method actually uses
the plane waves decomposition for the computed wave functions, the
required nodes number for the computational grid for each of the
dimensions, equal to the effective basis vector number on the same
degree of freedom, is determined in the same way as for the
corresponding basis in the matrix diagonalization method:
\[N_i\approx \left(\frac{n}{\eta_{PW}}\right)^{\frac1D},\]
where $\eta_{PW}$ is the rate of correctly calculable levels on the
plane waves basis. Therefore the computation time scales as
\[T_{SM}\approx N_T N_i^D\ln N_i\approx\frac{n^2}{\varepsilon\eta_{PW}}\ln\frac{n}{\eta_{PW}}\]
and memory usage scales as
\[M_{SM}\approx N_i^D\approx\frac{n}{\eta_{PW}}\]
As a result the spectral method is generally more efficient for both
CPU time and RAM usage criteria:
\[\frac{M_{SM}}{M_{MD}}\approx\frac{\eta^2}{n\eta_{PW}},\ \frac{T_{SM}}{T_{MD}}
\approx\frac{\eta^3}{n\varepsilon\eta_{PW}}\ln\frac{\eta}{\eta_{PW}}.\]

The last but not least advantage of the spectral method lies in the
fact that numerical simulation of wave packets temporal dynamics is
included into the method algorithm as an auxiliary procedure. In
some important applications this simulation represents the ultimate
goal of research, while the determination of energy levels and
stationary wave functions is not required. In such cases an
important advantage of the spectral method is the possibility of
achieving the goal by the shortest way, while it would require much
more computational efforts to reproduce the same results by the
matrix diagonalization technique.
\sat\chapter{Statistical Properties of Energy Spectra}\sat The
energy spectra represent historically the first object for
investigation of quantum signatures of chaos in Hamiltonian systems.
Substantial progress in the detection of quantum manifestations of
classical stochasticity in the 1980s was connected with the
transition to studies of statistical properties of energy spectra.
Up to that time there had been established the connection between
spectral properties of complex systems (for example, atomic nuclei)
and ensembles of random matrices, which created the background for
the understanding of the complicated pseudo-random nature of energy
spectra. This in turn stimulated the transition from investigation
of the behavior of separate levels to statistical characteristics of
the energy spectrum as a whole. In the research process it became
clear that only local (not averaged) statistical properties of
energy spectra were of interest from the point of view of
investigation of quantum manifestations of classical stochasticity.
Why should we address local characteristics of a spectrum? The fact
is that the global characteristics such as the numbers of states
$N(E)$ or the smoothed density of levels $\rho(E)$ are too crude. At
the same time, such local characteristics as the function of the
nearest-neighbour spacing distribution between levels (FNNS) are
very sensitive to the properties of potential or to the shape of a
billiard boundary. \sat\section{The standard semiclassical spectrum
unfolding method}\sat Already by definition statistical properties
imply analysis of sufficiently large numbers of states, which
necessarily lead us to the use of the semiclassical approach, which
is well known to have as long and rich a history as quantum
mechanics itself.

The most fundamental characteristics of quantum spectra are the
level density
\begin{equation}\label{rhoe}\rho(E)=\sum\limits_n\delta(E-E_n)\end{equation} and
staircase state number function
\begin{equation}\label{ne}n(E)=\int\limits_{-\infty}^E dE\rho(E)=\sum\limits_n\Theta(E-E_n),\end{equation}
where $E_n$ are stationary energy levels.

In the semiclassical approximation it is common to consider
separately so called smooth and oscillating components of the
functions (\ref{rhoe}) and (\ref{ne}):
\[\rho(E)=\bar{\rho}(E)+\tilde{\rho}(E)\]
\[n(E)=\bar{n}(E)+\tilde{n}(E).\]

The smooth component $\bar{\rho}(E)$ describes a gradual variation
of the averaged levels density with energy growth, while the
oscillating component describes the deviation of spectrum from the
"quasi-equidistant" one --- the hypothetical spectrum with levels
spacing equal to $\Delta(E)=1/\bar{\rho}(E)$. At the present time
there are very highly developed methods to obtain the semiclassical
approximations (usually in the form of decomposition in powers of
$\hbar$) for $\bar{\rho}(E)$. Among them are the widely known
Thomas-Fermi formulae, which are nothing but zero approximation for
corresponding values:
\begin{equation}\label{tf}\begin{array}{c}
\rho_{TF}(E)=\frac{1}{(2\pi\hbar)^D}\int dp dq \delta(E-H(p,q))\\
n_{TF}(E)=\frac{1}{(2\pi\hbar)^D}\int dp dq \Theta(E-H(p,q))
\end{array}\end{equation}
where $D$ is the dimensionality of the configuration space. As a
rule in studies of quantum chaos we deal with Hamiltonians of the
following type:
\[H(p,q)=\frac{p^2}{2}+U(q).\]
After integration on the momenta in (\ref{tf}), we obtain for the
case $D=2$ (further we will limit ourselves to the two-dimensional
systems only):
\begin{equation}\label{tf2}\begin{array}{c}
\rho_{TF}(E)=\frac{1}{2\pi\hbar^2}\int dx dy \delta(E-U(x,y))\\
n_{TF}(E)=\frac{1}{2\pi\hbar^2}\int dx dy (E-U(x,y))\Theta(E-U(x,y))
\end{array}\end{equation}
We can see that the Thomas-Fermi level density in two-dimensional
dynamical systems is a non-decreasing function --- constant for
billiards and increasing for smooth potential systems.

Quite often the Hamiltonian of the system is invariant to some
discrete symmetry transformation, for example it has a parity of
some kind\footnote{The presence of a continuous symmetry group in a
two-dimensional dynamical system is of no interest, because it leads
to the existence of an integral of motion additional to energy and
therefore integrability of the problem excludes the possibility of
dynamical chaos.}. Certain symmetry of the Hamiltonian allows us to
reduce integration in (\ref{tf2}) over the whole $(x,y)$ plane to
integration over some small region, which may significantly simplify
the task. It is even more important that, provided a certain
symmetry of the Hamiltonian, the quantum states of the system divide
on subsets
--- irreducible representations --- according to their symmetry
properties, for example even or odd states. States belonging to
different irreducible representations are statistically independent.
Therefore in order to research the statistical properties of such
systems it is necessary to consider the states of different symmetry
types separately. For that purpose it is necessary to know the
semiclassical expressions for "partial" densities $\rho^\lambda(E)$
and the number $n^\lambda(E)$ of states, were $\lambda$ labels the
symmetry type of the states. On that point there is an exhaustive
general semiclassical theory \cite{whelan}, but in many concrete
problems it is often easy to rediscover simple particular formulae.
For example, in the simplest case of discrete symmetry ---
invariance with respect to reflection $y\rightarrow -y$ --- it is
easy to obtain Thomas-Fermi expressions analogous to (\ref{tf}), for
odd states separately. Indeed, the wave functions odd on $y$
evidently satisfy the Schr\"odinger equation with potential, which
differs from the original one only by the presence of an infinitely
high potential barrier situated in the $y=0$ plane. Effectively such
impenetrable potential wall exactly halves the classically allowed
phase space, and we obtain:
\begin{equation}\label{tf3}\begin{array}{c}
\rho_{TF}(E)=\frac{1}{2\pi\hbar^2}\int\limits_{y>0} dx dy \delta(E-U(x,y))\\
n_{TF}(E)=\frac{1}{2\pi\hbar^2}\int\limits_{y>0} dx dy
(E-U(x,y))\Theta(E-U(x,y))\\
\rho_{TF}^{even}(E)=\rho_{TF}(E)-\rho_{TF}^{odd}(E),\
n_{TF}^{even}(E)=n_{TF}(E)-n_{TF}^{odd}(E).
\end{array}\end{equation}
For zero --- Thomas-Fermi --- approximation we get a trivial result:
\begin{equation*}\begin{array}{c}
\rho_{TF}^{even}(E)=\rho_{TF}^{odd}(E)=\rho_{TF}(E)/2\\
n_{TF}^{even}(E)=n_{TF}^{odd}(E)=n_{TF}(E)/2.
\end{array}\end{equation*}
However such a symmetry decomposition approach is applicable in any
order of $\hbar$-decomposition, where it already leads to
non-trivial corrections. In the same way, but with much more
difficulties, we can obtain the semiclassical approximations for
partial level densities and staircase numbers with rotational
symmetry --- in Hamiltonian, invariant under rotations on
$\varphi=2\pi/n$.

\begin{figure}
\includegraphics[width=0.5\textwidth,draft=false]{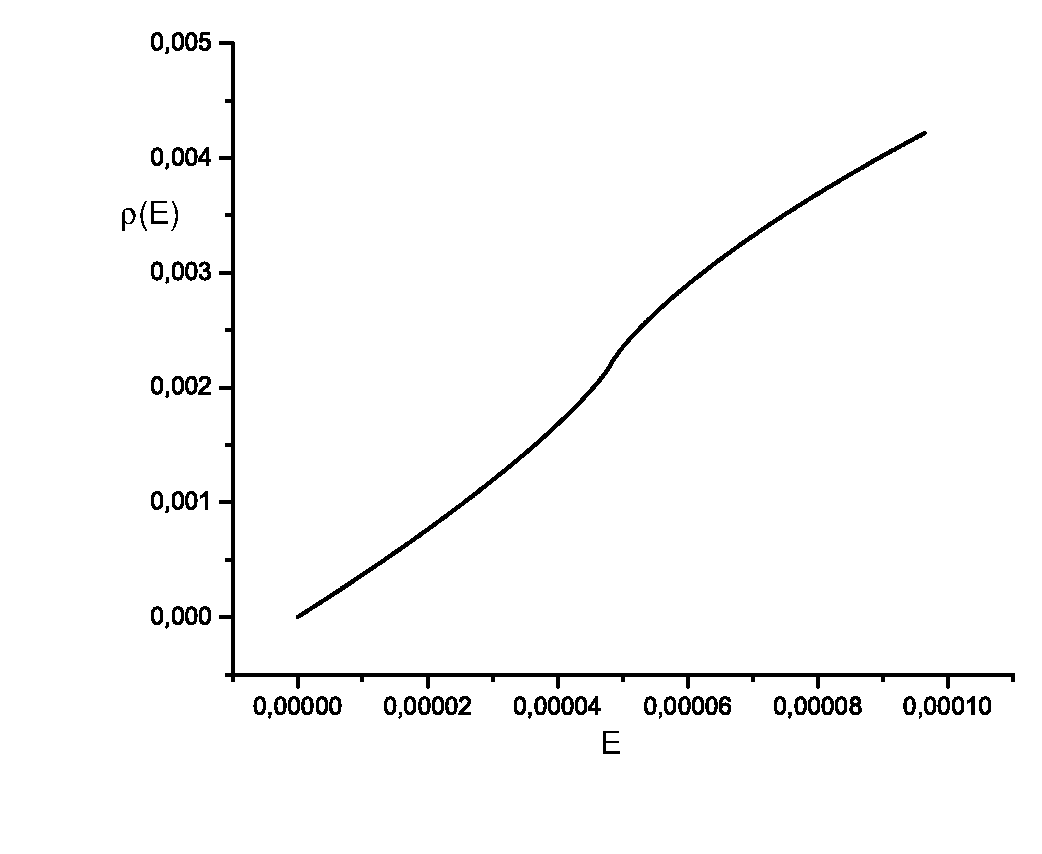}
\includegraphics[width=0.5\textwidth,draft=false]{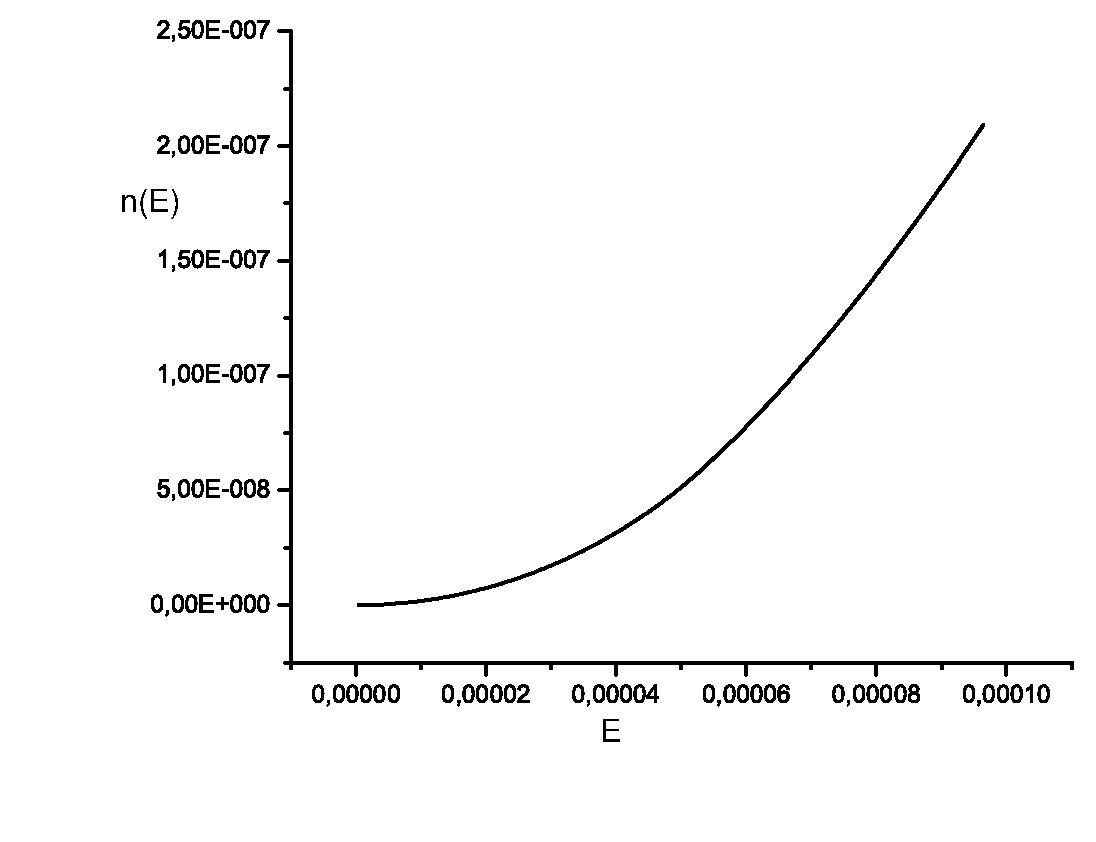}
\caption{\label{tf_qo} Density $\rho(E)$ (left) and staircase number
$n(E)$ (right) for $\hbar=1$ in Thomas-Fermi approximation in the
$QO$ potential (\ref{u_qo}).}
\end{figure}

Practically all two-dimensional models actively studied from the
point of view of quantum chaology, namely the above mentioned
potentials of quadrupole oscillation (\ref{u_qo}), H\'enon-Helis
(\ref{u_mu}), lower umbilic catastrophe $D_5$ (\ref{u_d5}), coupled
quartic oscillator (\ref{cqo}) and also Barbanis potential
\[U_B(x,y)=\frac{x^2+y^2}{2}+xy^2\]
are described by rather simple expressions --- polynomials of the
third or fourth order. As a rule it allows us to analytically solve
the equation $U(x,y)=E$ with respect to, for example, variable $y$,
and to obtain the explicit expression for the level line $y=f(x;E)$.
It gives us the possibility to reduce the Thomas-Fermi integrals
(\ref{tf3}), and similarly the higher order semiclassical
approximations, to one-dimensional integrals, which are evaluated if
not analytically, then numerically in a much faster way than the
original two-dimensional ones (Fig.\ref{tf_qo}).

One of the most important applications of the above discussed
semiclassical expressions is the so-called spectrum unfolding
procedure using $n(E)$:
\begin{equation}\label{unfold}\varepsilon_n=n(E_n).\end{equation}
As a result the original spectrum is transformed to
quasi-equidistant:
\[\rho(\varepsilon)=1.\]
Only for spectra transformed that way does it make sense to discuss
the quantum signatures of chaos in the statistical properties of
energy spectra.
\sat\section{Nearest neighbor spacing distribution function}\sat
In the 1960s of last century Wigner \cite{wigner}, Porter
\cite{porter} and Dyson \cite{dyson} built statistical theory of
complex systems spectra on the basis of the following hypothesis:
energy levels distribution in complex systems is equivalent to
distribution of the eigenvalues of random matrices ensemble with
certain symmetry. The ultimate result for FNNS obtained in that
theory has the following form:
\[P(S)\sim aS^\alpha e^{-bS^2},\]
where $a$ and $b$ are slowly varying functions of energy and
critical exponent $\alpha$; defining behavior of the distribution
function in the limit $S\rightarrow 0$, depends on symmetry
properties of the random matrices: $\alpha=1$ for orthogonal and
$\alpha=2$ for unitary ensemble of matrices.

Predictions of the statistical theory of energy spectra (mainly for
Gaussian orthogonal ensemble of random matrices) were carefully
compared with all available data on nuclear spectra. No considerable
contradictions between the theory and experiment were found. In
particular, the random matrices ensembles perfectly reproduced such
an important spectral characteristic as spectral rigidity,
describing small fluctuations of energy levels around averaged
values in a given interval. Measure of rigidity is the statistic
$\Delta_3$ of Dyson \cite{dyson} and Mehta \cite{mehta}
\begin{equation}\label{p_s}\Delta_3 (L;x)=\frac1L
\min\limits_{A,B}{\int\limits_x^{x+L}[n(\varepsilon)-A\varepsilon-B]^2d\varepsilon}\end{equation}
which determines the least-square deviation of the staircase
representing the cumulative density $n(\varepsilon)$ from the best
straight line fitting it in any interval $[x,x+L]$. The most
perfectly rigid spectrum is the picket fence with all spacing equal
(for instance, the one-dimensional harmonic oscillator spectrum),
therefore maximally correlated, for which $\Delta_3=1/12$, whereas,
at the opposite extreme the Poisson spectrum has a very large
average value of rigidity $(\Delta_3=L/15)$, reflecting strong
fluctuations around the mean level density.

Analogous comparisons were made also for atomic spectra. Good
agreement with theory was found for them too, although for much
poorer statistics.

A completely different approach to the problem of statistical
properties of energy spectra was developed on the basis of
non-linear theory of dynamical systems. As numerical simulations
show \cite{bohigas_prl,seligman,seligman_prl,bohigas_prl3},
confirmed by serious theoretical considerations
\cite{haake,gutzwiller,seligman,chirikov_pr}, the main universal
property of systems which have regular type of dynamics in the
classical limit, is the level clustering phenomenon, while for
systems that are chaotic in the classical limit level repulsion is
observed. This statement sometimes is called the hypothesis of
universal character of energy spectra fluctuations
\cite{bohigas_prl}.

In the case of regular motion any classical trajectory lies on a
surface topologically equivalent to a torus. Each such torus
corresponds to a certain set of integrals of motions (or quantum
numbers), with the total number equal to the number of degrees of
freedom. Different eigenfunctions of integrable quantum systems
correspond to different sets of quantum numbers, and therefore lie
on different tori. Their eigenenergies are not correlated, which
leads to Poissonian FNNS
\begin{equation}\label{poisson}P_P(S)=e^{-S}\end{equation}
responsible for level clustering.

In the transition to chaos some tori are destroyed, leading to
formation of chaotic regions in the phase space. Classical
trajectories in such regions diffuse between different (already
destroyed) tori. Therefore different chaotic modes can be
represented in the form of superposition of former regular modes,
which leads to repulsion between the corresponding levels. In the
limit of a completely chaotic system, all modes of initially
integrable system mix one with other, so that repulsion exists
between any pair of levels, changing FNNS from Poissonian to the
Wigner one:
\begin{equation}\label{wigner}P_W(S)=\frac\pi2Se^{-\frac\pi4S^2}.\end{equation}

The situation became more complicated for generic Hamiltonian
systems, where the phase space contains both regular and chaotic
components. How is it reflected in the energy spectra statistical
properties?

Historically the first was the phenomenological approach to the
solution of that problem, proposed by Brody \cite{brody}, who
considered FNNS of the form:
\begin{equation}\label{brody}P_B(S)=(\beta+1)bS^\beta e^{-bS^{\beta+1}},\
b=\left[\Gamma\left(1+\frac{1}{\beta+1}\right)\right]^{\beta+1}\end{equation}
where $\beta$ equals to the relative measure of classical phase
space, occupied by the chaotic trajectories: $\beta=0$ for regular
and $\beta=1$ for chaotic systems.

Berry and Robnik \cite{br} and independently Bogomolny
\cite{bogomolny} using on semiclassical arguments, showed that FNNS
in the case of mixed type dynamics represents the superposition of
Poisson and Wigner distributions with weights that are determined by
the relative phase space measure of regular and chaotic motion
respectively:
\begin{equation}\label{brb}\begin{array}{c}
P_{BRB}=\frac{d^2}{dS^2}\left[e^{-\rho
S}erfc\left(\frac{\sqrt\pi}{2}(1-\rho)S\right)\right]=\\
=\rho^2e^{-\rho S}erfc\left(\frac{\sqrt\pi}{2}(1-\rho)S\right) +
(1-\rho)\left(2\rho+\frac\pi2(1-\rho)^2 S\right)e^{-\rho S-\frac\pi4
(1-\rho)^2 S^2}
\end{array}\end{equation}
where $\rho$ is the relative phase volume occupied by regular
trajectories in the mixed system. The limit $\rho\rightarrow1$
corresponds to regular system, and $\rho\rightarrow0$ --- to
completely chaotic.

Further development theory of FNNS acquired in the works of
Narimanov and Podolskiy \cite{np}, who accounted for additional
effect of repulsion for close levels due to dynamical tunneling
(more details see Chapter \ref{ch_wp}):
\begin{equation}\label{np}\begin{array}{c}
P_{NP}=\rho^2 F\left(\frac{S}{\nu^2}\right) e^{-\rho
S}erfc\left(\frac{\sqrt\pi}{2}(1-\rho)S\right)+\\ +
(1-\rho)\left(2\rho F\left(\frac{S}{\nu^2}\right)
+\frac\pi2(1-\rho)^2 S\right)e^{-\rho S-\frac\pi4 (1-\rho)^2 S^2}\\
F(x)=1-\frac{1\sqrt{\frac\pi2}x}{e^x - x}.
\end{array}\end{equation}

\begin{figure}
\includegraphics[width=\textwidth,draft=false]{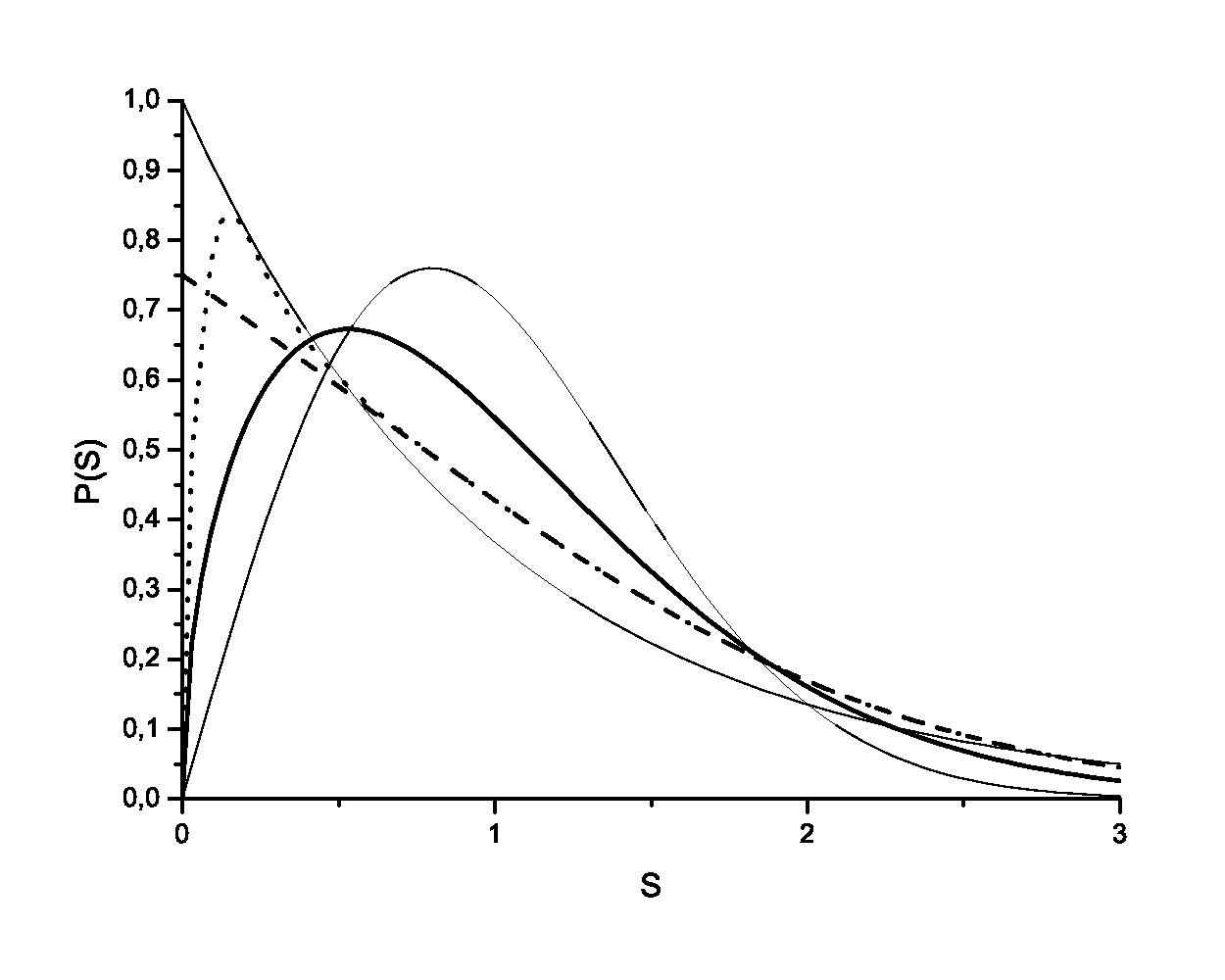}
\caption{\label{fnns}The distribution functions of Poisson and
Wigner (thin solid lines), Brody for $\beta=0.5$ (thick solid line),
Berry-Robnik-Bogomolny for $\rho=0.5$ (dashed line) and
Narimanov-Podolskiy for $\rho=0.5$ and $\nu=0.1$ (dotted line)}
\end{figure}

The $\rho$ parameter in the Narimanov-Podolskiy distribution
function (\ref{np}) has the same physical sense as in the
Berry-Robnik-Bogomolny distribution (\ref{brb}), and additional
parameter $\nu$ is connected to the intensity of dynamical
tunneling. Therefore the Berry-Robnik-Bogomolny distribution
function represents a particular case of the Narimanov-Podolskiy
distribution at $\nu=9$, which corresponds to the absence of the
tunneling (Fig.\ref{fnns}).

Among systems subject to extensive numerical analysis of the
spectral properties, two-dimensional billiards possess the central
place. The two-dimensional billiard represents a point particle
freely moving on the plane inside some region of arbitrary shape and
elastically reflecting from the boundary. Such systems attract
active interest in research for the following reasons:
\begin{enumerate}
\item the system has the lowest possible number of degrees of
freedom, allowing chaotic motion in a conservative system;
\item the simplicity of classical dynamics;
\item stochasticity criteria for billiards are formulated in
geometrical terms;
\item homogeneity of the phase space;
\item availability of efficient methods for solution of the
Schr\"odinger equation for billiards;
\item smooth component of level density is well known due to the Weyl
formula;
\item billiard dynamics reflect the real situation in many physical
systems (quantum dots and antidots, Josephson's junctions, nuclear
billiards)
\item at the present time there is efficient experimental
realization both for classical and quantum dynamics in microwave and
optical two-dimensional resonators.
\end{enumerate}

\begin{figure}
\includegraphics[width=\textwidth,draft=false]{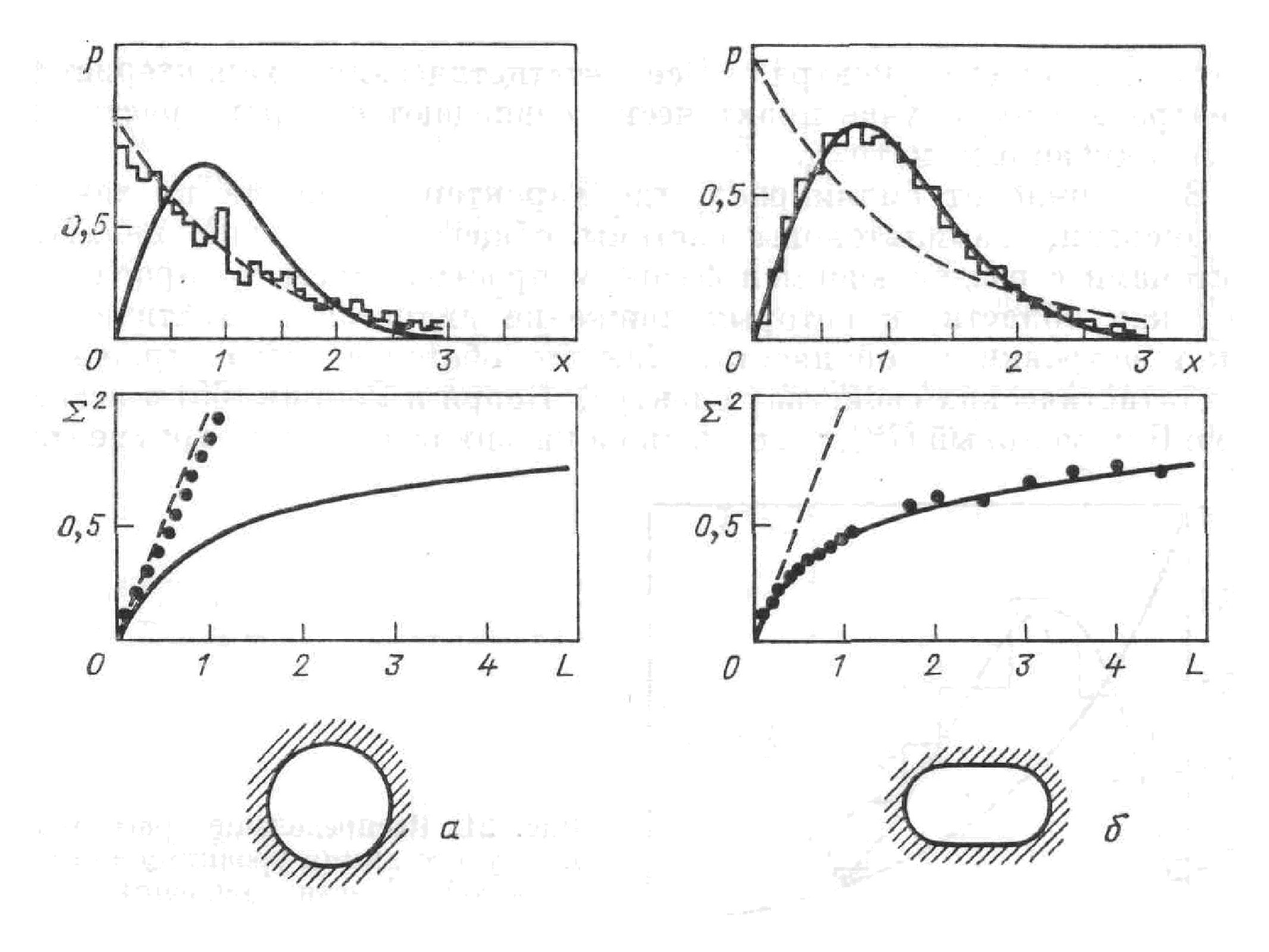}
\caption{\label{bill}Statistical characteristics of energy spectra
for circular (a) and "stadium" (b) billiards.}
\end{figure}

For billiards with certain boundary shape one of two limiting cases
is realized: exact integrability or absolute chaos. So in the
circular billiard (Fig.\ref{bill}a) angular momentum is the second
(after energy) integral of motion, and this system is integrable. A
billiard of "stadium" type (Fig.\ref{bill}b) is one of the simplest
stochastic systems. In Fig.\ref{bill} the statistical
characteristics of energy spectra (FNNS and dispersion) are
presented for both systems. In complete agreement with the
hypothesis of the universal character of spectral fluctuations, FNNS
for the circular billiard is perfectly approximated by Poissonian
distribution, and dispersion is a linear function of the length of
the considered interval. In the non-integrable case the level
repulsion effect is pronouncedly manifested, leading to Wigner
distribution, and dispersion grows much more slowly because of the
higher rigidity of the considered spectrum.
\sat\subsection{Uniform potential case}\sat
Let us note that an important advantage of billiards is the
possibility of spectrum unfolding (\ref{unfold}) with high precision
and independence of relative measures of regular and chaotic
components from the energy of particles. Both features equally share
homogeneous potentials, in particular coupled quartic oscillators
potential (\ref{cqo}), where we consider for simplicity only
$A_1$-type states --- even on both coordinates and symmetric with
respect to exchange $x\leftrightarrow y$.

As for quantum billiards, for the coupled quartic oscillators
potential (\ref{cqo}) there is a quite precise semiclassical
expression for the smooth component of staircase state number
function up to the third order in energy \cite{zhao_du}:
\begin{equation}\label{n_a1}\bar{n}_{A_1}(E)\simeq\frac{E^\frac32}{48}\left[F\left(\frac12,\frac12,1,\frac{2-\alpha}{4}\right)
+ 6\frac{\frac{\Gamma\left(\frac54\right)}{2^{\frac54}}+\frac{1}{\alpha+2}}{\Gamma\left(\frac74\right)\sqrt{2\pi
E^{\frac32}}}+\frac{9}{2E^{\frac32}}\right],\end{equation}
which allows us to perform spectrum unfolding with sufficiently high
accuracy in a wide energy range. However because of technical
difficulties in numerical calculation of hypergeometric function in
(\ref{n_a1}), in practice it is much more convenient to use the
corresponding Thomas-Fermi approximation for (\ref{n_a1}) in the
form:
\begin{equation}\label{n_a1_tf}\bar{n}_{A_1}(E)\approx\frac{E^\frac32}{12}\int\limits_0^{\frac\pi2}
\frac{d\varphi}{\sqrt{1+\frac{\alpha-2}{4}\sin^2\varphi}}.\end{equation}

\begin{figure}
\includegraphics[width=0.5\textwidth,draft=false]{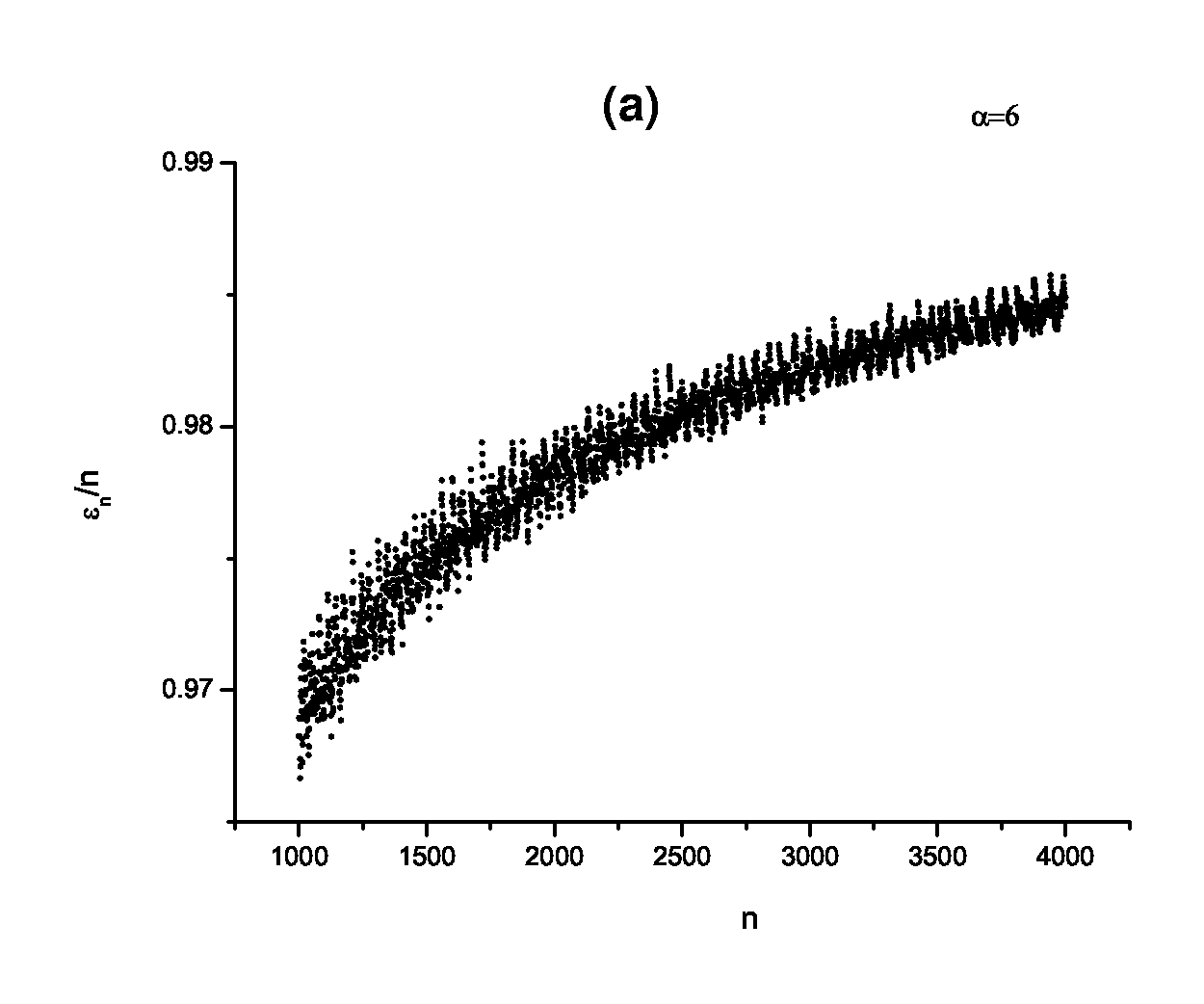}
\includegraphics[width=0.5\textwidth,draft=false]{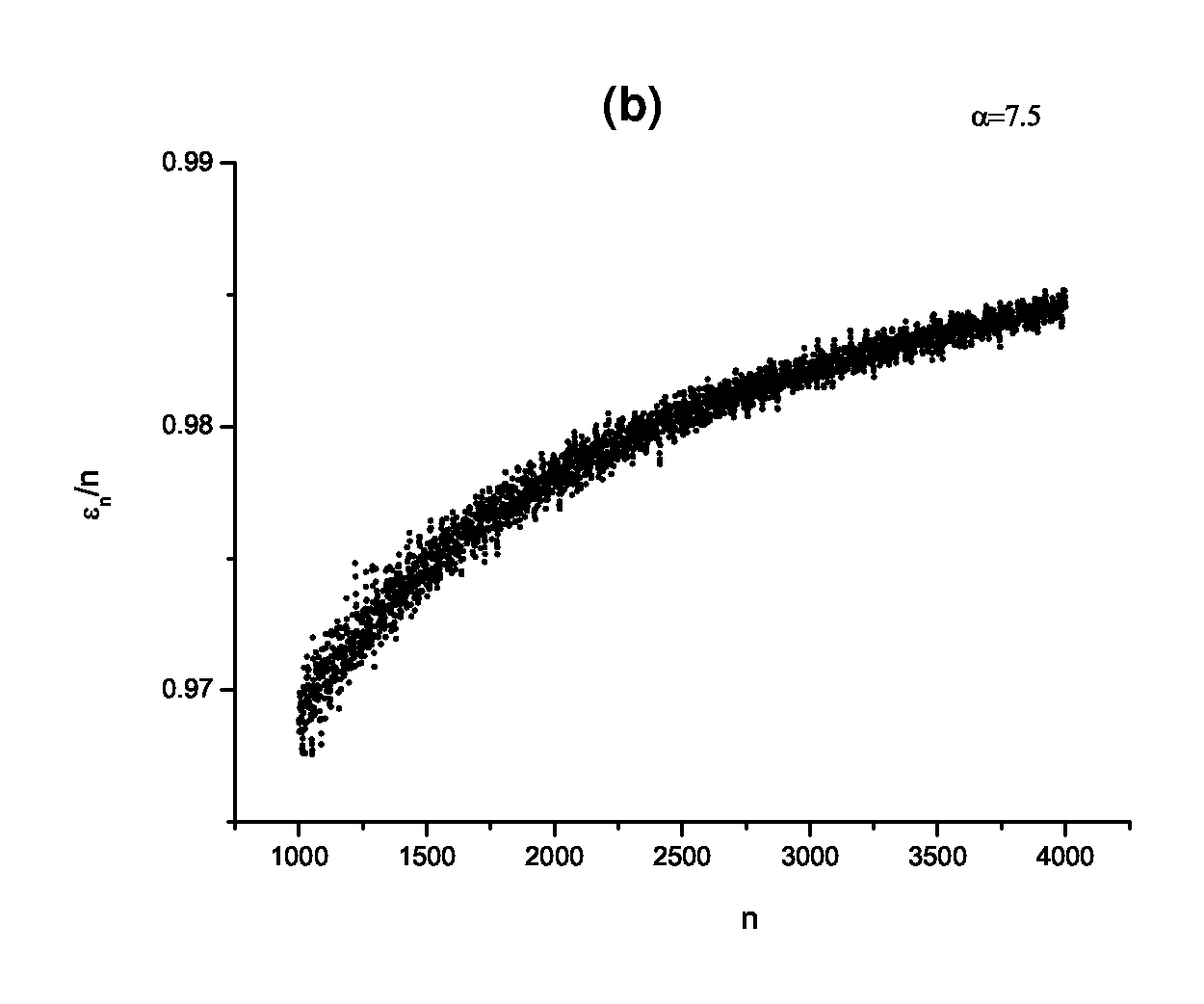}
\includegraphics[width=0.5\textwidth,draft=false]{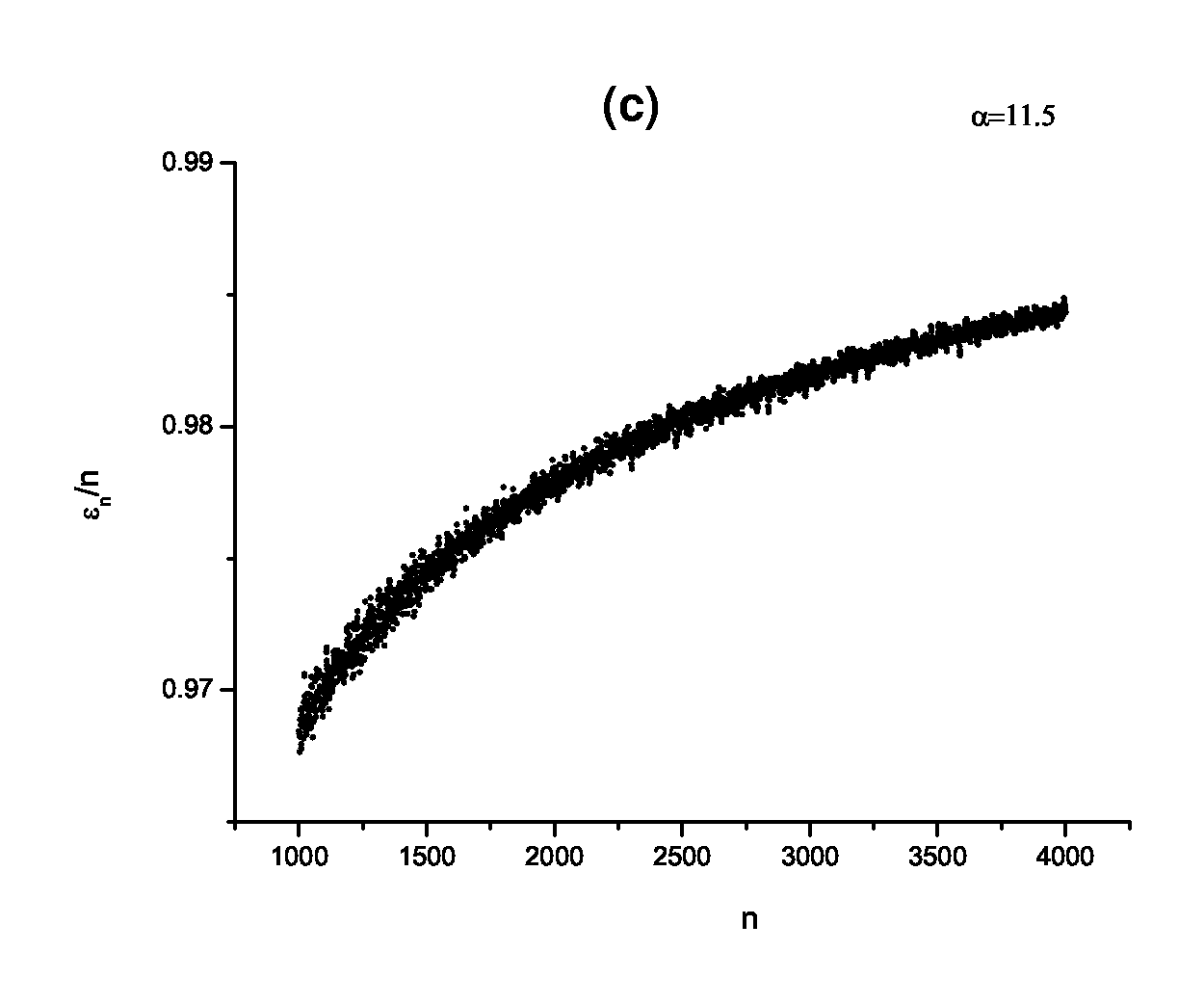}
\includegraphics[width=0.5\textwidth,draft=false]{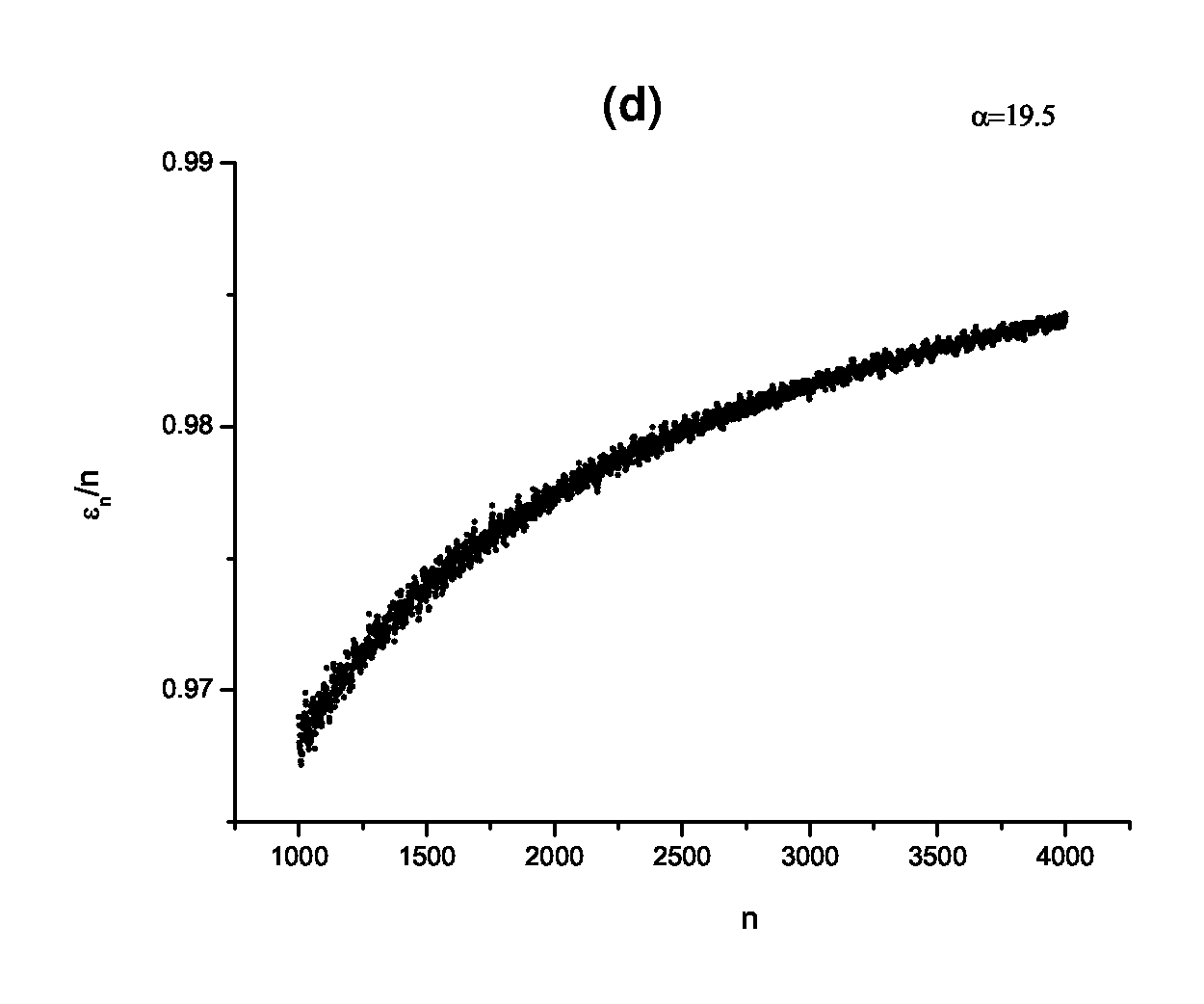}
\caption{\label{cqo_unfold}Accuracy estimation for the unfolding
with (\ref{n_a1}) in the spectrum of coupled quartic oscillator
potential (\ref{cqo}) for $\alpha=6$(a), $\alpha=7.5$(b),
$\alpha=11.5$(c), $\alpha=19.5$(d).}
\end{figure}

It is convenient to estimate the accuracy of the spectrum unfolding
with a diagram of variable $\varepsilon_n/n$ plotted for all levels
in the spectral series under consideration. For a spectrum
standardly unfolded by (\ref{unfold}), this variable should be
sufficiently close to unity in the whole energy interval. As can be
seen from Fig.\ref{cqo_unfold}, the Thomas-Fermi
approximation(\ref{n_a1_tf}) gives slightly understated values for
$\bar{n}_{A_1}(E)$, but nevertheless allows us to achieve $1\%$
accuracy of spectrum unfolding, and this accuracy improves with
energy growth.

\begin{figure}
\includegraphics[width=0.5\textwidth,draft=false]{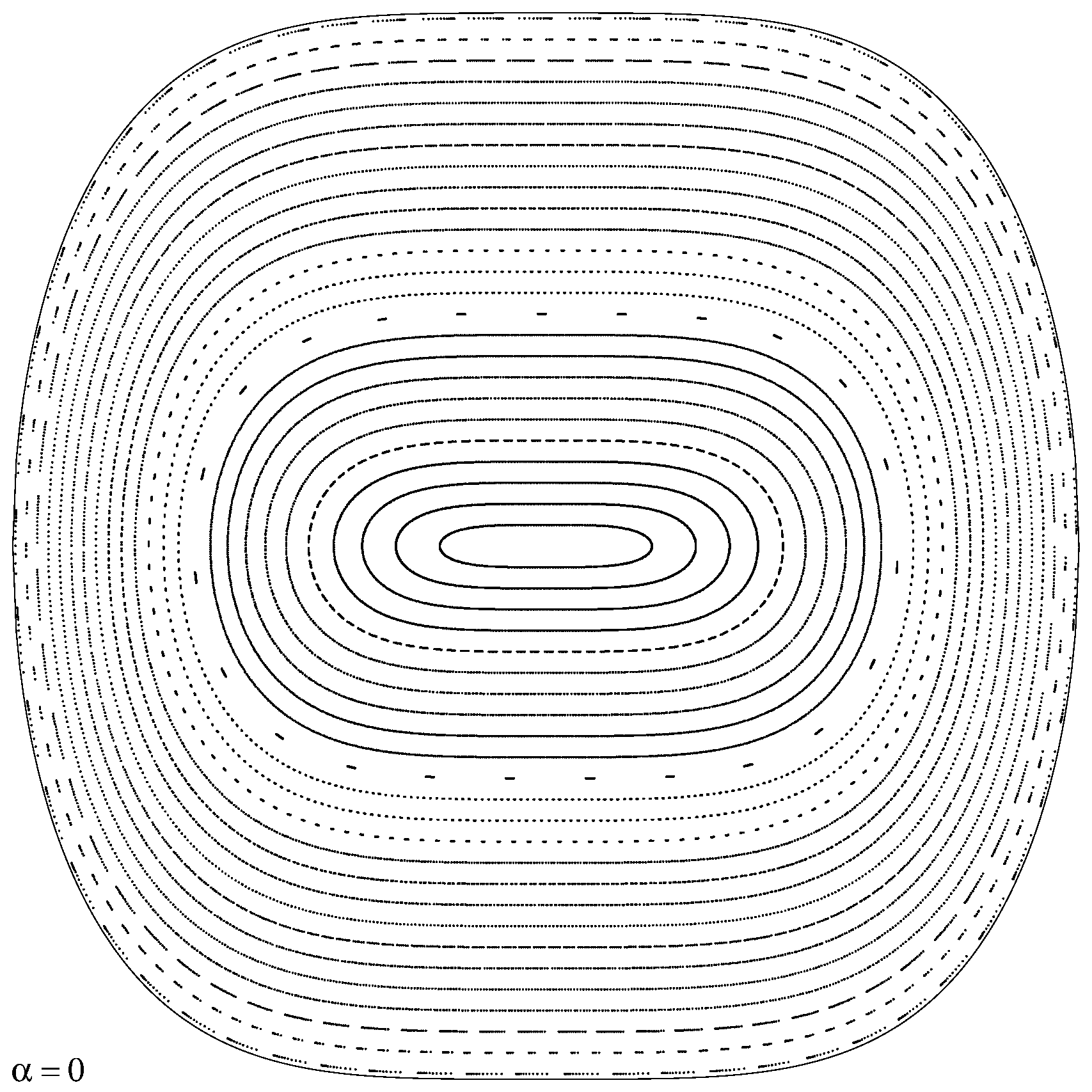}
\includegraphics[width=0.5\textwidth,draft=false]{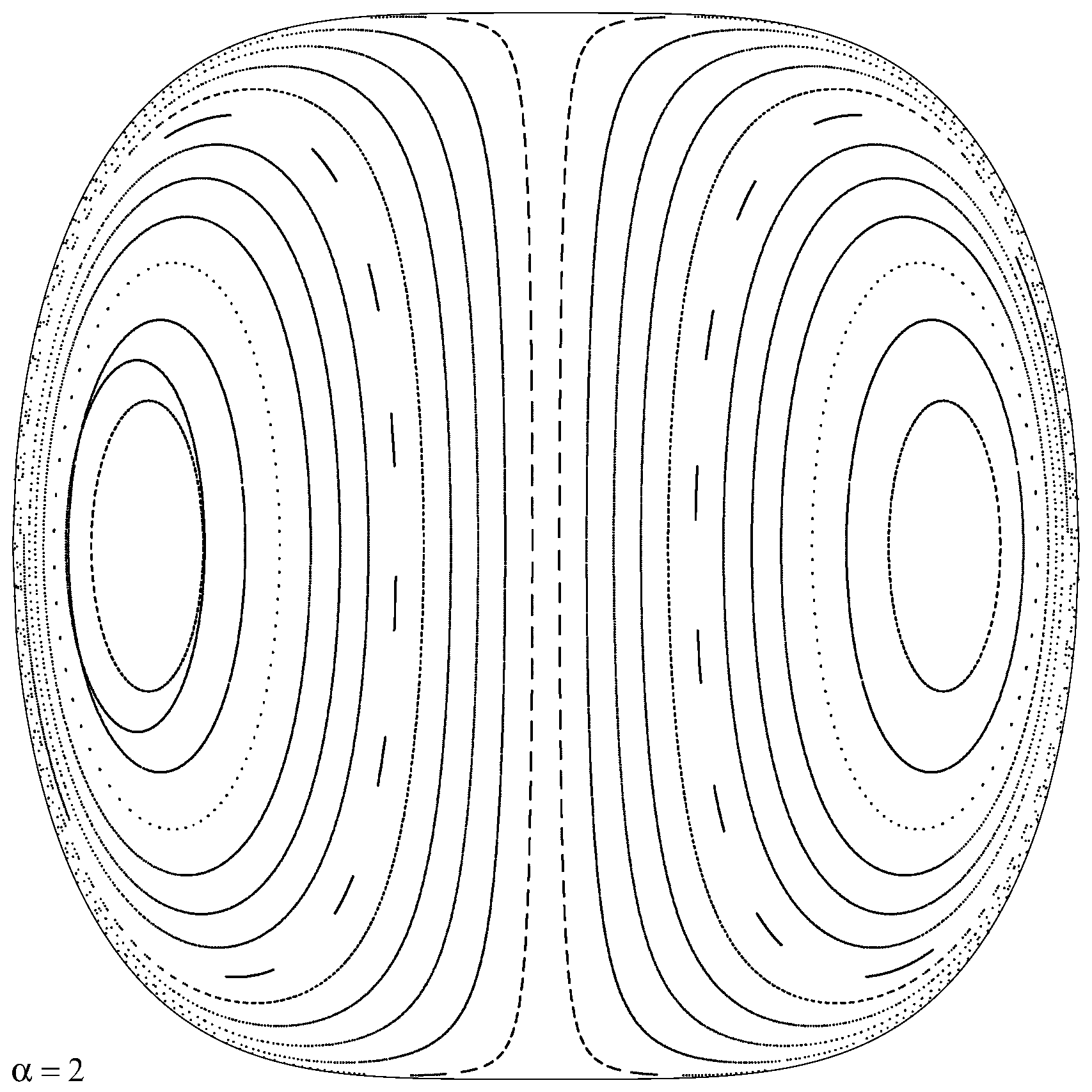}
\includegraphics[width=0.5\textwidth,draft=false]{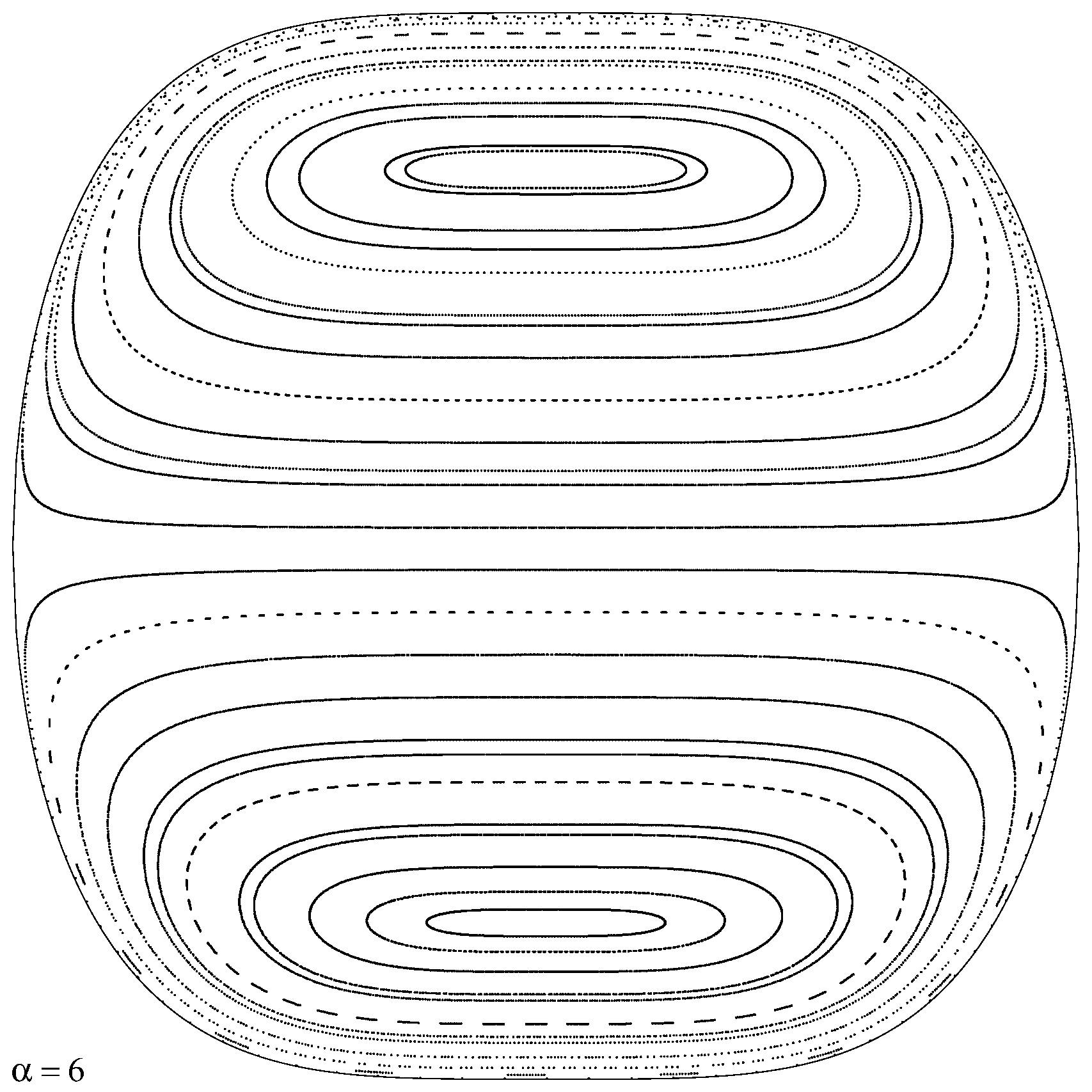}
\includegraphics[width=0.5\textwidth,draft=false]{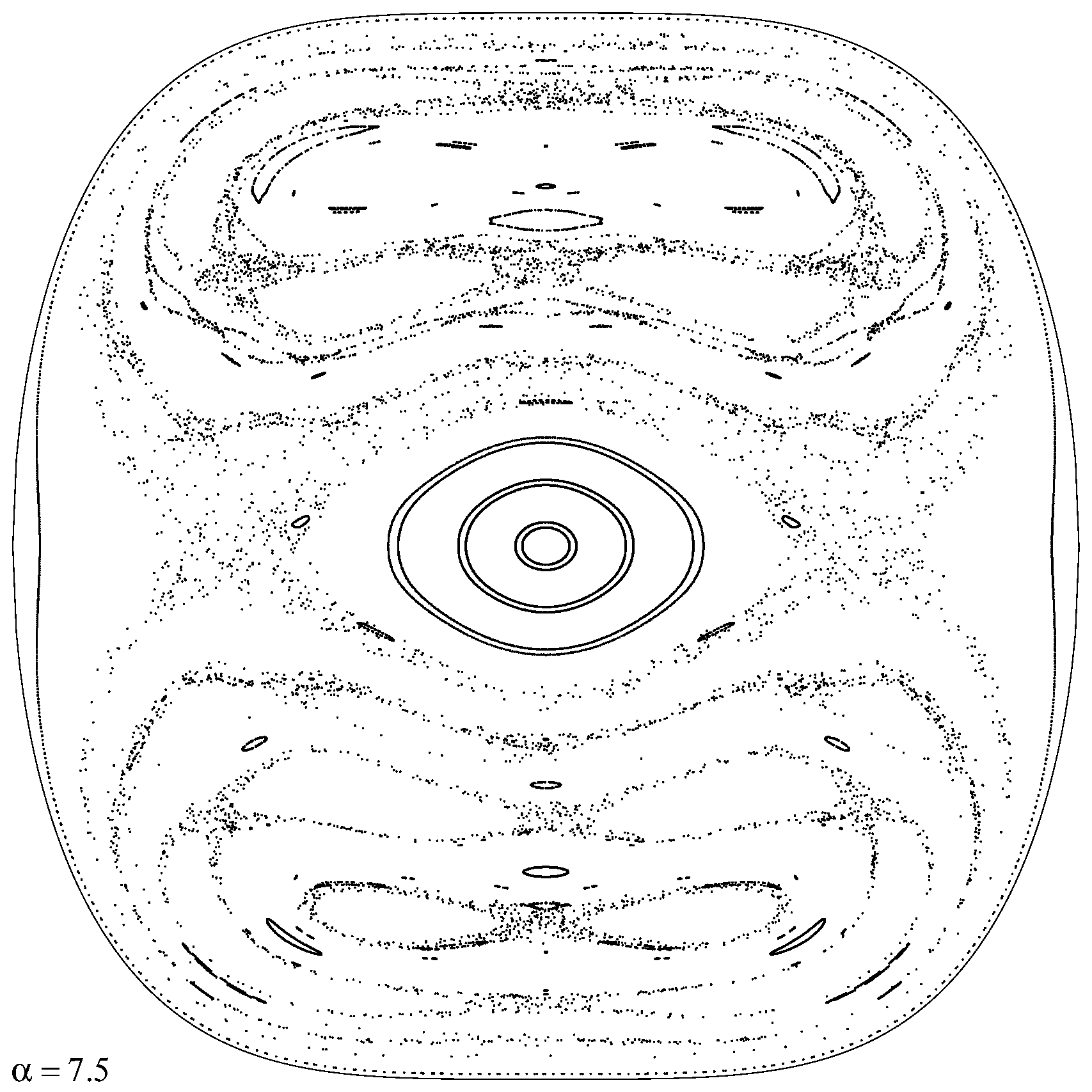}
\includegraphics[width=0.5\textwidth,draft=false]{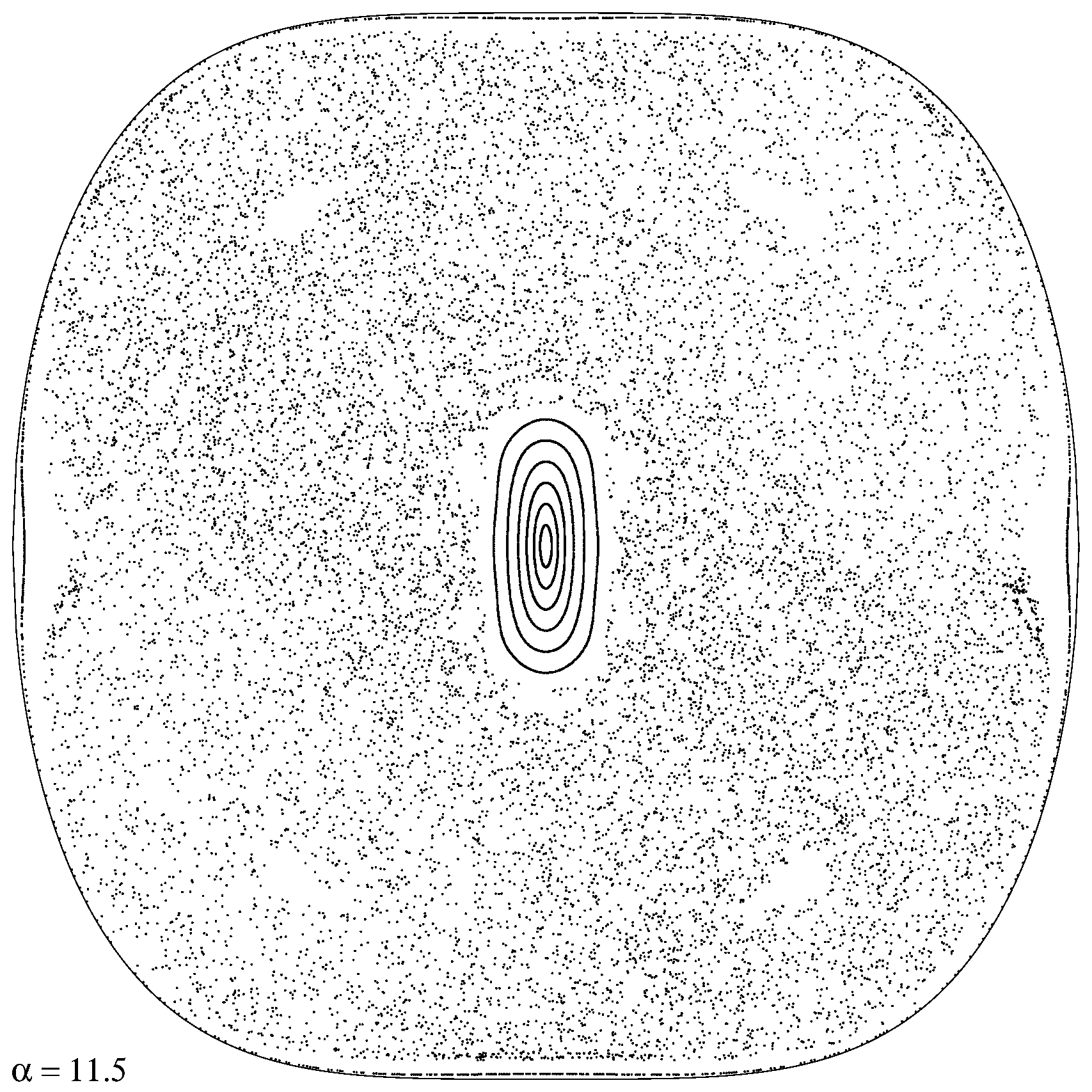}
\includegraphics[width=0.5\textwidth,draft=false]{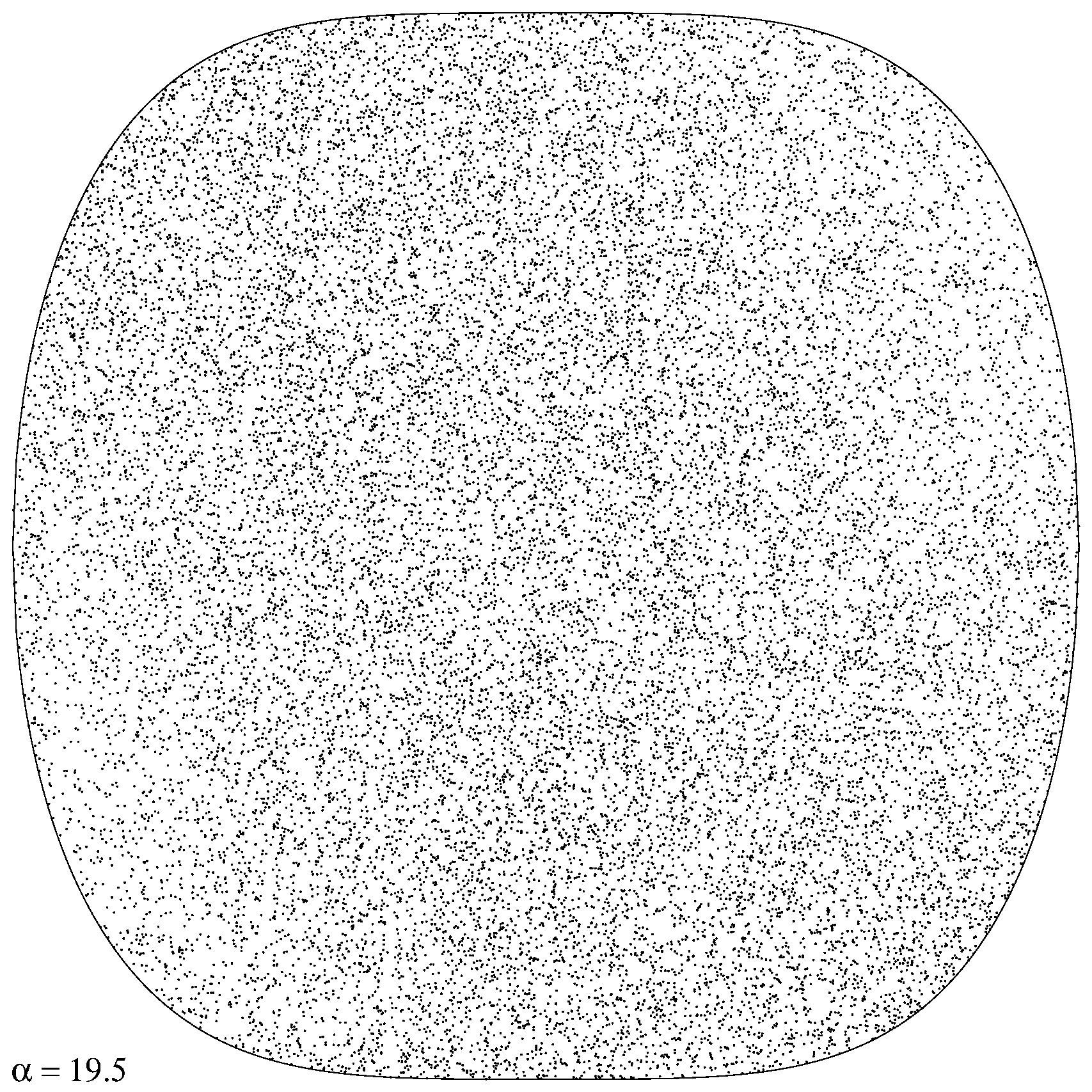}
\caption{\label{cqo_pss}Poincar\'e surfaces of section in the
$(x,p_x)$ plane for the coupled quartic oscillator potential with
coupling constant $\alpha=0,2,6,7.5,11.5,19.5$. The solid line
limits the classically allowed region of the phase space.}
\end{figure}

In Fig.\ref{cqo_unfold} one can see gradual decrease of maximum
amplitude of the spectral fluctuations with growth of chaoticity
measure (see Fig.\ref{cqo_pss}), which already qualitatively points
to gradual changes of the fluctuations character from Poissonian
(Fig.\ref{cqo_unfold}a) to Wigner one (Fig.\ref{cqo_unfold}d).

\begin{figure}
\includegraphics[width=0.5\textwidth,draft=false]{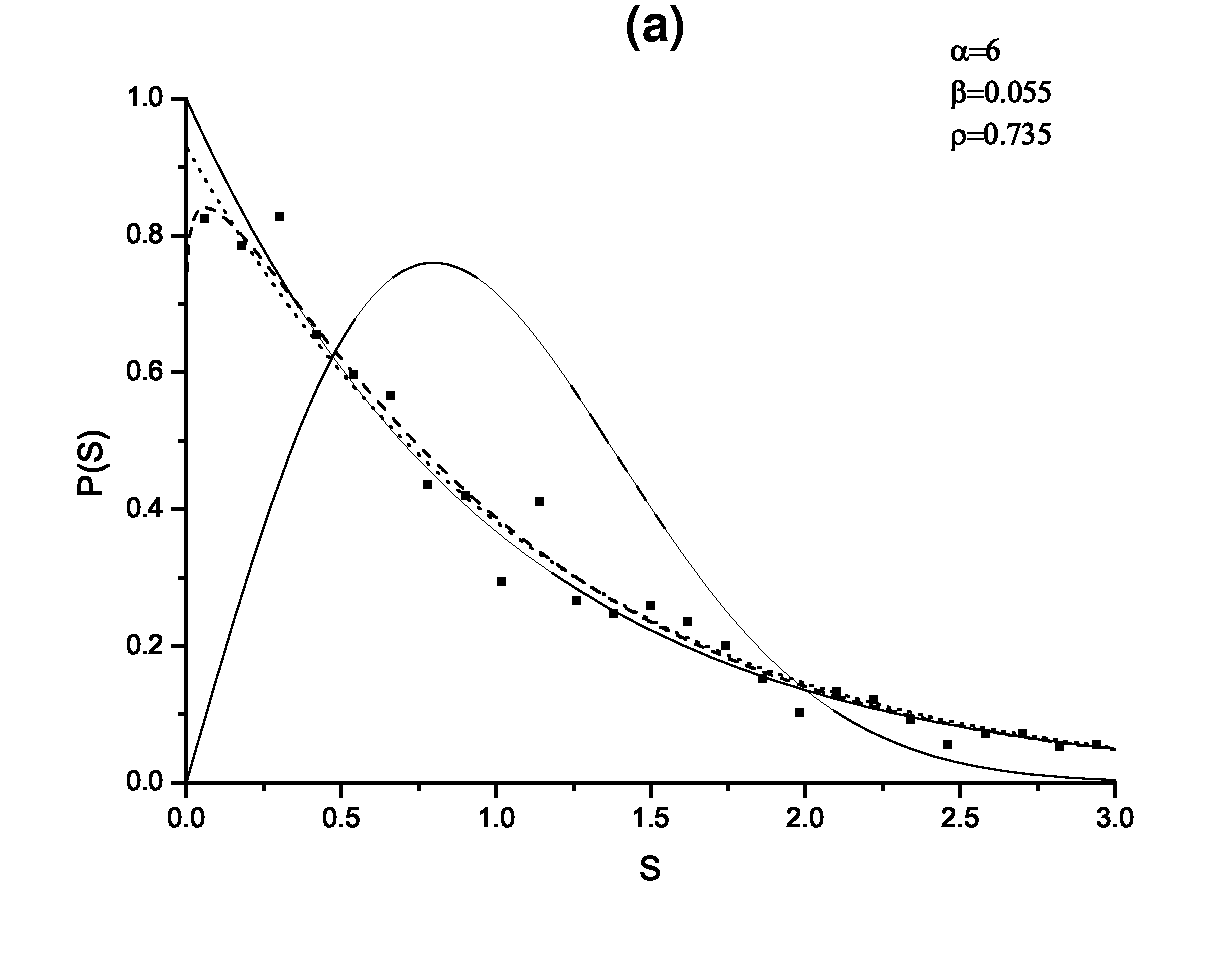}
\includegraphics[width=0.5\textwidth,draft=false]{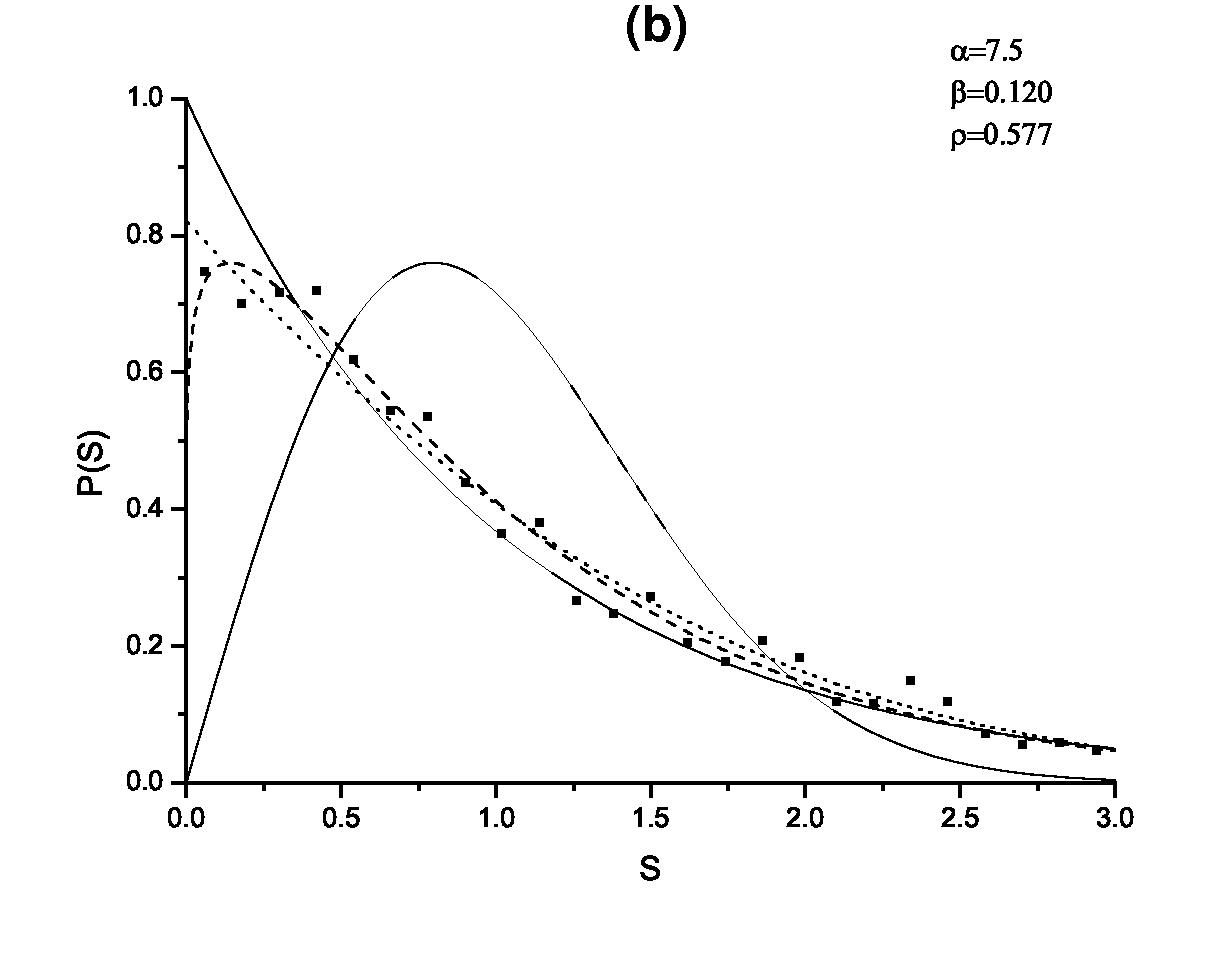}
\includegraphics[width=0.5\textwidth,draft=false]{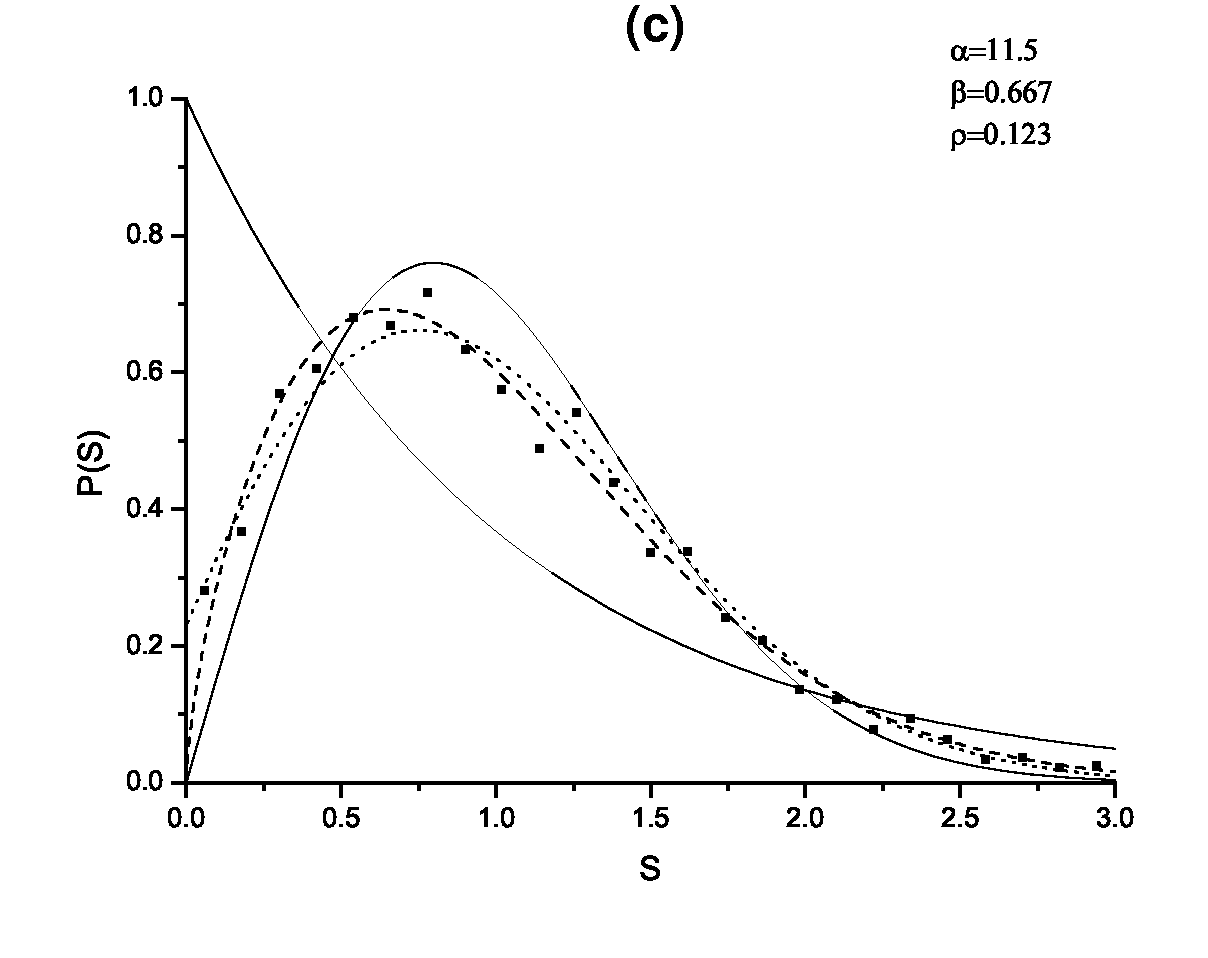}
\includegraphics[width=0.5\textwidth,draft=false]{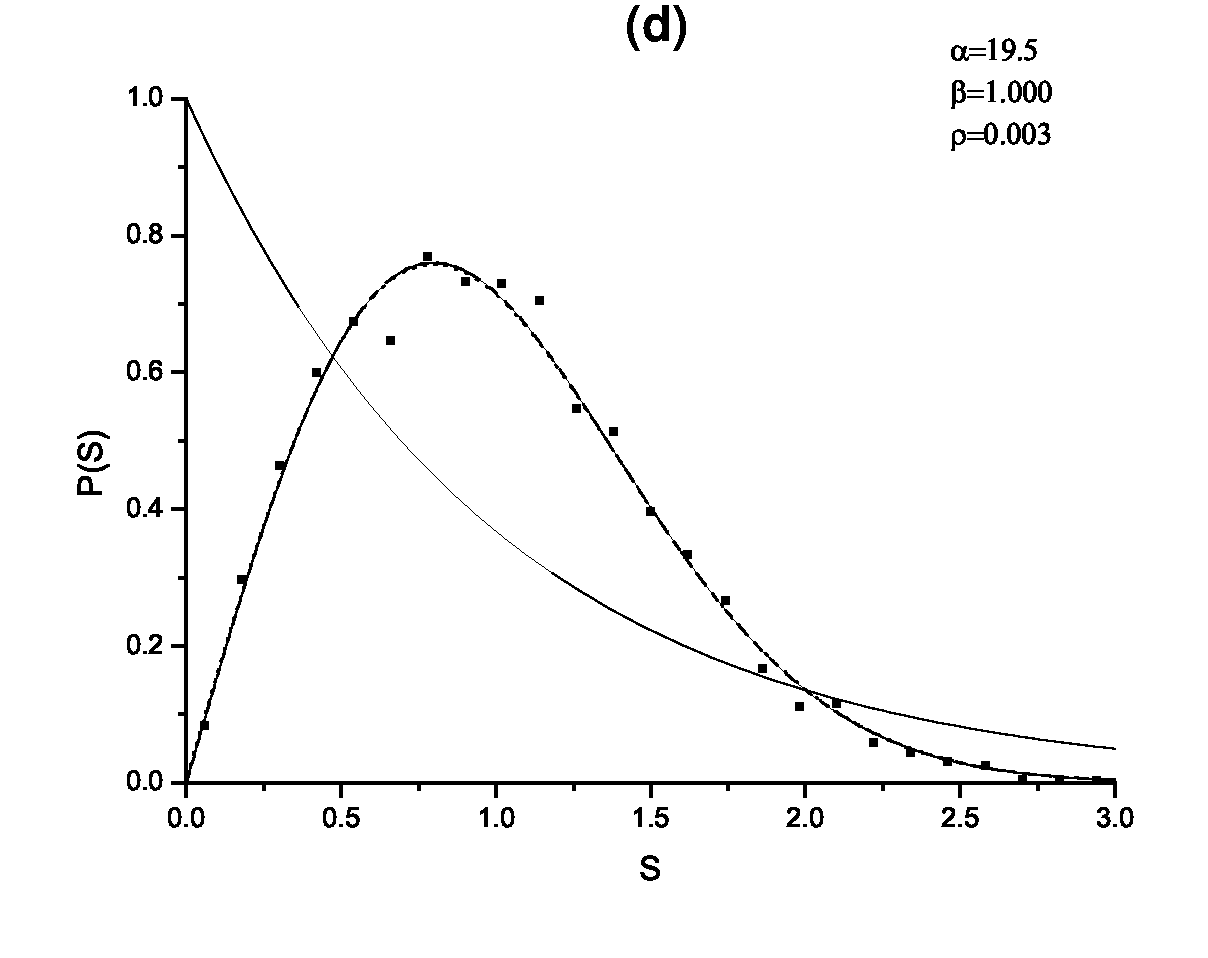}
\caption{\label{cqo_p}FNNS for the coupled quartic oscillator
potential (\ref{cqo}) for $\alpha=6$(a), $\alpha=7.5$(b),
$\alpha=11.5$(c), $\alpha=19.5$(d). Points represent numerical data,
solid lines --- Poisson (\ref{poisson}) and Wigner (\ref{wigner})
distributions, dashed and dotted lines --- the best fits by the
Brody (\ref{brody}) and the Berry-Robnik-Bogomolny (\ref{brb})
distributions respectively.}
\end{figure}

\begin{figure}
\includegraphics[width=0.5\textwidth,draft=false]{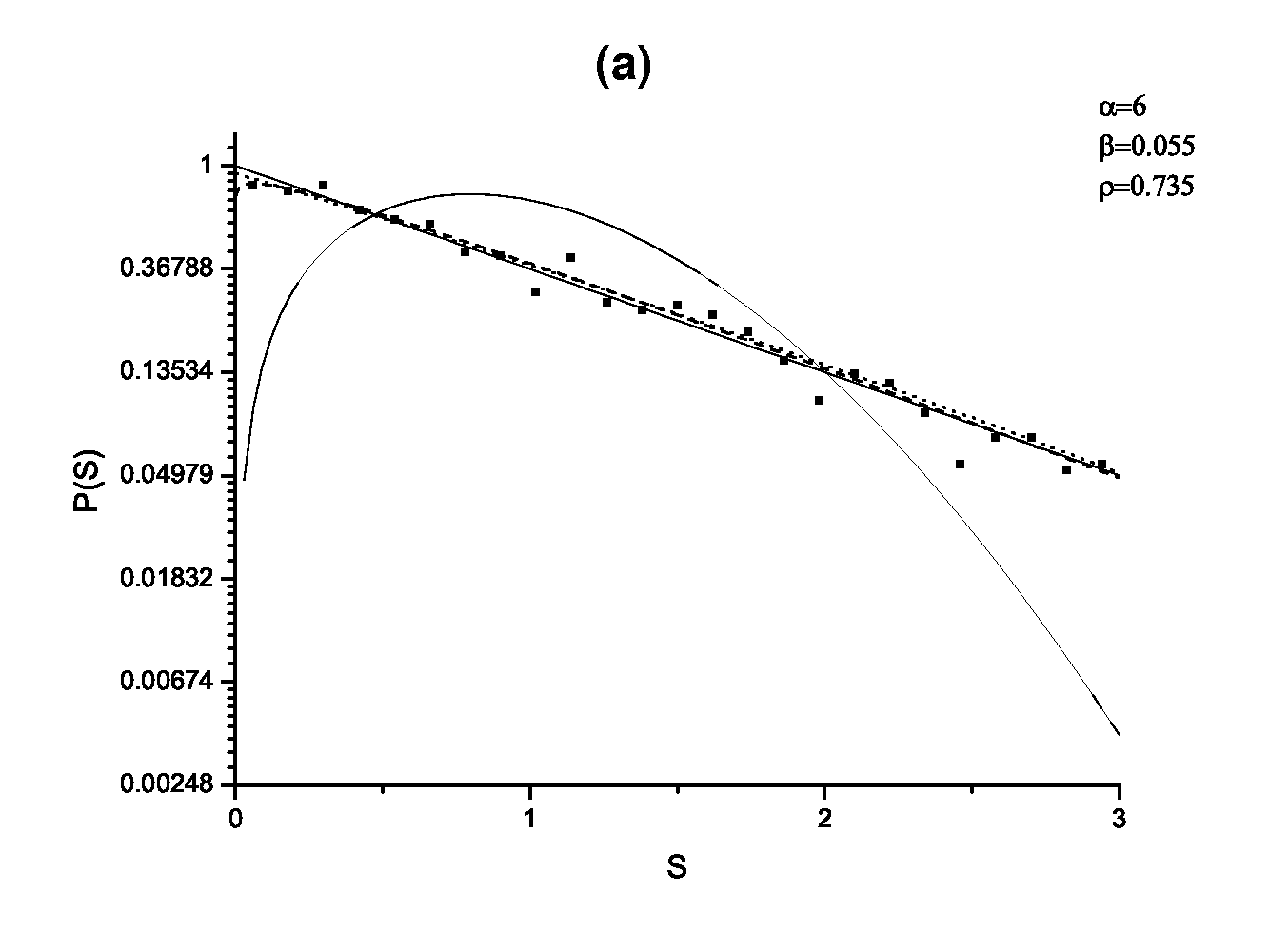}
\includegraphics[width=0.5\textwidth,draft=false]{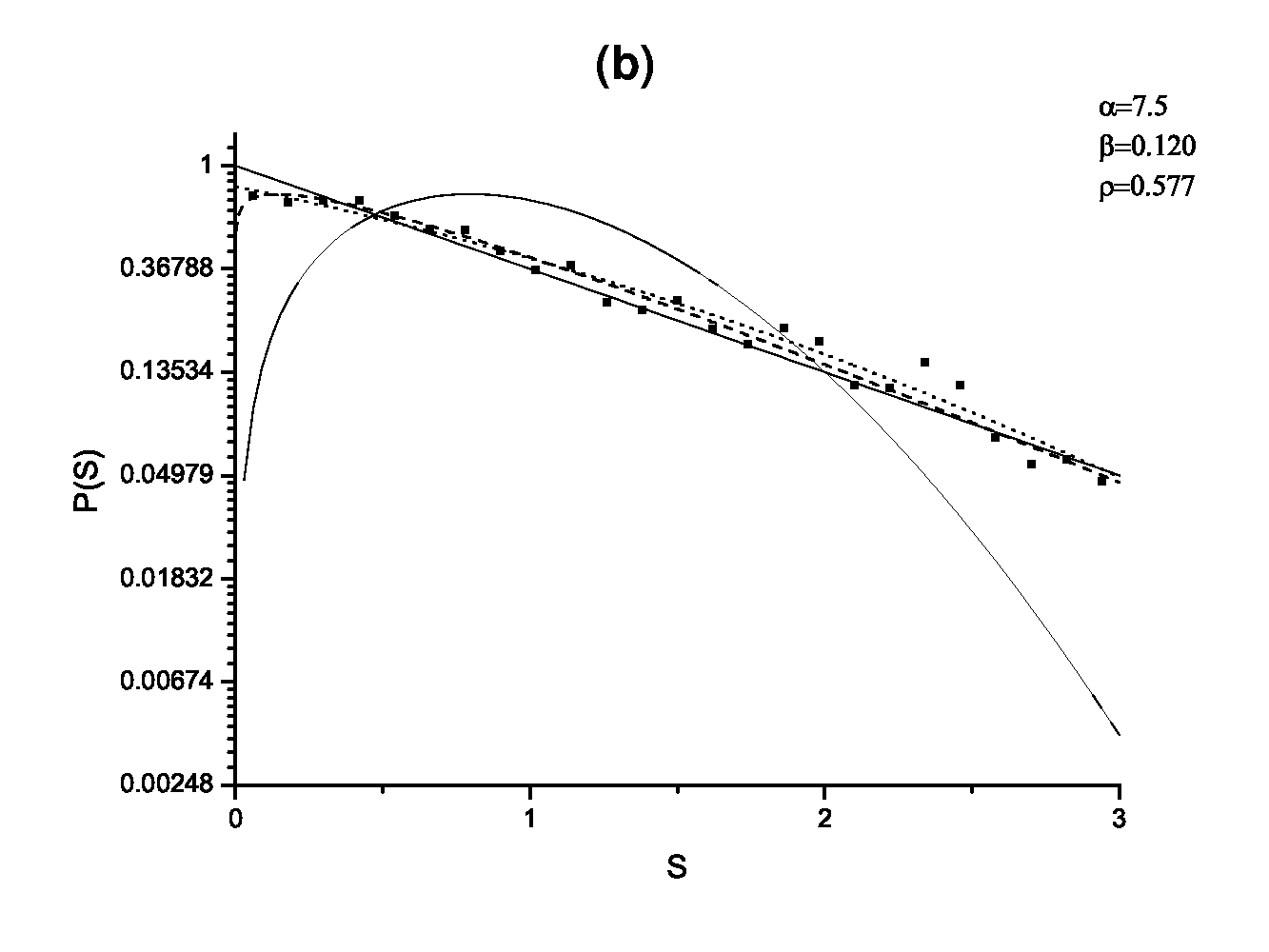}
\includegraphics[width=0.5\textwidth,draft=false]{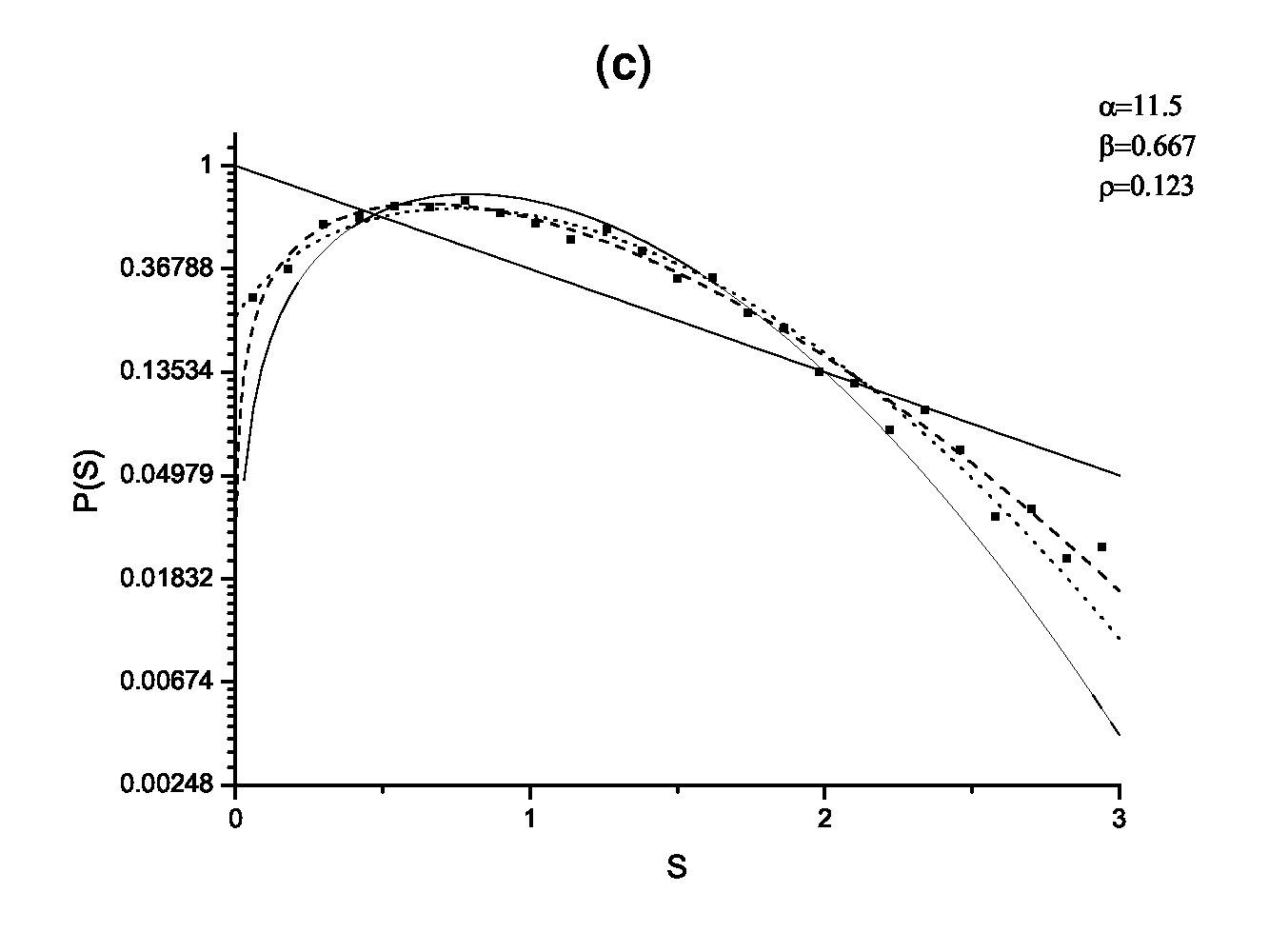}
\includegraphics[width=0.5\textwidth,draft=false]{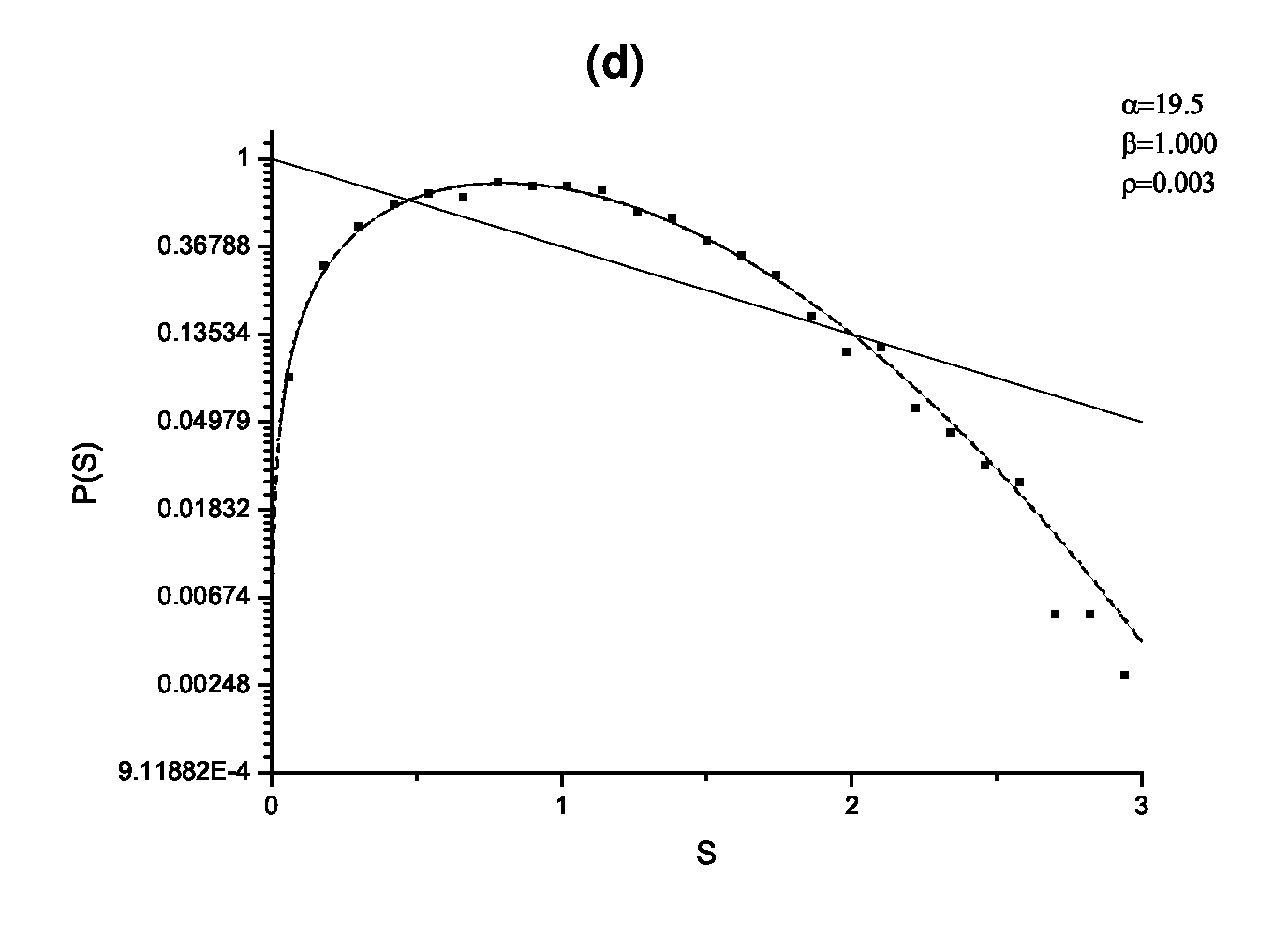}
\caption{\label{cqo_pl}The same as on Fig.\ref{cqo_p} but in
logarithmic scale.}
\end{figure}

The FNNS for coupled quartic potential (\ref{cqo}) together with the
best fits by Brody (\ref{brody}) and Berry-Robnik-Bogomolny
(\ref{brb}) distributions are presented in Fig.\ref{cqo_p} in normal
and on Fig.\ref{cqo_pl} in logarithmic scale. The latter is
especially convenient because in the logarithmic representation the
Poissonian distribution turns into linear function:
\[\ln P_P(S)=-S,\]
and proximity of numerically obtained distribution to Poissonian one
in the case of regular motion (Fig.\ref{cqo_pl}) becomes especially
clear.

According to the hypothesis of univeral character of spectral
fluctuations the FNNS transforms from Poissonian at $\alpha=6$ to
Wigner one at $\alpha=19.5$. In the intermediate cases the numerical
distribution function is satisfactorily described by the
Berry-Robnik-Bogomolny distribution as well as Brody distribution,
and the latter gives even better accuracy regardless of its
phenomenological nature.

Distributions on Fig.\ref{cqo_p} and Fig.\ref{cqo_pl} are obtained
from analysis of spectral series consisted of $3000$ levels each,
but even these statistics appears too low to give preference to
Brody distribution over the Berry-Robnik-Bogomolny distribution. In
the case of too pure statistics which is common in smooth potentials
it is better to use integral, or cumulative distribution:
\[W(S)=\int\limits_0^SP(S)dS\]

It is easy to obtain cumulative distributions corresponding to Brody
and Berry-Robnik-Bogomolny distributions:
\begin{equation}\label{w_brb}\begin{array}{c}
W_B(S)=1-e^{-bS^{\beta+1}}\\
W_{BRB}(S)=1+\frac{d}{dS}\left[e^{-\rho
S}erfc\left(\frac{\sqrt\pi}{2}(1-\rho)S\right)\right]
\end{array}\end{equation}

Cumulative Poisson and Wigner distributions, as well as the usual
ones, are limiting cases of the Brody distribution at $\beta=0$ and
$\beta=1$ and of the Berry-Robnik-Bogomolny one at $\rho=1$ and
$\rho=0$ respectively:
\begin{equation}\label{w_pw}\begin{array}{c}
W_B(S)=1-e^{-S}\\
W_{BRB}(S)=1-e^{-\frac{\pi}{4}S^2}
\end{array}\end{equation}

\begin{figure}
\includegraphics[width=0.5\textwidth,draft=false]{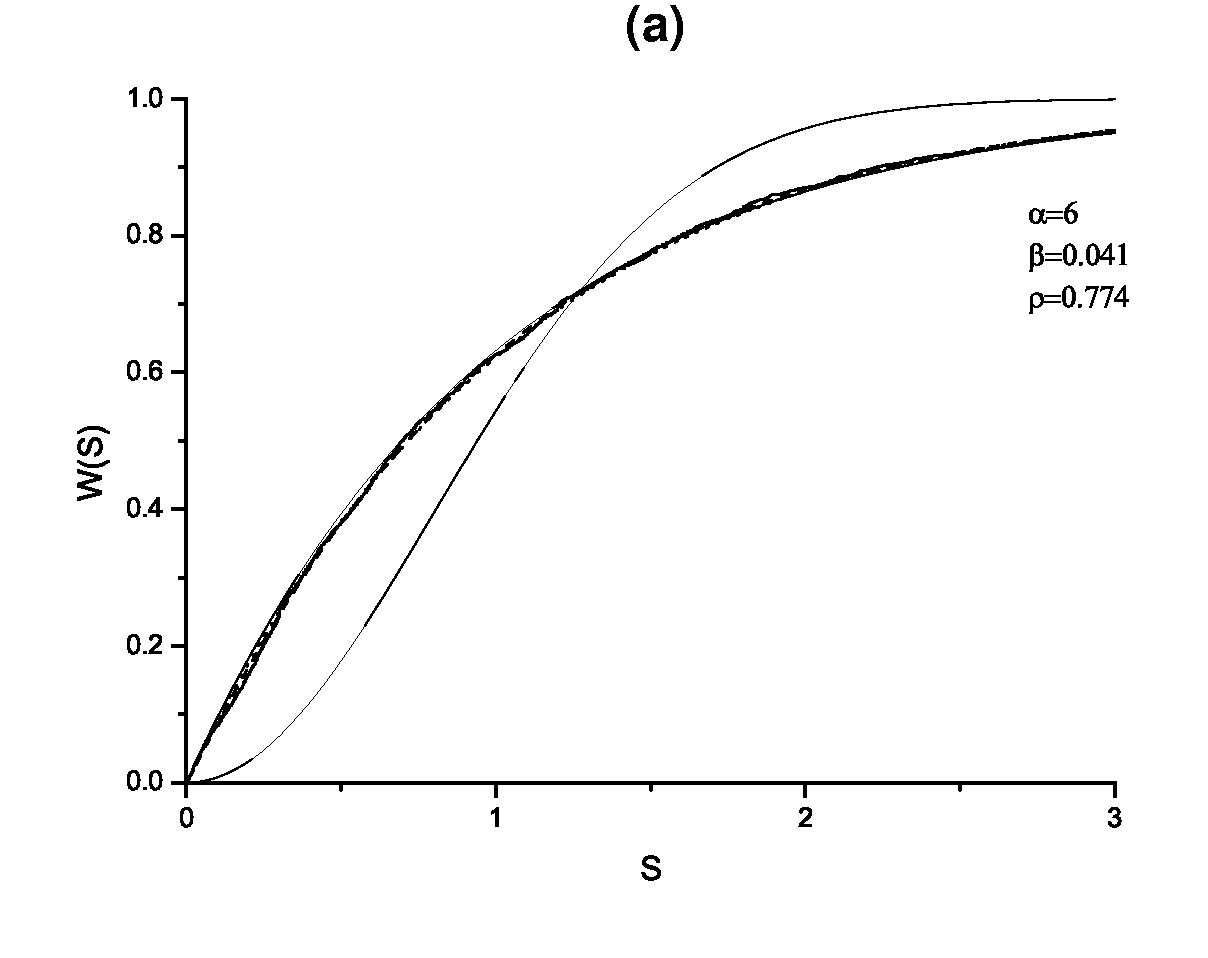}
\includegraphics[width=0.5\textwidth,draft=false]{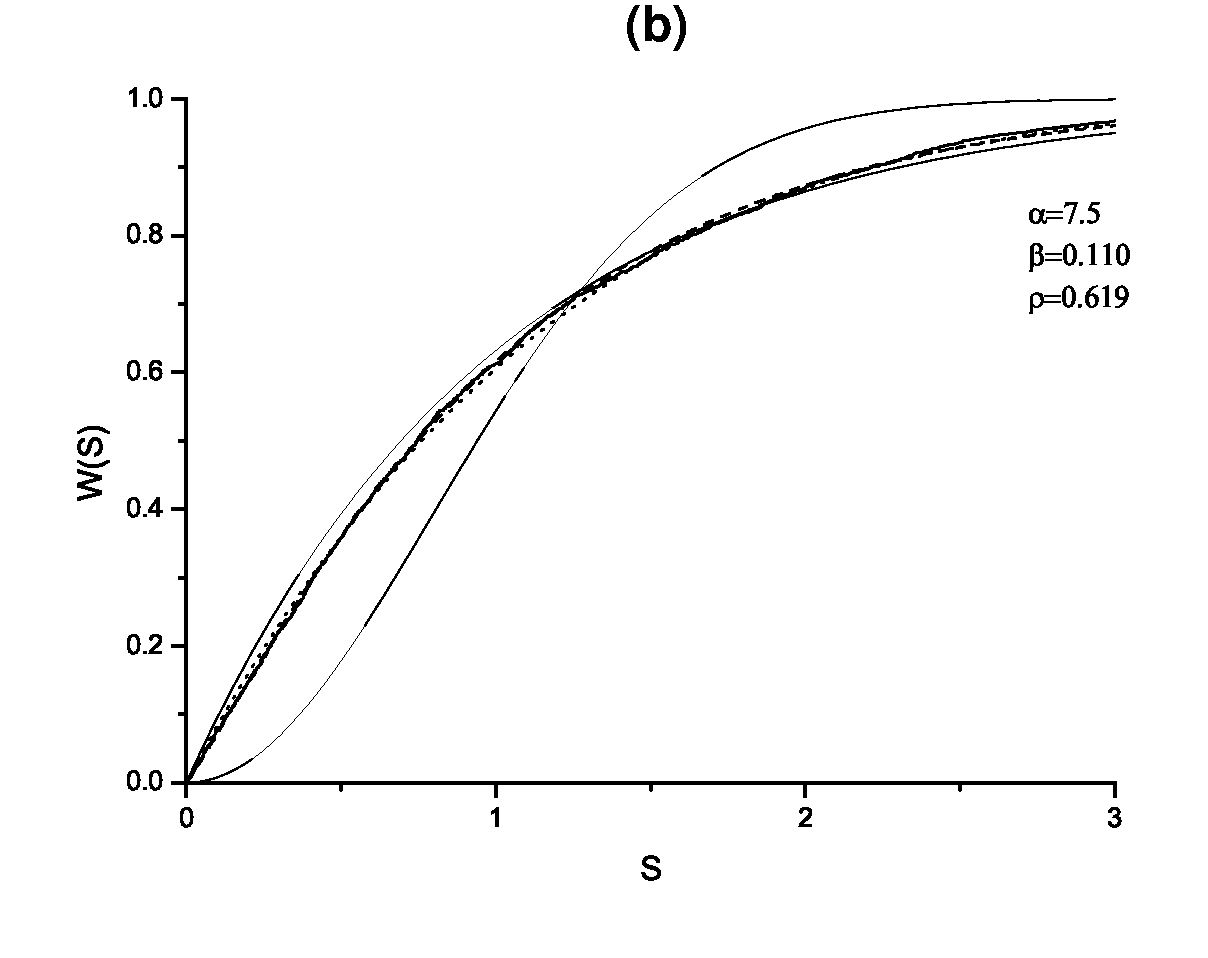}
\includegraphics[width=0.5\textwidth,draft=false]{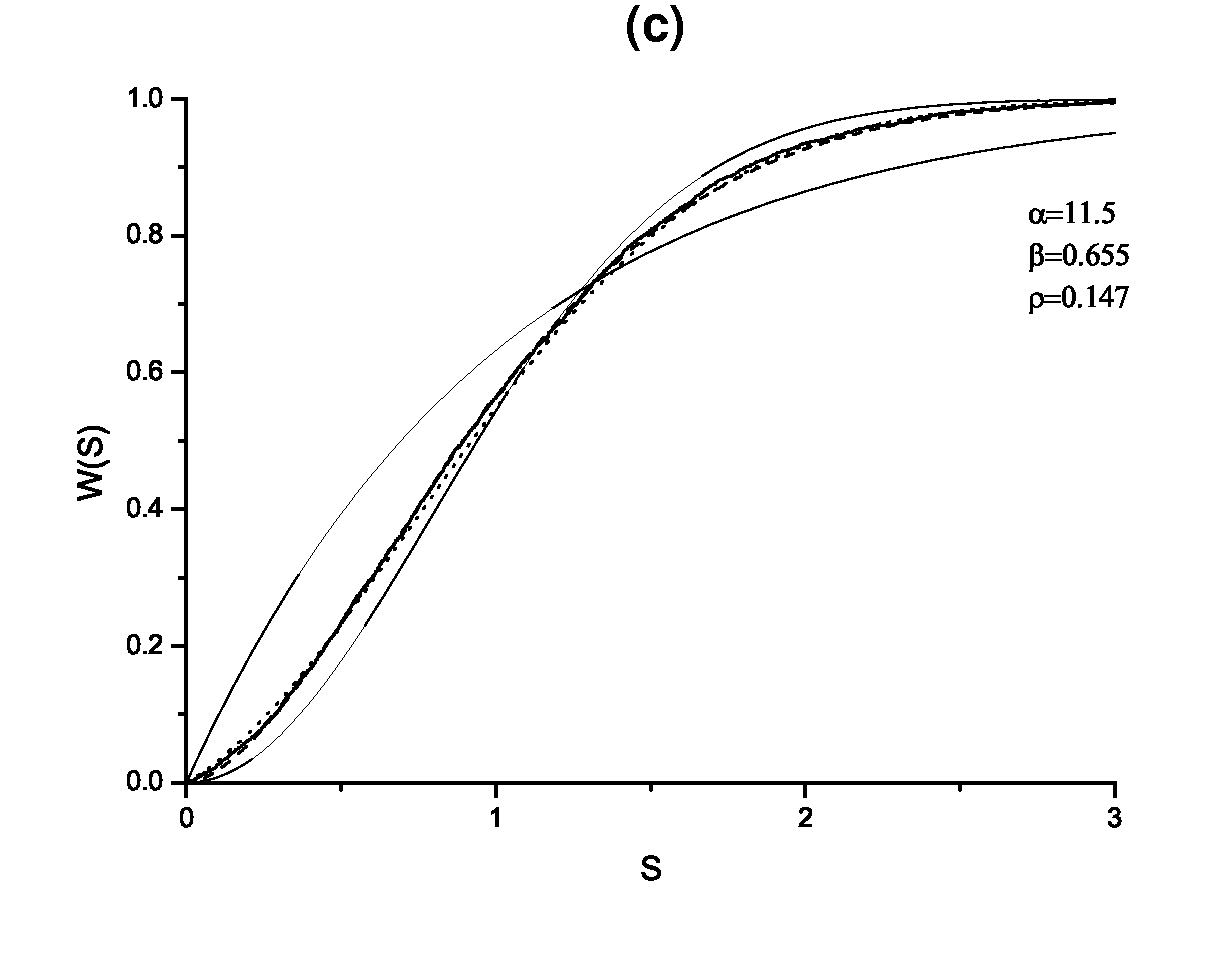}
\includegraphics[width=0.5\textwidth,draft=false]{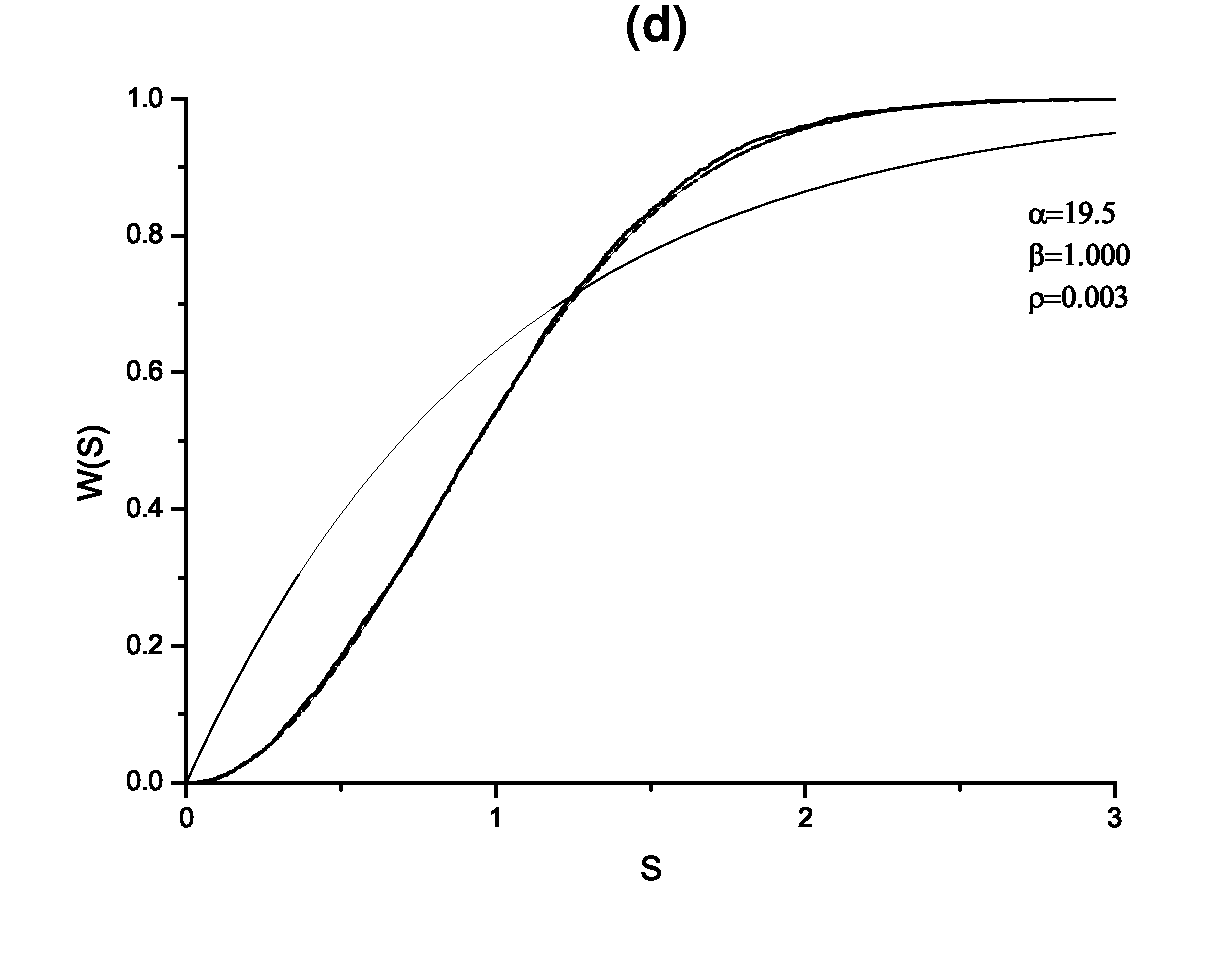}
\caption{\label{cqo_w}The cumulative FNNS for the coupled quartic
oscillator potential (\ref{cqo}) for $\alpha=6$(a), $\alpha=7.5$(b),
$\alpha=11.5$(c), $\alpha=19.5$(d). Points represent numerical data,
solid lines --- Poisson and Wigner (\ref{w_pw}) distributions,
dashed and dotted lines --- the best fits by the Brody and the
Berry-Robnik-Bogomolny (\ref{w_brb}) distributions respectively.}
\end{figure}

The numerically obtained cumulative FNNS for coupled quartic
potential (\ref{cqo}) together with the best fits by the Brody and
Berry-Robnik-Bogomolny (\ref{w_brb}) distributions are presented in
Fig.\ref{cqo_w}. It should be noted that parameters  of the best fit
made independently by usual (\ref{brody},\ref{brb}) and cumulative
(\ref{w_brb}) distributions agree one with another very well.
However cumulative distributions demonstrate better agreement with
numerical data exactly for the Brody distribution rather than for
the Berry-Robnik-Bogomolny distribution. The usual explanation of
that fact states that the Berry-Robnik-Bogomolny distribution is a
theoretically well-grounded universal high-energy asymptote for FNNS
but it has limited applicability for insufficiently high energies.
Therefore level fluctuations in the spectral series available for
most potential systems are better described by phenomenological
Brody distribution (\ref{brody}).

The most striking distinction between the Brody and
Berry-Robnik-Bogomolny distributions is displayed in the so-called
$T$-representation:
\[T(S)=\ln\{-\ln[1-W(S)]\}.\]

In the variables $(T,\ln T)$ the Brody distribution with any
parameter $\beta$, as well as Poisson and Wigner distributions,
become linear functions:
\begin{equation}\label{t_b}
T_B(S)=\ln b +(\beta+1)\ln S\\
\end{equation}
\begin{equation}\label{t_pw}\begin{array}{l}
T_P(S)=\ln S\\
T_W(S)=\ln \frac\pi4 +2\ln S,
\end{array}\end{equation}
which is not the case for the Berry-Robnik-Bogomolny distribution:
\begin{equation}\label{t_brb}T_{BRB}=\ln\left\{-\ln\left(\frac{d}{dS}\left[e^{-\rho S}
erfc\left(\frac{\sqrt\pi}{2}(1-\rho)S\right)\right]\right)\right\}.\end{equation}

\begin{figure}
\includegraphics[width=0.5\textwidth,draft=false]{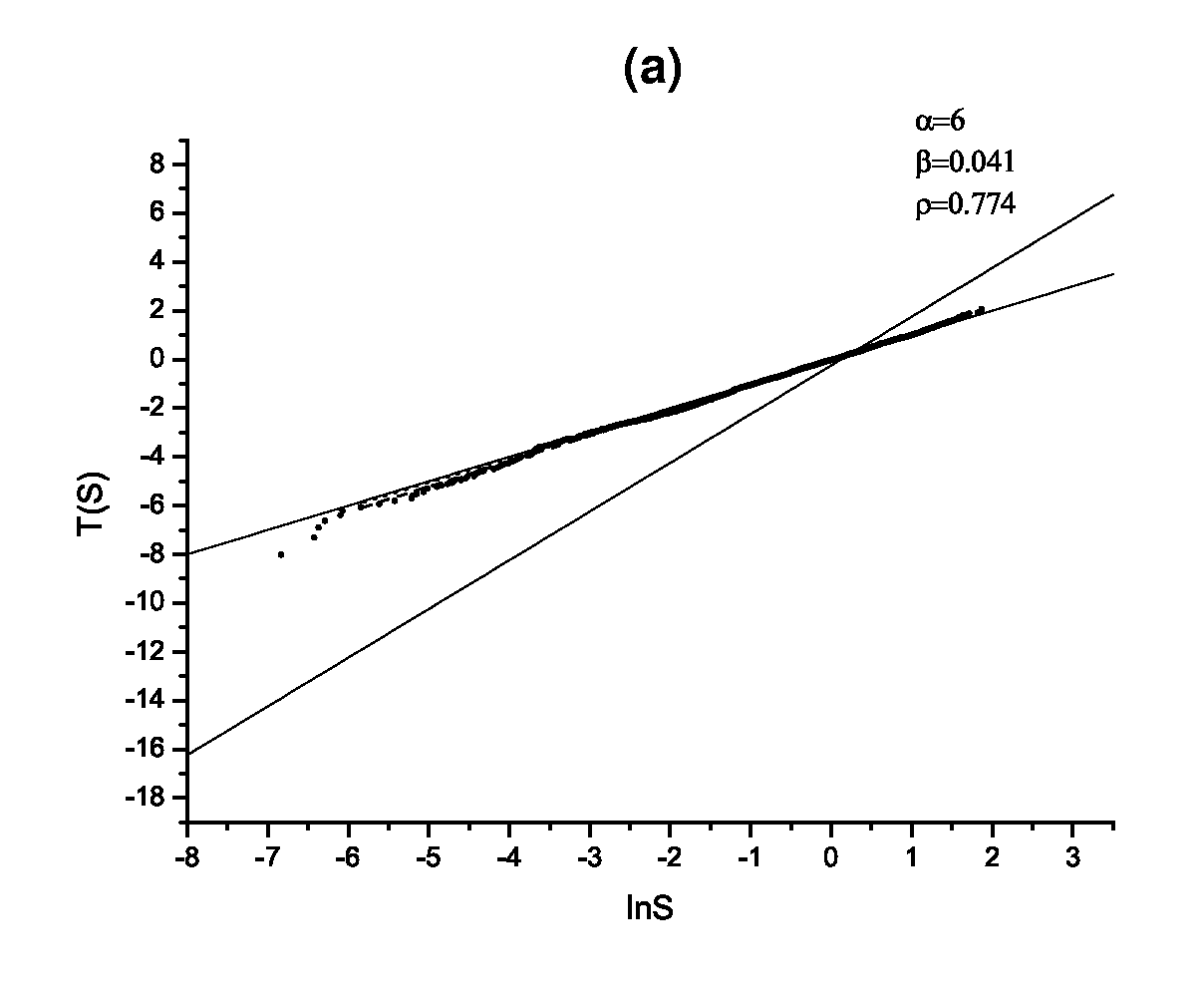}
\includegraphics[width=0.5\textwidth,draft=false]{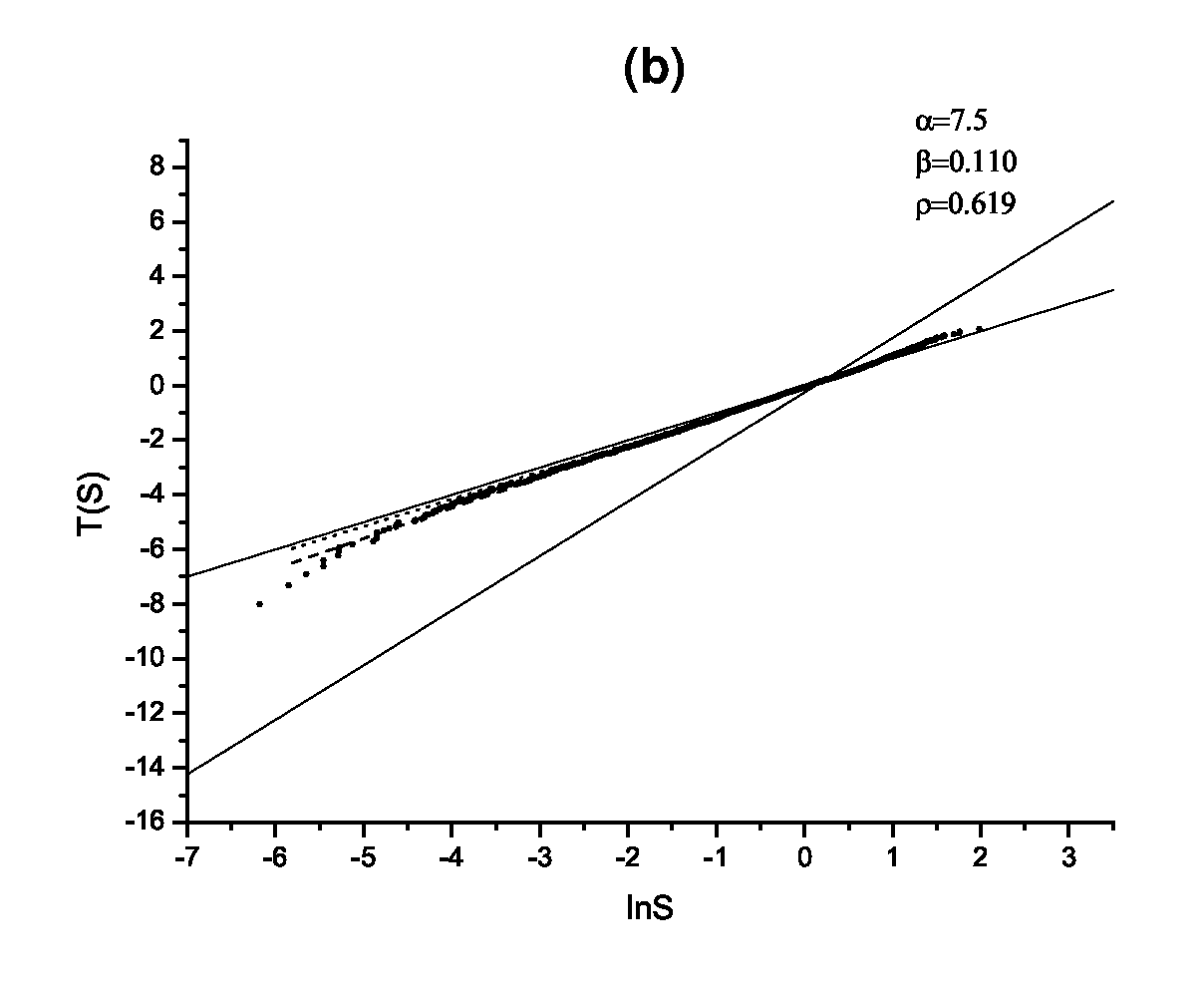}
\includegraphics[width=0.5\textwidth,draft=false]{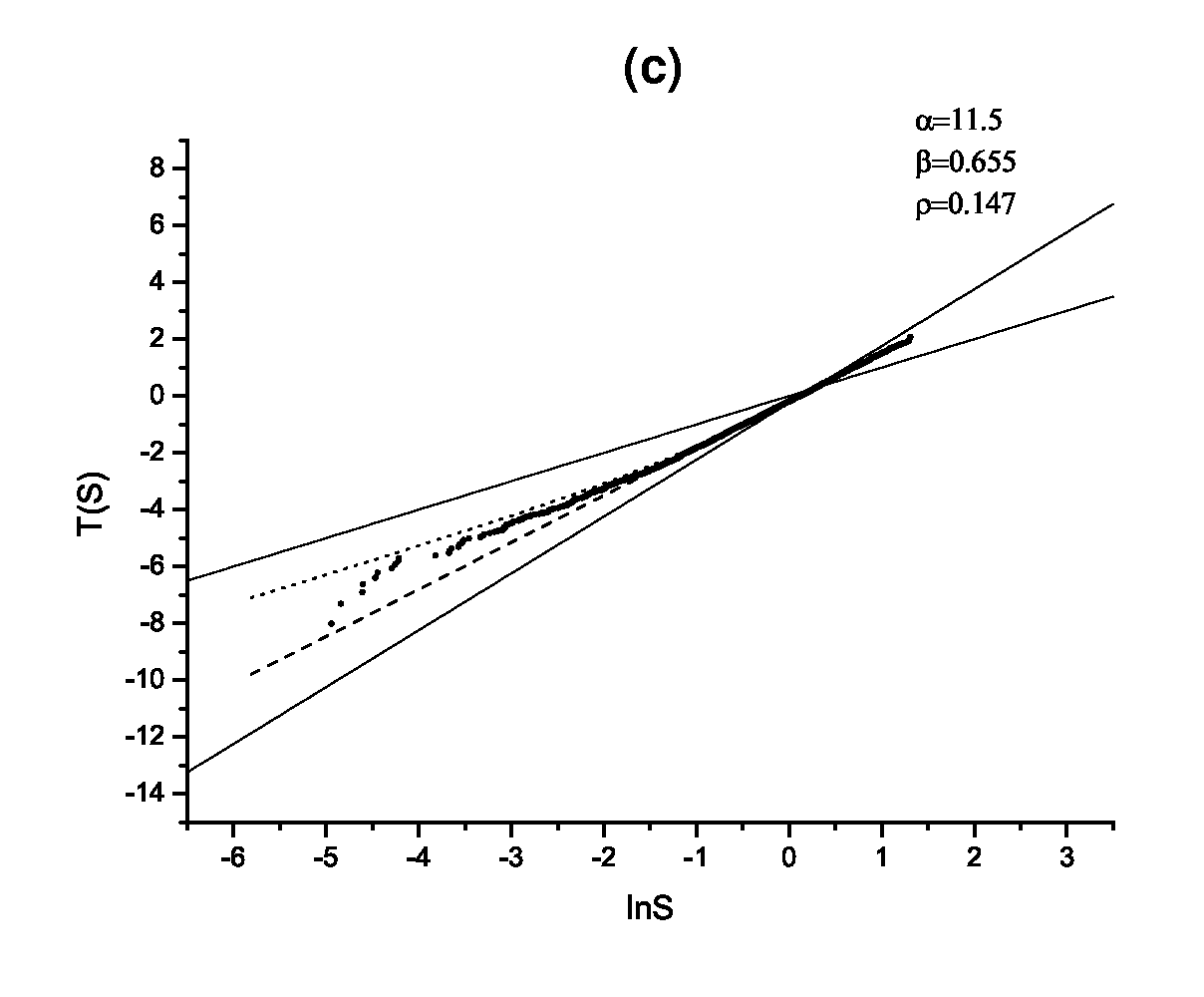}
\includegraphics[width=0.5\textwidth,draft=false]{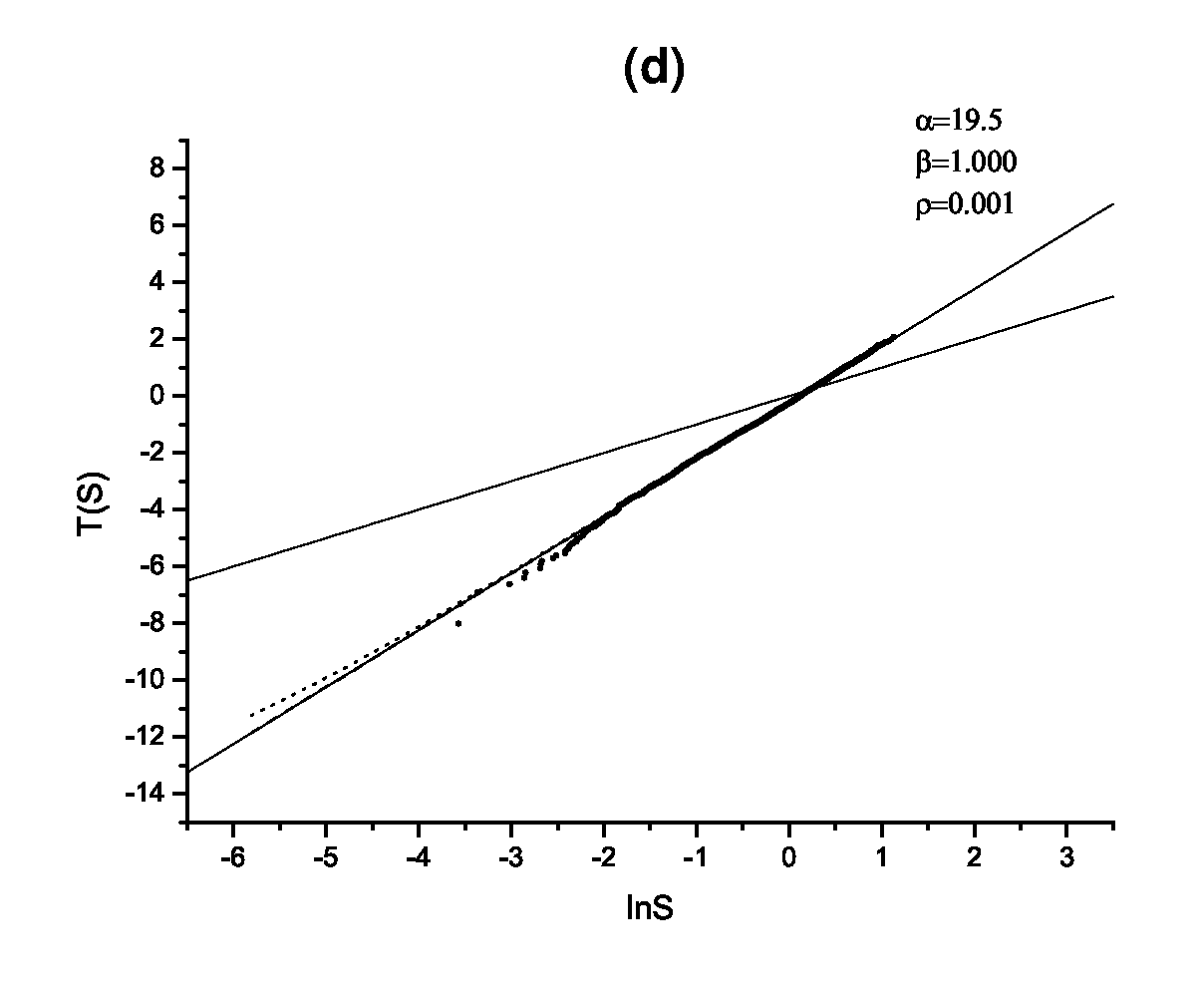}
\caption{\label{cqo_t}The cumulative FNNS in the $T$-representation
for the coupled quartic oscillator potential (\ref{cqo}) for
$\alpha=6$(a), $\alpha=7.5$(b), $\alpha=11.5$(c), $\alpha=19.5$(d).
Points represent numerical data, solid lines --- Poisson and Wigner
(\ref{t_pw}) distributions, dashed and dotted lines --- the best
fits by the Brody (\ref{t_b}) and the Berry-Robnik-Bogomolny
(\ref{t_brb}) distributions respectively.}
\end{figure}

The slope of the $T$-distribution on the $(T,\ln T)$ determines the
repulsion rate in the model (\ref{p_s}): Wigner distribution
corresponds to linear repulsion and Poisson distribution --- to
zero, or absence of repulsion. Therefore to the Brody distribution
we can assign so-called fractional levels repulsion with the
exponent continuously changing from $0$ to $1$. At the same time the
Berry-Robnik-Bogomolny distribution describes only integer repulsion
with the exponent equal to $1$. As can be seen from Fig.\ref{cqo_t},
the fractional level repulsion effect actually takes place in smooth
potential systems and is correctly described by the Brody
distribution.

\sat\subsection{Generic potential case}\sat

\begin{figure}
\includegraphics[width=\textwidth]{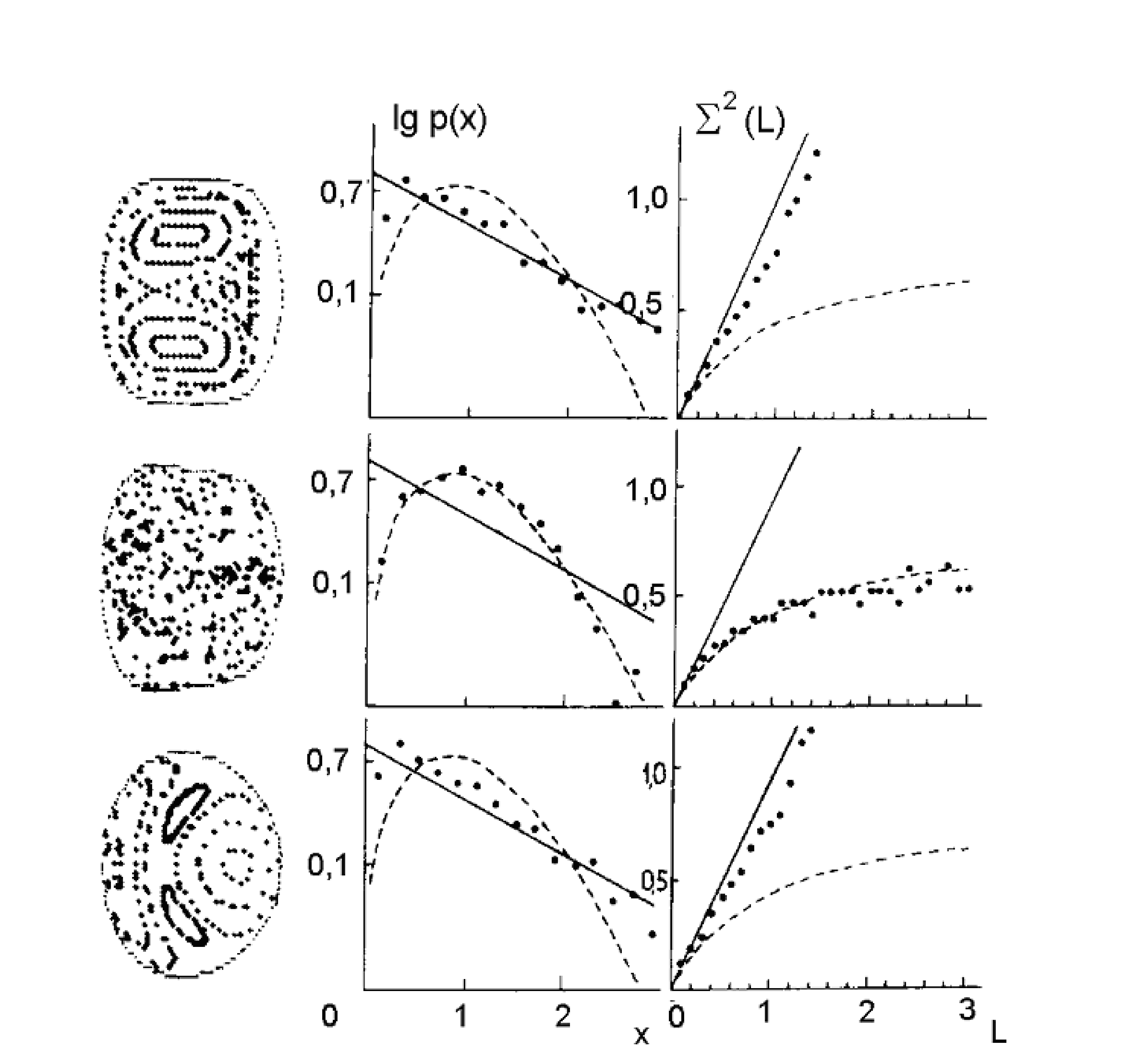}
\caption{\label{rcr}Correlation between the character of classical
motion and statistical properties of energy spectra in $R-C-R$
transition for the Hamiltonian of quadrupole oscillations
(\ref{qo_ham}). On the left --- Poincar\'e sections, in the middle ---
FNNS distribution  $p(x)$, on the right --- the variance $\Sigma^2$.
From the bottom up: the first regular range $R_1$, the chaotic range
$C$, the second regular range $R_2$}
\end{figure}

Let us pass now from the special case of uniform potentials to the
more generic situation. Let us note at the beginning that, even in
the case of non-uniform potentials in the presence of $R-C$ or
$R-C-R$ transitions, it is possible to build rather representative
series of energy levels corresponding to a definite type of
classical motion. As an example let us consider the transformation
of statistical properties of the energy spectrum of the quadrupole
oscillations Hamiltonian (\ref{qo_ham}) in the $R-C-R$ transition
for one-well case $(W<16)$. At the fixed topology of potential
surface $(W=const)$ the unique free parameter of the Hamiltonian is
the scaled Planck's constant $\bar{\hbar}$. In the study of the
concrete energetic interval ($R_1$, $C$ or $R_2$), corresponding to
a definite type of classical motion, the choice of $\bar{\hbar}$ is
dictated by the possibility of attainment of the necessary
statistical assurance (the required number of levels in the
investigated energy interval) with conservation of precision of
spectrum calculation (restrictions to possibility of diagonalization
of matrices of large dimension). The numerical results are presented
in Fig.\ref{rcr}. Both the FNNS $p(x)$ and the average value of
$\Delta_3$ well correspond to the predictions of GOE for the chaotic
$(C)$ region. The logarithmic scale for $p(x)$ is suitable to trace
this correspondence at large $x$. For regular regions ($R_1$ and
$R_2$) the distribution function, in the same scale, according to
the hypothesis of the universal character of fluctuations of energy
spectra, must be represented by a straight line (the logarithm of
Poisson's distribution). The results demonstrate the agreement with
this hypothesis, though small-sized deviations are observed for
small distances between levels. Such a tendency to the rise of some
repulsion in the regular region, apparently, is connected with a
small admixture of chaotic components. At the construction of
statistical characteristics, a purity of sequence is provided by
using only those levels, which are relative to definite irreducible
representation of $C_{3v}$ group (the levels of $E$-type were used
for results represented in Fig.\ref{rcr}; the statistical
characteristics of levels of $A_1$ and $A_2$-types have similar
form).

Spectral ranges of multi-well potentials realizing the mixed state
open new possibilities for investigation of the intermediate
statistics. At those energies chaotic and regular components are
separated not only in Phase space as in the usual case of mixed type
motion), but already in the configuration space. A priori the FNNS
form in the mixed state is not necessarily reduced to definitely
weighted superposition of Poisson and Wigner distribution. In that
case we deal not with statistics of admixture of two level series
with different FNNS, but with statistics of spectrum series, where
each level cannot be attributed to definite (Poisson or Wigner)
statistics. Statistical properties of such systems have not been
under study up to now although they are the systems corresponding to
the common case situation.

\begin{figure}
\includegraphics[width=0.5\textwidth]{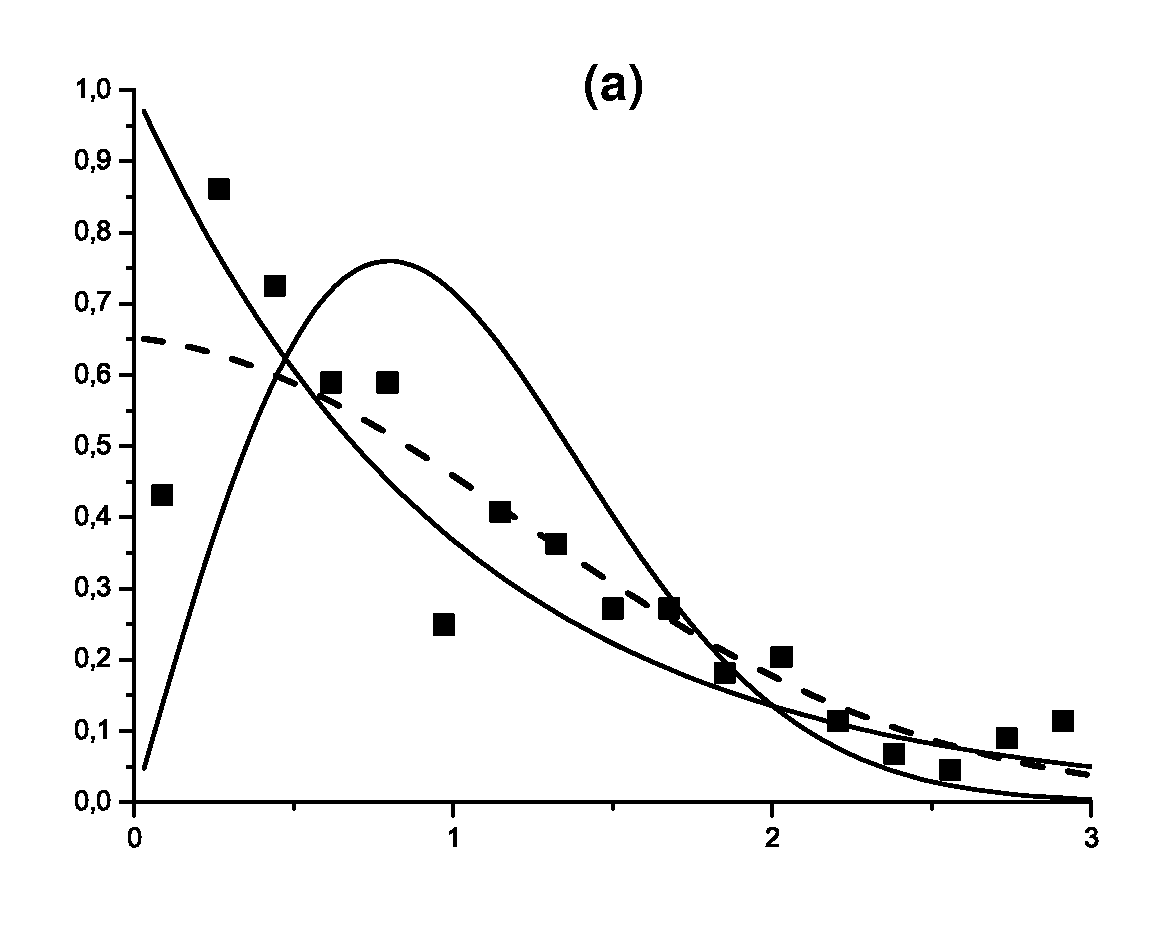}
\includegraphics[width=0.5\textwidth]{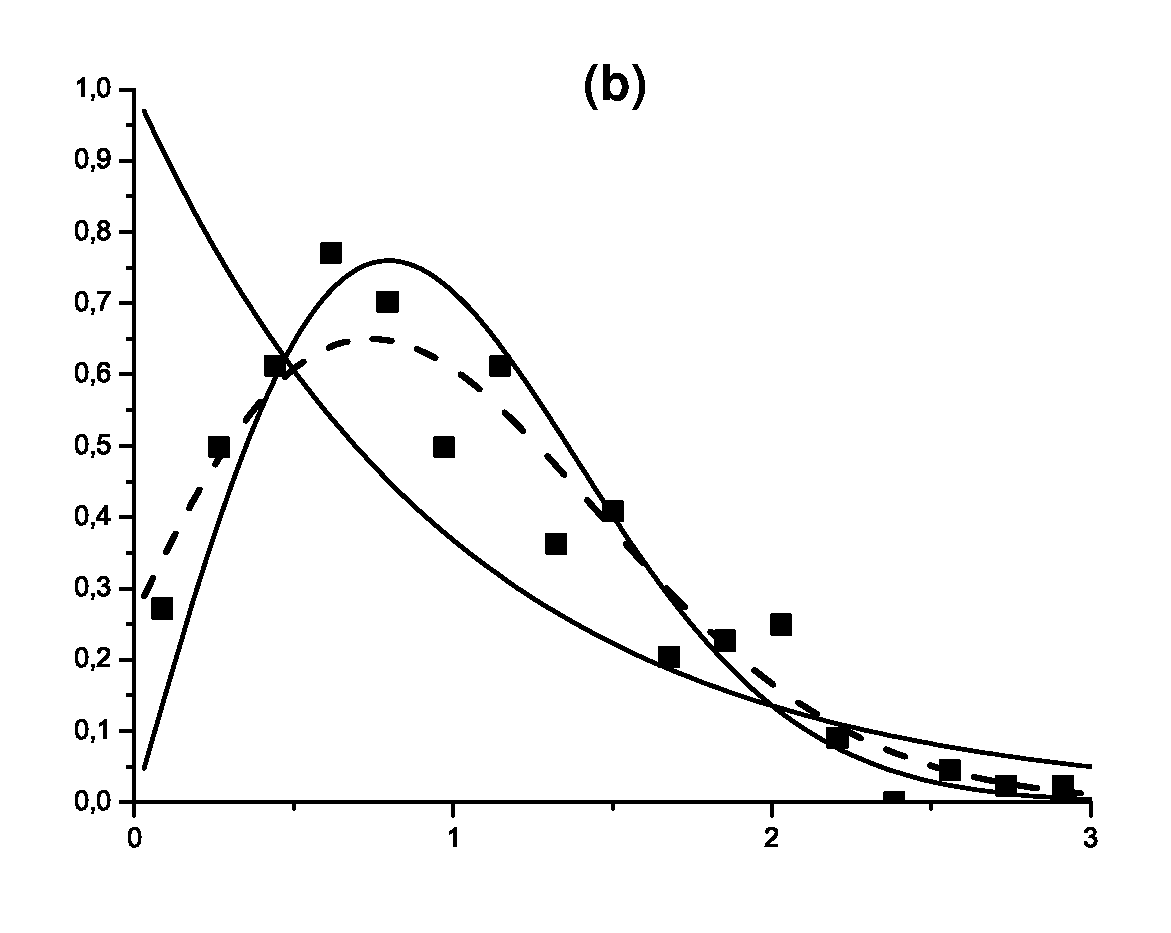}
\caption{\label{p_d5} FNNS for the $D_5$ potential (\ref{u_d5}) for
$E<E_{cr}$(a) and $E>E_{cr}$(b). Points represent numerical data,
solid lines --- Poisson (\ref{poisson}) and Wigner (\ref{wigner})
distributions, dashed line --- the best fit by the
Berry-Robnik-Bogomolny (\ref{brb}) distribution.}
\end{figure}

Let us consider the simplest potential realizing the mixed state ---
two-well potential of lower umbilic catastrophe $D_5$ (\ref{u_d5}).
In order to describe the statistical properties of the corresponding
energy spectrum, let us try to use the Berry-Robnik-Bogomolny
distribution (\ref{brb}), where the $\rho$ parameter equals the
relative phase volume occupied by regular trajectories. In the case
of $D_5$ potential $\rho\sim1$ for $E<E_{cr}$ and $\rho\sim1/3$ for
$E>E_{cr}$, which qualitatively agree with numerical data for energy
levels statistics Fig.\ref{p_d5}

The Berry-Robnik-Bogomolny distribution function by its construction
describes the energy level fluctuations in the absence of
interaction between the regular and chaotic components. Therefore
the best agreement between the Berry-Robnik-Bogomolny spectral
fluctuations theory and experimental data we should expect exactly
in the mixed state spectra, where interaction of chaotic and regulat
states each with other is additionally suppressed by the energy
barrier separating the corresponding local minima. In practice
however such agreement is never observed for several reasons.

\begin{figure}
\includegraphics[width=0.5\textwidth,draft=false]{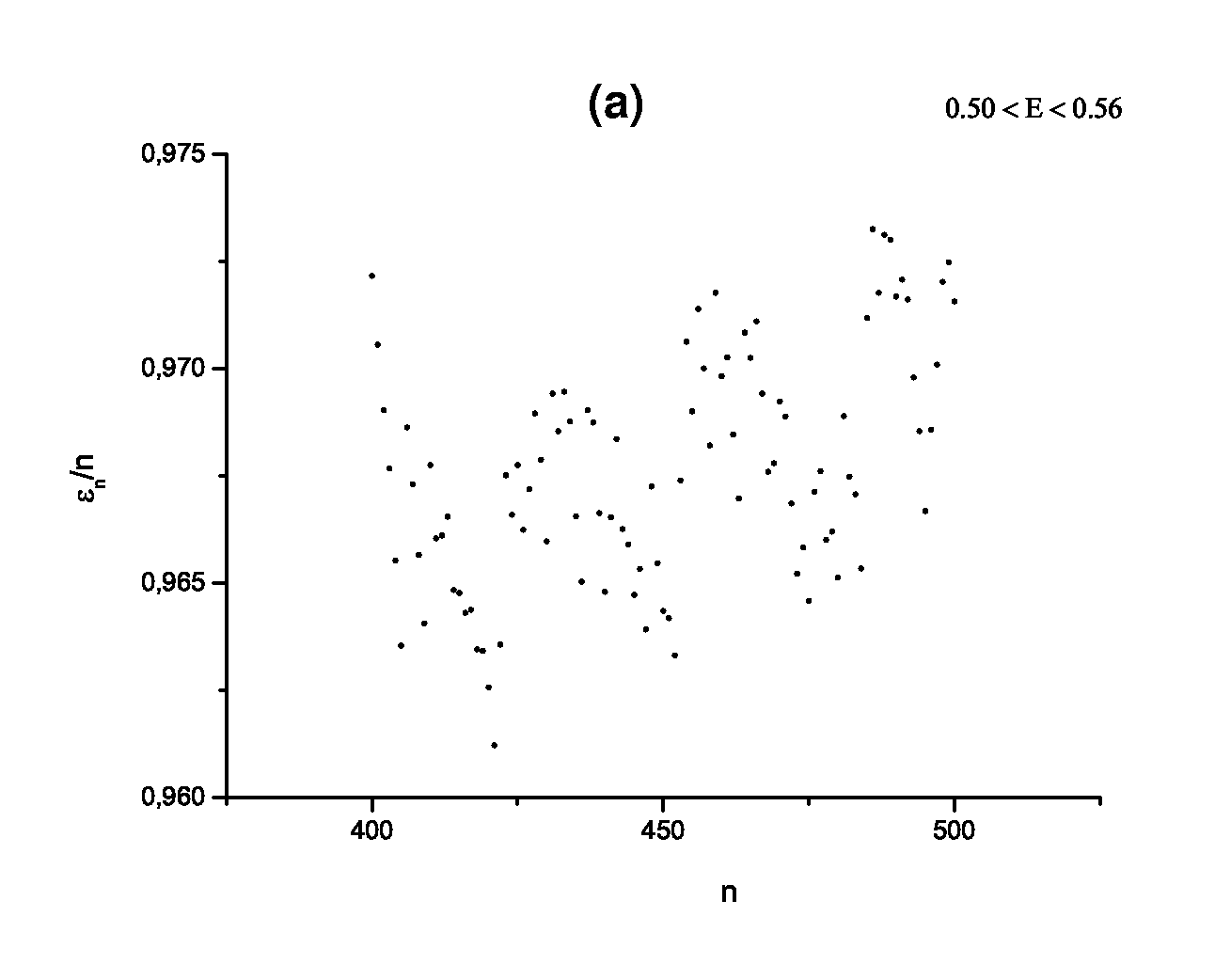}
\includegraphics[width=0.5\textwidth,draft=false]{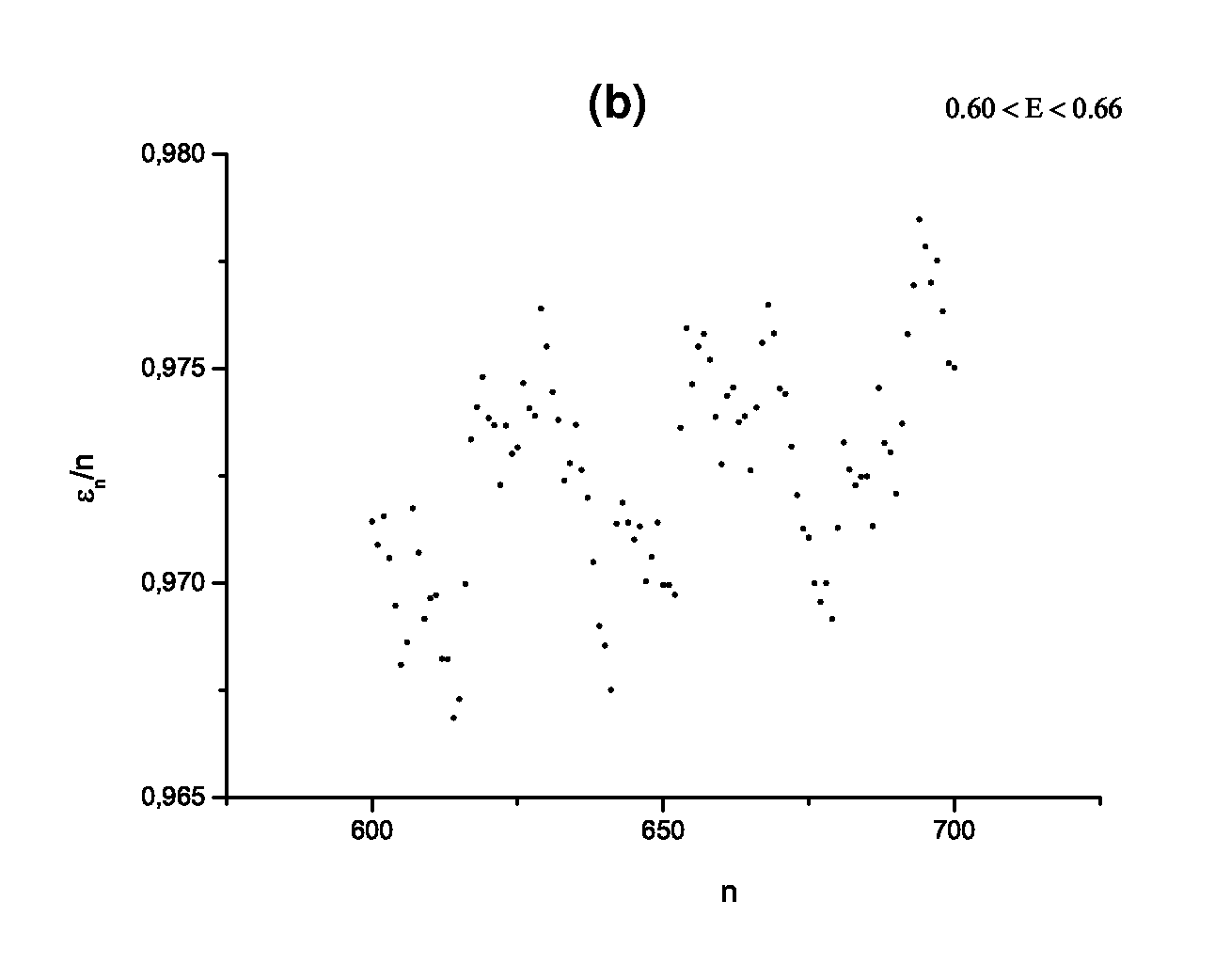}
\includegraphics[width=0.5\textwidth,draft=false]{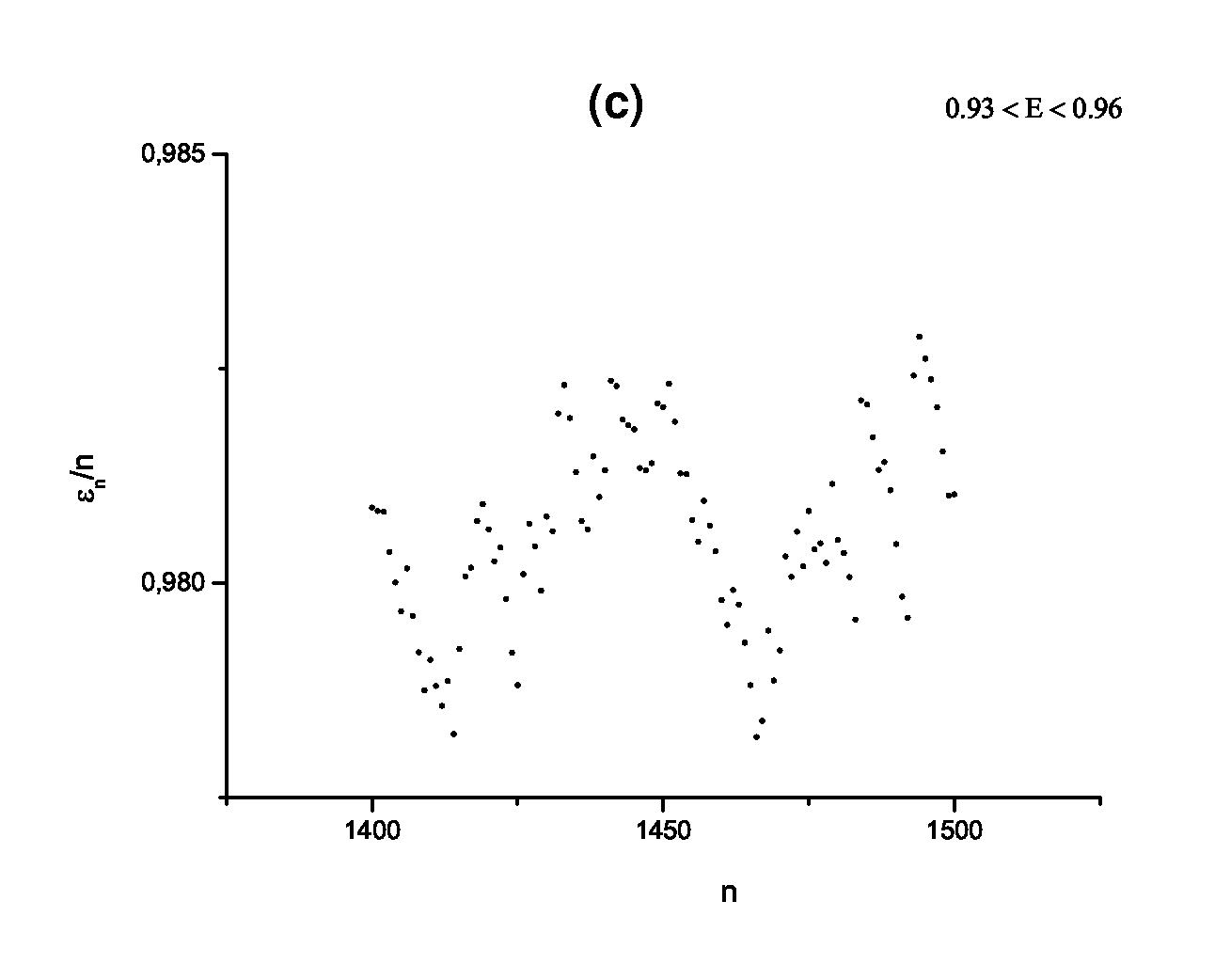}
\includegraphics[width=0.5\textwidth,draft=false]{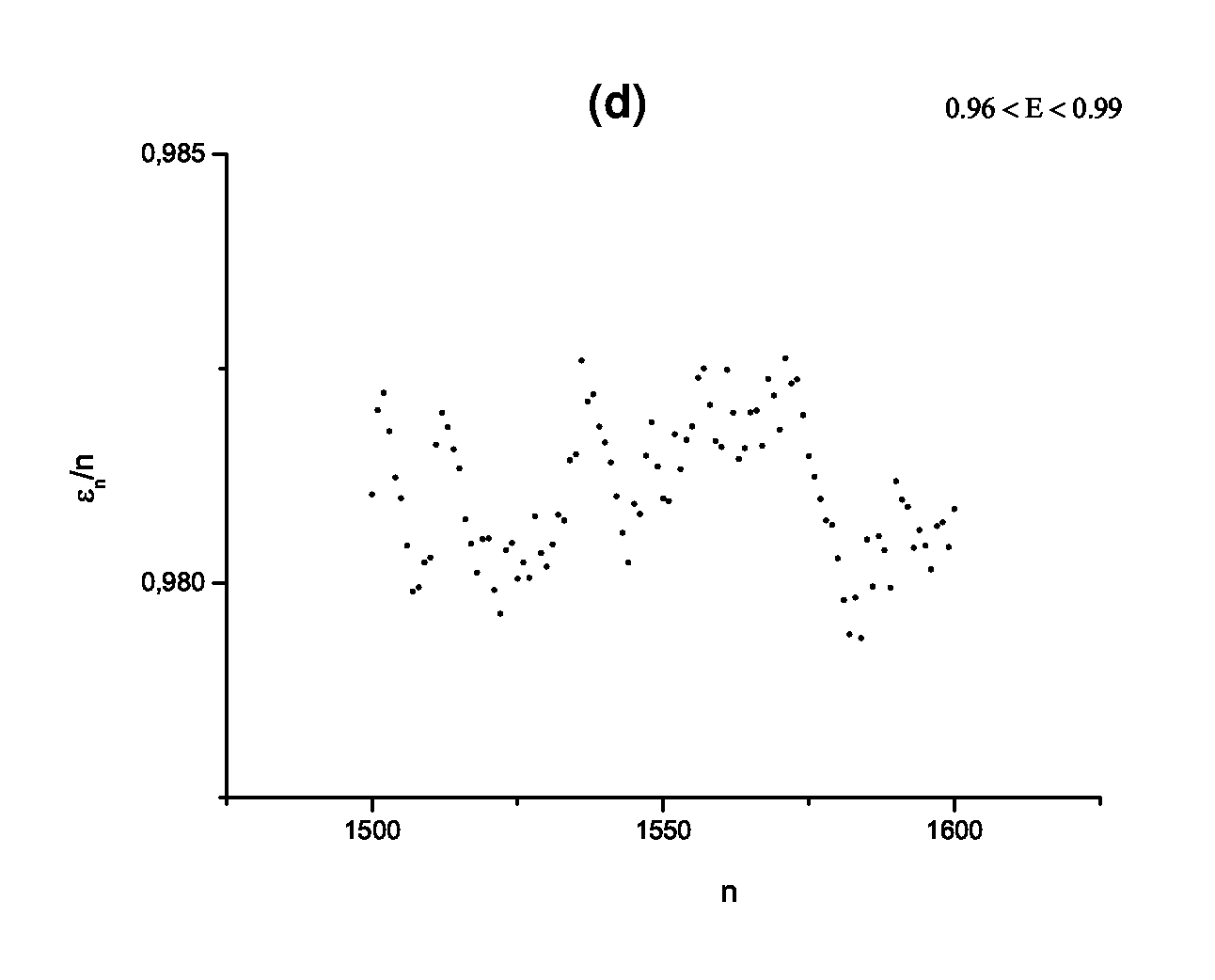}
\caption{\label{d5_unfold}Accuracy estimation for the unfolding with
(\ref{ne_d5}) in the spectrum of lower umbilic catastrophe $D_5$
potential (\ref{u_d5}) for $0.50<E<0.56$(a), $0.60<E<0.66$(b),
$0.93<E<0.96$(c), $0.93<E<0.99$(d).}
\end{figure}

The first reason is not connected immediately with the mixed state
properties and lies in technical difficulties of calculation of long
spectral series in potentials with several local minima. In practice
one usually should be happy with a series of several hundred levels
at most, which is definitely insufficient to obtain accurate
distribution functions for spectral fluctuations. Another difficulty
of a technical nature is connected with the fact that the
multiplicity of generic smooth potentials does not allow any
systematic semiclassical analysis, because in particular there are
no sufficiently precise semiclassical approximations for the
smoothed component of staircase states number function, required for
the standard unfolding procedure (\ref{unfold}). Therefore in order
to unfold the spectra we should be happy with the zero semiclassical
approximation, i.e. usual Thomas-Fermi approximation, which for even
states in the $D_5$ potential takes the following form:
\begin{equation}\label{ne_d5}\begin{array}{c}
\bar{n}(E)=\bar{n}_R(E)+\bar{n}_C(E)\\
\bar{n}_{R,C}(E)\approx\frac{1}{3\pi\hbar^2}\int\limits_{\sqrt{2(1-\sqrt
E)}}^{\sqrt{2(1+\sqrt
E)}}\frac{dx}{\sqrt{x\pm2}}\left[E-\left(\frac{x^2}{2}-1\right)^2\right]^{\frac32},
\end{array}\end{equation}
where $\bar{n}_R(E)$ and $\bar{n}_C(E)$ represent partial
contributions from regular and chaotic minima respectively. Such a
simple approximation nevertheless allows us to perform the spectrum
unfolding with maximum errors of order $1\%$ in the whole energy
interval of interest (see Fig.\ref{d5_unfold}).

Therefore in potentials with multiple local minima, the
smoothed semiclassical state number function can be calculated as
the sum of partial contributions from each local minimum. In the
case of mixed state it gives a convenient way to estimate the
relative density of states of regular and chaotic type.

The second reason for disagreement between Berry-Robnik-Bogomolny
distribution and actual numerical data is common for all potential
systems (excluding uniform potentials like the coupled quartic
oscillator one (\ref{cqo})), and it consists of the fact that in
smooth potential systems the classical chaoticity measure
essentially changes with energy; therefore upper and lower levels in
sufficiently long spectral series will correspond to different
relations of regular and chaotic components, while all known
theoretical distribution functions assume the relation to be
constant. It forces us to consider spectral series that are
sufficiently narrow in energy, and this makes levels statistics even
poorer.

Finally, the specific feature of $D_5$ potential, which it shares
with $D_7$, Barbanis and H\'enon--Heilis potential, consists of the
fact that in all the above mentioned potentials the discrete energy
spectrum, strictly speaking, is absent. It is due to the fact that
all those models allow infinite motion for all energies.
Nevertheless many researchers discuss spectral fluctuations in such
systems, implying the spectrum of quasistationary states, localized
in the corresponding potential well and extremely slowly decaying
due to tunneling into the continuous spectrum. Any numerical
calculation of such states practically contains implicit
reformulation of the original model in order to make it more correct
from the point of view of quantum mechanics. The quasi-stationarity
of such states becomes especially manifest near the saddle energies
--- the most interesting energy region for investigation of the
mixed state.

\begin{figure}
\includegraphics[width=0.5\textwidth,draft=false]{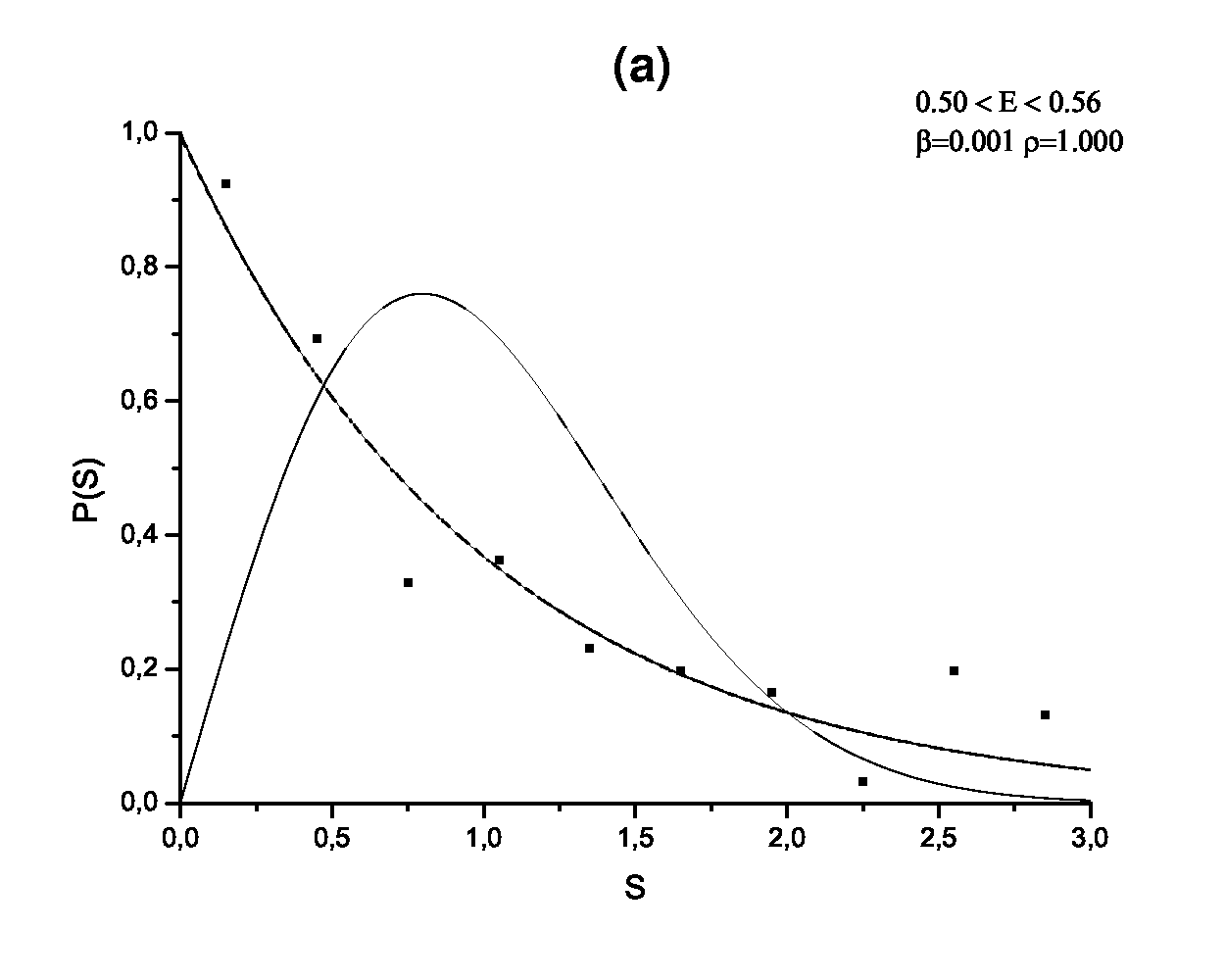}
\includegraphics[width=0.5\textwidth,draft=false]{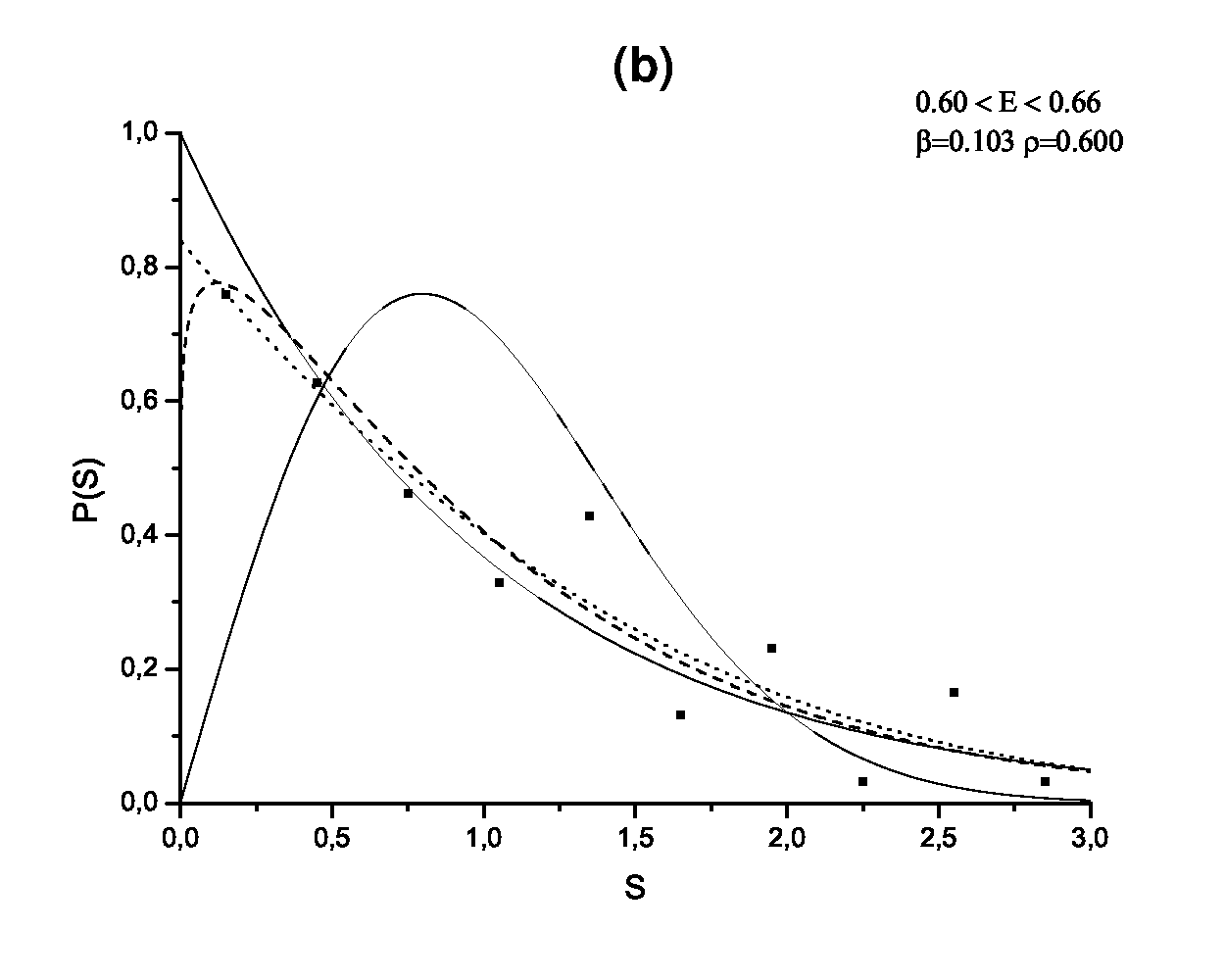}
\includegraphics[width=0.5\textwidth,draft=false]{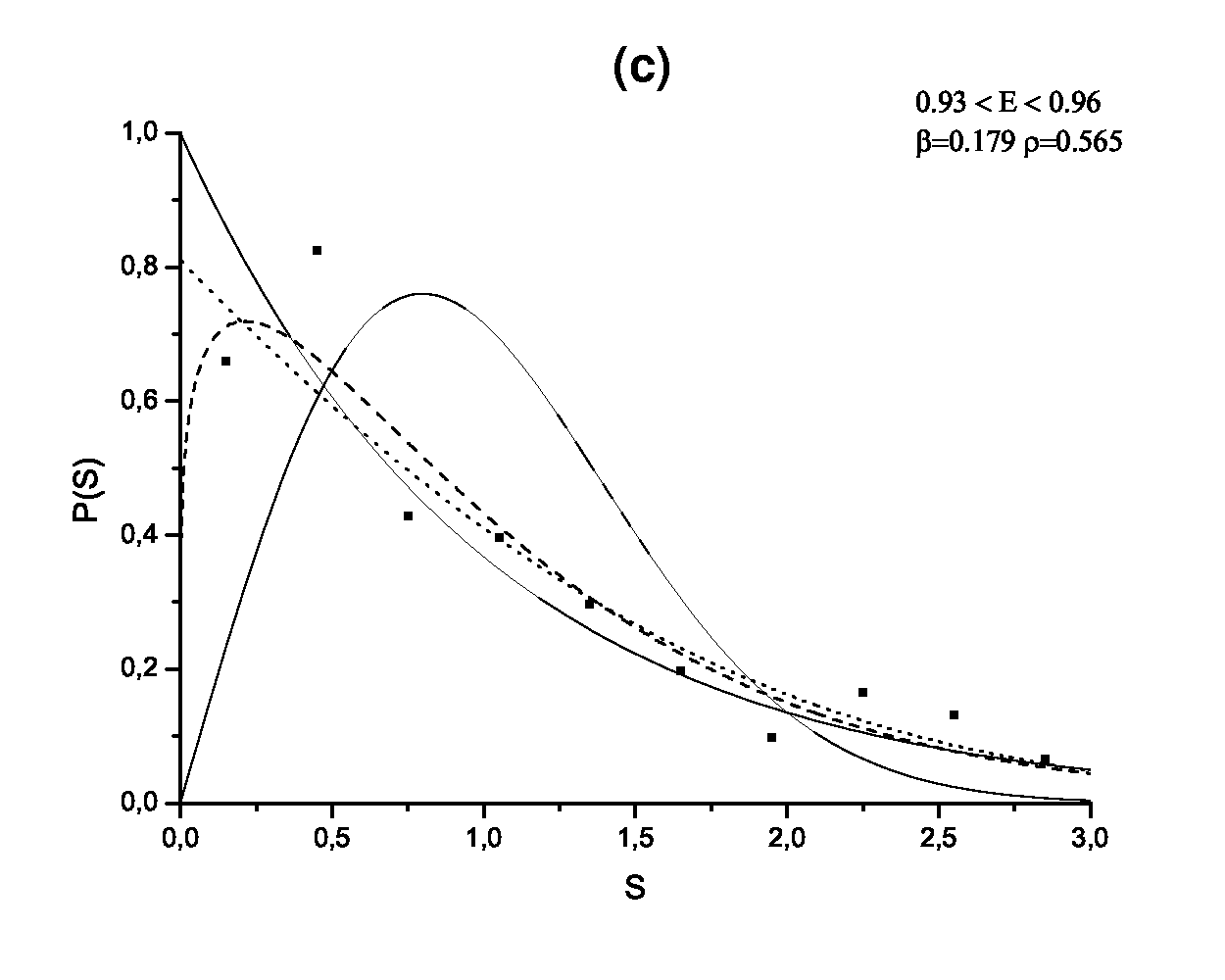}
\includegraphics[width=0.5\textwidth,draft=false]{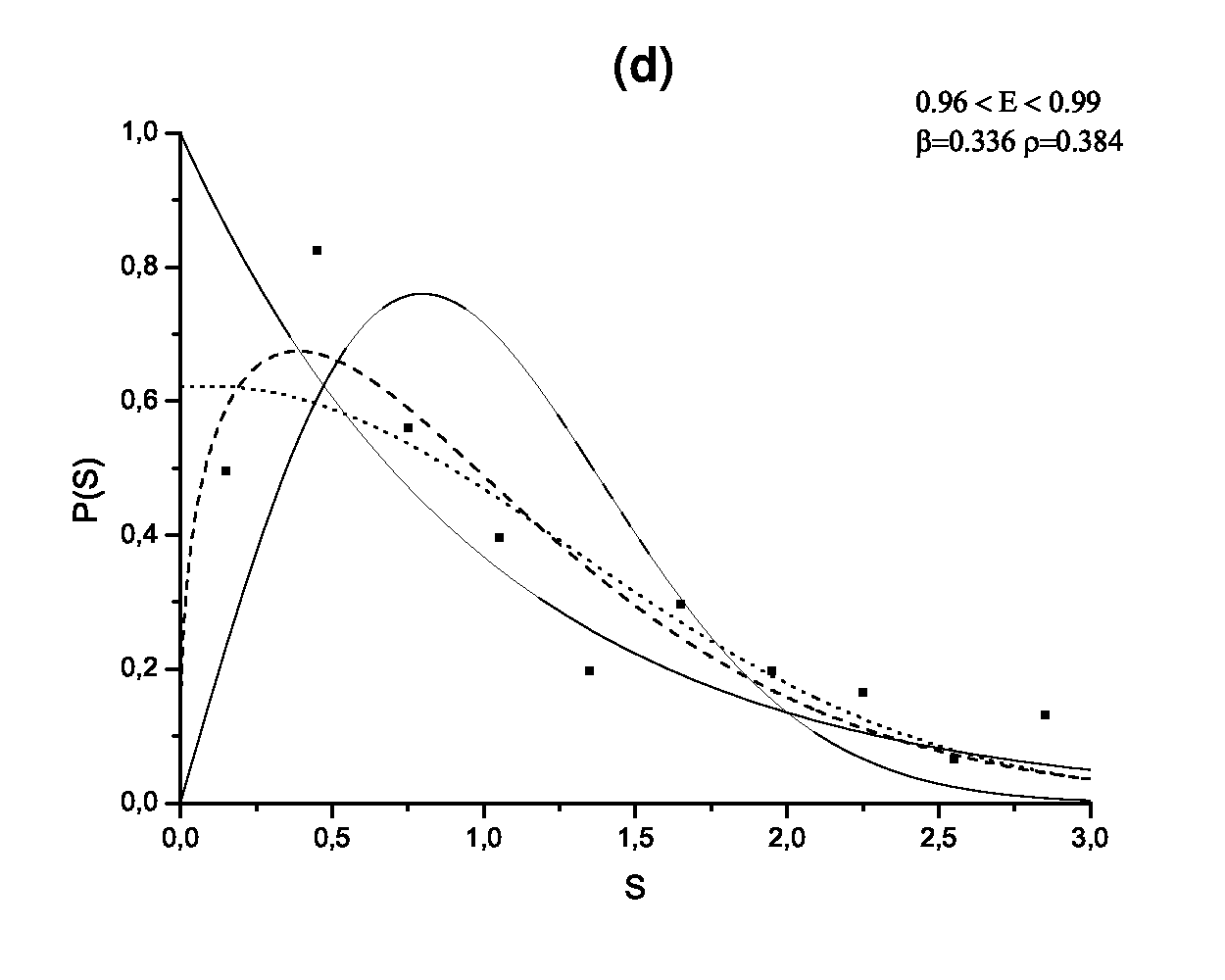}
\caption{\label{d5_p} FNNS for lower umbilic catastrophe $D_5$
potential (\ref{u_d5}) for $0.50<E<0.56$(a), $0.60<E<0.66$(b),
$0.93<E<0.96$(c), $0.93<E<0.99$(d). Points represent numerical data,
solid lines --- Poisson (\ref{poisson}) and Wigner (\ref{wigner})
distributions, dashed and dotted lines --- the best fits by the
Brody (\ref{brody}) and the Berry-Robnik-Bogomolny (\ref{brb})
distributions respectively.}
\end{figure}

\begin{figure}
\includegraphics[width=0.5\textwidth,draft=false]{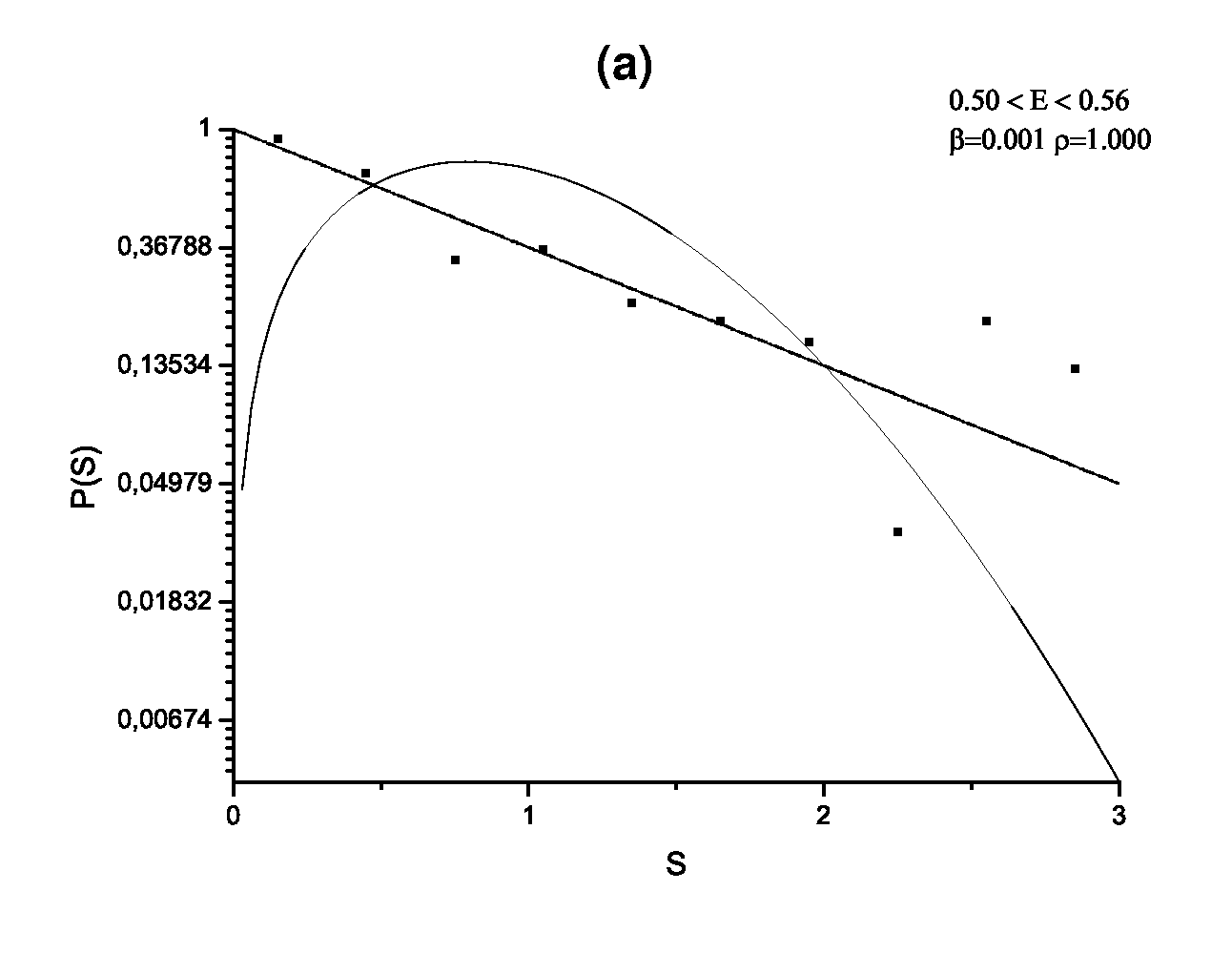}
\includegraphics[width=0.5\textwidth,draft=false]{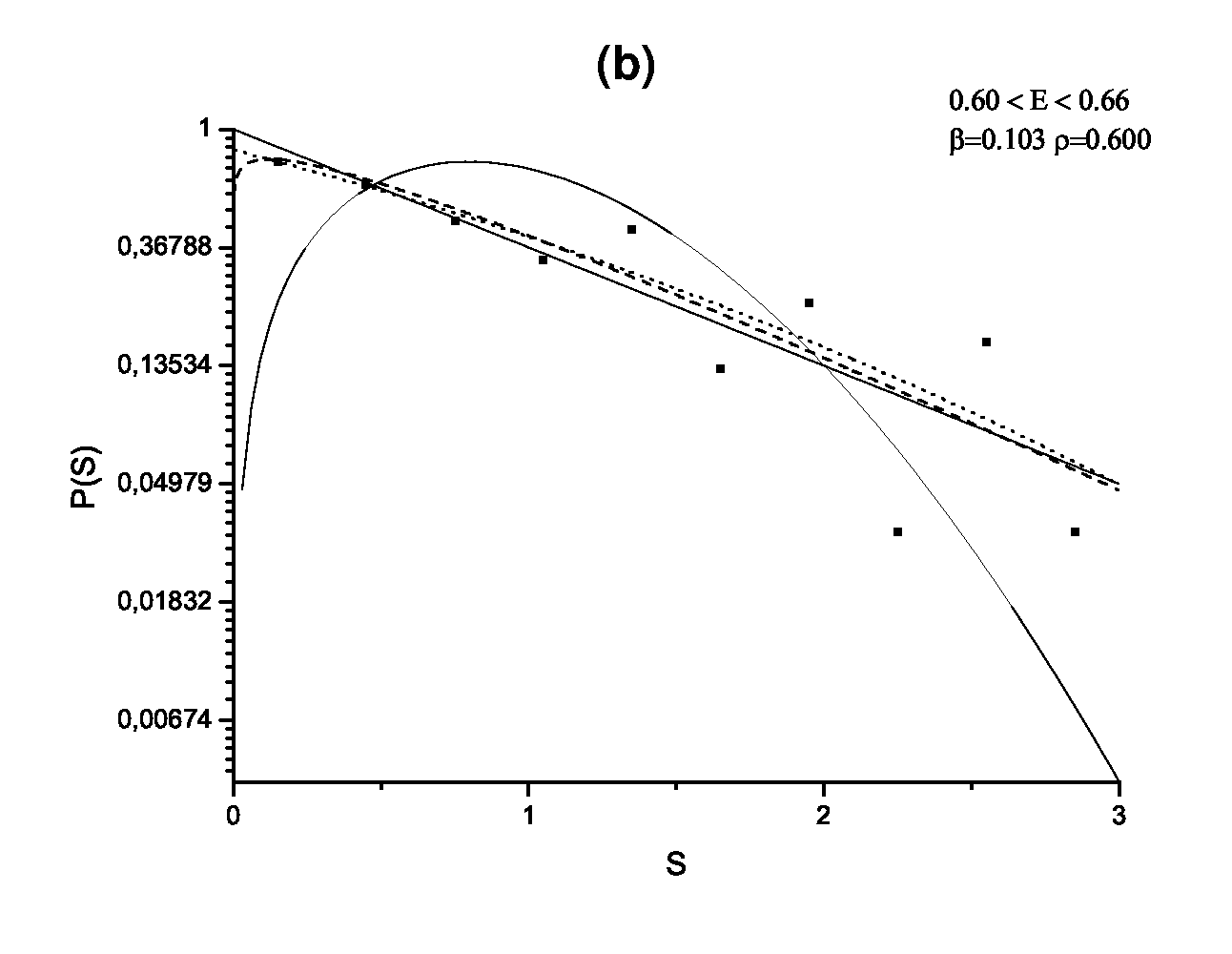}
\includegraphics[width=0.5\textwidth,draft=false]{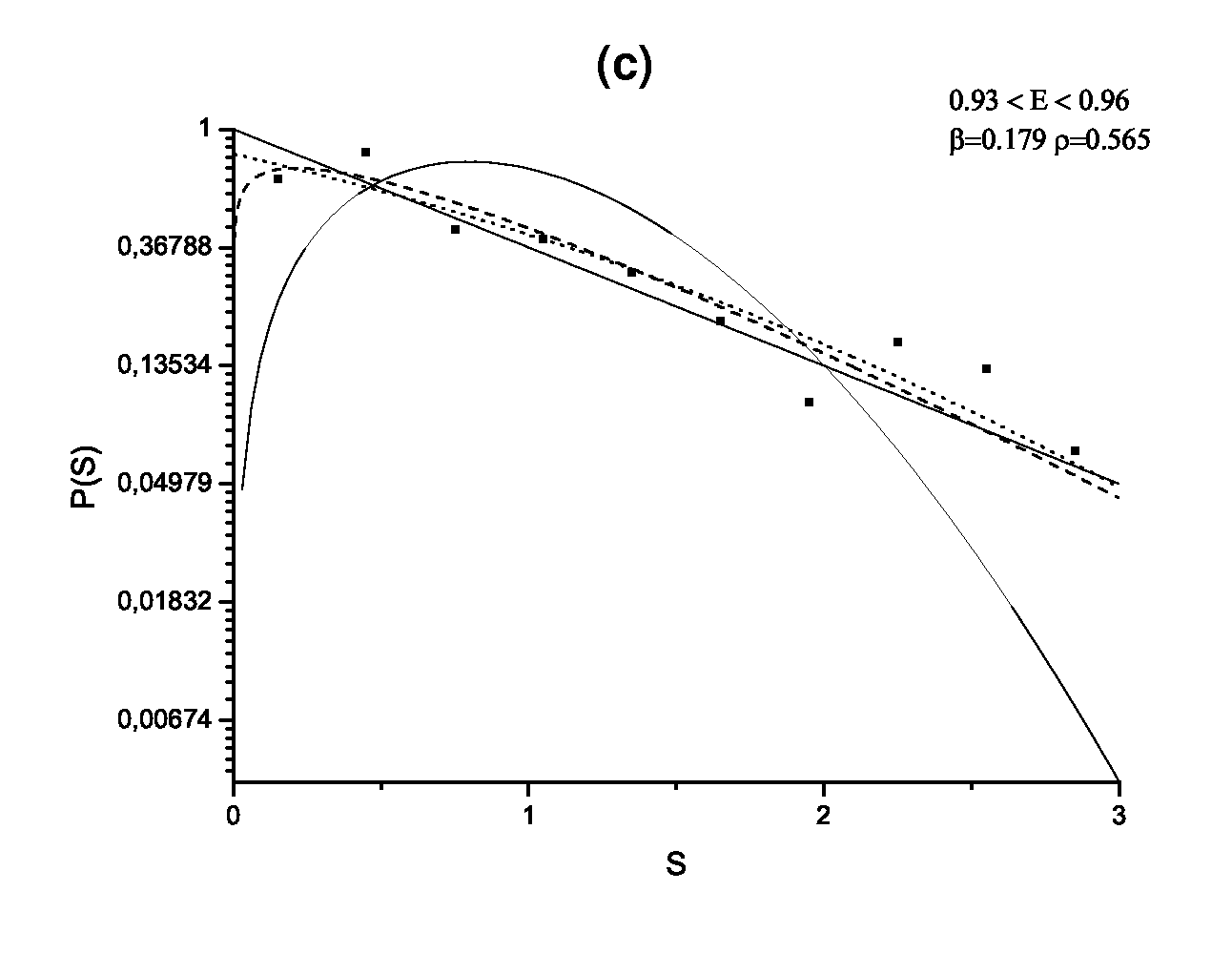}
\includegraphics[width=0.5\textwidth,draft=false]{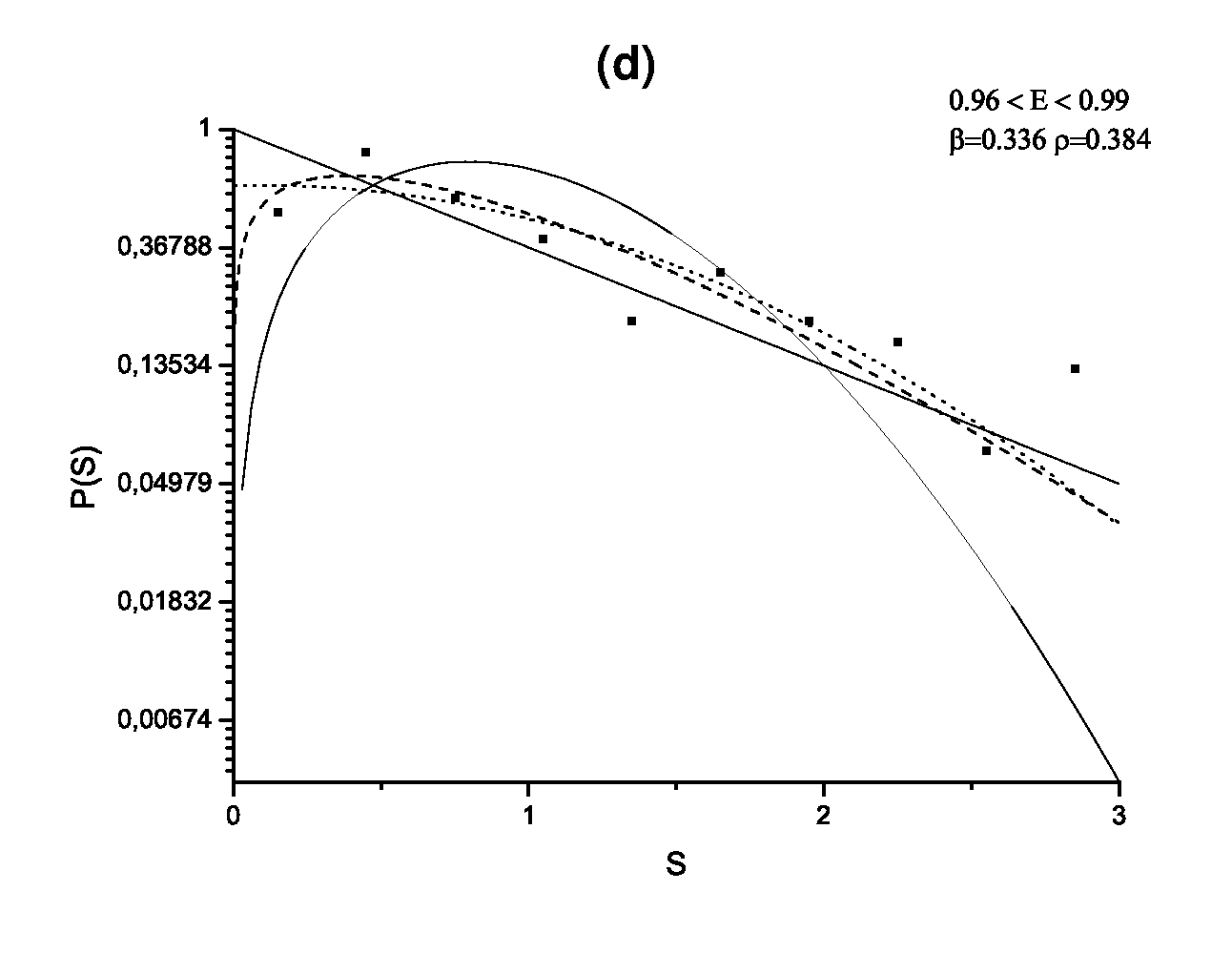}
\caption{\label{d5_pl} The same as on Fig.\ref{d5_p} but in
logarithmic scale.}
\end{figure}

\begin{figure}
\includegraphics[width=0.5\textwidth,draft=false]{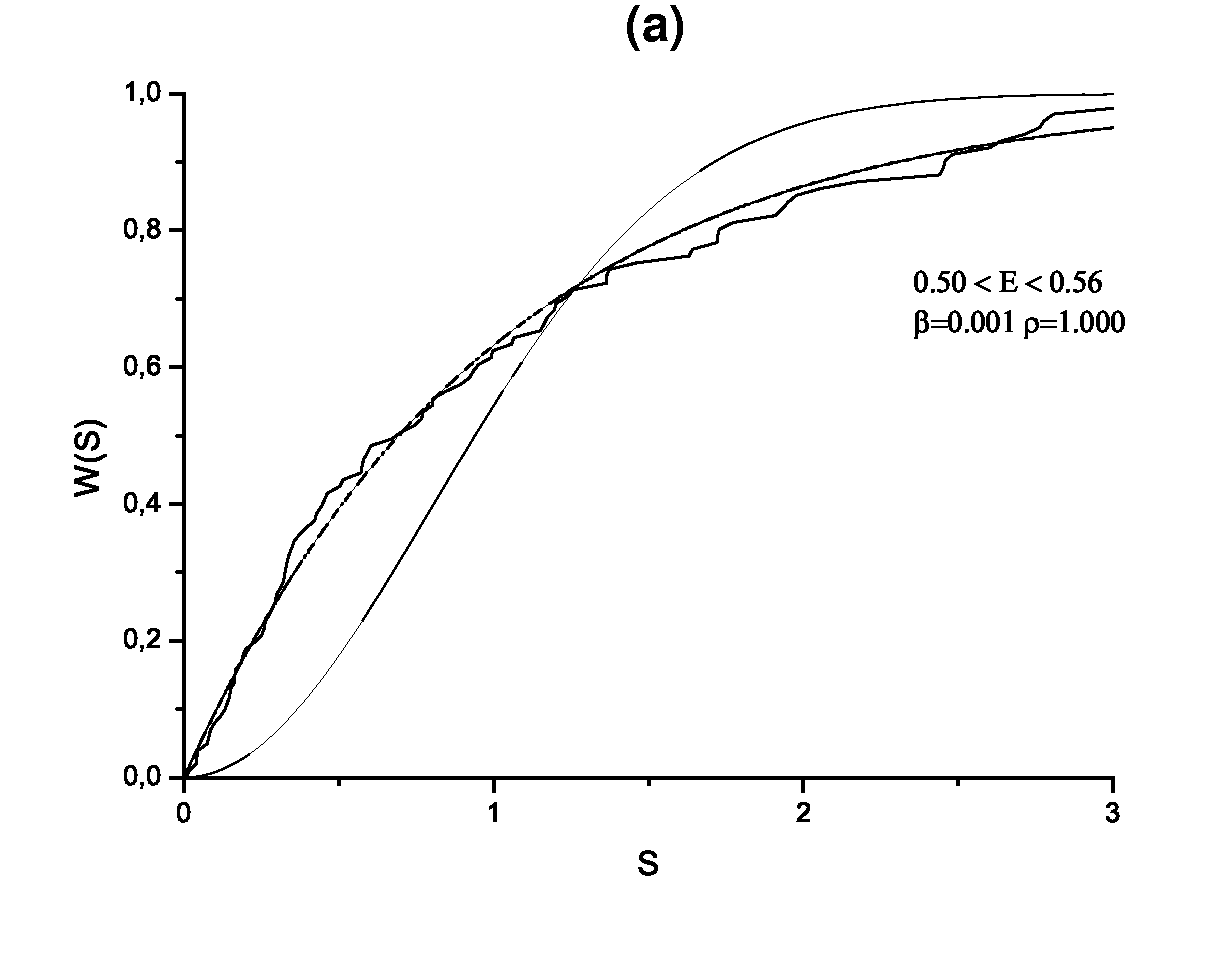}
\includegraphics[width=0.5\textwidth,draft=false]{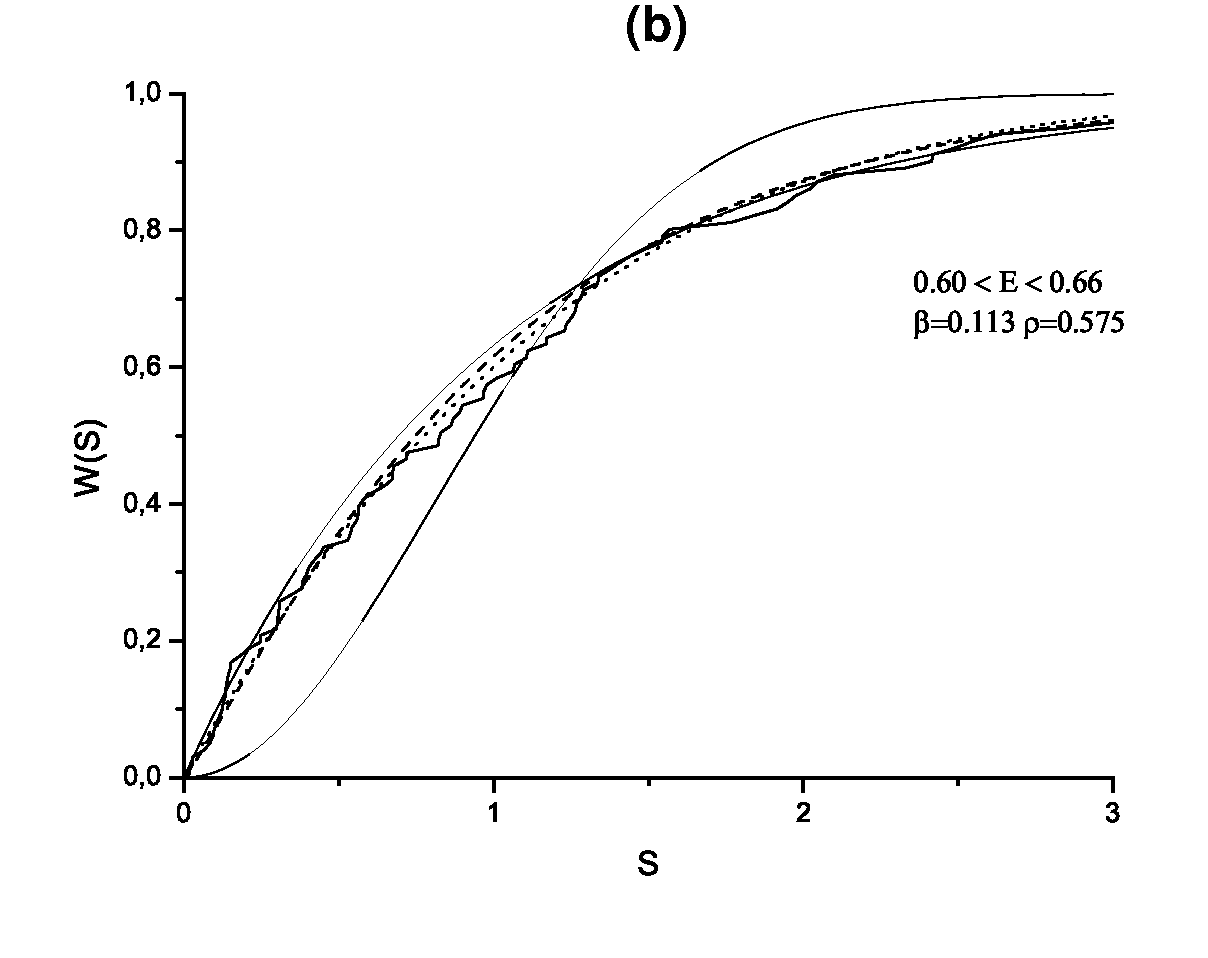}
\includegraphics[width=0.5\textwidth,draft=false]{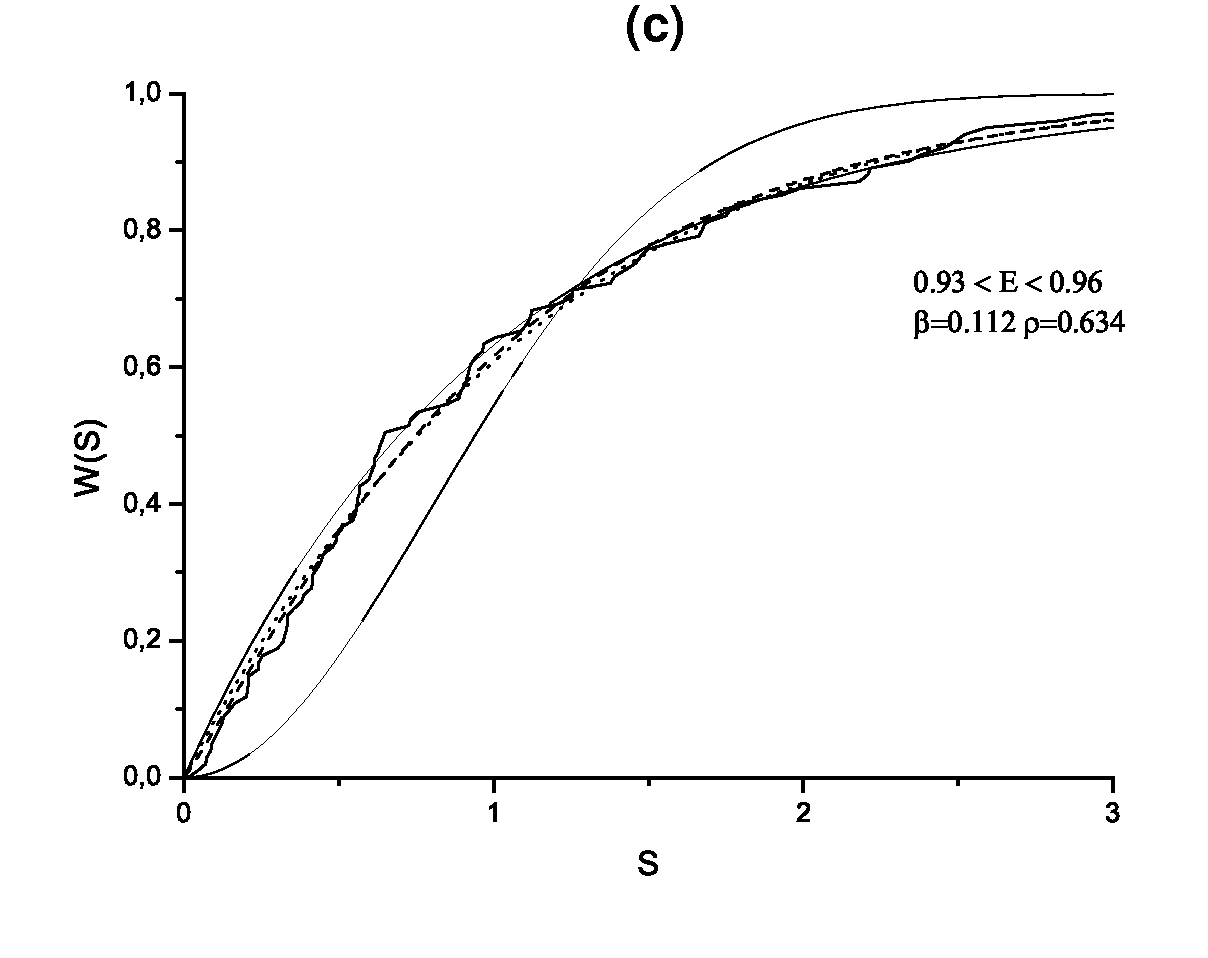}
\includegraphics[width=0.5\textwidth,draft=false]{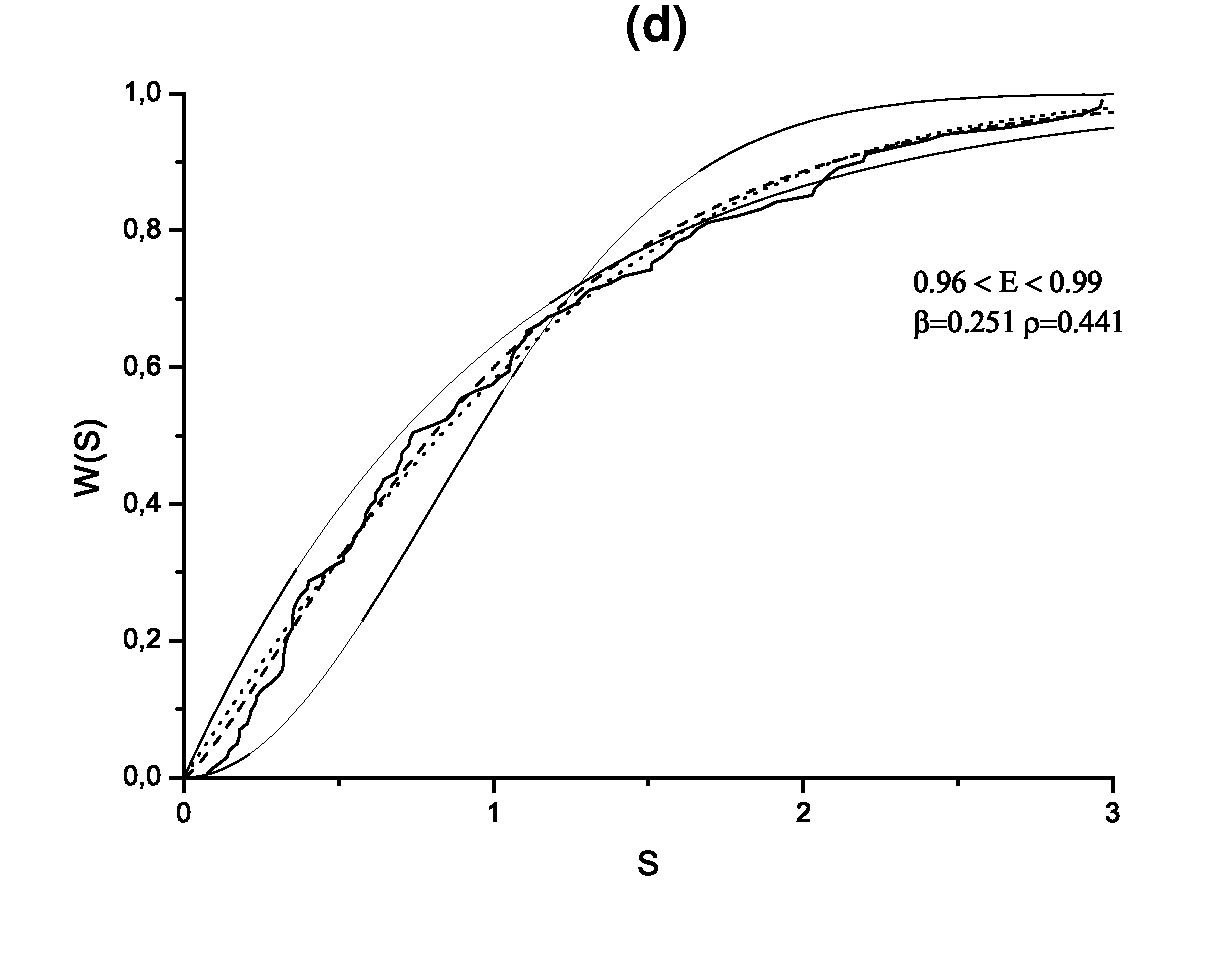}
\caption{\label{d5_w} Cumulative FNNS for lower umbilic catastrophe
$D_5$ potential (\ref{u_d5}) for $0.50<E<0.56$(a), $0.60<E<0.66$(b),
$0.93<E<0.96$(c), $0.93<E<0.99$(d). Points represent numerical data,
solid lines --- Poisson and Wigner (\ref{w_pw}) distributions,
dashed and dotted lines --- the best fits by by Brody and
Berry-Robnik-Bogomolny (\ref{w_brb}) distributions respectively.}
\end{figure}

\begin{figure}
\includegraphics[width=0.5\textwidth,draft=false]{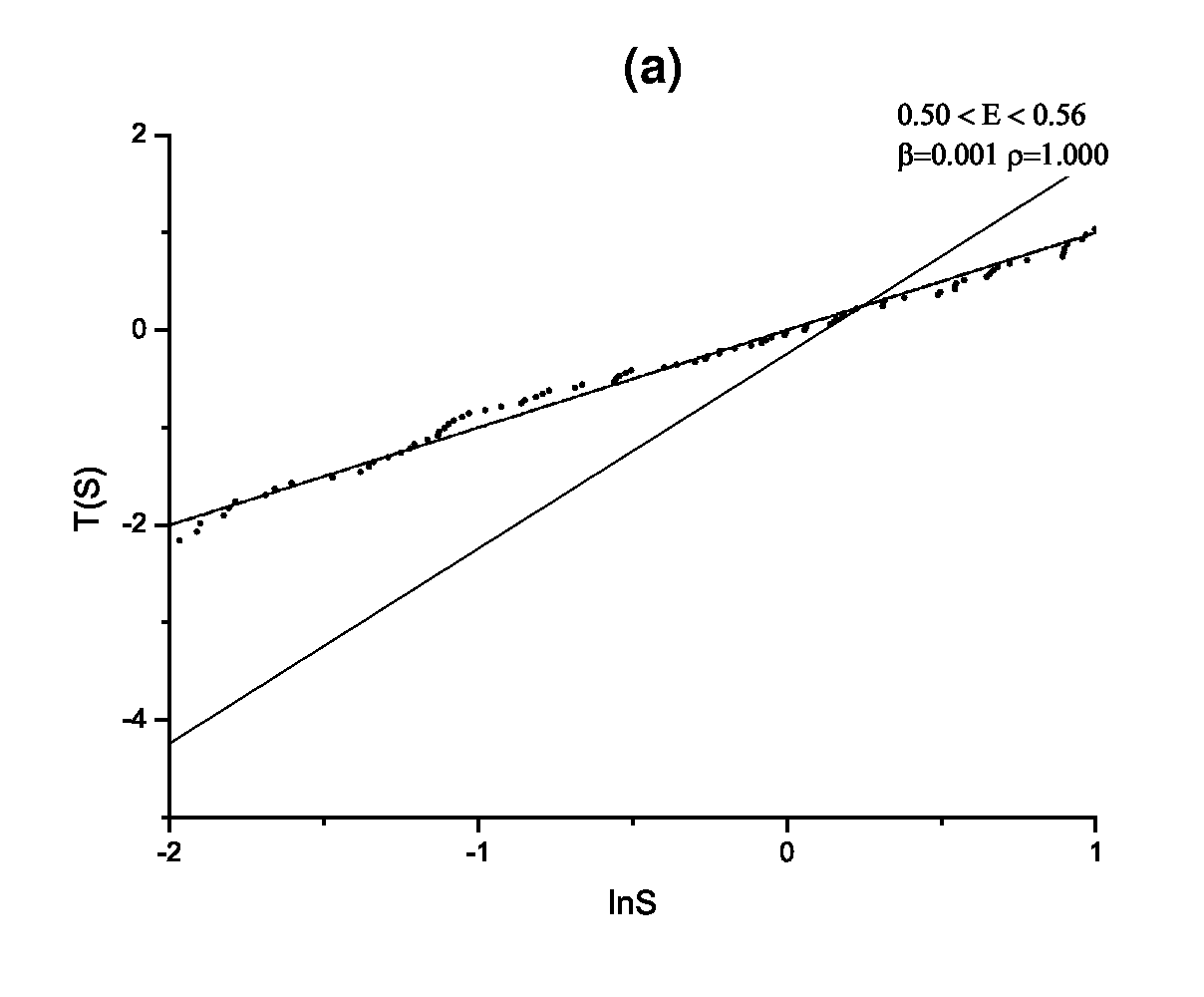}
\includegraphics[width=0.5\textwidth,draft=false]{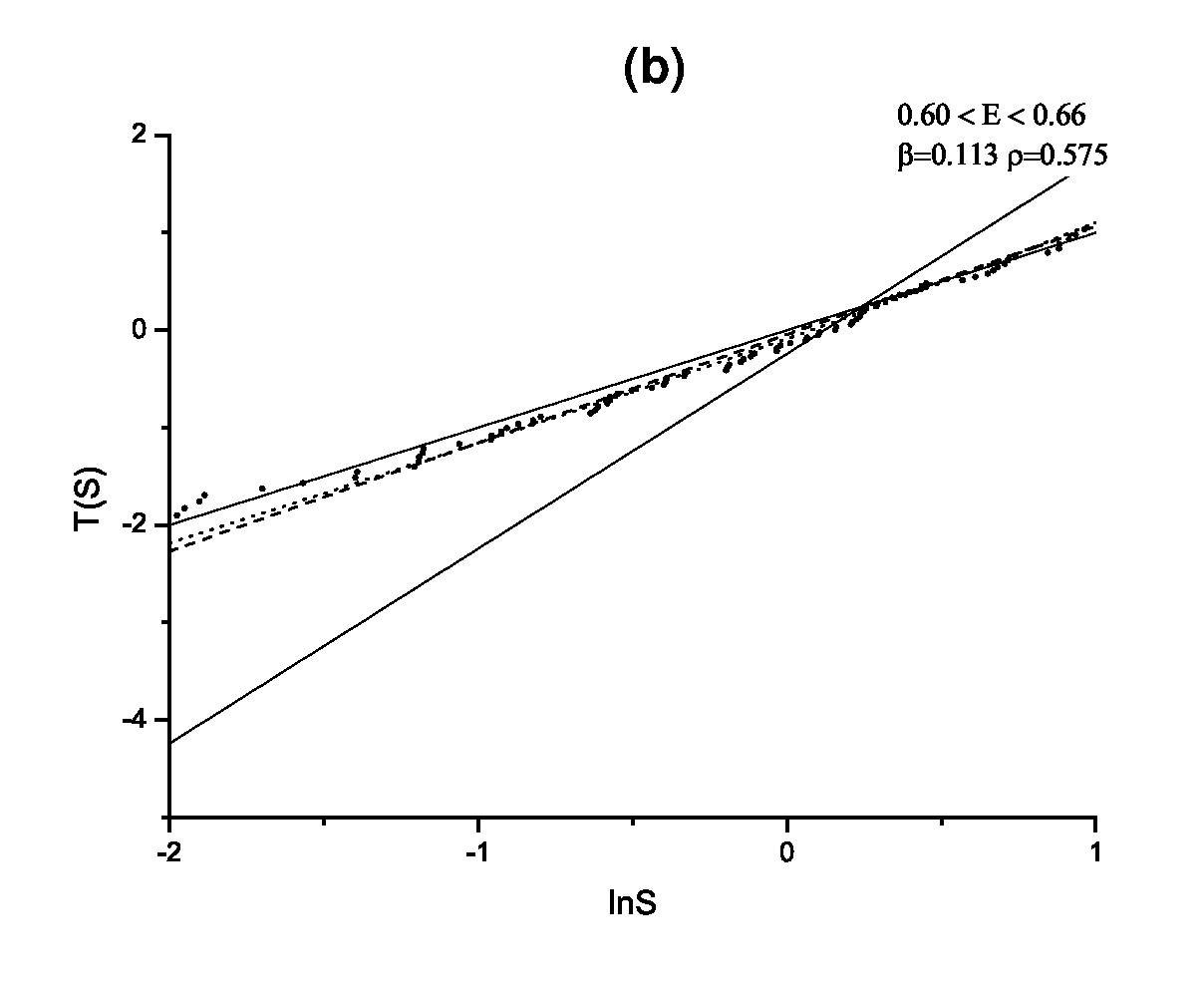}
\includegraphics[width=0.5\textwidth,draft=false]{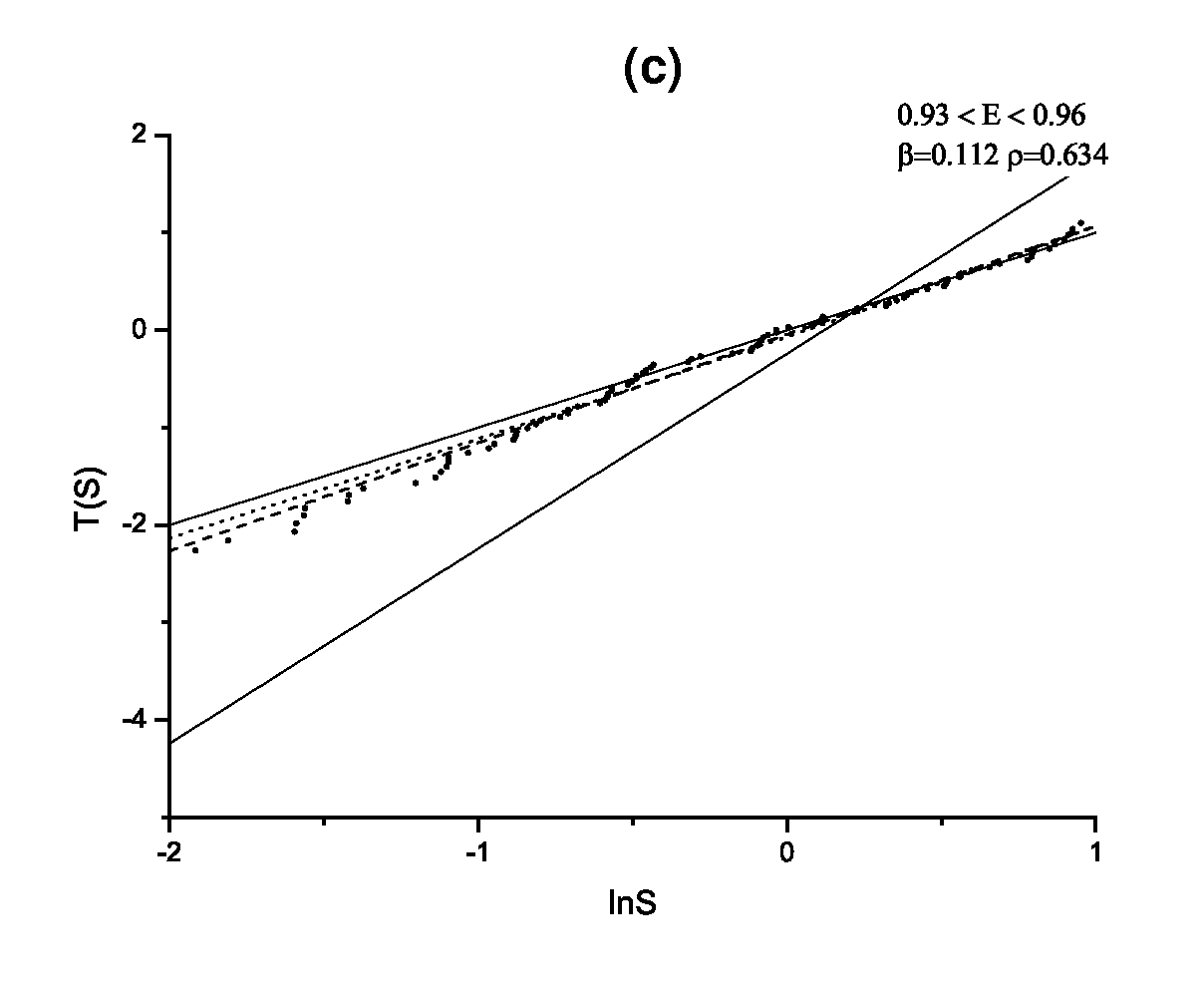}
\includegraphics[width=0.5\textwidth,draft=false]{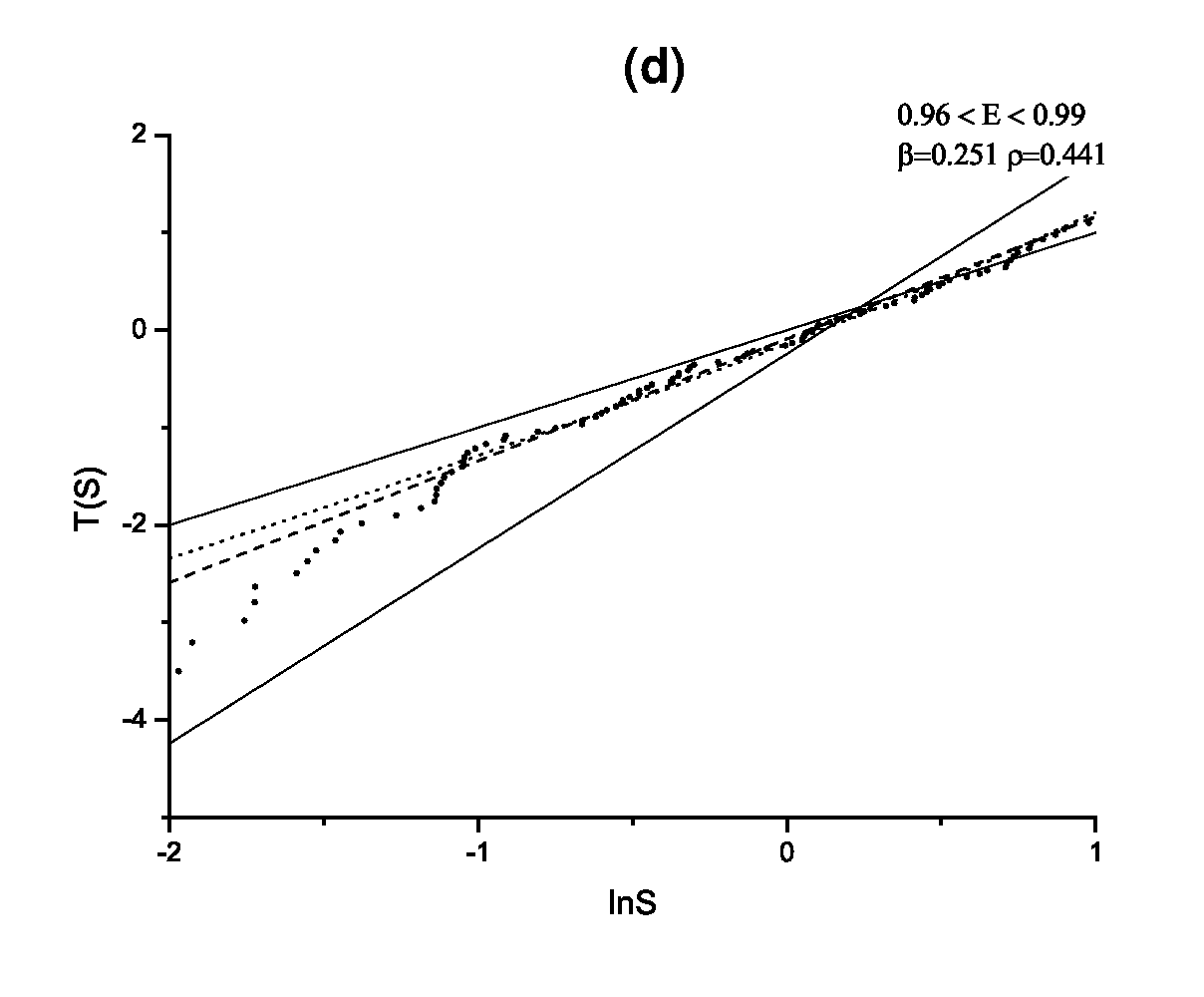}
\caption{\label{d5_t} Cumulative FNNS in the $T$-representation for
lower umbilic catastrophe $D_5$ potential (\ref{u_d5}) for
$0.50<E<0.56$(a), $0.60<E<0.66$(b), $0.93<E<0.96$(c),
$0.93<E<0.99$(d). Points represent numerical data, solid lines ---
Poisson and Wigner (\ref{t_pw}) distributions, dashed and dotted
lines --- the best fits by the Brody (\ref{t_b}) and the
Berry-Robnik-Bogomolny (\ref{t_brb}) distributions respectively.}
\end{figure}

\begin{figure}
\includegraphics[width=0.5\textwidth,draft=false]{d5_3.png}
\includegraphics[width=0.5\textwidth,draft=false]{d5_2.png}
\includegraphics[width=0.5\textwidth,draft=false]{d5_1.png}
\includegraphics[width=0.5\textwidth,draft=false]{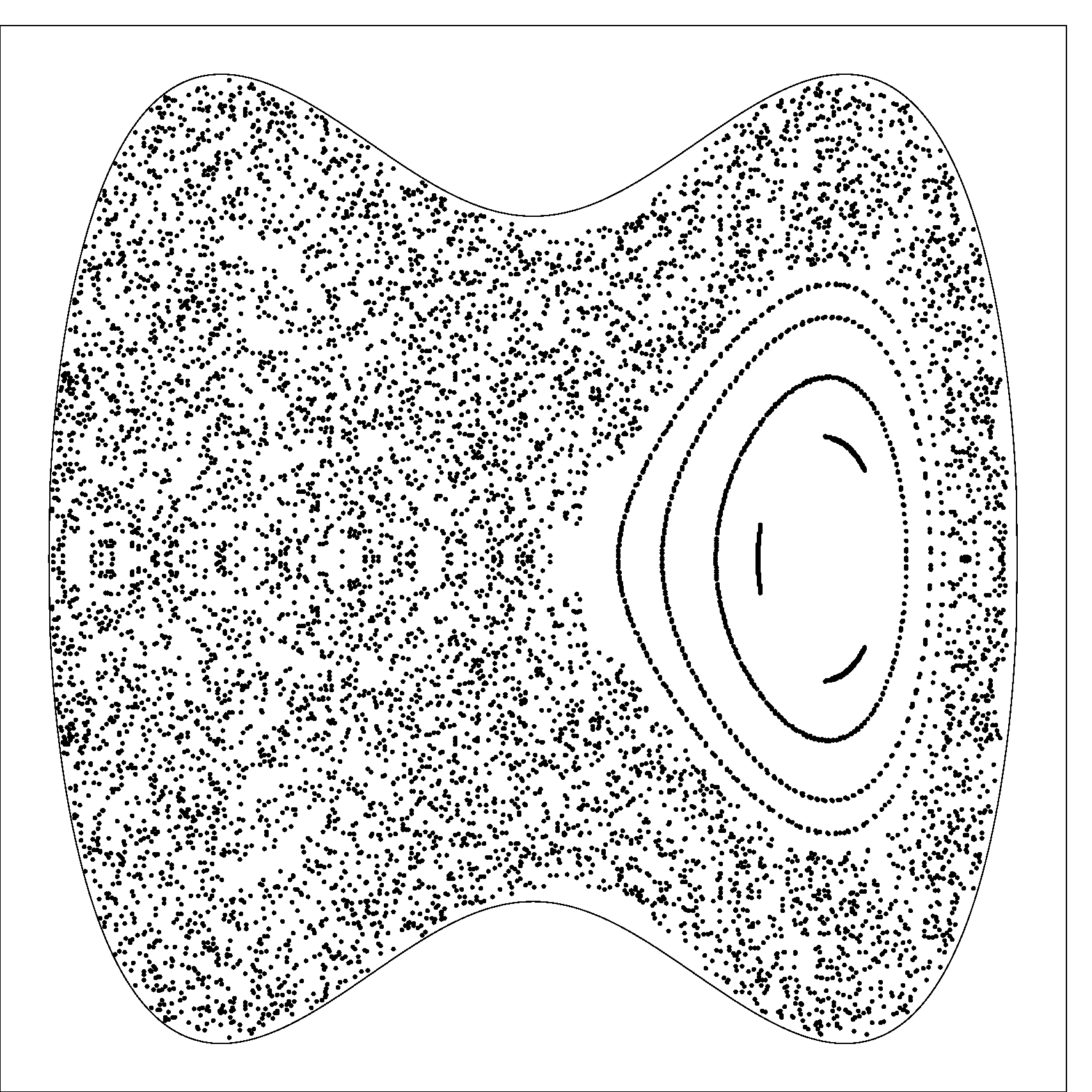}
\caption{\label{d5_pss} Poincar\'e surfaces of section for lower
umbilic catastrophe $D_5$ potential (\ref{u_d5}) for $E=0.5$(a),
$E=0.6$(b), $E=0.9$(c), $E\lesssim1$(d). Solid lines limit
classically allowed region of the phase space.}
\end{figure}

The FNNS for $D_5$ potential obtained for narrow energy intervals
are presented in Fig.\ref{d5_p} in normal and in Fig.\ref{d5_pl} in
logarithmic scale, and the corresponding cumulative distributions
are presented in Fig.\ref{d5_w} in $W$-representation and on
Fig.\ref{d5_t} in $T$-representation. Regardless of the rather poor
statistics (only $100$ levels in each spectral series), we should
note quite good qualitative agreement between the fitting parameters
of the Brody and the Berry-Robnik-Bogomolny distributions on the one
hand and relative phase volumes of regular and chaotic regions in
classical phase space (see the Poincar\'e surfaces of section on
Fig.\ref{d5_pss}).

As we can see, the Brody distribution gives better agreement than
the Berry-Robnik-Bogomolny distribution with numerical FNNS
distributions in the $D_5$ potential. It is manifested both in the
accuracy for each energy interval separately and in the variation of
$\beta$ parameter with energy growth; in contrast to the $\rho$
parameter in the fitting by Berry-Robnik-Bogomolny distribution, it
grows monotonously with growth of chaoticity measure of classical
motion and coincides in order of magnitude with the relative volume
of the classical chaotic component.

The unsatisfactory applicability of Berry-Robnik-Bogomolny the
distribution for description of spectral fluctuations in the mixed
state is explained first of all by insufficient quasiclassicality of
the considered states. However a more important reason for such
disagreement of theory and experiment appears to be connected with
the specific features of energy spectra in the mixed state. It lies
in the fact that that many of the energy levels in the potentials
with multiple local minima actually appear to be very close to those
energy levels that correspond to each local minimum separately ---
neglecting the small influence of other minima. Those levels are in
their turn very well described by spectral series obtained in the
harmonic oscillator approximation, especially for not too high
energies. In the case of the mixed state especially close to the
harmonic oscillator levels appear to be those states that correspond
to minima with regular motion in the classical limit.

\begin{figure}
\includegraphics[width=0.9\textwidth,draft=false]{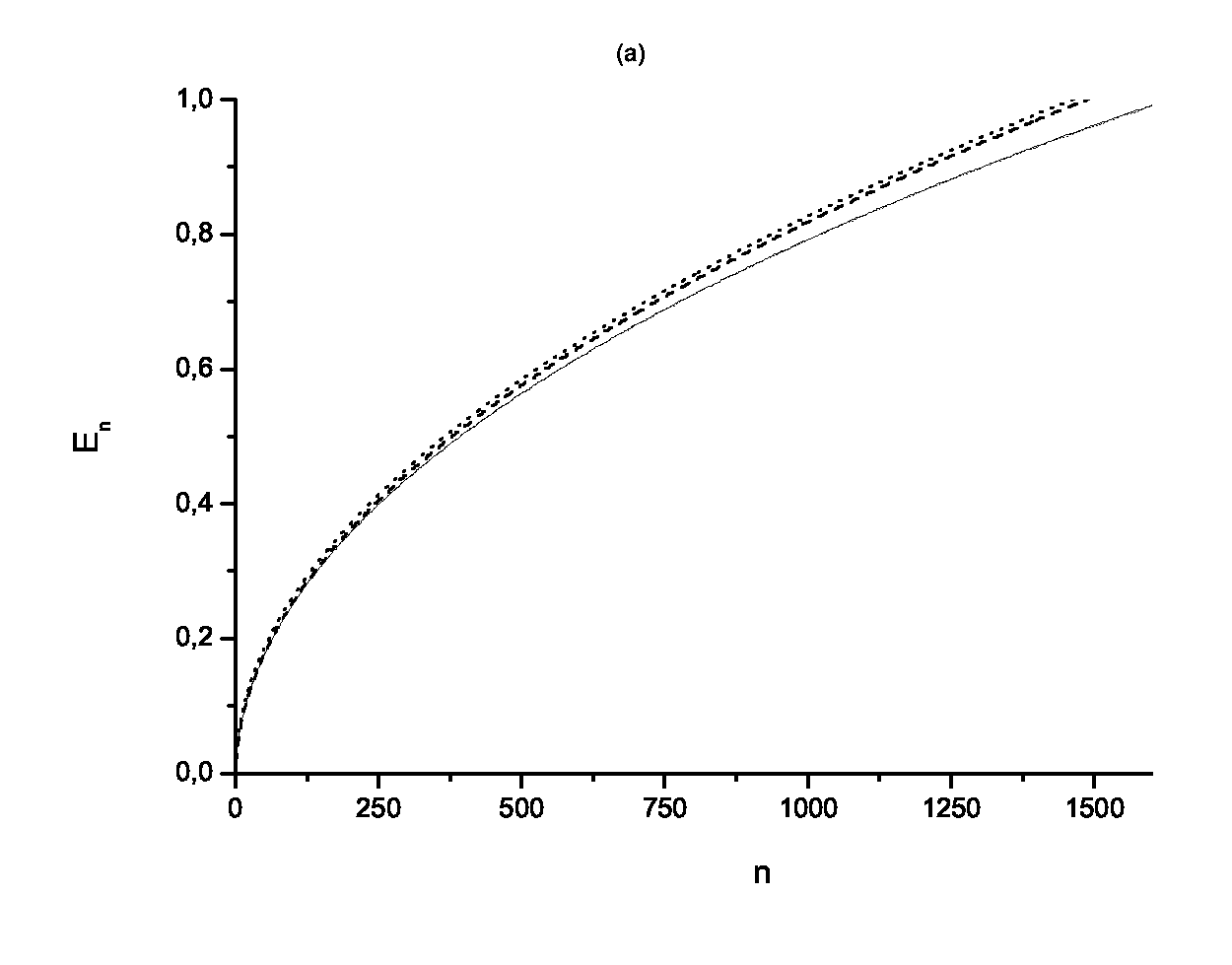}
\includegraphics[width=0.45\textwidth,draft=false]{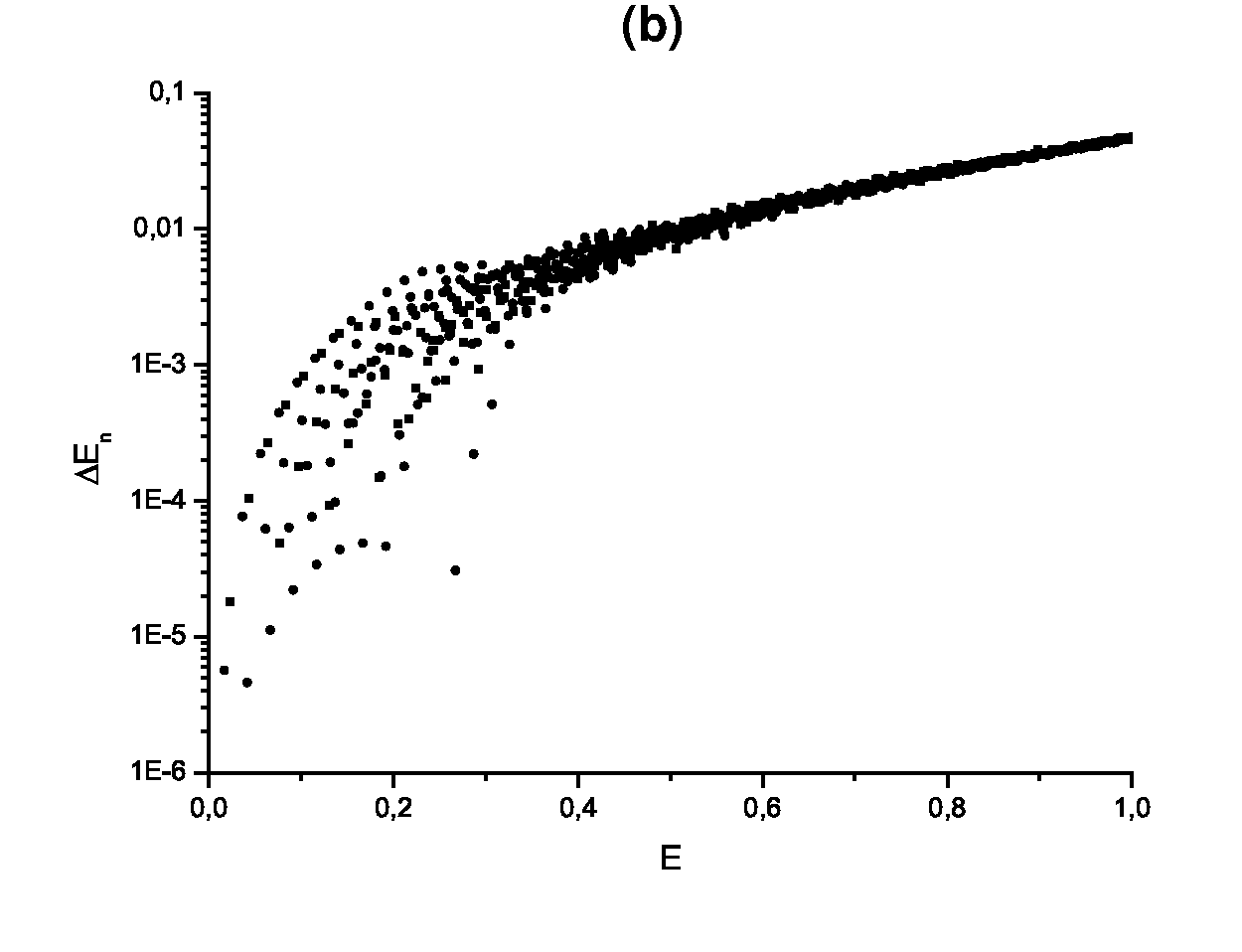}
\includegraphics[width=0.45\textwidth,draft=false]{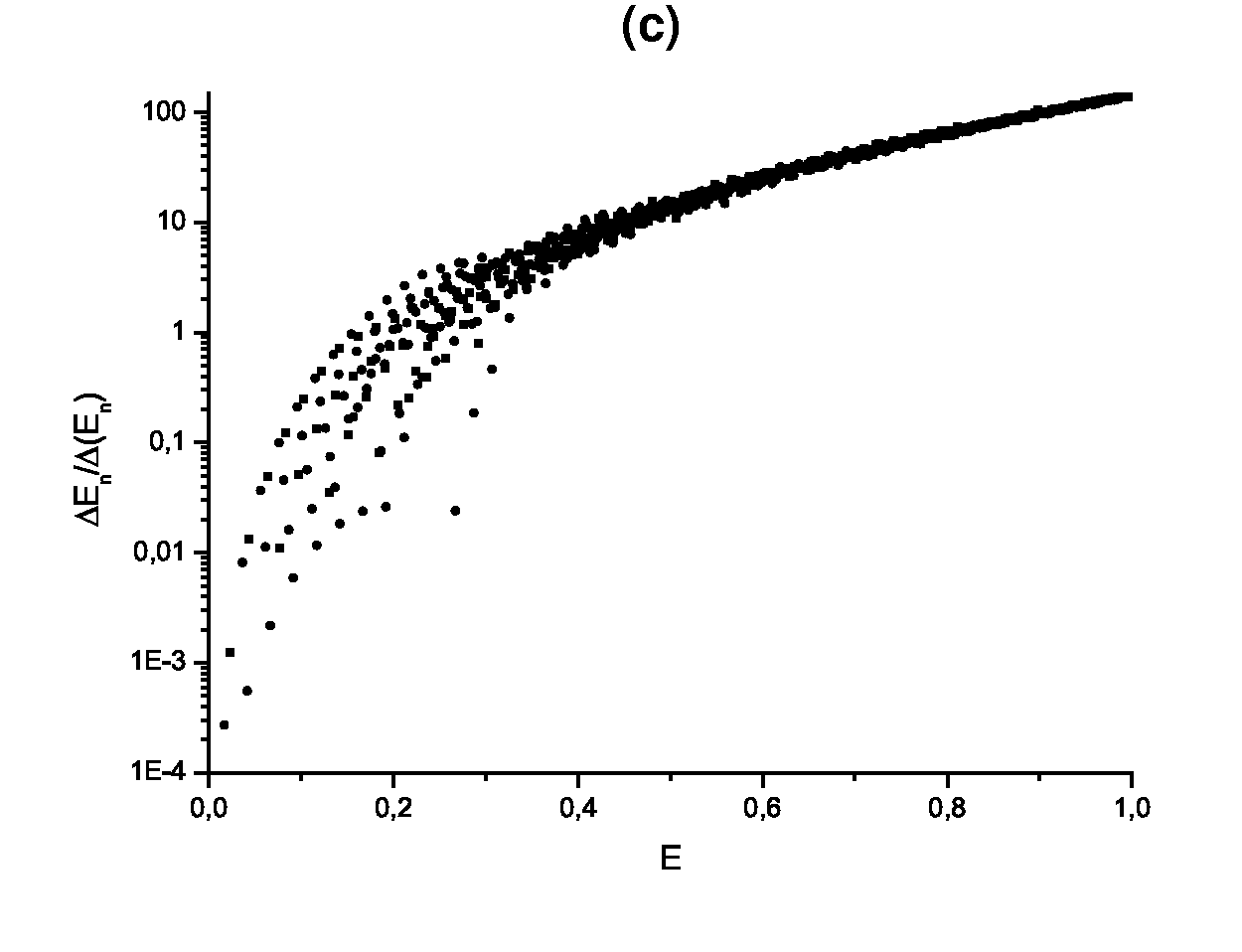}
\caption{\label{d5_ho} Energy levels of lower umbilic catastrophe
$D_5$ potential (\ref{u_d5}): (a) numerically exact spectrum (solid
line), harmonic oscillator (dashed line) and the semiclassical
(dotted line) approximations; (b) absolute value of difference
between the numerically exact  energy levels and harmonic oscillator
approximation; (c) absolute value of deviance of numerically exact
energy levels from the harmonic oscillator approximation referred to
mean level spacing.}
\end{figure}

In $D_5$ potential harmonic approximation is given by a mixture of
two spectral series of two-dimensional harmonic oscillator levels
with eigenfrequencies
\[\omega_x=2\]
\[\omega_y^{(R,C)}=2\sqrt{1\pm\frac{1}{\sqrt2}}\]
where the eigenfrequencies $\omega_y^{(R,C)}$ are different for
regular and chaotic local minima. Fig.\ref{d5_ho} demonstrates
unexpectedly good agreement (in the limits of $1\%$ of mean level
spacing for not too high energies) between the harmonic
approximation and actual energy levels in the $D_5$ potential, and
this agreement is better for the regular minimum.

\begin{figure}
\includegraphics[width=0.3\textwidth,draft=false]{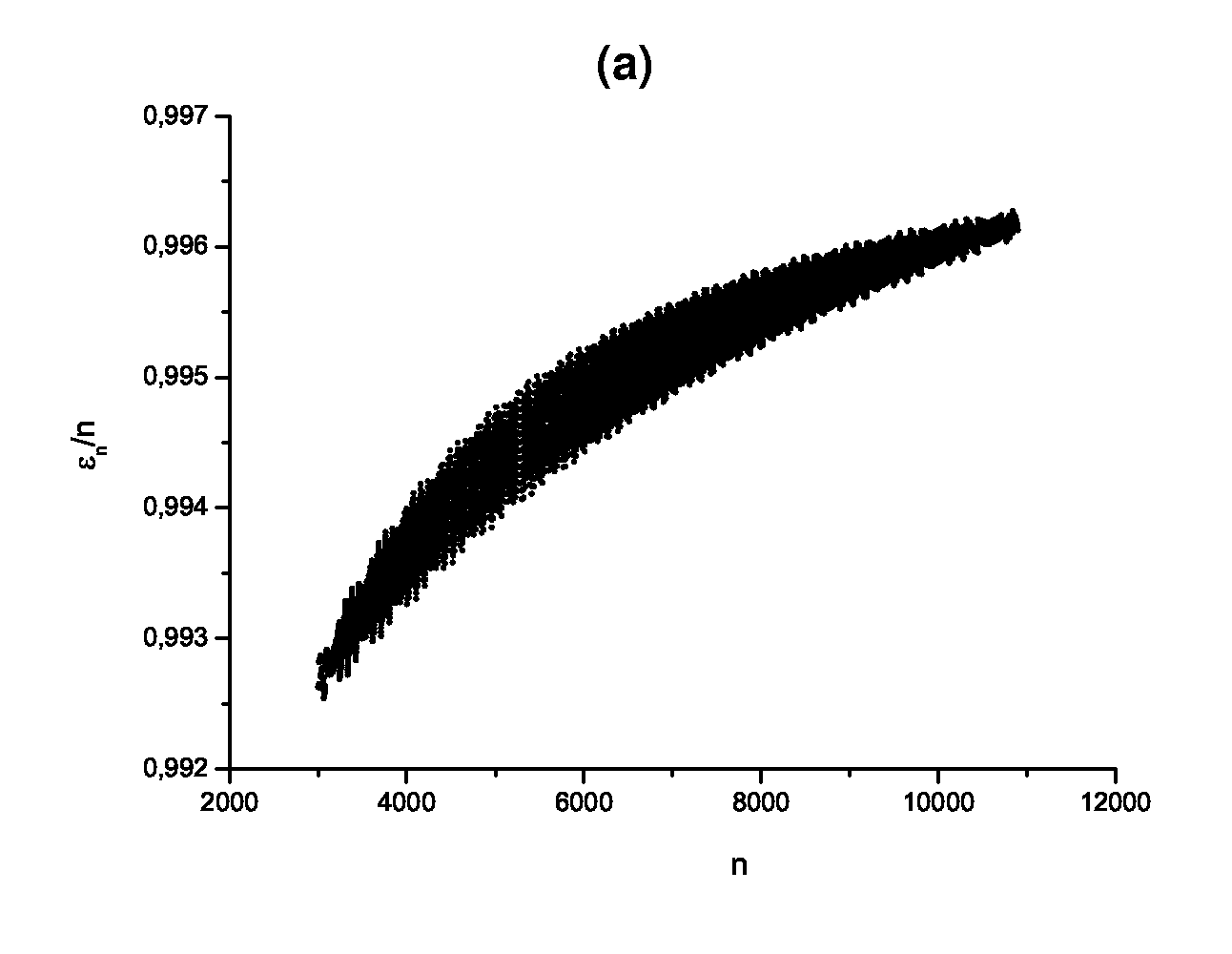}
\includegraphics[width=0.3\textwidth,draft=false]{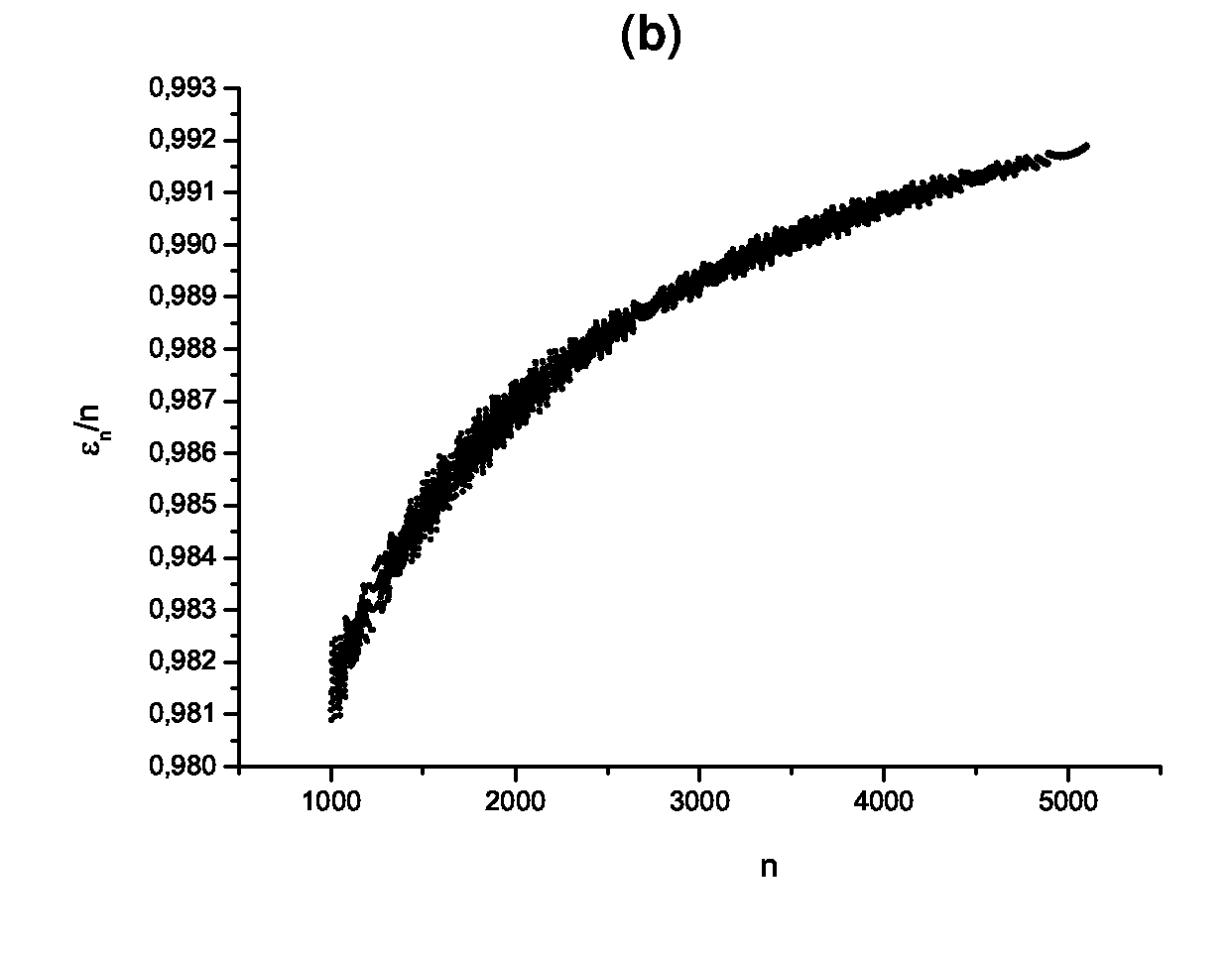}
\includegraphics[width=0.3\textwidth,draft=false]{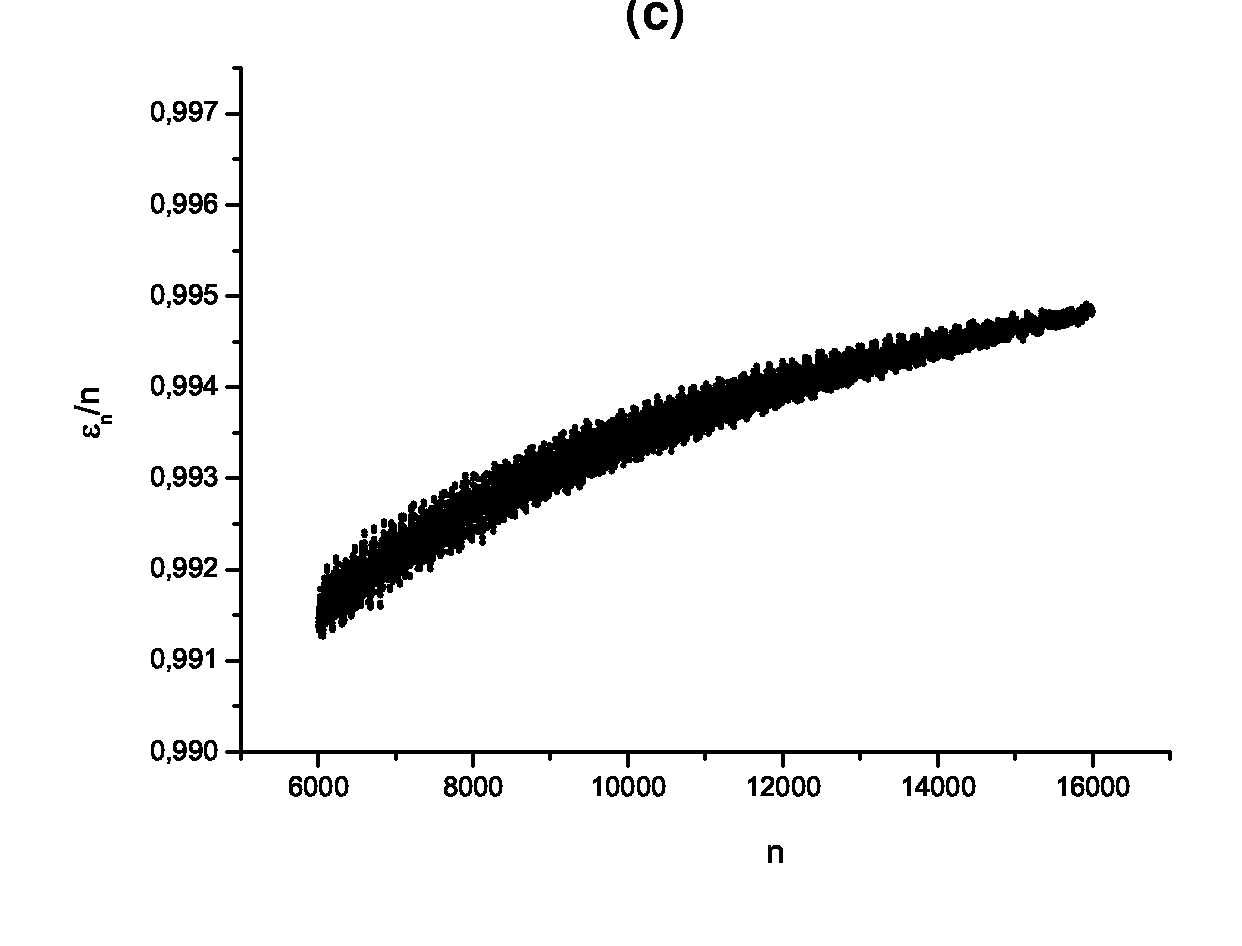}
\caption{\label{rc_unfold}Accuracy estimation for the spectrum
unfolding of harmonic approximation in lower umbilic catastrophe
$D_5$ potential (\ref{u_d5}): (a) for chaotic minimum ($x<0$); (b)
for regular minimum ($x>0$);(c)for both minima together.}
\end{figure}

\begin{figure}
\includegraphics[width=0.3\textwidth,draft=false]{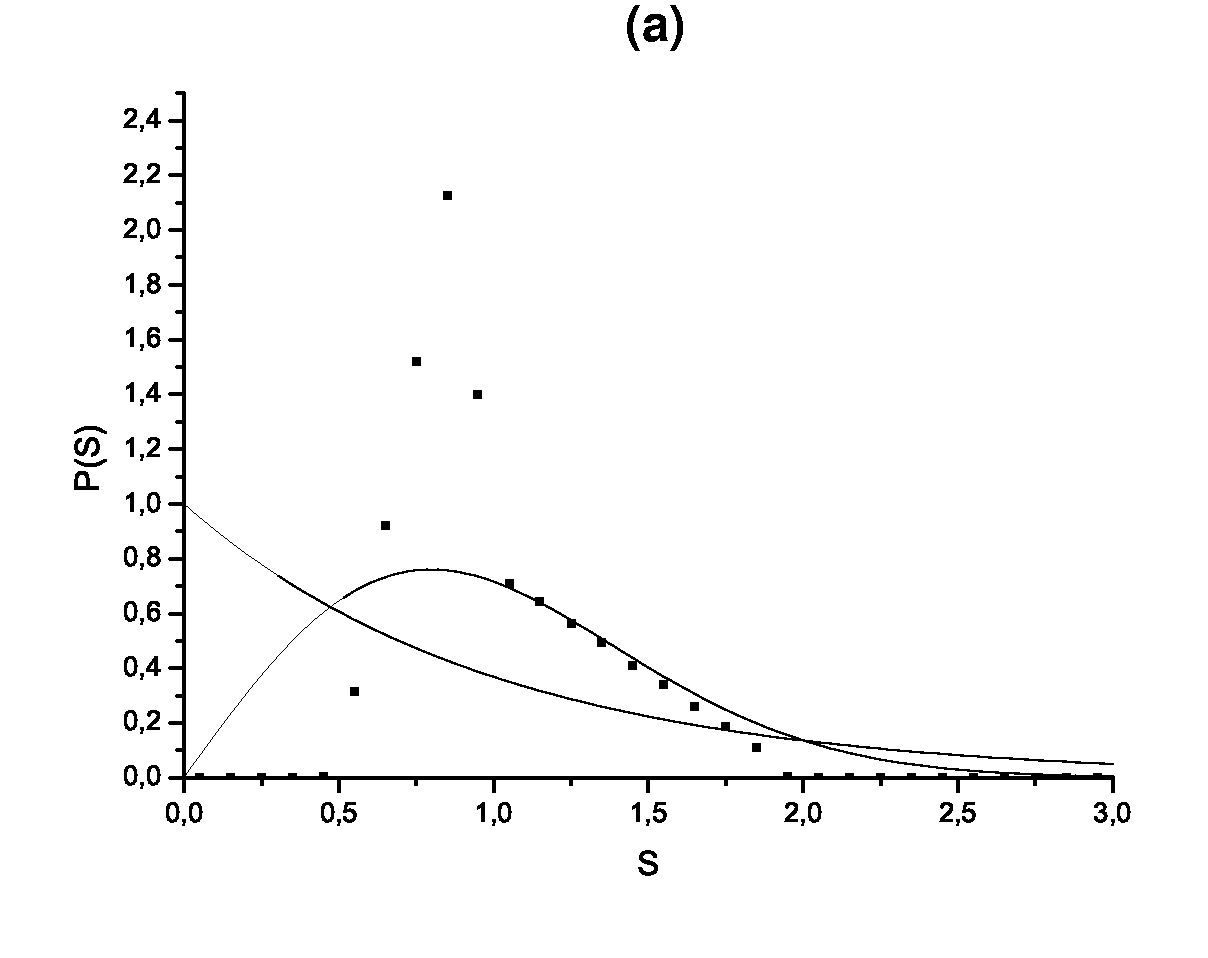}
\includegraphics[width=0.3\textwidth,draft=false]{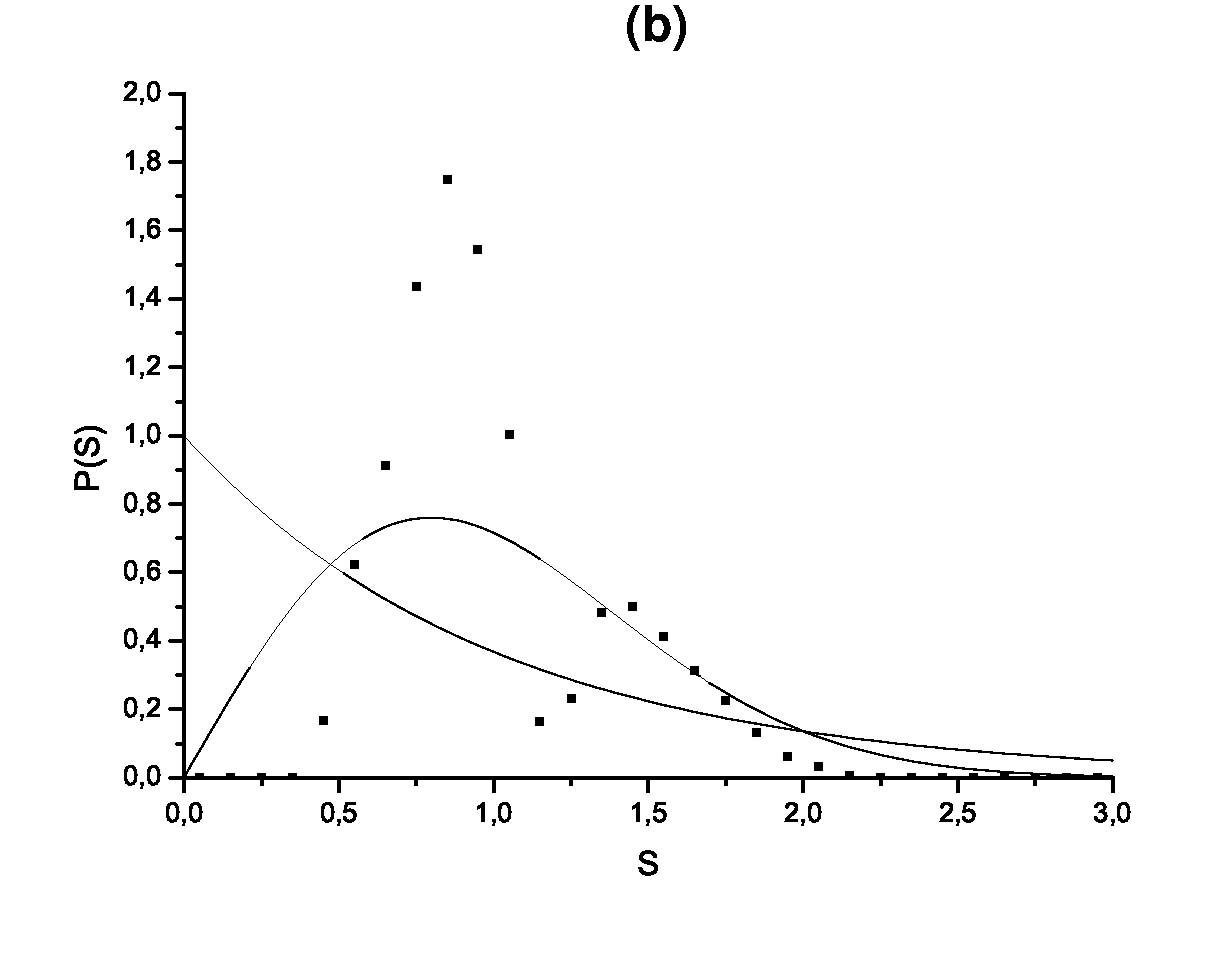}
\includegraphics[width=0.3\textwidth,draft=false]{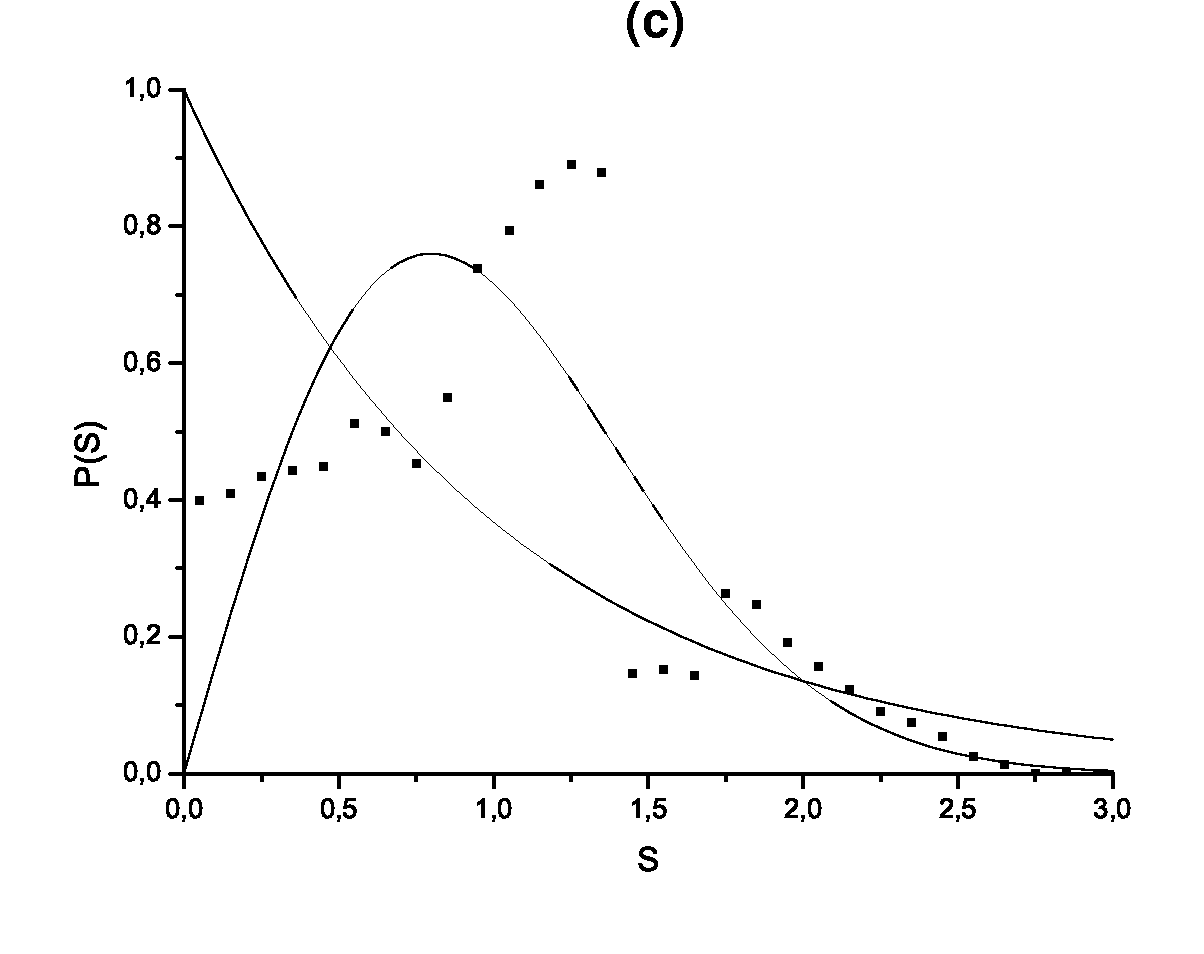}
\caption{\label{rc_p}FNNS for harmonic approximation in lower
umbilic catastrophe $D_5$ potential (\ref{u_d5}): (a) for chaotic
minimum ($x<0$); (b) for regular minimum ($x>0$);(c)for both minima
together. Points represent numerical data, solid lines --- Poisson
(\ref{poisson}) and Wigner (\ref{wigner}) distributions.}
\includegraphics[width=0.3\textwidth,draft=false]{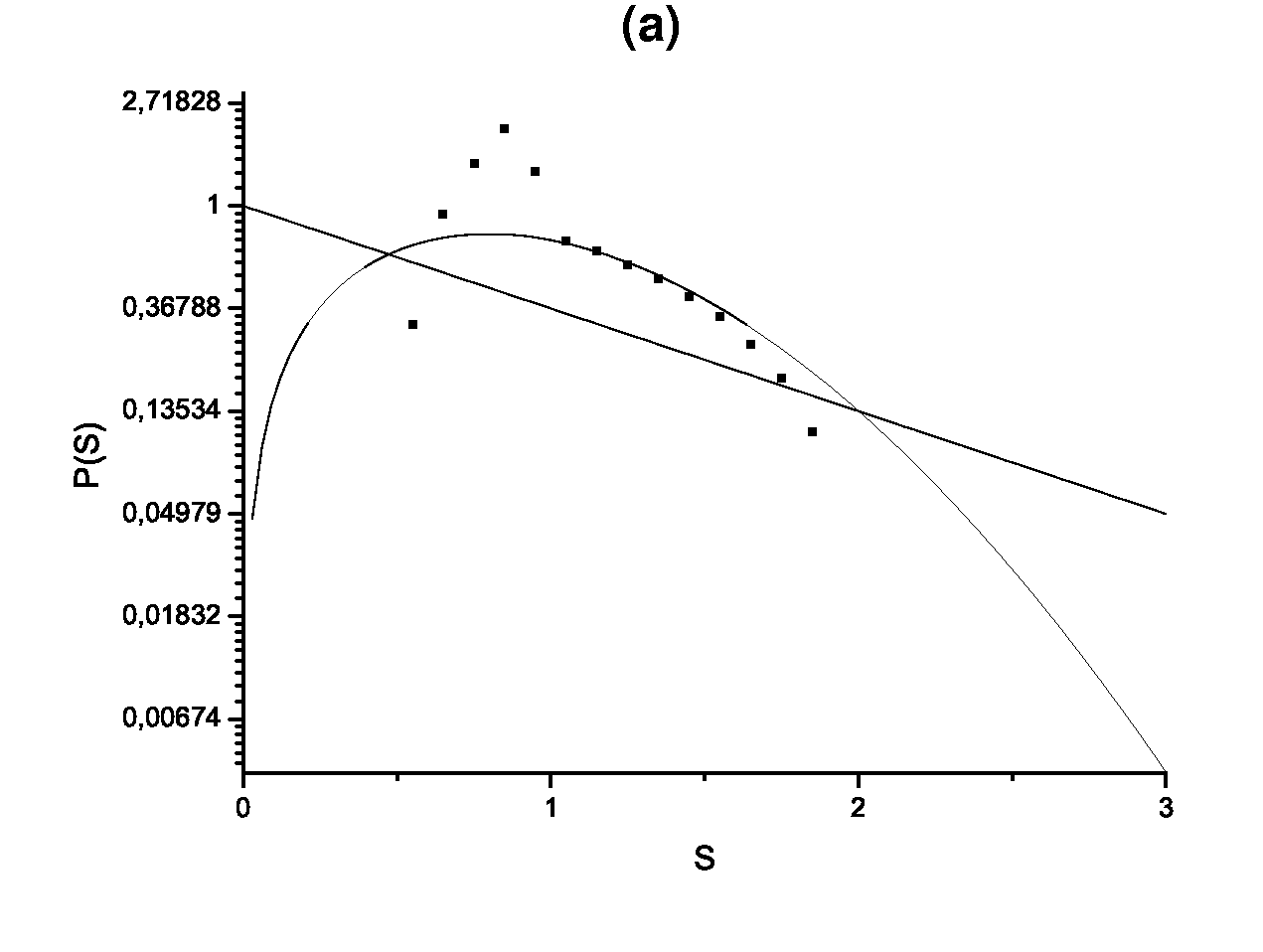}
\includegraphics[width=0.3\textwidth,draft=false]{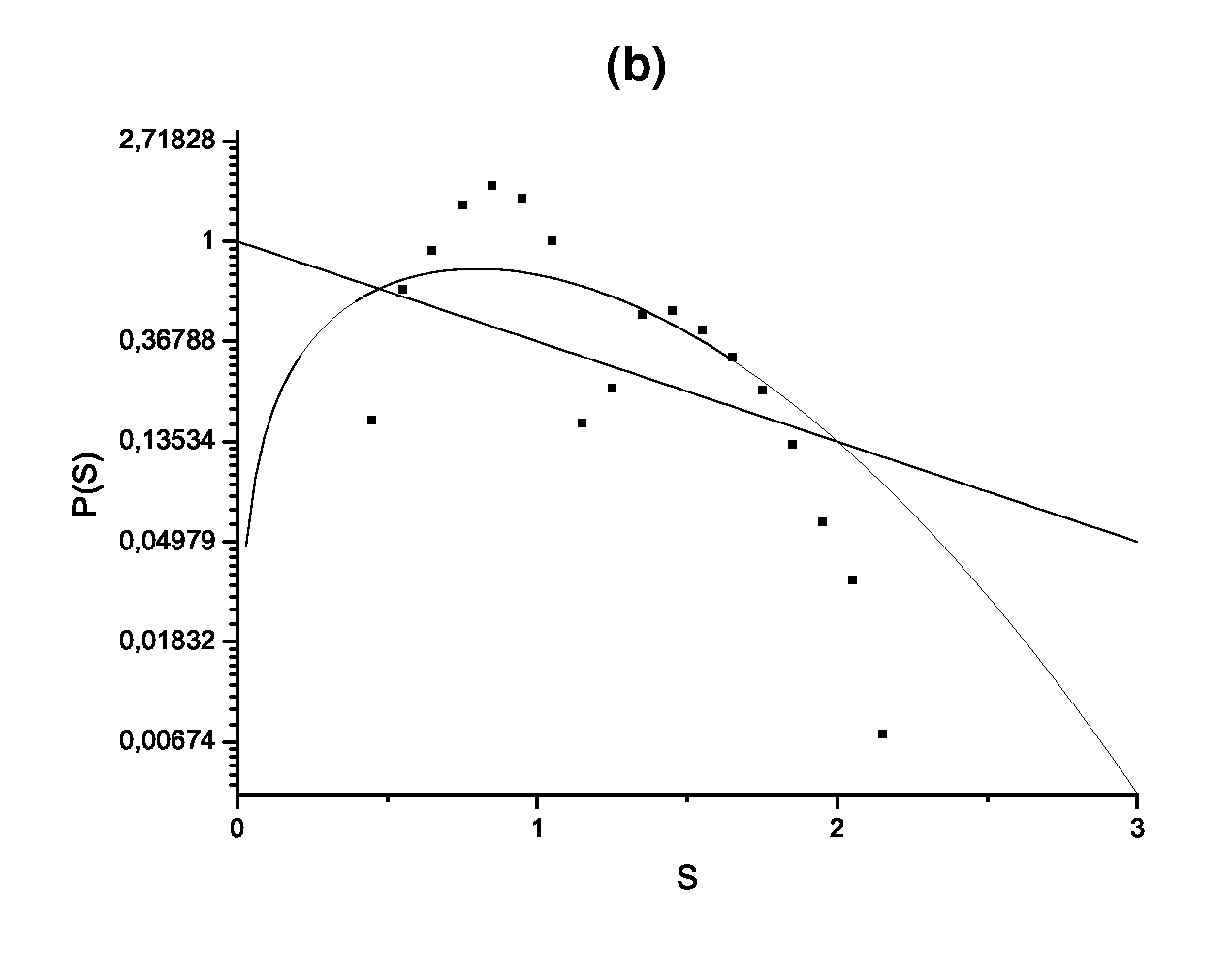}
\includegraphics[width=0.3\textwidth,draft=false]{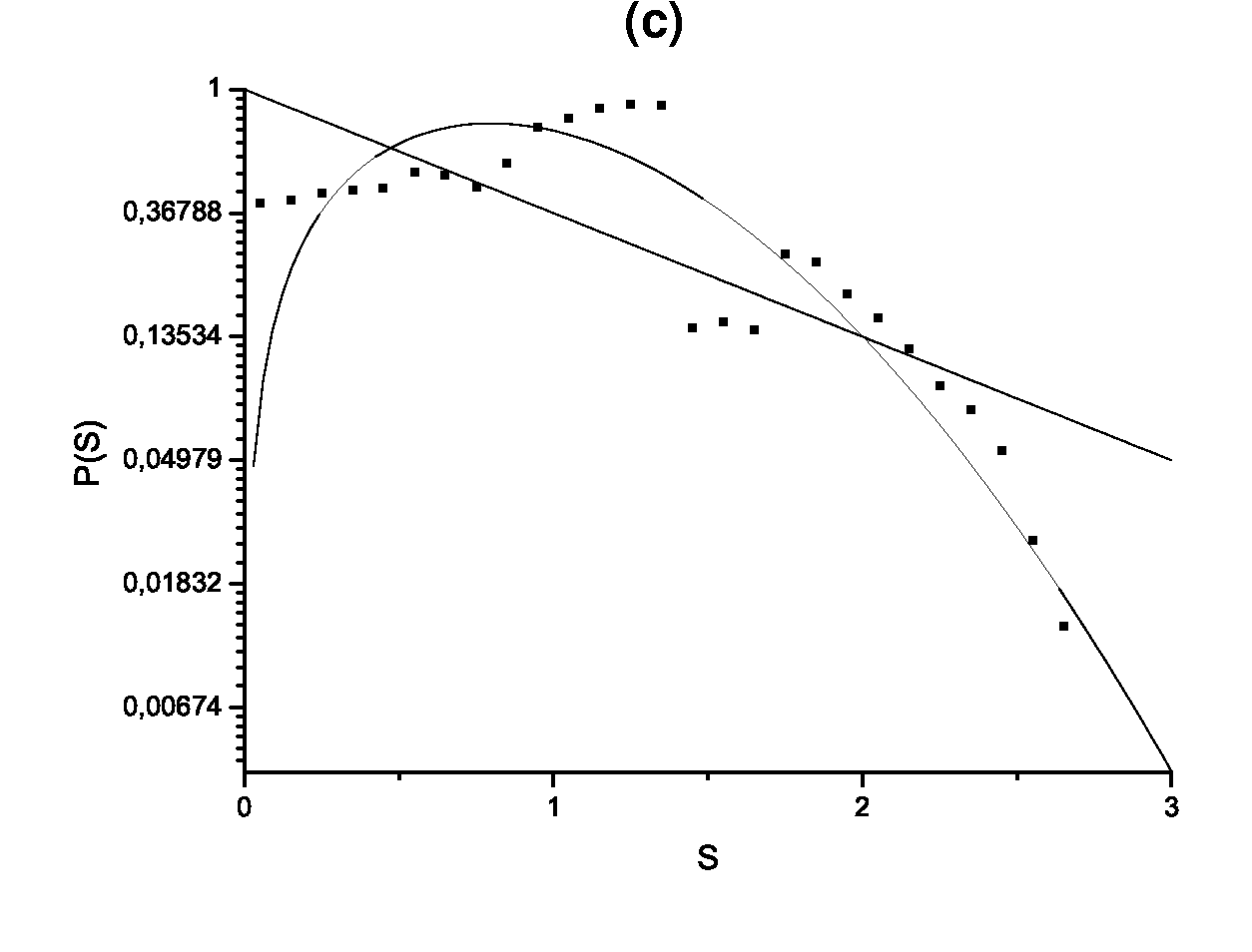}
\caption{\label{rc_pl}The same as Fig.\ref{rc_p} but in logarithmic
scale.}
\end{figure}

\begin{figure}
\includegraphics[width=0.3\textwidth,draft=false]{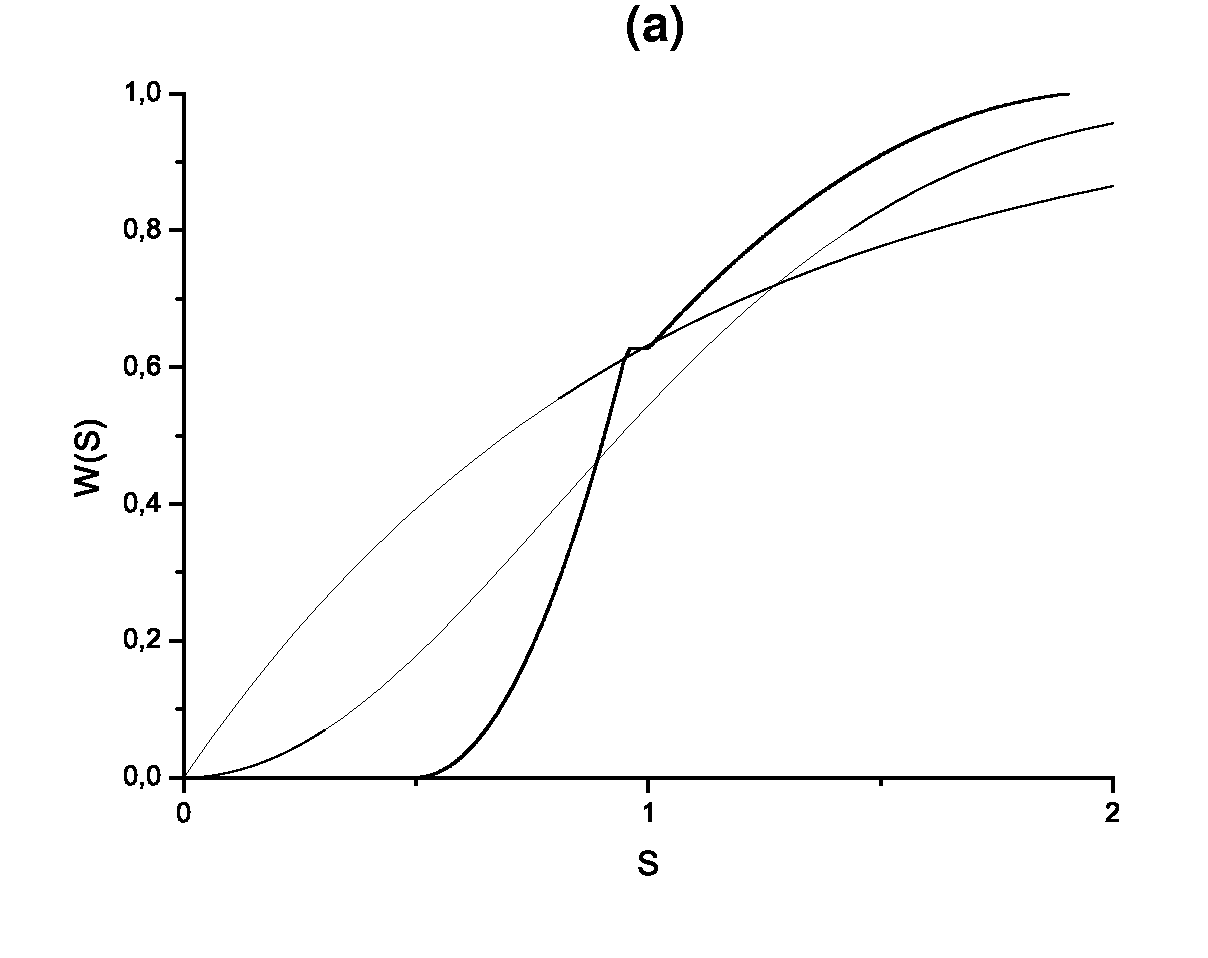}
\includegraphics[width=0.3\textwidth,draft=false]{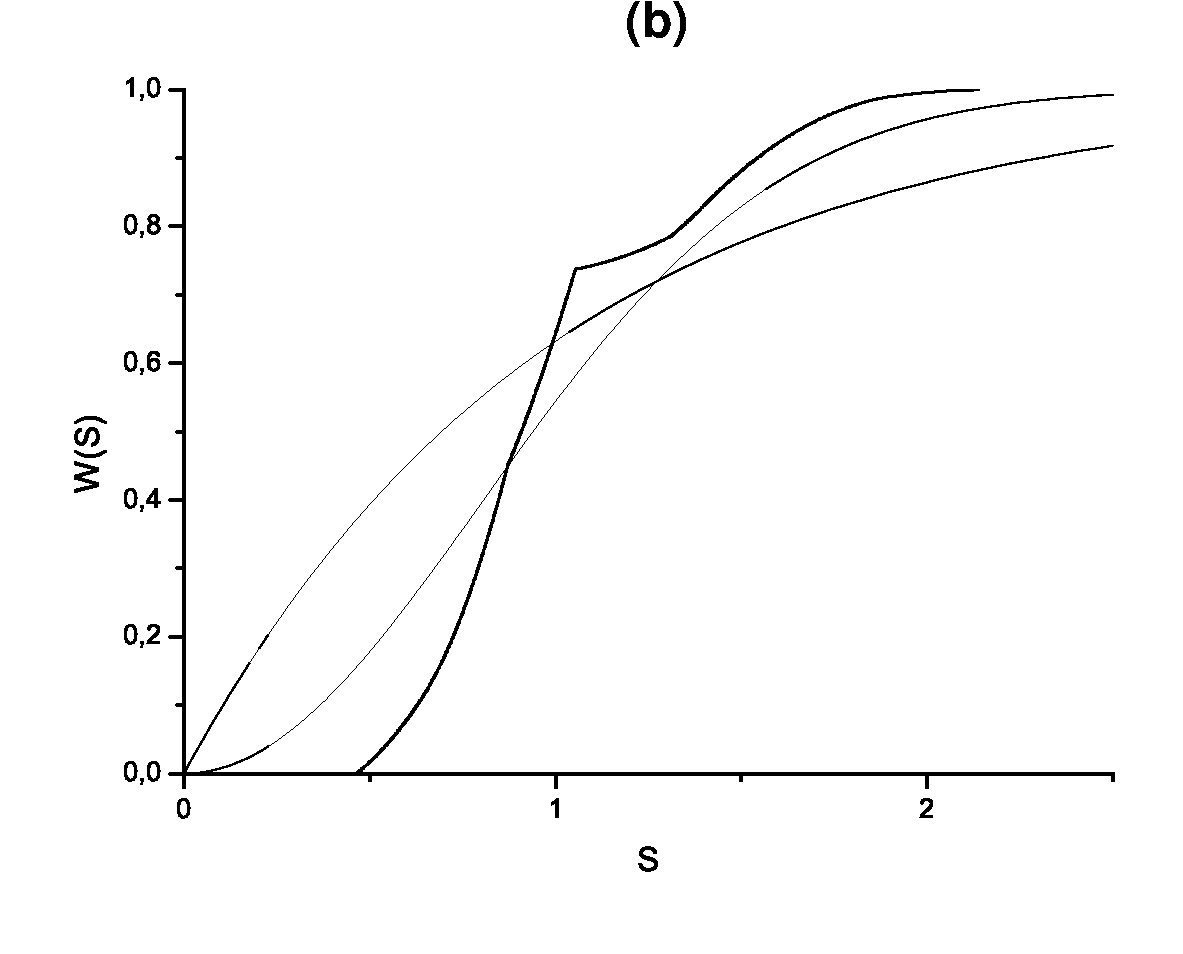}
\includegraphics[width=0.3\textwidth,draft=false]{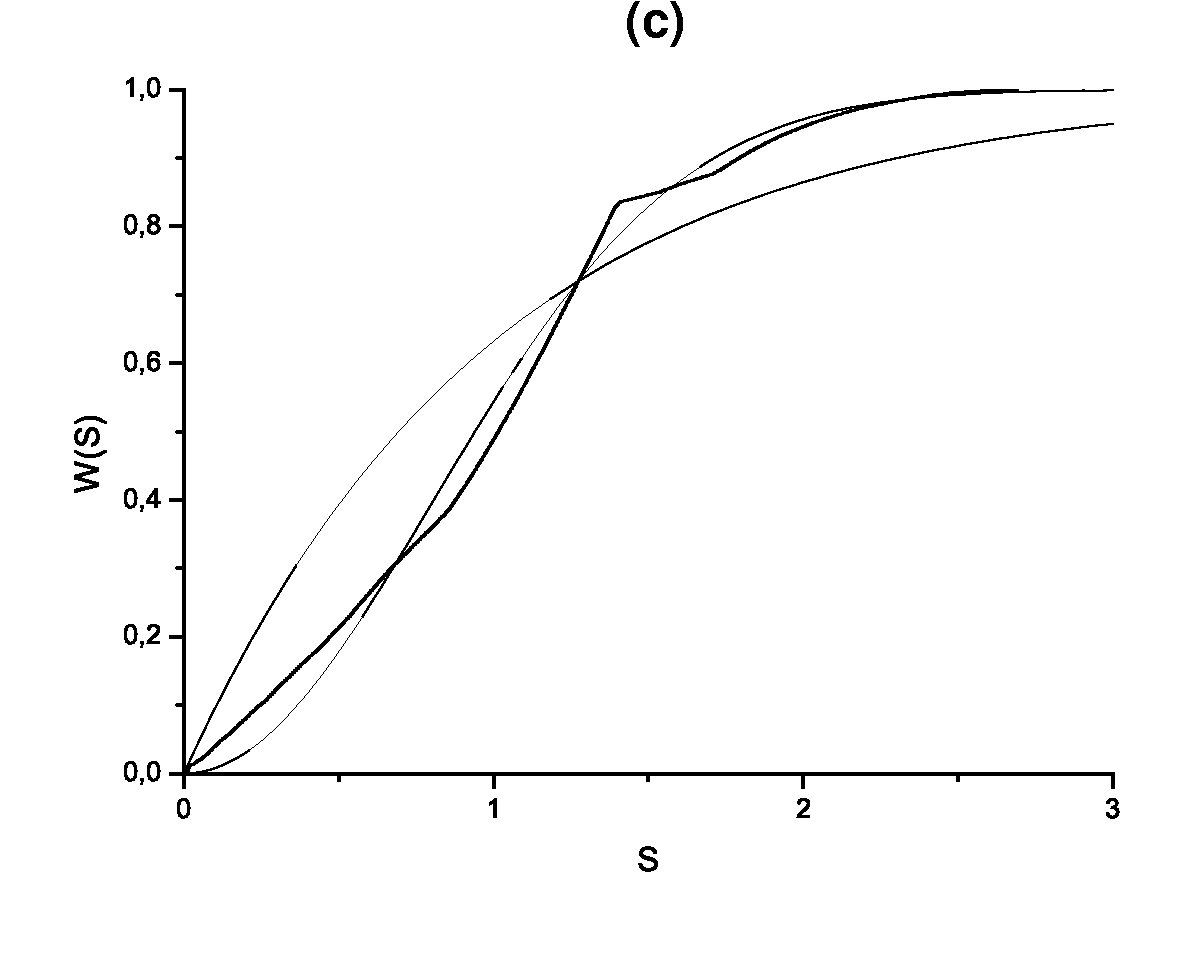}
\caption{\label{rc_w} Cumulative FNNS for harmonic approximation in
lower umbilic catastrophe $D_5$ potential (\ref{u_d5}): (a) for
chaotic minimum ($x<0$); (b) for regular minimum ($x>0$);(c)for both
minima together. Points represent numerical data, solid lines ---
Poisson and Wigner (\ref{w_pw}) distributions.}
\includegraphics[width=0.3\textwidth,draft=false]{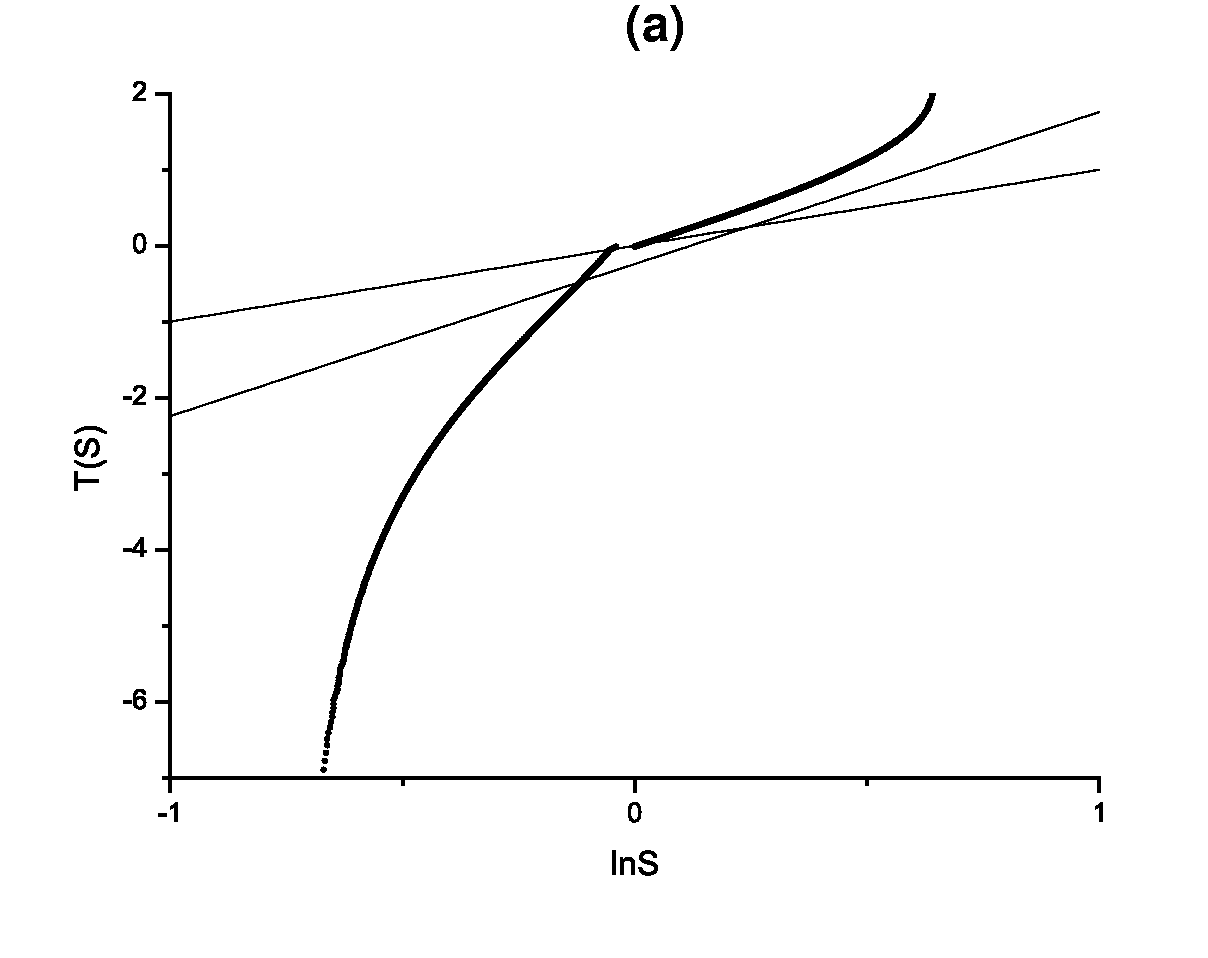}
\includegraphics[width=0.3\textwidth,draft=false]{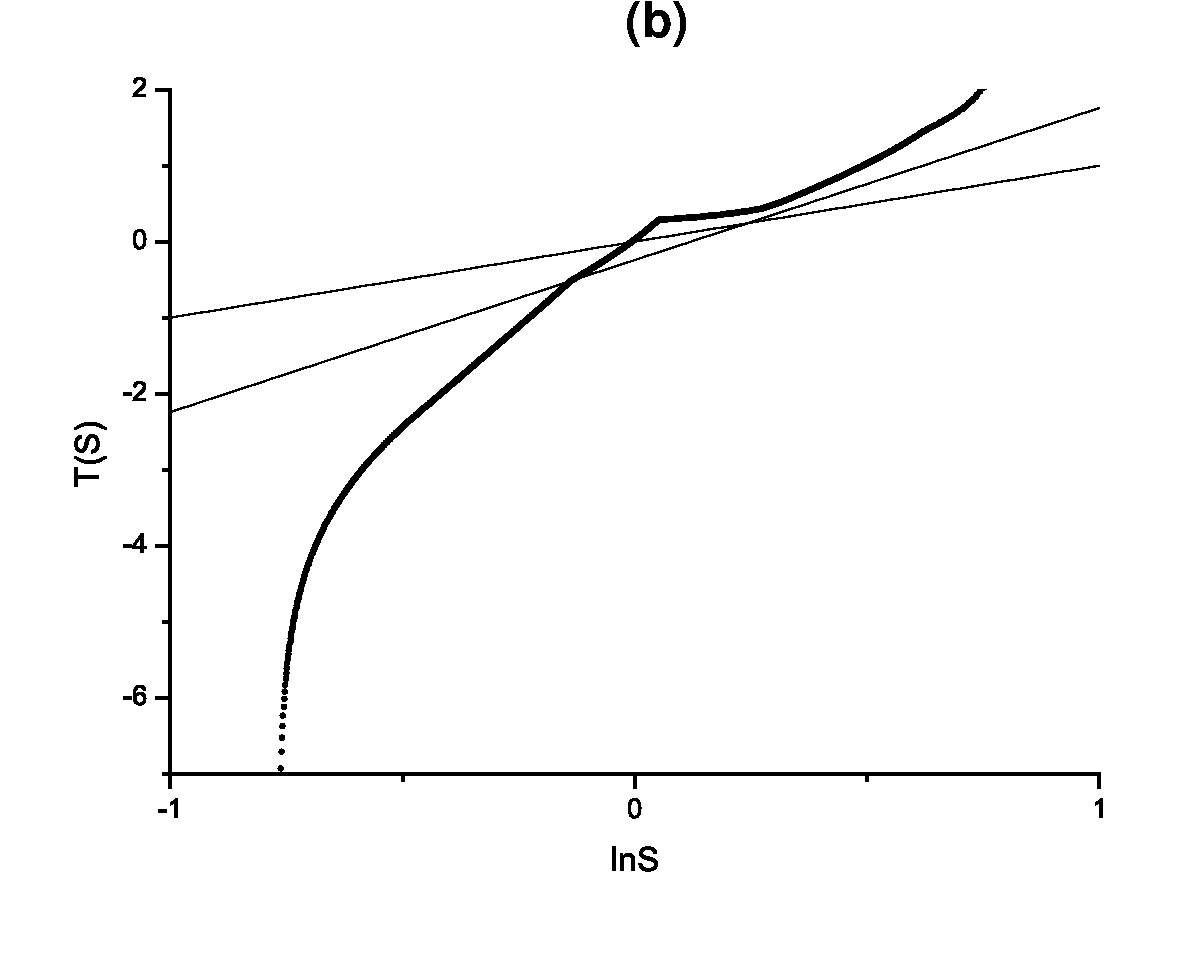}
\includegraphics[width=0.3\textwidth,draft=false]{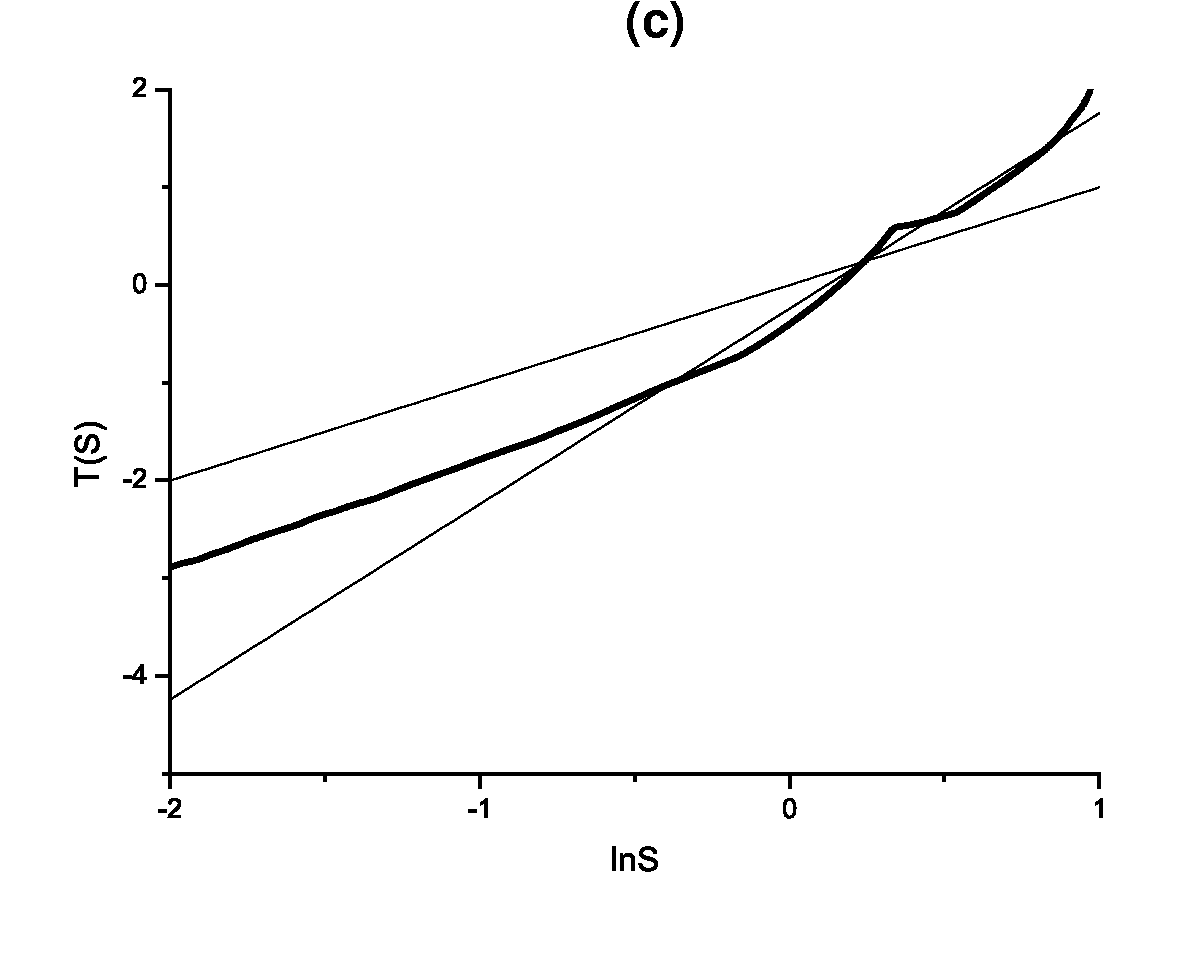}
\caption{\label{rc_t} Cumulative FNNS for harmonic approximation in
lower umbilic catastrophe $D_5$ potential (\ref{u_d5}) in the
$T$-representation: (a) for chaotic minimum ($x<0$); (b) for regular
minimum ($x>0$);(c)for both minima together. Points represent
numerical data, solid lines
--- Poisson and Wigner (\ref{t_pw}) distributions.}
\end{figure}

Such proximity of the considered spectrum to the harmonic one
automatically implies also the mutual proximity of corresponding
FNNS distributions. As is well known, FNNS distribution for the
harmonic oscillator has a pathological nature --- it does not have
any universal form, and, therefore, is not described by any of the
known distribution functions. Fig.\ref{rc_unfold}-\ref{rc_t} present
the results of analysis of spectral fluctuations in the harmonic
oscillator approximation in lower umbilic catastrophe $D_5$
potential (\ref{u_d5}) for chaotic and regular minima separately,
and also for a spectrum obtained by mixing of spectral series of
both minima.

The FNNS for harmonic oscillator spectrum has multiple maxima,
determined by its eigenfrequencies. Therefore the FNNS distribution
for the harmonic approximation spectrum in the mixed state has many
maxima, corresponding to all eigenfrequencies of the approximations.
That dependence of the FNNS distribution form on the
eigenfrequencies of the system is the main reason for the
pathological character of level fluctuations in the harmonic
oscillator spectrum. In the case of the mixed state, energy levels,
close to the harmonic ones, will give essential (and pathological)
contribution to FNNS distribution. The theoretical distribution
function, accounting for such a contribution, should be constructed
as mixed statistics from components of three types: not only chaotic
or regular, but also harmonic ones, carrying pathological spectral
fluctuations. The construction of such a distribution function is a
problem for the future.

\sat\section{Other statistical characteristics of spectra
--- $1/f$-noise}\sat

FNNS is far from being the unique tool for investigation of quantum
manifestations of classical stochasticity in the structure of
quantum energy spectra in the systems whose classical analogues
demonstrate chaos. One of the alternative approaches \cite{relano}
to quantum chaos description is based on traditional time series
analysis methods. Let us consider the energy spectrum as a discrete
signal, and a sequence of $N$ energy levels as a time series. Let us
characterize the spectral fluctuations with the
$\delta_n$-statistics, defined as the following:
\[\delta_n=\sum\limits_{i=1}^n(s_i-\langle s\rangle)=\sum\limits_{i=1}^n w_i,\]
where index $n$ runs from $1$ to $N-1$. Quantities $w_i$ give the
fluctuation of the $i$-th spacing between neighboring levels with
respect to its mean value $\langle s\rangle=1$. We assume that the
standard procedure of spectrum unfolding (\ref{unfold}) was made in
advance, i.e. the original spectrum was mapped into another spectrum
$(E_i\rightarrow\varepsilon_i)$ with mean level spacing equal to
unity, then $s_i=\varepsilon_{i+1}-\varepsilon_i$, $i=1,\ldots,N$.

Our goal is to study transformation of the above introduced
characteristics $\delta_n$ in transition from regular systems to
chaotic ones. One uf the useful methods is to calculate the power
spectrum $S(k)$ of the discrete and finite series $\delta_n$
according to the following:
\[S(k)=|\hat{\delta}_k|^2\]
where $\hat{\delta}_k$ is the Fourier transform of $\delta_n$,
\[\hat{\delta}_k=\frac{1}{\sqrt N}\sum\limits_n\delta_n e^{-\frac{2\pi ikn}{N}}\]
and $N$ is the number of terms in the series.

As an example of a chaotic system authors \cite{relano} have chosen
the atomic nucleus at high excitation energies, where level density
is very high. In order to obtain the energy spectrum the authors
performed calculations in the shell model with realistic interaction
reproducing the experimental data well. Hamiltonian matrices for
differen values of angular momentum were diagonalized and the
unfolding operation was accurately done for the obtained spectra.
After that the authors selected sets from $256$ consecutive levels
each with the same $J^\pi,T$ from high level density regions. In
order to characterize the statistical properties of the signal
$\delta_n$ the authors calculated the ensemble average of its power
spectrum. The averaging was performed in order to decrease the
statistical fluctuations (errors) and to clarify the general
tendency. Average $\langle S(k)\rangle$ was calculated over the
ensemble of $25$ sets.

\begin{figure}
\includegraphics[width=\textwidth]{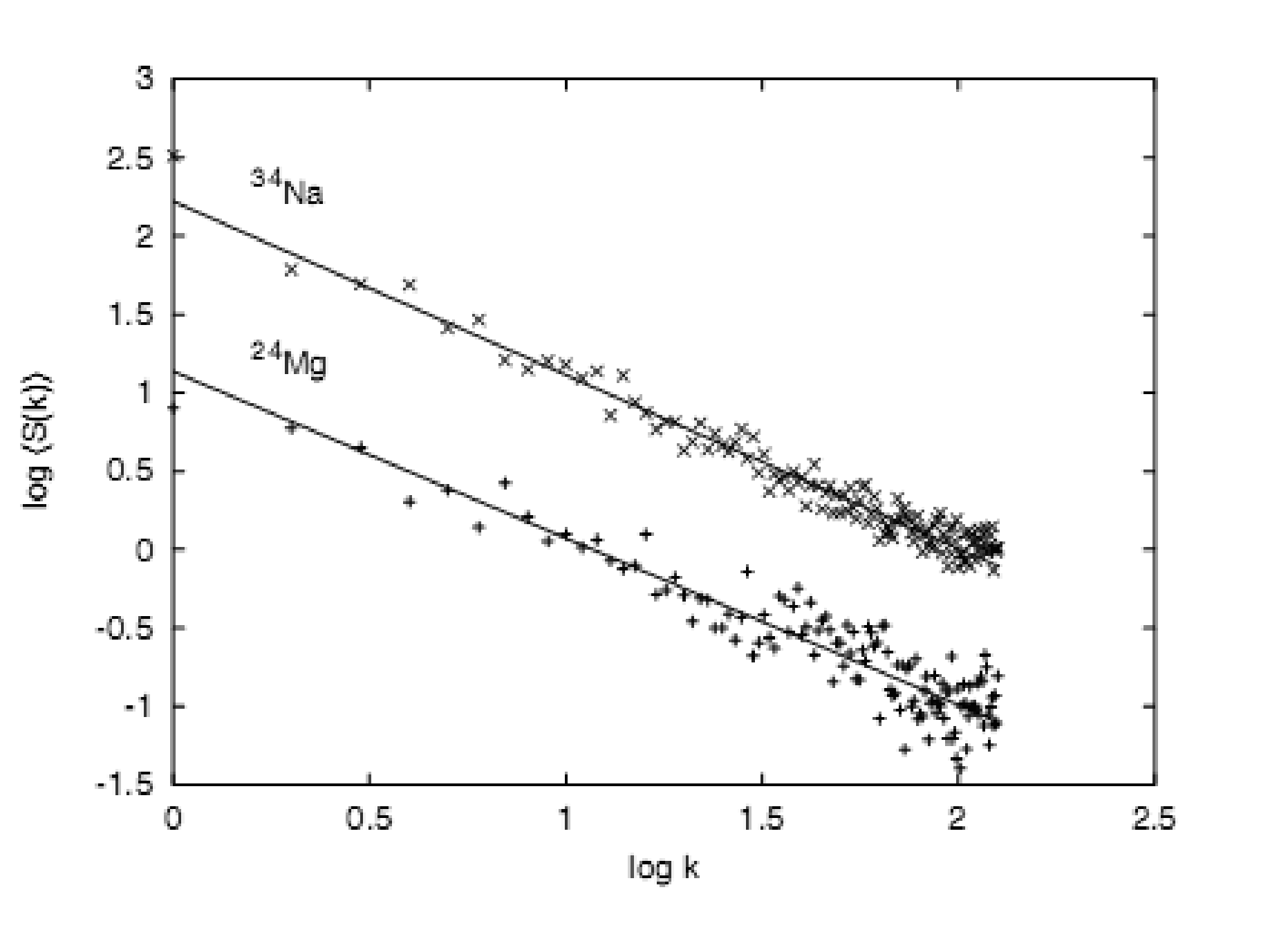}
\caption{\label{s_k}Average power spectrum of the $\delta_n$
function for ${}^{24}$Mg and ${}^{34}$Na, using $25$ sets of $256$
levels from the high level density region. The plots are displaced
to avoid overlapping.}
\end{figure}

Fig.\ref{s_k} represents the results for a typical stable $sd$ shell
of ${}^{24}$Mg nucleus (the spectrum was obtained by diagonalization
of a $2000\times2000$ martrix) and for a very exotic ${}^{34}$Na
nucleus --- $sd$ proton shell and $pf$ neutron one ($5000\times5000$
matrix diagonalization). It can be clearly seen that the power
spectrum for $\delta_n$ is close to power law. It is possible to
assume simple functional form
\[\langle S(k)\rangle\sim\frac{1}{k^\alpha}.\]

The least squares fit gives $\alpha=1.11\pm0.03$ for ${}^{34}$Na
$\alpha=1.06\pm0.05$ for ${}^{24}$Mg. A natural question arises:
does there exist an universal relation between peculiarities of
quantum spectrum, connected with the character of classical motion,
and the power spectrum of $\delta_n$ fluctuations?

\begin{figure}
\includegraphics[width=\textwidth]{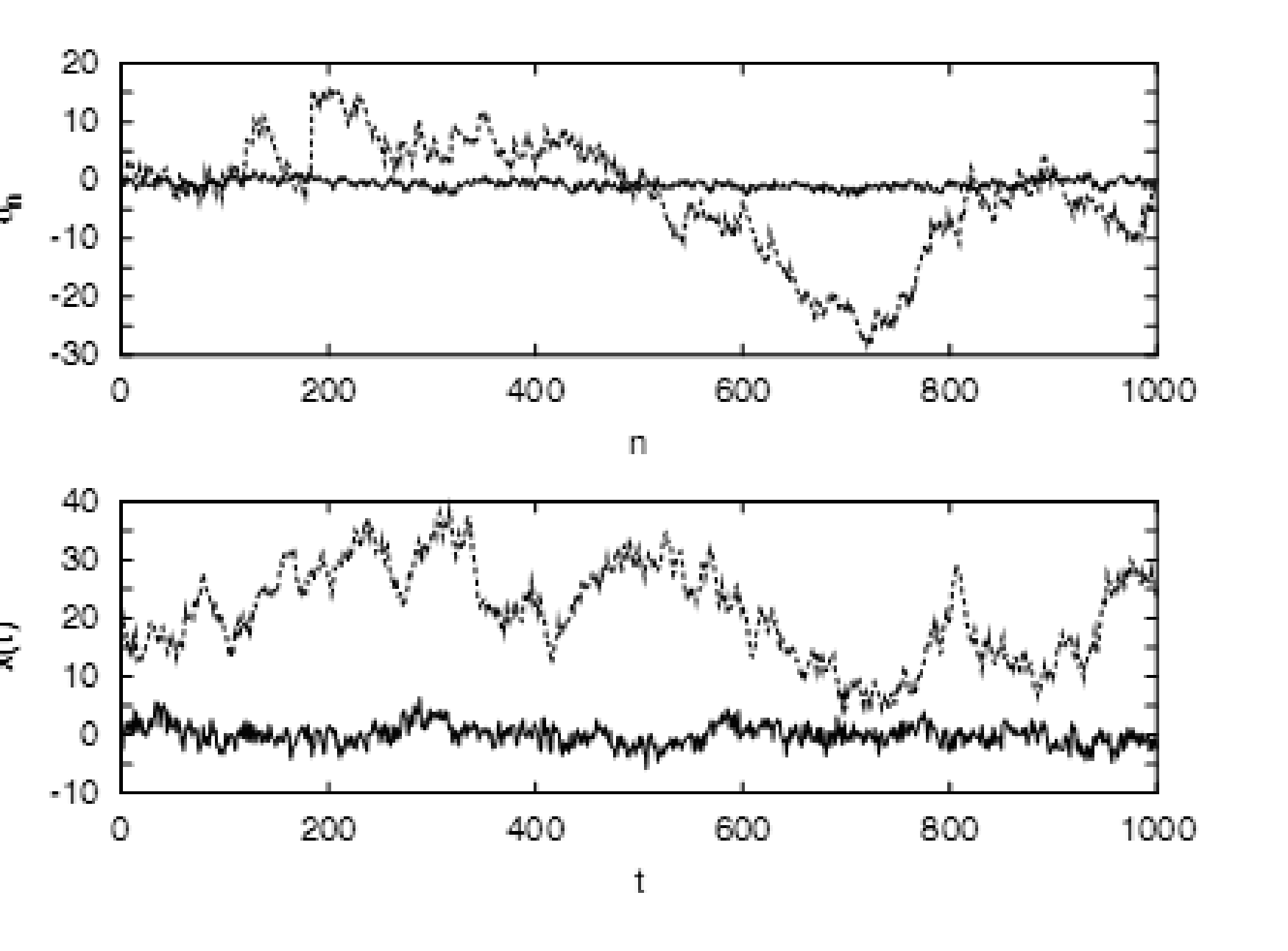}
\caption{\label{d_n_x_t_compare}Comparison of the $\delta_n$
function for Poisson (dashed line) and GOE spectra (solid line),
with a standard time series $x(t)$ with $1/k^\alpha$ power spectrum,
for $\alpha=2$ (dashed line) and $\alpha=1$ (solid line).}
\end{figure}

Probably the simplest way to answer that question is to compare
$\delta_n$ and $\langle S(k)\rangle$ for level sequences
corresponding to Poissonian statistics with the results for
different random matrices ensembles. Fig.\ref{d_n_x_t_compare}
presents the $\delta_n$ signal for the Poissonian and GOE spectra of
dimensionality $1000$. It is clear that the signals are absolutely
different. We compare them with discrete time series $x(t)$
corresponding to power spectra $1/k$ and $1/k^\alpha$. The proximity
of the $\alpha=2$ time series to the Poissonian spectrum and of
$\alpha=1$ --- to to GOE one is evident.

\begin{figure}
\includegraphics[width=\textwidth]{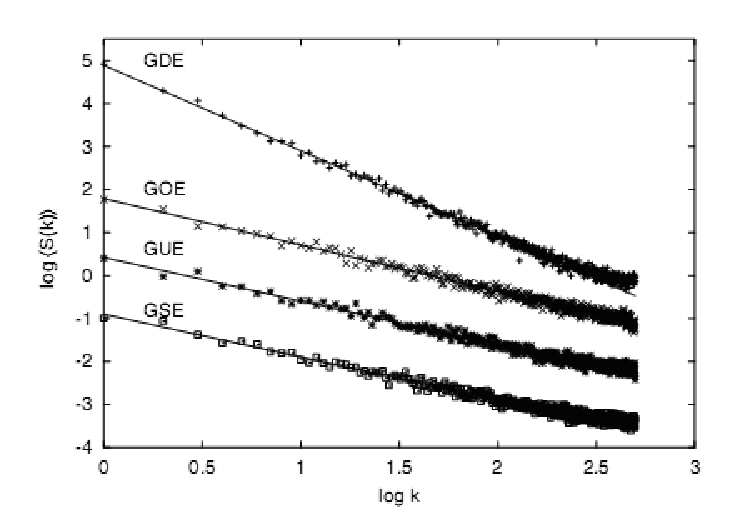}
\caption{\label{4_ensembles}Power spectrum of the $\delta_n$
function for GDE (Poisson) energy levels, compared to GOE, GUE and
GSE. The plots are displaced to avoid overlapping.}
\end{figure}

In order to calculate $\langle S(k)\rangle$ there were built $30$
different matrices of dimensionality $1000\times1000$ for random
matrix ensembles of different types mentioned above.
Fig.\ref{4_ensembles} presents the results of the calculations in
double logarithmic scale. In all the considered cases the general
tendency is linear with the exception of the very high energy
region, where some deviations are observed which are probably due to
the finiteness of dimensionality of the considered matrices.
Ignoring frequencies with $k>2.2$, the fitting gives $\alpha=1.99$
for Poissonian spectrum with uncertainty about $2\%$.

In contrast to the last example, the GOE of high dimension matrices
is usually considered as a paradigm for the chaotic quantum
spectrum. It demonstrates level-level correlations on all scales.
The same refers also to unitary (GUE) and simplectic (GSE) ones, but
with increasing level repulsion rate.

Power spectra of $\delta_n$ for all the three ensembles are
presented on Fig.\ref{4_ensembles}. For the exponents we obtained
the following values: $\alpha_{GOE}=1.08$, $\alpha_{GUE}=1.02$,
$\alpha_{GSE}=1.00$. All three ensembles agree with the same power
law $\alpha\simeq1$. It is evident that the power spectrum $\langle
S(k)\rangle$ behaves as $1/k^\alpha$ both for regular and chaotic
spectra, but level correlations decrease from limiting $\alpha=2$
for a regular uncorrelated spectrum to the minimal value $\alpha=1$
for quantum chaotic systems.

The above result for the $\delta_n$ statistics power spectrum gives
rise to the following hypothesis: energy spectra of chaotic systems
are characterized by $1/f$ noise. Let us recall that $1/f$ (flicker
noise) was discovered in the 1920s: slow flicks (from which the name
comes) of emission capacity of cathode lamps (as a result of current
fluctuations) superposed on fast thermal (fraction) current
fluctuations. For white noise $S(\omega)=const$ in a wide frequency
range, while for the flicker noise $S(\omega)\propto1/\omega$.

Flicker noise was discovered in many physical and biological
systems. A common property of all such systems is the fact that
$1/f$ noise is a satellite of stationary irreversible processes: its
contribution disappears in the absence of irreversible flows and the
system reaches the thermodynamic equilibrium. In reality $1/f$ noise
is always weak in the sense that it carries only a small part of
average fluctuation squared: the main contribution is due to white
noise. But at low frequencies $1/f$ noise may exceed white noise by
several orders of magnitude and becomes the main harmful factor for
devices that work on such frequencies. An important feature of
flicker noise is the increase of its spectral density with frequency
decrease, which takes place up to minimal measurable frequencies of
about $10^{-7}$ Hertz without any tendency to saturation. Thus the
main question is: should one seek for a universal flicker nose
mechanism common for all media and systems, or it is better to
assume the existence of many flicker noises of different characters.
This question is unanswered up to now.

The hypothesis proposed in \cite{relano} has several attractive
features. The considered properties characterize the chaotic
spectrum immediately without reference to properties of other
systems (such as random matrices ensembles). It is universal for all
types of chaotic system and does not depend on such properties as
time reversal invariance and whether spins are integer or
half-integer. Also the $1/f$ characteristics unites a rich diversity
of quantum chaotic systems from many different fields of science ---
flicker noise is everywhere! Therefore the energy spectra of chaotic
systems demonstrate the same type of fluctuations as many other
systems. However there is no indication that $1/f$ noise in spectral
fluctuations of quantum systems implies $1/f$ noise in the
corresponding classical analogues.
\sat\chapter{Signatures of Quantum Chaos in Wave Function
Structure}\sat

\sat\section{QMCS in wave functions structure}\sat

As we have seen in the previous section, the statistical properties
of spectra have been rigidly correlated with the type of classical
motion. It is natural to try to discover the analogous
correlations in the structure of wave functions, i.e. to assume
that the form of the wave function for a semiclassical quantum state
depends on whether it is associated with classical regular or
chaotic motion. Furthermore it should be pointed out that in the
analysis of quantum manifestations of classical stochasticity at the
level of energy spectra, the principal role was given to statistical
characteristics, i.e., quantum chaos was treated as a property of a
group of states. The choice of a stationary wave function of the
quantum system, which is chaotic in the classical limit, as the
basic object of investigation relates the phenomenon of quantum
chaos to an individual state. Joint investigation of both the
possibilities is not a contradiction but reflects the universality of
the considered phenomenon.

In contrast to spectrum, the form of wave functions depends on the
basis on which they are determined. For QMCS studies the following
three representations are most often used:
\begin{enumerate}
\item The so-called $H_0$-representation is the representation of eigenfunctions $\{\varphi_n\}$ of the
integrable part $H_0$ of total Hamiltonian $H=H_0+V$. The main
objects of investigation in this case are the coefficients of
expansion $C_{mn}$ of stationary functions $\psi_m$ on basis
$\{\varphi_n\}$. $H_0$-representation is natural in the realization of
numerical calculations, as the diagonalization of Hamiltonian $H$ is
realized more often in this representation.
\item Coordinate representation $\Psi(\mathbf{x})$ is the most
convenient from the point of view of visual clarity as it allows direct
comparison with classical motion along the trajectories in
coordinate space. Indeed, according to Schnirelman theorem
\cite{shnirelman}, the mean probability density $|\Psi(\mathbf{x})|^2$
in the semiclassical limit $\hbar\rightarrow1$ coincides with
projection of micro-canonical distribution onto the coordinate
space:
\[\langle|\Psi(q)|^2\rangle\rightarrow\rho_0(q)\ with\ \hbar\rightarrow1\]
\[\rho_0(q)=\frac{\int d^n p\delta(E-H(p,q))}{\int d^n p' d^n q' \delta(E-H(p',q'))}\]
Actually this theorem transmits the correspondence principle on wave
functions. What can we expect from the wave function structure
basing on the theorem? It is concentration of the probability
density in the regions of coordinate space covered by
(quasi-)periodic trajectories for regular wave functions, in
contrast to almost uniformly smeared probability density for
chaotic wave functions.
\item Representation with the help of
Wigner functions \cite{wfr1,wfr2} has a set of properties in common
with the classical function of distribution in phase space.
\end{enumerate}

As early as 1977 Berry \cite{berry77} assumed that the form of the
wave function $\psi$ for a semiclassical regular quantum state
(associated with classical motion on a  $N$-dimensional torus in the
$2N$-dimensional phase space) was very different from the form of
$\psi$ for an irregular state (associated with stochastic classical
motion on all or part of the $(2N-1)$-dimensional energy surface in
phase space). For regular wave functions, the average probability
density in the configuration space was seen to be determined by the
projection of the corresponding quantized invariant torus onto the
configuration space, which implies global order. The local
structure is implied by the fact that the wave function is locally a
superposition of a finite number of plane waves with the same wave
number as determined by classical momentum. In the opposite case
for chaotic wave functions the averaged (over small intervals of
energy and coordinates) $|\psi_n|^2$ in the semiclassical limit
$\hbar\rightarrow0$ coincides with the projection of the classical
micro-canonical distribution onto the coordinate space. Its local
structure is spanned by the superposition of infinitely many plane
waves with random phases and equal wave numbers. The random phases
might be justified by classical ergodicity, and this assumption
immediately predicts locally the Gaussian randomness for the
probability of amplitude distribution. Such a structure of wave
function is in good agreement with the picture of classical phase
space: the classical trajectory homogeneously fills the isoenergetic
surface in the case of chaotic motion. By contrast, from the
consideration of regular quantum state as an analogue of classical
motion on a torus, a conclusion should be done about the singularity
(in the limit $\hbar\rightarrow0$) of the wave function near caustics
(boundaries of the region of classically allowed motion in a
coordinate space).

Berry's hypothesis was subjected to the most complete test for
billiards of different types and, in particular, for a billiard
stadium type \cite{mcdonald}. The amplitude of a typical wave function
of an integrable circular billiard is negligible in the classically
prohibited region (from conservation of angular momentum in a circular
billiard it should be clear that the arbitrary trajectory is
enclosed between the external and a certain internal circle, the radius of
which is determined by the angular momentum magnitude) , whereas
near the caustics it is maximal. Distribution of $|\psi_n|^2$ for
the case of a stadium billiard, (classical dynamics is stochastic) is
strongly distinguished from the integrable case. However, this
distribution is not so homogeneous, as we could expect starting from
the ergodicity of classical motion \cite{scars_heller}.

The above considered $R-C-R$ transition represents an attractive
possibility to check Berry's hypothesis for potential systems
with a localized instability region. Let us start from the
$H_0$-representation, or more exactly the representation of linear
combinations of wave functions of a two-dimensional harmonic
oscillator with equal frequencies
\begin{equation}\label{wf_expansion}\varphi_k=\sum\limits_{N,L}C^k_{NLj}|NLj\rangle,\end{equation}
where
\[|NLj\rangle=\frac{P_{L,j}}{\sqrt2}(|N,L\rangle+j|N,-L\rangle),\ j=\pm1;\]
\[N=0,1,2,\ldots;\ L=N,N-2,\ldots,1\ or\ 0;\ P_{L,j}=j^{Mod(L,3)},\]
\[\langle NLj|N'L'j'\rangle=\frac{1}{\sqrt2} 2^{\delta_{L,0}}\delta_{NN'}\delta_{LL'}\delta_{jj'}.\]

\begin{figure}
\includegraphics[width=\textwidth]{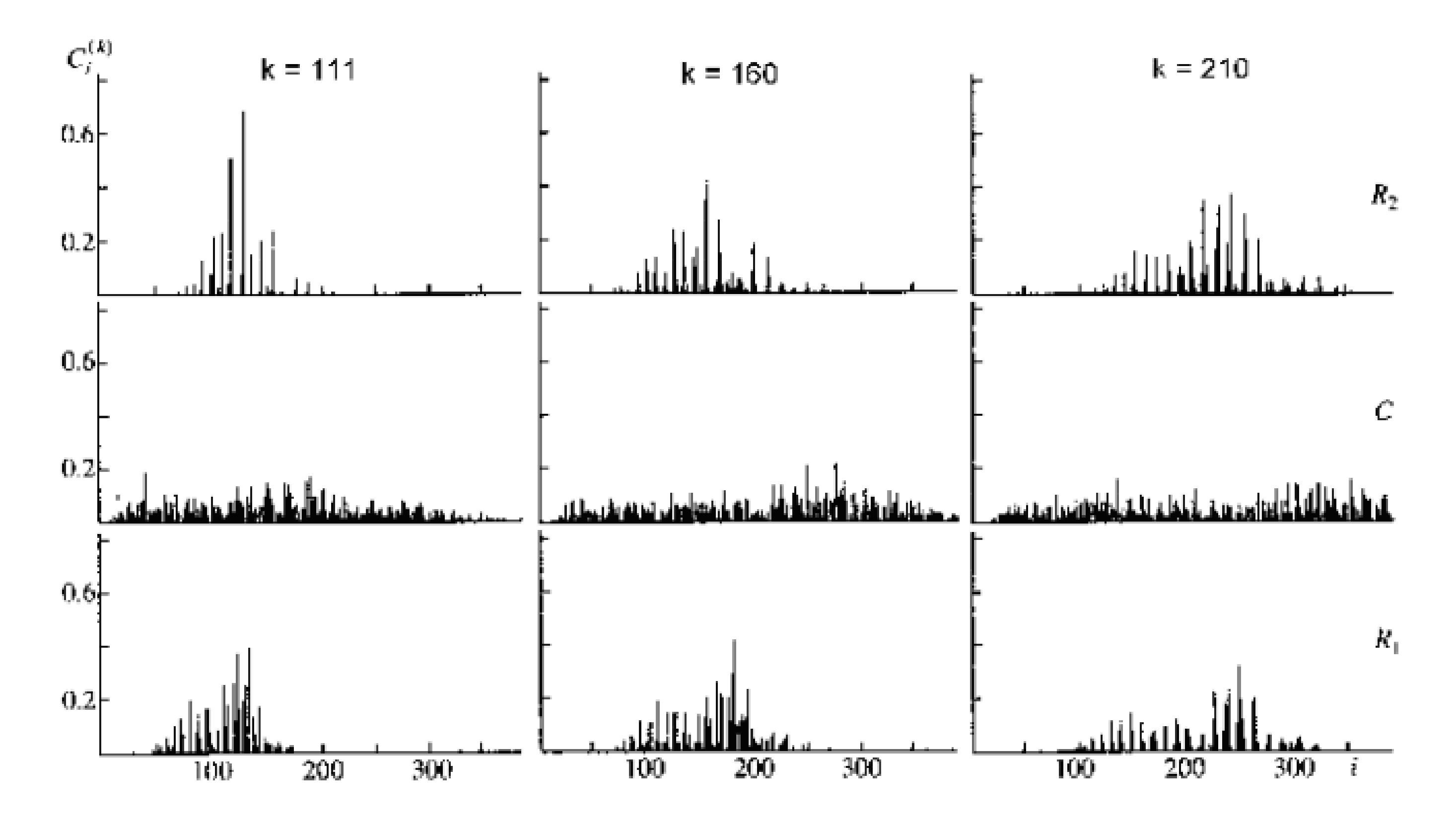}
\caption{\label{wf_distribution} Distribution of the coefficients
$C_i^{(k)}$ on the number $i=\{N,L,j\}$ of the basis states. The
numbers $k$ of the corresponding levels are shown over the figures.}
\end{figure}

If we introduce the notion of distributivity of wave function in
basis, then the criterion of stochasticity formulated by Nordholm
and Rise \cite{rice} states that the average degree of
distributivity of wave functions arises with the degree of
stochastization in the system. It is clear that this criterion is a
direct analog of Berry's hypothesis for $H_0$-representation, if
one interprets the number of the basis state $i_{NLj}$ as a discrete
coordinate. Fig.\ref{wf_distribution} qualitatively confirms this
criterion. It can be seen from this figure that the states that
correspond to regular motion (regions $R_1$ and $R_2$) are
distributed in a relatively small number of basis states. At the
same time, states corresponding to chaotic motion (region $C$) are
distributed in a considerably larger number of basis states. In the
latter case, the contributions from a large number of basis states
in expansion (\ref{wf_expansion}) interfere; this results in a
complicated spatial structure of the wave function $\psi(x,y)$.

\begin{figure}
\includegraphics[width=\textwidth]{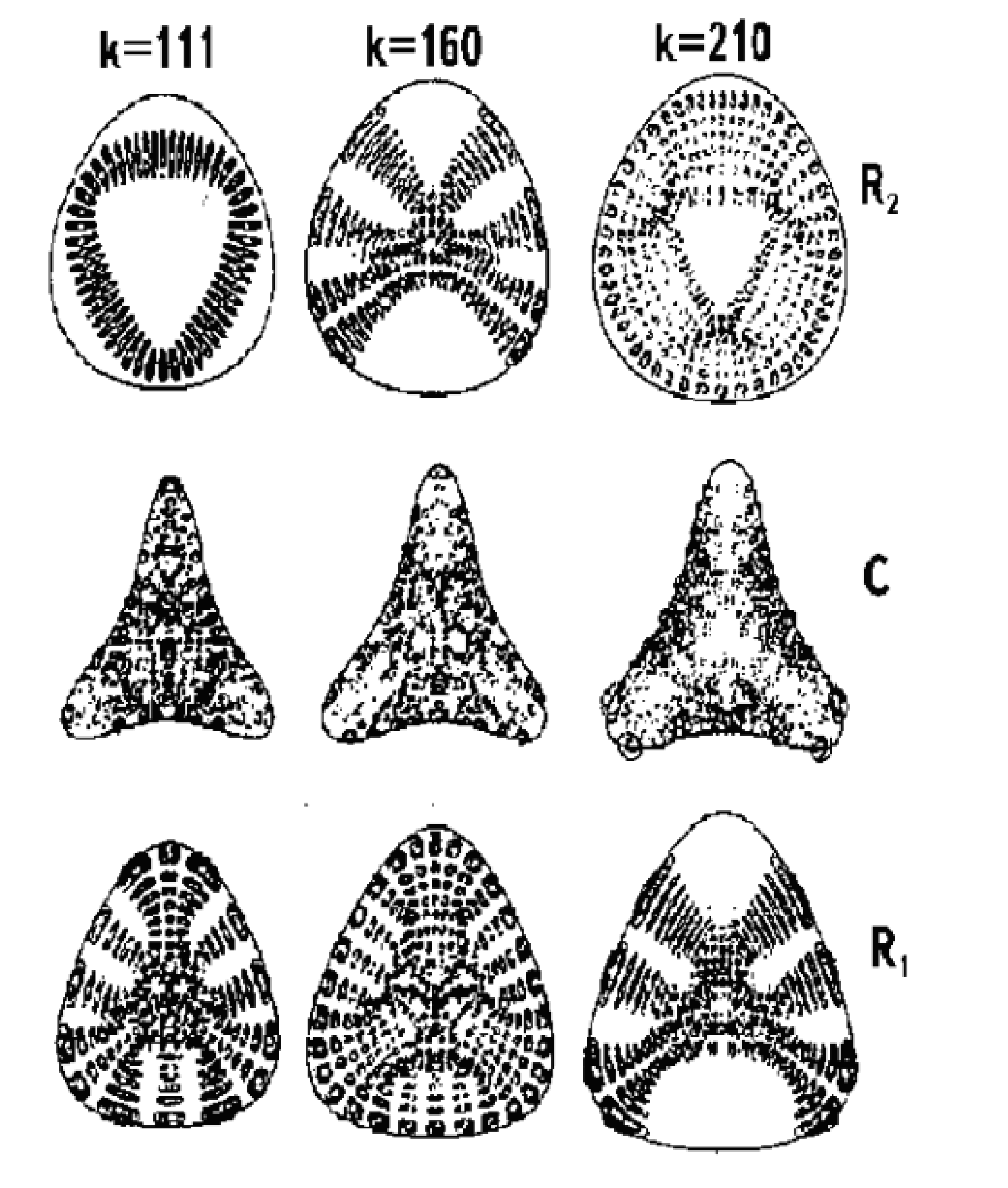}
\caption{\label{rcr-psi2} Isolines of probability density
$|\psi_k(x,y)|^2$. The step between the isolines is constant. The
numbers $k$ of the corresponding levels are shown over the figures.}
\end{figure}

Correlations between the structure of wave function and the type of
classical motion are also demonstrated in fig.\ref{rcr-psi2}, where
the probability density $|\psi_k(x,y)|^2$ for states with numbers
111, 160 and 210 is represented \cite{bolotin-87}. The squared
module of the wave functions reproduces rather well the transition from
functions with a clear internal structure (region $R_1$) to an
irregular distribution (region $C$) and the restoration of structure
in the second regular region ($R_2$). For the chosen technique, in
which transition is traced for the wave function with a fixed number (
scaled Planck constant), a change in the wave function is associated
only with $R-C-R$ transition.

The specific character of stationary wave functions corresponding to
a definite type of classical motion manifests also in the level lines
structure, in particular in the structure of manifolds
$\Psi(\mathbf{q})=0$. Depending on the configuration space
dimensionality, those manifolds are called nodal points ($N=1$),
lines ($N=2$), or surfaces ($N>2$).

Already in 1979, Stratt, Handy and Miller \cite{shm} had proposed a
quantum chaos criterion directly connected with nodal geometry:
for regular states $\Psi_n$ the system of nodal lines
$\Psi(\mathbf{q})=0$ represents a grid of quasi-orthogonal curves
(or is very close to it),while for chaotic states there is no such
representation.

\begin{figure}
\includegraphics[width=\textwidth,draft=false]{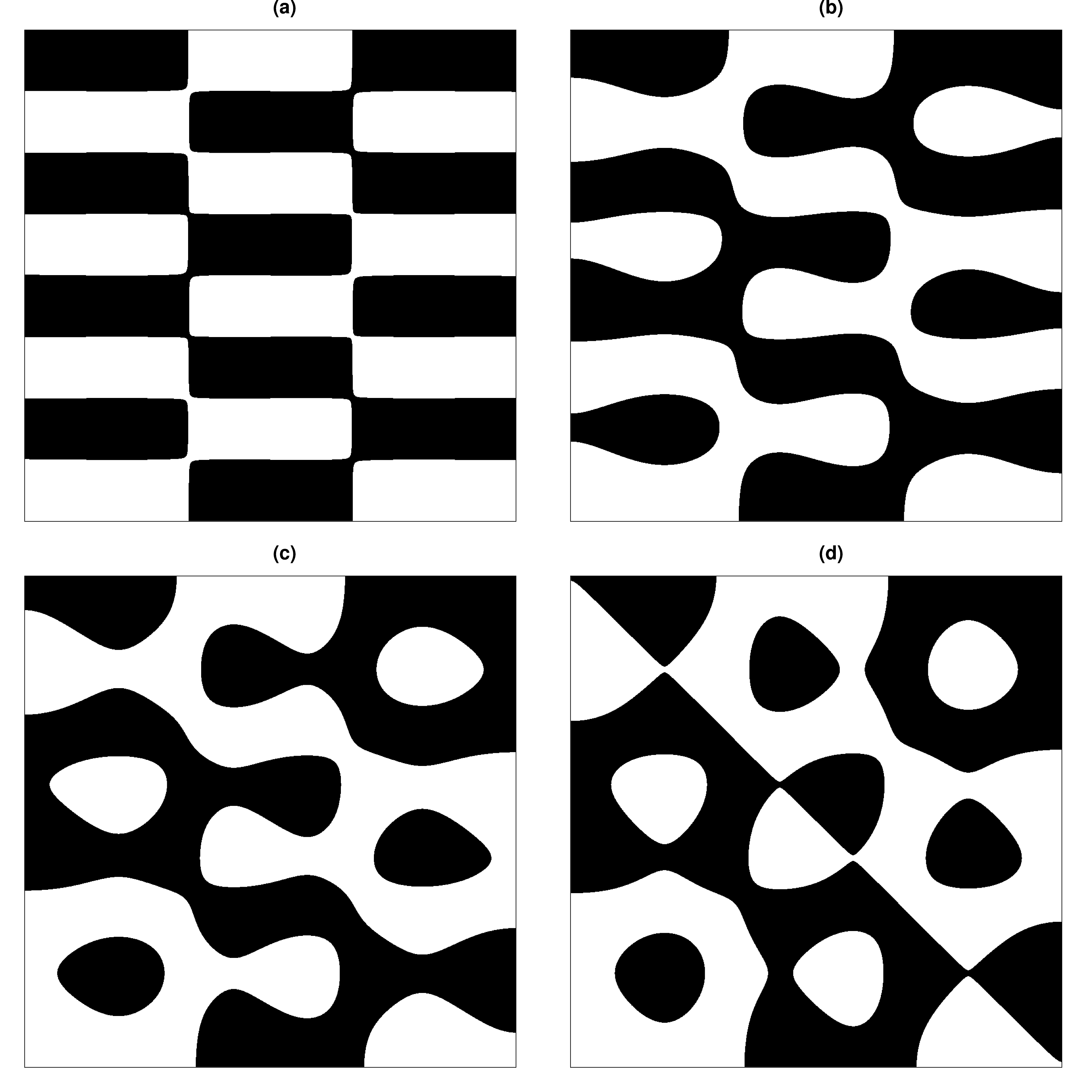}
\caption{\label{rbnd}The nodal domain structure for the rectangular
billiard stationary wave functions $\Psi(x,y)=\sin3\pi x\cos8\pi y +
\varepsilon\sin8\pi x\cos3\pi y$ for $\varepsilon=0.01$ (a),
$\varepsilon=0.3$ (b), $\varepsilon=0.6$ (c) and $\varepsilon=0.99$
(d).}
\end{figure}

For two-dimensional systems it is more convenient to describe not
the nodal lines themselves, but the regions of constant sign of the
wave function --- the so-called nodal domains, and then the borders
between the nodal domains coincide with the nodal lines. As can be
easily seen from fig.\ref{rbnd}, the stationary wave functions nodal
structure in integrable systems may have properties usually assigned
to chaotic ones --- it is an irregular picture of nodal domains and
avoids intersections of nodal lines.

\begin{figure}
\includegraphics[width=\textwidth]{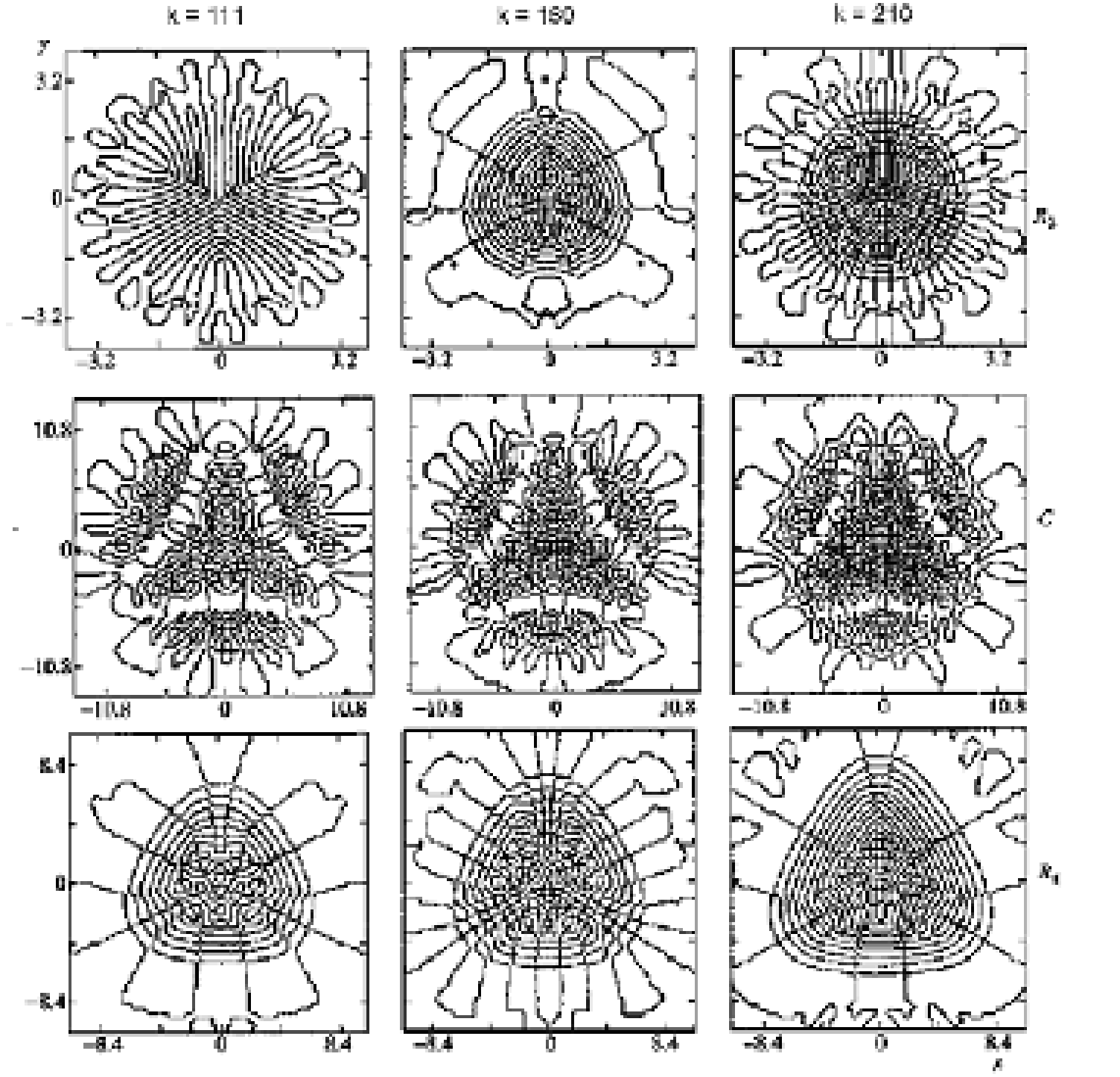}
\caption{\label{qonl}The nodal curves for the wave functions
$\Psi_k(x,y)$.  The numbers $k$ of the corresponding levels are
shown over the figures.}
\end{figure}

Fig.\ref{qonl} confirms that the structure of the lattice of nodal
curves of $QO$ wave functions undergoes a change in the $R-C-R$
transition. The spatial structure of nodal curves for states from
 regions $R_1$ and $R_2$ of regular classical motion is considerably
simpler than this structure for states from the chaotic region $C$.

The considered possibility corresponds to the traditional approach
in search of QMCS in wave function structure: it is the
investigation of changes in the wave functions structure with energy
increasing. Such an approach adresses the difficulties connected with
the necessity to separate the QMCS from the trivial modifications in
the wave functions structure due to quantum numbers changes.

\begin{figure}
\includegraphics[width=\textwidth]{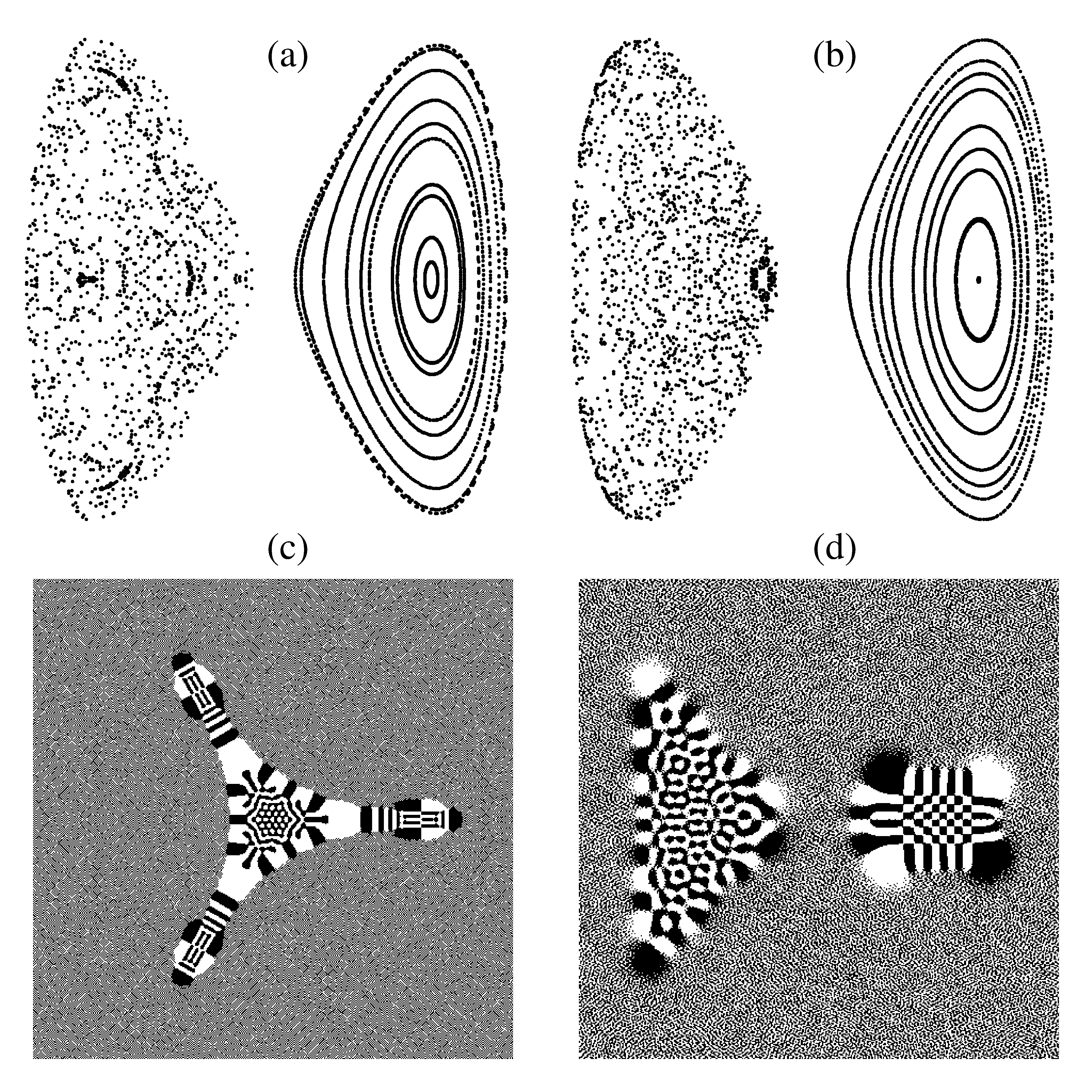}
\caption{\label{qod5}QMCS in the wave functions structure for
$QO$(a,c) and $D_5$(b,d) potentials.}
\end{figure}

Let us now turn to the multi-well case in the potentials, for
example, $QO$ with $W>16$ and $D_5$. The stationary wave functions
corresponding to the energies at which the classical mixed state is
observed, allow us to see the QMCS in comparison of different parts of
one and the same wave function \cite{we_ptp,we_pla}. As in the mixed
state, different types of classical motion are realized in different
local minima; it is manifested in the wave functions structure. The
parts of a wave function corresponding to different local minima
show the above described features, characteristic for regular or
chaotic motion respectively. Comparing the structure of the
eigenfunction in central and peripheral minima of the $QO$ potential
(Fig.\ref{qod5}a,c), or in left and right minima of the $D_5$
potential (Fig?\ref{qod5}b,d), it is evident that the nodal
structures of the regular part and the chaotic part of the
eigenfunction are clearly different.
\begin{description}
\item[i)] within the classically allowed region the nodal domains of
the regular part of the wave function form a well recognizable
checkerboard-like pattern \cite{davis}, while nothing similar can be
observed for the chaotic part;
\item[ii)] the nodal lines of the
regular part exhibit crossings or very tiny quasi-crossings; in the
chaotic part the nodal lines quasi-crossings have significantly
larger avoidance ranges;
\item[iii)] while crossing the classical turning line $U(x,y)=E_n$, the nodal
lines structure of the regular part immediately switches to the
straight nodal lines, going to infinity, which makes the turning
point line itself easily locatable in the nodal domains structure;
in the chaotic part, an intermediate region exists around the turning
line, where some of the nodal lines pinch-off, making transition to
the classically forbidden region more gradual and not so
manifest in the nodal structure.
\end{description}

Let us note an additional advantage in nodal structure
investigations for QMCS search in the mixed state case. In generic
multi-well potentials, stationary wave functions as a rule are
strongly localized in one or a few local minima. The localization
rate is conveniently characterized by the dimensionless variable
(for example in the case of two-well potential $D_5$)
\[\eta_n=\frac{P_n^R-P_n^C}{P_n^R+P_n^C},\]
where
\[P_n^R=\int\limits_{x>0}|\psi_n|^2 dxdy\]
\[P_n^C=\int\limits_{x<0}|\psi_n|^2 dxdy\]

\begin{figure}
\includegraphics[width=0.5\textwidth]{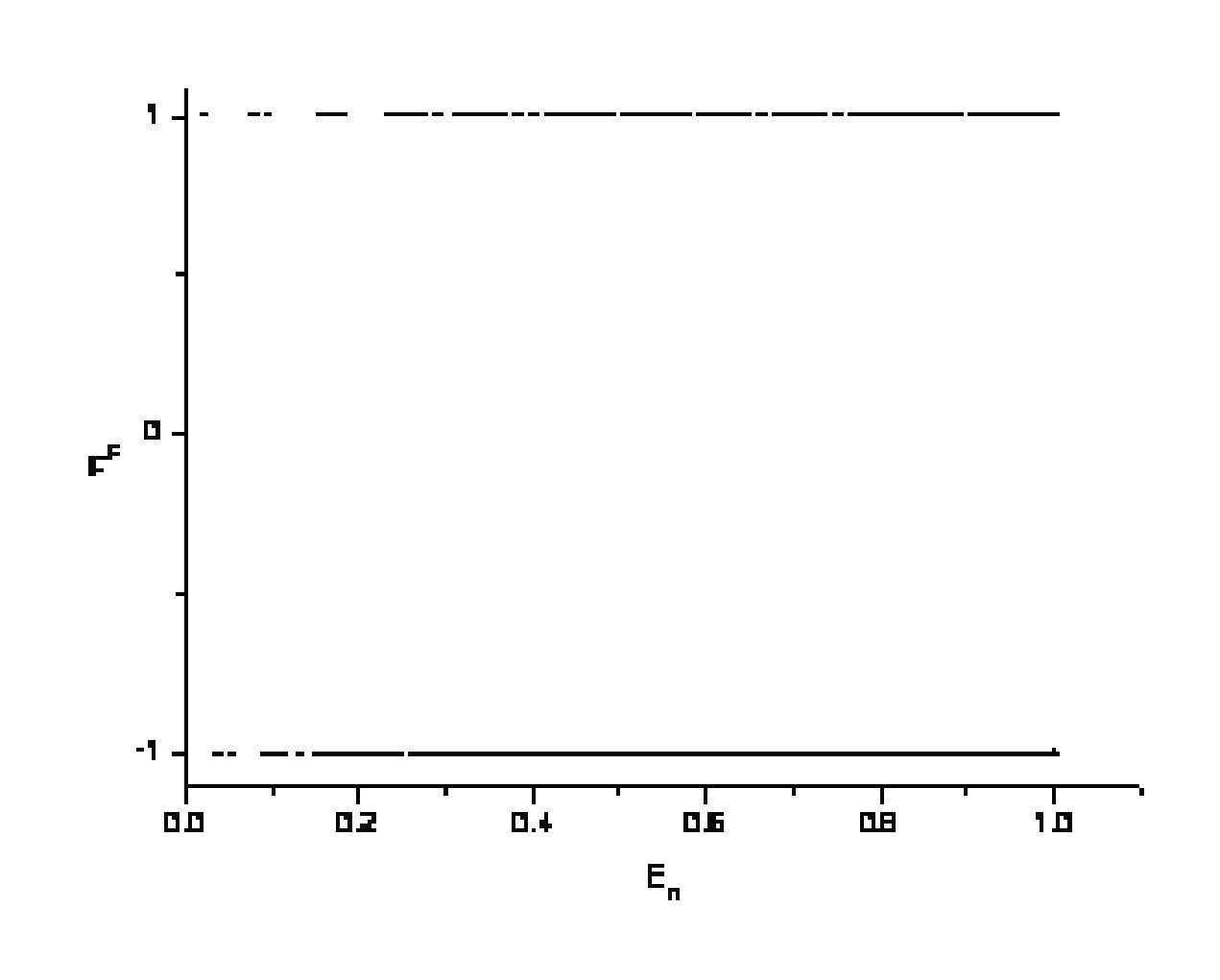}
\includegraphics[width=0.5\textwidth]{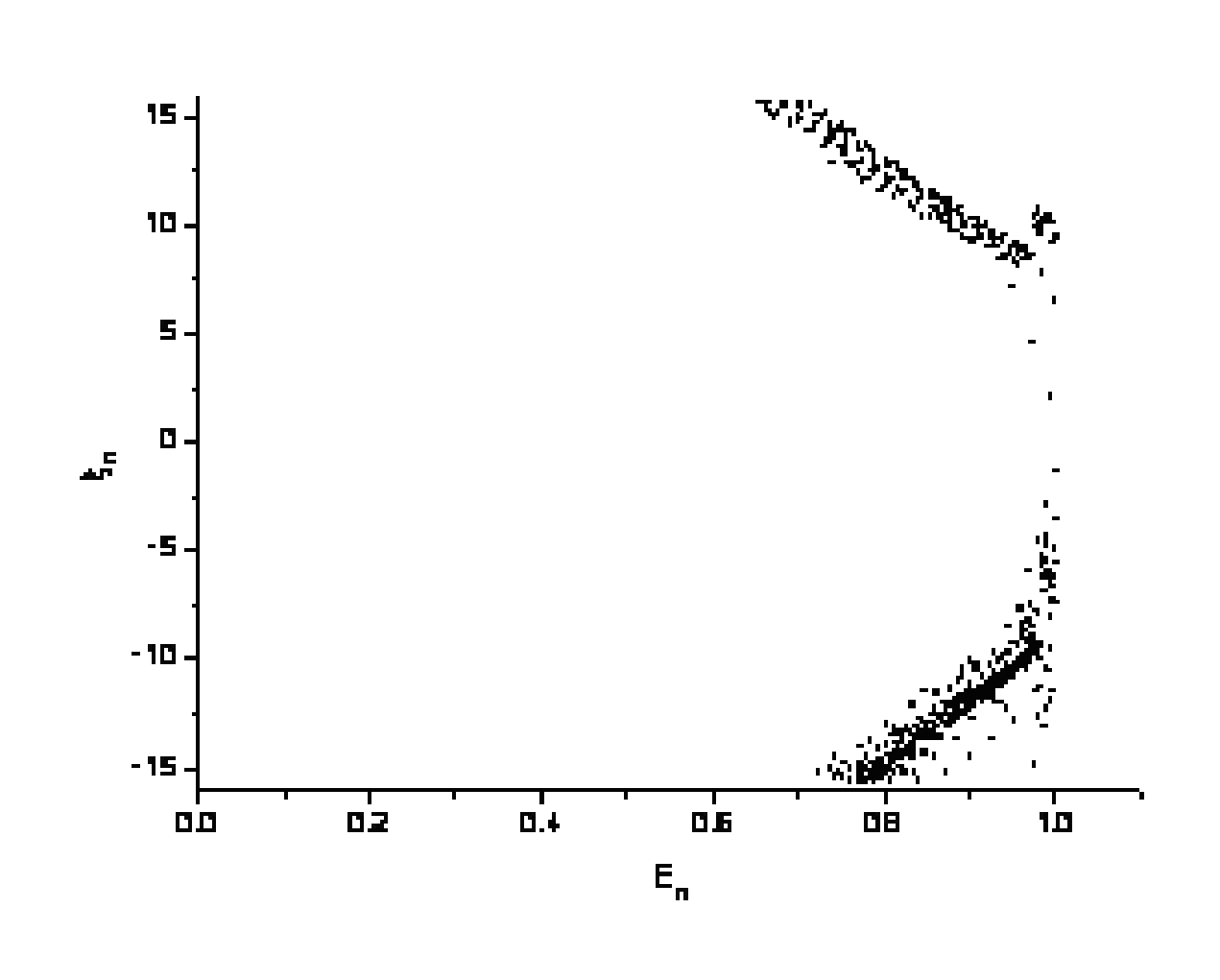}
\caption{\label{local}Localization of the wave functions in $D_5$
potential.}
\end{figure}

Evidently $\eta=1$ for regular and $\eta=-1$ for chaotic states.
Proximity of $\eta$ to unity is characterized by the variable
\[\xi_n=-sign(\eta_n)\log_{10}(1-|\eta_n|)\]
Therefore $\xi_n=1$ corresponds to a wave function for $90\%$
localized in the regular minimum, and $\xi_n=-1$ --- for $90\%$ in
the chaotic one. In the mixed state case, the amplitude of overwhelming
majority of wave functions in different local minima differs by
several orders of magnitude (fig.\ref{local}), which makes
difficulties for analysis of the probability density distribution, but those difficulties are however inessential for the analysis of
wave functions nodal structure.

\sat\section{Evolution of shell structure in the process of $R-C-R$
transition}\sat

Introduction of the stochasticity concept in nuclear theory made it
possible to take a fresh look at the old paradox: how one could
reconcile the liquid drop --- short mean free path --- model of the
nucleus with the independent particles --- gas-like shell model. To
solve this paradox it is sufficient to assume the following:
\begin{enumerate}
\item When the nucleonic motion inside the nucleus is integrable,
we expect to see strong shell effects in nuclear structure, quite
well reproducible, for example, by the model of independent
particles in the potential well.
\item If in the nuclear dynamics
a chaotic component is dominant, it is necessary to expect, that the droplet model or Thomas-Fermi approximation will appear more
useful.
\end{enumerate}

With such an approach, the elucidation of the mechanism of
destruction of shell effects in the process of the $R-C$ transition
plays the key role. More appropriately the problem can be formulated
in the following way: how does the shell dissolve with deviations
from integrability or, conversely, how do incipient shell effects
emerge as the system first begins to feel its proximity to an
integrable situation?

As has been mentioned above, the finite motion of integrable
Hamiltonian system with  $N$-degrees of freedom, in general, is
conditionally periodic, and the phase trajectories lie on
$N$-dimensional tori. In the action-angle variables $(I,\theta)$ the
Hamiltonian is cyclic with respect to angle variables, $H=H_0(I)$.
Poincar\'e called the main problem of dynamics the problem of the
perturbation of conditionally periodic motion in the system defined
by the Hamiltonian
\[H=H_0(I)+\varepsilon V(I,\theta)\]
where $\varepsilon$ is a small parameter. The essential step in the
solution of this problem was the KAM theorem, asserting, that at turning on non-integrable perturbation, the majority of
non-resonance tori, i.e. the tori for which
\[\sum\limits_{i=1}^N n_i\frac{\partial H_0}{\partial I_i}\ne0\]
are conserved, distinguishing from the unperturbed cases only by
small (to the extent of smallness of $\varepsilon$) deformation. As
was noted before, at definite conditions the KAM formalism allows us
to remove the terms depending on the angle out of the Hamiltonian,
using the convergent sequence of canonical transformations. When it
succeeds, we find that perturbed motion lies on rather deformed
tori, so that trajectories, generated by perturbed Hamiltonian
remain quasi-periodic. In other words, the KAM theorem reflects an
important peculiarity of classical integrable systems: they conserve
regular behavior even at rather strong non-integrable perturbation.
In the problem of our interest, concerning the destruction of the
shell structure of quantum spectrum, the KAM theorem also will be
able to play an important role. Considering the residual
nucleon-nucleon interaction as non-integrable addition to the
self-consistent field, obtained, for example, in the Hartree-Fock
approximation, we can try to connect the destruction of shell with
the deviation from integrability. The existence of shell structure
at rather strong residual interaction (or at large deformation) can
be due to the rigidity of KAM tori, contributing to survival of
regular behavior. Such an assumption seems rather natural,
especially if we take into account that the procedure of
quasi-classical quantization \cite{morse_1,morse_2} itself as well
as KAM theorem are based on convergence of the same sequence of
canonical transformations.

The aim of this section is to trace the evolution of the shell
structure of $QO$ Hamiltonian. In numerical calculations of this
section it will be more convenient to use non-scaled version of the
$QO$ Hamiltonian (\ref{qo_ham}) . The unperturbed Hamiltonian
$H_0$ we will understand as the Hamiltonian of two-dimensional harmonic
oscillator with equal frequencies:
\[H_0\equiv H(a=1,b=0,c=0).\]
Its degenerate equidistant spectrum is well known. At switching on
perturbation, the degeneration disappears and the shell structure
forms. The number of states, for example the states of $E$-type
(the numerical results, represented below, are relative to the
states of this type, while analogous results are observed for the
states belonging to another irreducible representations of
$C_{3v}$-group) for the given quantum number $N$ is equal to
\[N_0=\frac12(N_1+Mod(N,2)),\]
where
\[N_1=\frac13(2N+Mod(N,3)),\]
and $Mod(N,M)$ is the remainder of division of $N$ by $M$.

It is obvious, that eigenfunctions of exact Hamiltonian $QO$ are no
longer the eigenfunctions of operators $\hat{N}$ and $\hat{L}$.
Nevertheless, as numerical calculations show, one can use the
quantum numbers $N$ and $L$ for the classification of wave functions
even at rather large nonlinearity. The measure of nonlinearity, at
which such classification (i.e. the existence of shell structure)
remains reasonable, is connected with quasi-crossing of neighboring
levels. Quasi-crossing we shall understand as the approach of
levels up to the distances of order of numerical calculations error.

\begin{figure}
\includegraphics[width=\textwidth]{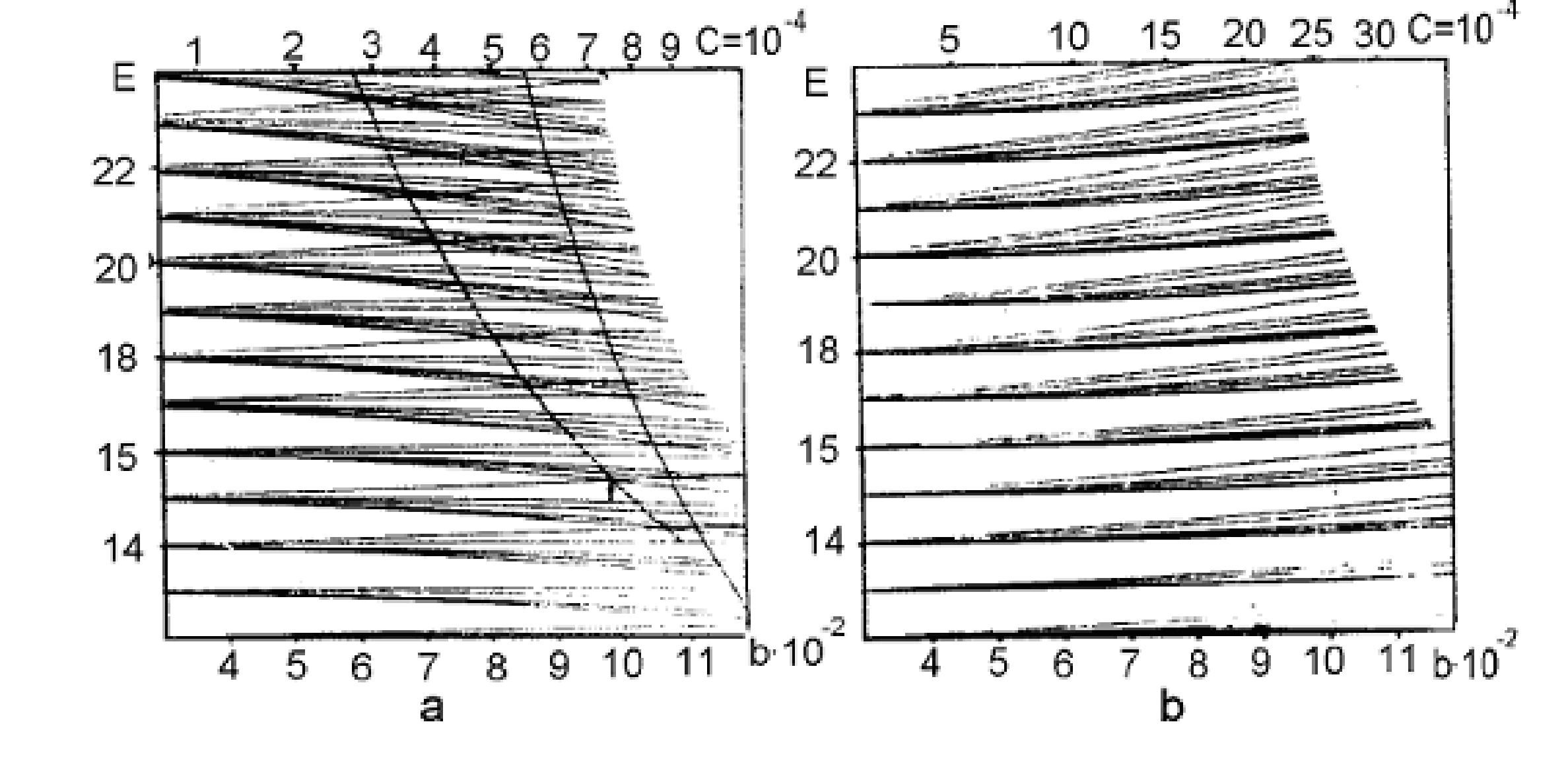}
\caption{\label{level_qc}a) Energy spectra for QO Hamiltonian
(\ref{qo_ham}) depending on values of parameter $b$ for region $C$
$W=13$. Points mark the quasi-crossings of levels. Bold solid line
shows the dependence of classical critical energy $E_{cr}$ on
parameter $b$. The dashed line shows the quasi-crossing region
border. Arrow shows the point of quasi-crossing of levels with
numbers $k=40$ and $k=41$. b) Energy spectra for QO Hamiltonian
(\ref{qo_ham}) depending on values of parameter $b$ for region $C$
$W=3.9$. There are no quasi-crossings of levels.}
\end{figure}

The dependence of energy spectra of $QO$ Hamiltonian on the
parameter $b$ for the values $W=3.9$ and $W=13$ are represented in
Fig.\ref{level_qc}. As can be seen from Fig.\ref{level_qc}a,  for
the PES with $W=13$ at the approach to the line of critical
energy of the transition to chaos, defined according to the negative
curvature criterion, the destruction of shell structure, which we understand as the beginning of multiple quasi-crossings, takes
place. At the same time, for PES with $W=3.9$ on
Fig.\ref{level_qc}b \footnote{as we have shown in Section \ref{ms},
the local instability is absent for that case and the classical
motion is regular at all energies}  quasi-crossings are absent
even at larger nonlinearity than for $W=13$.

\begin{figure}
\includegraphics[width=\textwidth]{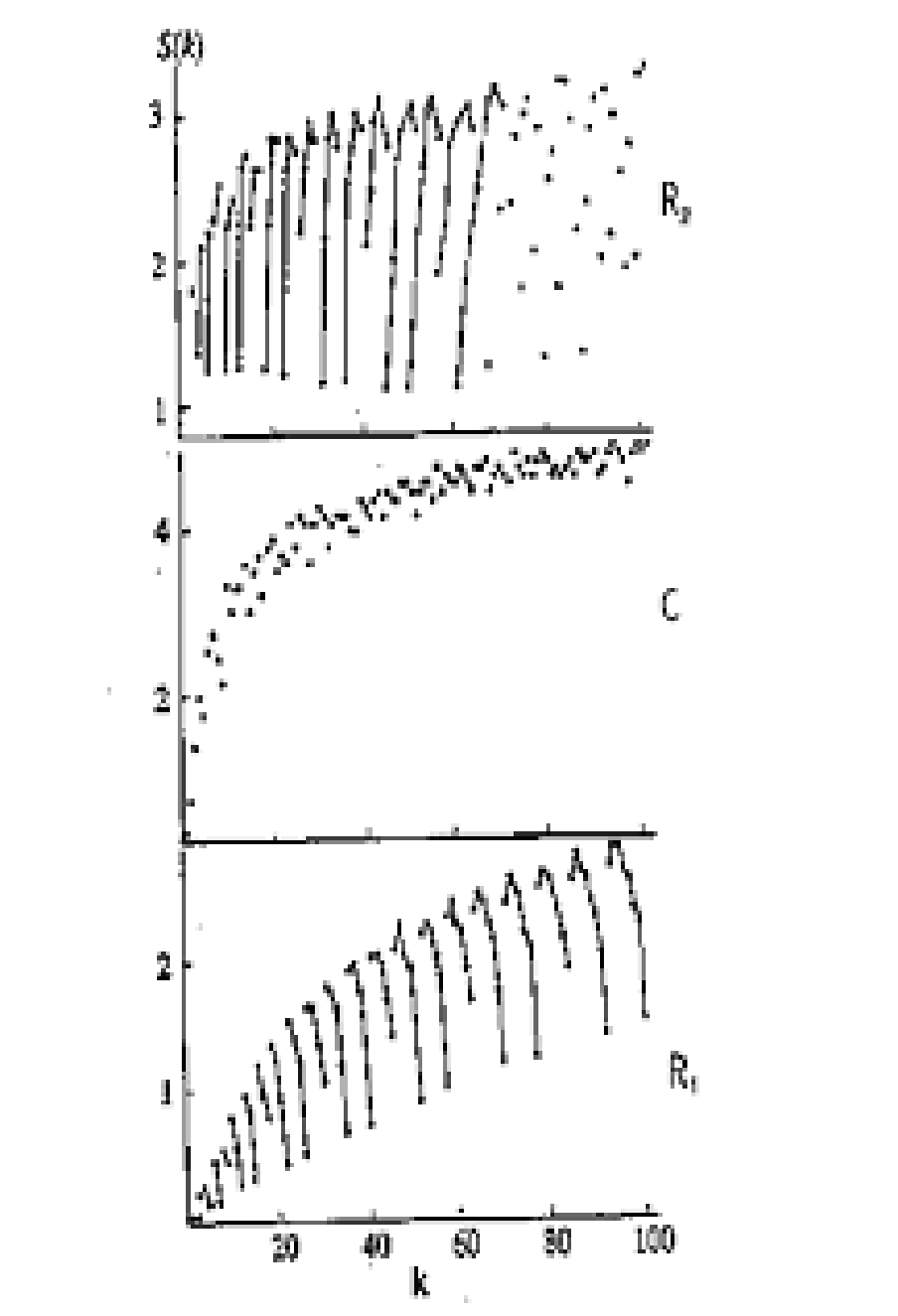}
\caption{\label{shell} Entropy $S$ as function of the state number.
The straight lines connect points corresponding to shell
classification on the basis of $N$.}
\end{figure}

The destruction of shell structure can be traced for the wave
functions, using the analogue of thermodynamic entropy $S_k$
\cite{yonezava,reichl_s},
\[S_k=-\sum|C^k_{NLj}|^2\ln|C^k_{NLj}|^2\]
The character of changes of entropy in regions $R_1$ and $R_2$,
corresponding to regular classical motion, correlates with the
transition from shell to shell (see Fig.\ref{shell}). Two effects
are observed in the region $C$ corresponding to chaotic classical
motion. Firstly, quasi-periodical dependence of entropy on energy is
violated, which testifies to the destruction of shell structure.
Secondly, monotone growth of entropy is observed on average with
going out on the plateau corresponding to the entropy of random sequence
at the energies essentially exceeding the critical energy.

\sat\chapter{Wave Packet Dynamics\label{ch_wp}}\sat

The investigation of the time evolution of non-stationary states,
i.e. of wave packets in quantum systems with classical analogues
that admit chaotic behavior, provides important information about
QMCS. The localized quantum wave packet (QWP) is the closest
analogue of the point in phase space, which describes the state of a
classical system. However, such correspondence between localized QWP
and the classical point particle in the chaotic region is broken
down for a very short time interval. Let us explain this fact by
using Takahashi arguments \cite{takahashi}. We take two localized
QWP $\Psi_1(\mathbf{x})$ and $\Psi_2(\mathbf{x})$ which are put in
the chaotic region at an initial time to be slightly different from
each other so that the difference between
$\langle\Psi_1|\mathbf{\hat{x}}|\Psi_1\rangle$ and
$\langle\Psi_2|\mathbf{\hat{x}}|\Psi_2\rangle$ (or
$\langle\Psi_1|\mathbf{\hat{p}}|\Psi_1\rangle$ and
$\langle\Psi_2|\mathbf{\hat{p}}|\Psi_2\rangle$) is very small. We
assume that in chaotic region the localized QWP does not either
extend in a certain time interval of the order $1/\sqrt\hbar$ like
that in regular region. Hence, following the Ehrenfest theorem, the
packets $\Psi_1(\mathbf{x})$ and $\Psi_2(\mathbf{x})$ move as
classical particles and the distance between
$\langle\Psi_1|\mathbf{\hat{x}}|\Psi_1\rangle$ and
$\langle\Psi_2|\mathbf{\hat{x}}|\Psi_2\rangle$ (or
$\langle\Psi_1|\mathbf{\hat{p}}|\Psi_1\rangle$ and
$\langle\Psi_2|\mathbf{\hat{p}}|\Psi_2\rangle$) is increasing
exponentially in time. Let us consider the superposition
\begin{equation}\label{wp2}\Psi(\mathbf{x})=\Psi_1(\mathbf{x})+\Psi_2(\mathbf{x}),\end{equation}
which also becomes a localized QWP in the initial state. From that
assumption, we can expect that the QWP (\ref{wp2}) does not extend
in a certain time interval of the order $1/\sqrt\hbar$ . However,
considering the exponential increment of the distance between
$\langle\Psi_1|\mathbf{\hat{x}}|\Psi_1\rangle$ and
$\langle\Psi_2|\mathbf{\hat{x}}|\Psi_2\rangle$, $\Psi(\mathbf{x})$
(\ref{wp2}) extends exponentially in the chaotic region and does not
behave as a classical particle. This result is inconsistent with
initial assumption and implies that in the chaotic region localized
QWP s (i.e. $\Psi_1(\mathbf{x})$, $\Psi_2(\mathbf{x})$ and
$\Psi(\mathbf{x})$) extend exponentially as does the classical
probability distribution in the first stage of time development.

In order to describe such unusual behavior of QWP we should address
to the concept of quantum-mechanical phase space. There are a few
well-established schemes to introduce phase-space variables in
quantum mechanics. In the present study we shall follow the
procedure proposed by Weissman and Jortner \cite{jortner}. Let us
consider an initially localized QWP $\Psi$ characterized by
coordinates $\mathbf{q}$ and momentum $\mathbf{p}$,
\[\mathbf{q}=\langle\Psi|\mathbf{\hat{q}}|\Psi\rangle,\ \mathbf{p}=\langle\Psi|\mathbf{\hat{p}}|\Psi\rangle.\]
Now we introduce the coherent states
$|\mathbf{p},\mathbf{q}\rangle$, which in the coordinate
$x$-representation are given by Gaussian wave packets
\begin{equation}\label{gwp}\langle\mathbf{x}|\mathbf{p},\mathbf{q}\rangle=\prod\limits_{j=1}^N
\left(\pi\sigma_j^2\right)^{\frac14}
e^{-\frac{(x_j-q_j)^2}{2\sigma_j^2}+\frac{ip_j}{\hbar}\left(x_j-\frac{q_j}{2}\right)}.\end{equation}

In the study of a system of $N$ coupled harmonic oscillators, it is
convenient to choose for constants $\sigma_j$ the rms zero-point
displacements
\[\sigma_j=\sqrt{\frac{\hbar}{m_j\omega_j}}\]
where $m_j$ are masses and $\omega_j$ are frequencies of the
uncoupled oscillators. With this choice of $\sigma_j$, the
coherent states $\alpha\equiv|\mathbf{p},\mathbf{q}\rangle$ become
the eigenstates of the harmonic oscillator annihilation operators
\begin{equation}\label{annihilation}a_j|\alpha\rangle=\alpha_j|\alpha\rangle=\end{equation} where
\[\alpha_j=\frac{1}{\sqrt2}\left(\frac{q_j}{\sigma_j}+i\frac{\sigma_j}{\hbar}p_j\right)\]
is a complex number.

Using these coherent states, it is possible to introduce the
following quantum-mechanical phase-space density
\[\rho_\Psi(\mathbf{q},\mathbf{p})=|\langle\alpha|\Psi\rangle|^2,\]
where $\Psi$ is any general wave packet. This quantum-mechanical
coherent-state phase-space density may be regarded as a quantum
analogue of classical phase-space density, since it satisfies an
equation of motion where the leading term (when expanded in powers
of $\hbar$), corresponds to the classical Liuville equation. The
stationary phase space densities are the following
\begin{equation}\label{rho_e}\rho_E(\mathbf{q},\mathbf{p})=|\langle\alpha|E\rangle|^2,\end{equation}
where $|\alpha\rangle$ is given in terms of Gaussian wave packet
(\ref{gwp}), while the eigenstate $|E\rangle$ is given by squares of
the projections of the eigenstates on the coherent state,
Eq.(\ref{annihilation}). Using the well-known expressions for scalar
products $\langle\alpha|NL\rangle$ we finally obtain
\[\rho_E(\mathbf{q},\mathbf{p})=\frac12 e^{-\frac12(|\alpha_+|^2+|\alpha_-|^2)}
\sum\limits_{N,L}\frac{C_{NL}}{\sqrt{n_+!n_-!}}(\alpha_+^{*n_+}\alpha_-^{*n_-}+j\alpha_+^{*n_-}\alpha_-^{*n_+}),\]
where
\[\alpha_\pm=\frac{1}{\sqrt2}(\alpha_2\mp i\alpha_1)\]
with
\[\alpha_{1,2}=\frac{1}{\sqrt2}(q_{1,2}+ip_{1,2}),\ n_+=\frac{N+L}{2},\ n_-=\frac{N-L}{2},\ j=\pm1.\]
The phase-space density $\rho_E(\mathbf{q},\mathbf{p})$ is a
function of the four real variables $p_1,q_1$  and $p_2,q_2$. We can
get the contour maps of $\rho_E(p_1,q_1;p_2,q_2)$ in the $(p_2,q_2)$
plane, taking $q_1=0$ and calculating $p_1$ from the relation
\[H(p_1,q_1=0,p_2,q_2)=E\]

Quantum Poincar\'e maps (QPM), obtained this way, constitute the
quantum analogues of the classical Poincar\'e maps and can be used
for the search for QMCS both in the structure of wave functions of
stationary states and in the dynamics of wave packets.

Let us consider next the time evolution of a wave packet, which is
initially in coherent state
\[|\Psi(t=0)\rangle=|\alpha\rangle\]
The time evolution of such initially coherent wave packet is given
by
\[|\Psi(t)\rangle=\sum\limits_k|E_k\rangle\langle E_k|\alpha\rangle e^{-iE_kt}\]
The probability $p(t)$ of finding the system in its initial state
--- the survival probability --- is given by
\begin{equation}\label{p_t_g_a}p(t)=|g_\alpha(t)|^2\end{equation}
where $g_\alpha(t)$ is the overlap of $\Psi(t)$ with the initial
state
\[g_\alpha(t)=\langle\alpha|\Psi(t)\rangle=\sum\limits_k|\langle E_k|\alpha\rangle|^2 e^{-iE_kt}.\]
Using the definition of stationary phase-space density
(\ref{rho_e}), we can write
\begin{equation}\label{g_a}g_\alpha(t)=\sum\limits_k\rho_{E_k}
e^{-iE_kt}.\end{equation} Equation (\ref{g_a}) implies that dynamics
can be specified by the spectrum of the initial coherent state
$|\alpha\rangle$.

\begin{figure}
\includegraphics[width=\textwidth]{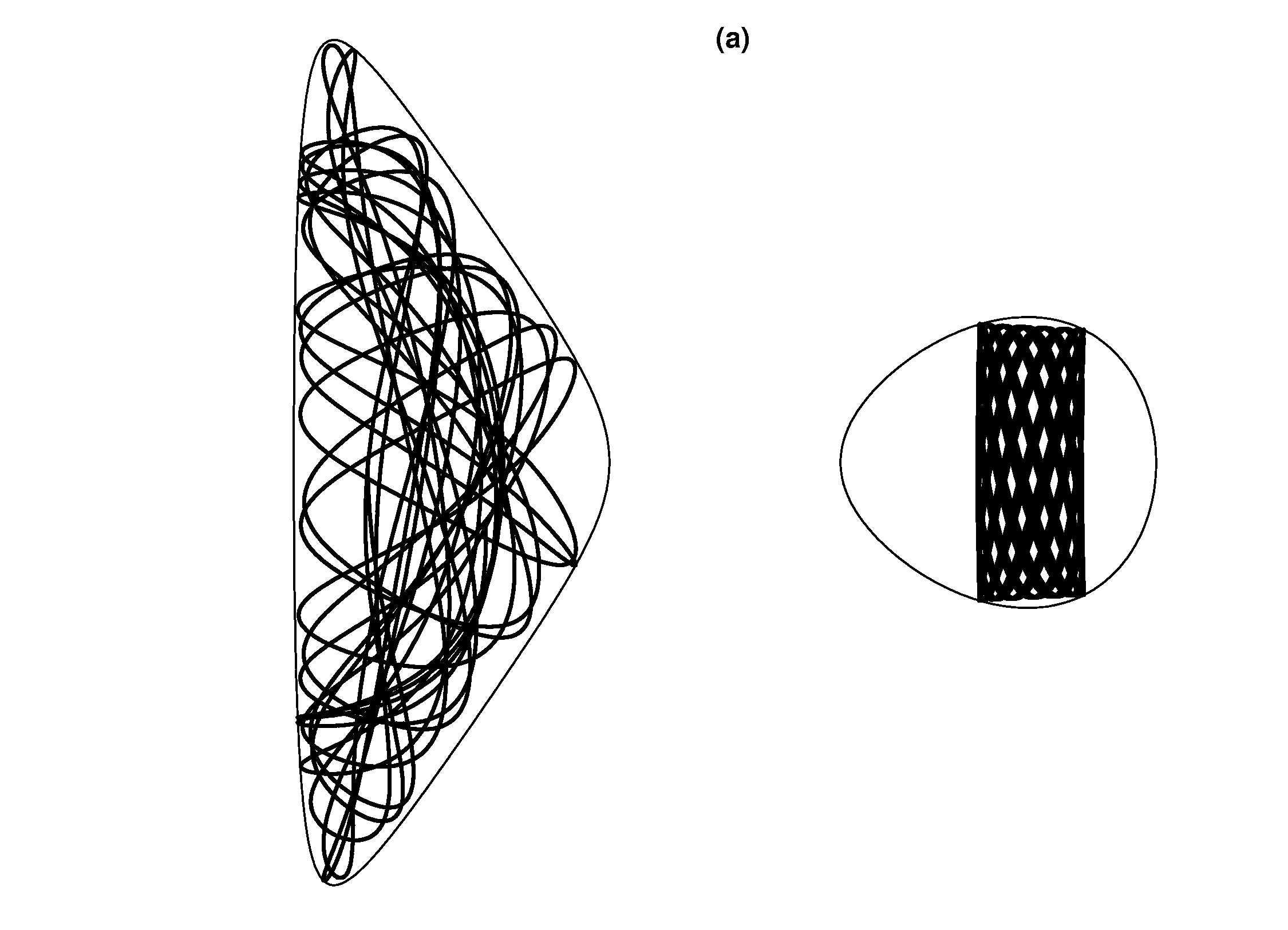}
\includegraphics[width=0.5\textwidth]{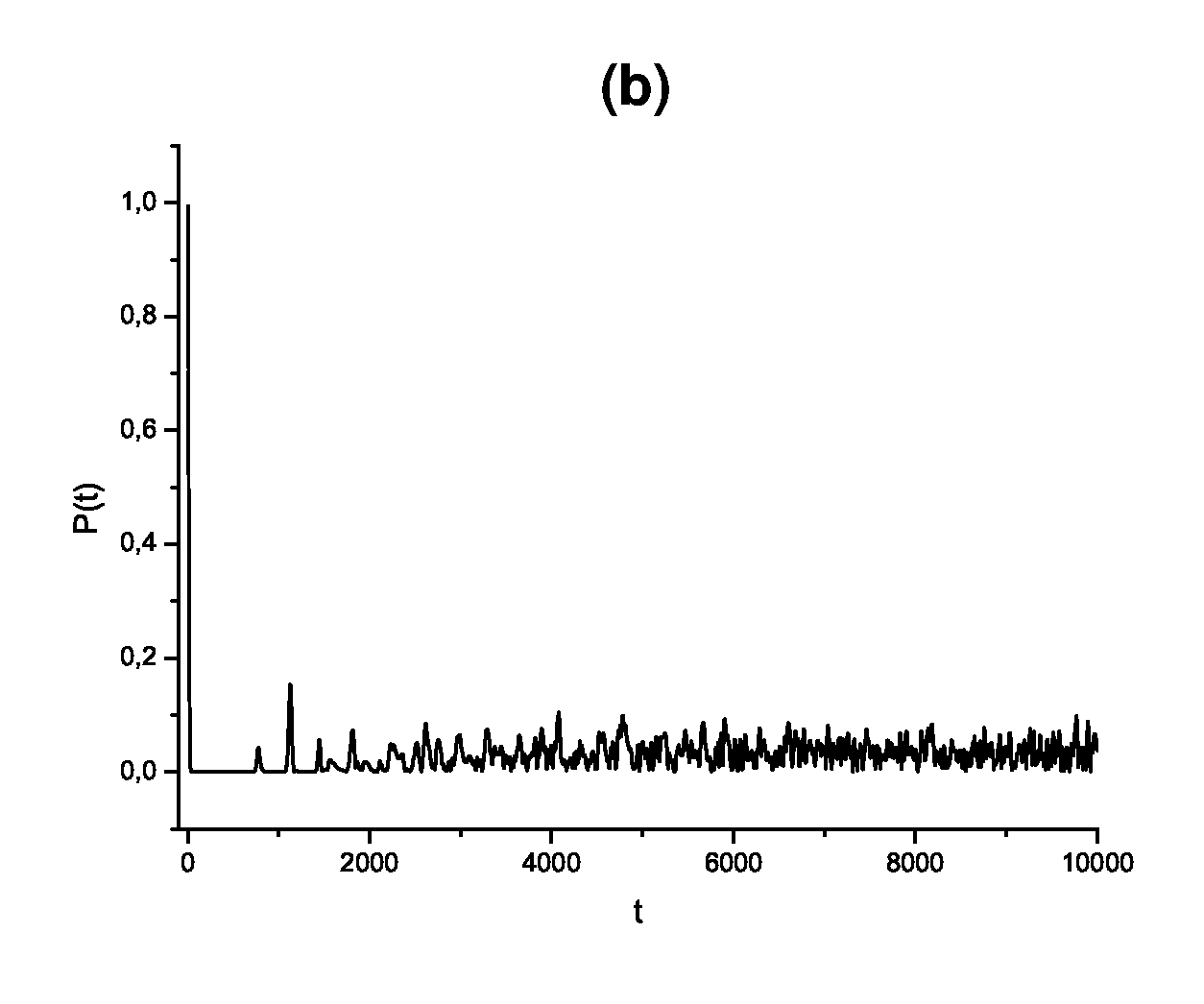}
\includegraphics[width=0.5\textwidth]{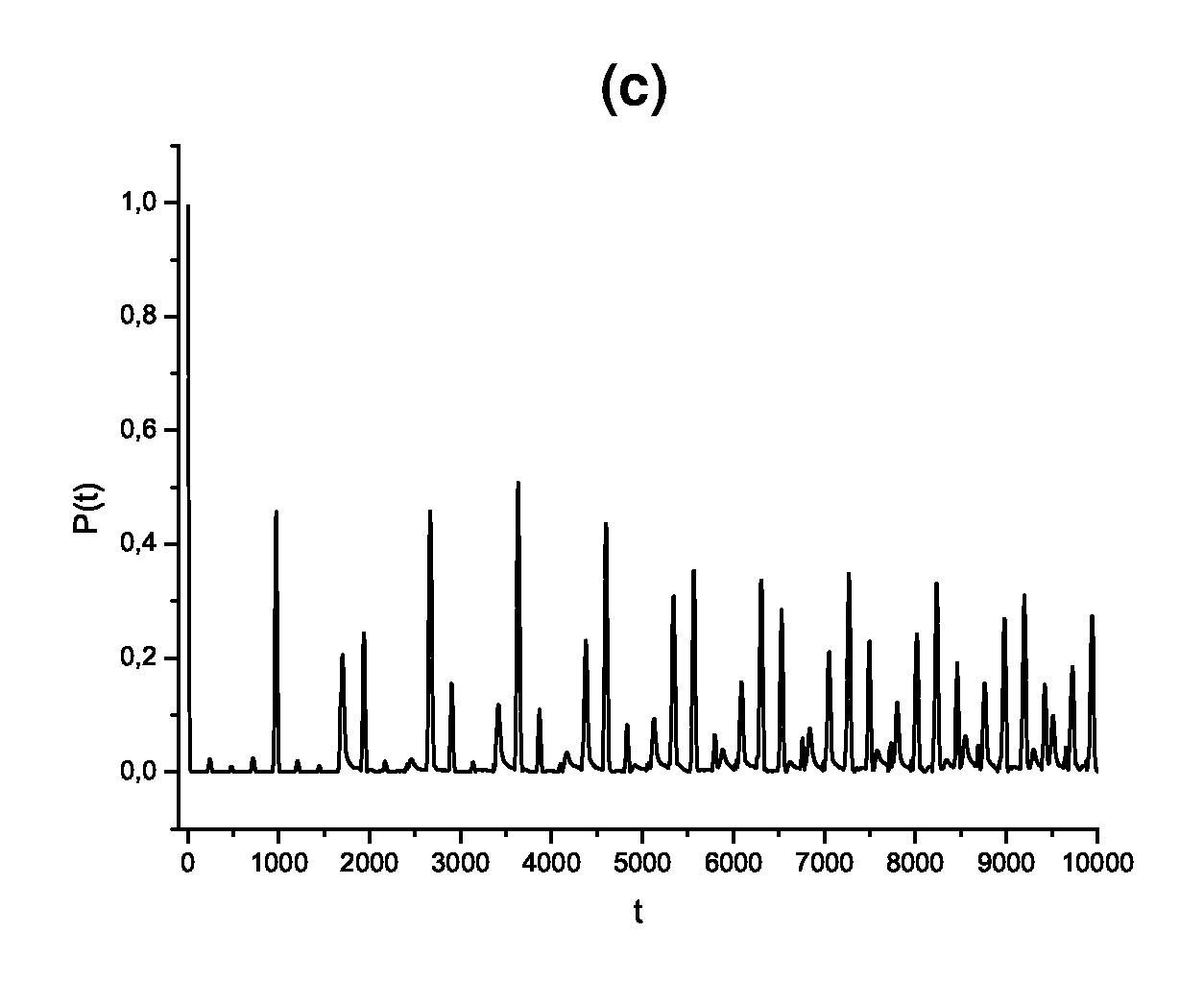}
\caption{\label{correlations_d_5} Typical trajectories for energies
corresponding to the mixed state (a) and correlation functions
(\ref{p_t_g_a}) for equivalent Gaussian wave packets in chaotic (b)
and regular (c) local minima of potential $D_5$ (\ref{u_d5}).}
\end{figure}

Weissman and Jortner \cite{jortner} have observed for H\'enon-Heiles
Hamiltonian two limiting types of QWP dynamics of initially coherent
Gaussian wave packets, which correspond to quasiperiodic time
evolution and to rapid decay of the initial state population
probability. Quasiperiodic time evolution is exhibited by wave
packets initially located in a regular region, while rapid decay of
the initial state population probability is revealed by those ones
that are initially placed in irregular regions. An analogous
situation takes place also in the multi-well potentials of our
primary interest, in the energy region where the mixed state is
realized. The quantum correlator $p(t)$ for two-well potential $D_5$
reproduces well the characteristic properties of classical
trajectories (see Fig.\ref{correlations_d_5}): quasi-periodic motion
in the regular minimum manifests itself in oscillations of the
correlator, and chaotic motion in another local minimum --- in fast
decay of correlations.

Let us turn to the consideration of the dynamics of QWP in the PES
with a few local minima (QO with $W>16$) \cite{bolotin95}.
Transitions between different local minima can be divided into
induced (the excitation energy exceeds the value of potential
barrier) and tunnel transitions. The latter are subdivided into
transitions from discrete spectrum into continuous spectrum (for
example,  $\alpha$-decay, spontaneous division), and from discrete
spectrum into discrete spectrum (for example, transitions between
isomeric states). Up to now the process of tunneling across a
multidimensional potential barrier, when initial and final states
are in discrete spectrum, has been the most complicated problem.

Most often the time evolution of wave packet is studied by two
methods: either by direct numerical integration of the Schr\"odinger
time-dependent equation with corresponding initial condition
$\Psi(\mathbf{r},t=0)$, or by expansion of the packet
$\Psi(\mathbf{r},t)$ in the eigenfunctions of the stationary
problem. The first one has some shortcomings, e.g., the complication
relative to the interpretation of the obtained results and the necessity
of separating the contributions from sub-barrier and tunnel
transitions for the packet of an arbitrary shape. These difficulties
can be avoided if the sub-barrier part of the spectrum
$E_n<U_0$ ($U_0$ is the height of the barrier) and the corresponding
stationary wave functions $\psi_n(\mathbf{r})$ are known. Pure
tunnel dynamics will take place for the packets representable in the
form
\[\Psi(\mathbf{r},t)=\sum\limits_n C_n e^{-\frac i \hbar E_n t}\psi_n(\mathbf{r}),\ E_n<U_0,\]
\[C_n=\int\Psi(\mathbf{r},t=0)\psi_n(\mathbf{r})d\mathbf{r}.\]

The probability $p^R(t)$ of finding the particle at the moment of time
$t$ in certain local minimum $R$ is
\[p^R(t)=\int\limits_R |\Psi(\mathbf{r},t)|^2d\mathbf{r}=\sum\limits_{m,n}C_m^* C_n e^{\frac{i(E_m-E_n)t}{\hbar}}
\int\psi_m^*(\mathbf{r})\psi_n(\mathbf{r})d\mathbf{r}\] or in the
two level approximation
\[p^R(t)=p^R(0) - 4C_1 C_2
\sin^2\left(\frac{i(E_m-E_n)t}{2\hbar}\right)
\int\psi_1^*(\mathbf{r})\psi_2(\mathbf{r})d\mathbf{r}.\]

Let us introduce the value
\[\bar{p}^R=\max_{\forall t}p^R(t)\]
which is the maximum probability of finding the particle in certain local minimum $R$, if initially it was localized to an arbitrary minimum. If the number of local minima is more than two,
then of separateinterest is
\[\bar{\bar{p}}^{R_0}=\min_{\forall t}p^{R_0}(t)\]
i.e., the minimum probability to find the wave packet in the minimum
$R_0$ corresponding to its initial localization.

\begin{figure}
\center{\includegraphics[width=0.5\textwidth]{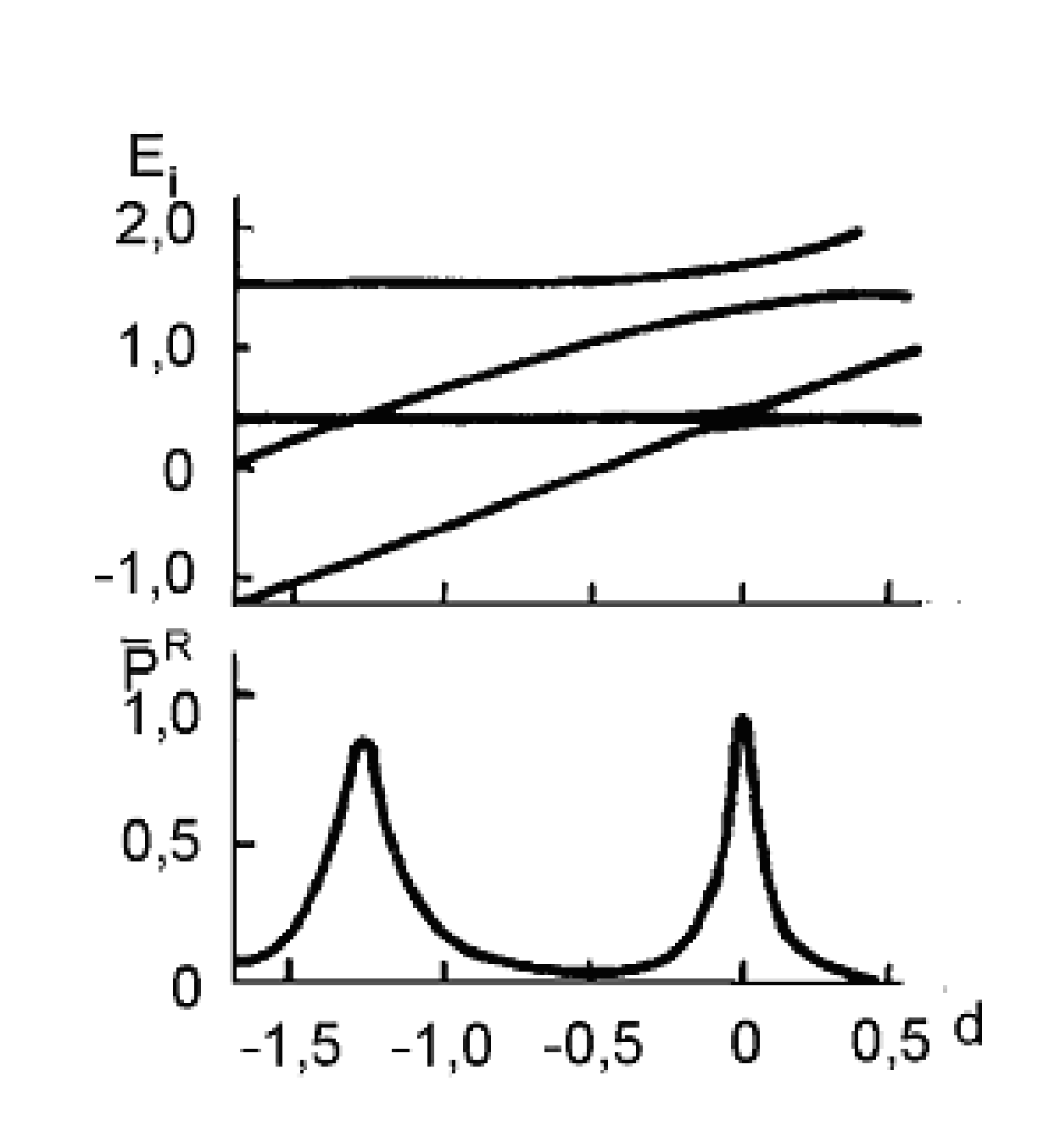}}
\caption{\label{level_crossing} Dependence of sub-barrier energy
levels $E_i$ for a double asymmetric one-dimensional rectangular well
with infinite external walls (above) and $\bar{p}^R$ on $d$ (below).
Width of the well equals 3, width of the barrier --- 1, barrier
height --- 2.}
\end{figure}

Intuitively, we may suggest that $\bar{p}^R\approx1$ if the initial
minimum is local, and the final one is absolute. However, the
results of \cite{nieto} obtained for the simplest one-dimensional
models (asymmetric double wells of different shapes) are
inconsistent with the intuitive expectations. The probability of
tunneling from the local minimum to the absolute one depends
resonantly on the potential parameters. Fig.\ref{level_crossing}
gives the dependence of $\bar{p}^R$ on the well depth displacement
$d$. It can be seen that at an arbitrary asymmetry $\bar{p}^R\ll1$.

The resonant behavior of $\bar{p}^R$ becomes more clear if one
considers the spatial structure of the sub-barrier wave functions.
For a sufficiently wide barrier in the case of an arbitrary
asymmetry, the sub-barrier wave functions are largely localized in
separate minima. The delocalization takes place only in the vicinity
of the level quasi-crossing. The degree of this delocalization
directly depends on the distance between the interacting levels.
Obviously, the QWP, in which the components localized in the certain
minimum are dominating, cannot  tunnel effectively to the
neighboring minimum. It is precisely this reason that explains the
stringent correlation between the $\bar{p}^R$ minima and the level
quasi-crossing (see Fig.\ref{level_crossing}).

\begin{figure}
\includegraphics[width=\textwidth]{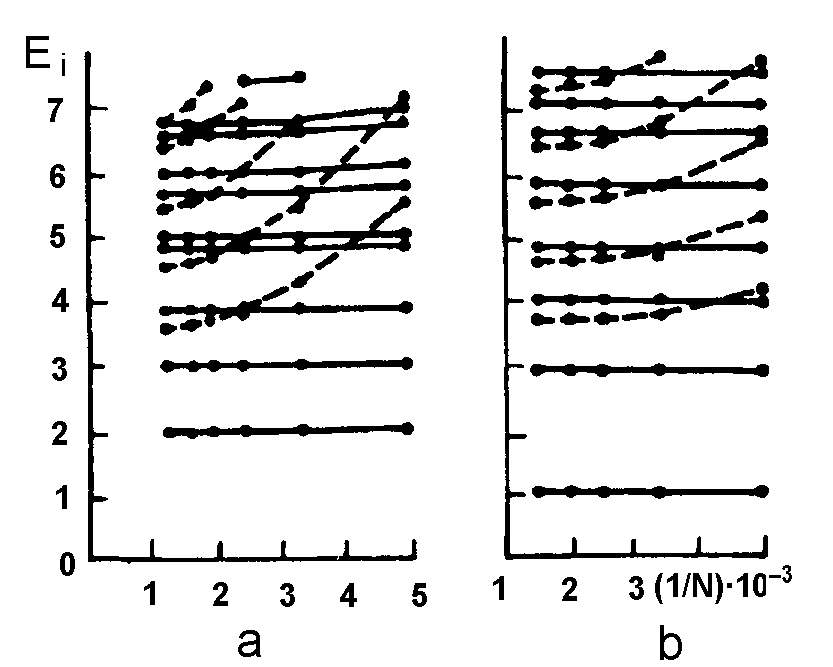}
\caption{\label{fig24} Dependence of the energy spectrum  of
$E$-type (a) at $b=0.173798$ and  $A_1$-type (b) at $b=0.17$ on the
dimension of sub-matrices $N$ for $W=17.8$. Dashed lines correspond
to the states localized in peripheral minima.}
\end{figure}

Now the question arises if a similar correlation between the level
quasi-crossings, the delocalization of wave functions and resonant
tunneling persists in the two-dimensional case. To give an answer to
this question let us turn to $QO$ Hamiltonian (\ref{qo_ham}) in the
region $W>16$. Recall that three ($C_{3v}$-symmetric) identical
additional minima appear at $W>16$ apart from the central minimum.
The central minimum exceeds the lateral ones in depth in the region
$16<W<18$. At $W>18$, the central minimum becomes the local one. In
this region of parameters the procedure of diagonalization of $QO$
Hamiltonian in oscillator basis becomes essentially complicated. It
is connected with the fact that the basis of Hamiltonian, the
potential of which has the unique minimum, is used for the
diagonalization of Hamiltonian with complex topology of the PES. In
addition to the large dimension of the basis, it is necessary for the
basis wave functions to have sufficient value in the region of
lateral minima. It has been achieved by the optimization of
frequency $\omega_0$ of the oscillator basis. For the values of
parameters used in the calculations ($W=17.8,\ b=0.17$) the value of
the optimal oscillator frequency was $\omega_0=0.2$. Fig.\ref{fig24}
gives lower eigenvalues of $E$ and $A_1$ types depending on the
basis dimension. We can see that it is possible to get saturation in the basis at the dimension of sub-matrices $N\sim10^3$
even for low states localized in the lateral minima (dashed lines).

\begin{figure}
\includegraphics[width=\textwidth]{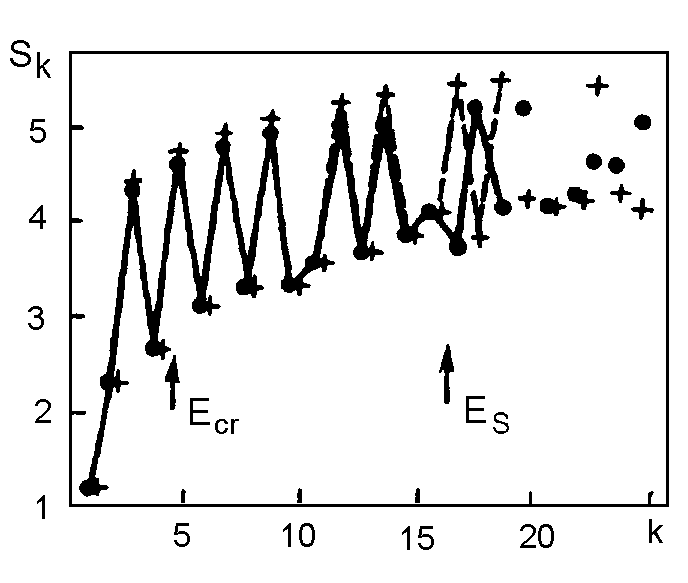}
\caption{\label{fig25} Dependence of $S_k$ on the state number $k$
($A_1$-type) for the dimension of sub-matrices $408$ (dark points)
and $690$ (crosses)  at $W=17.8$ and $b=0.17$. $E_S$ corresponds to
the saddle energy. and $E_{cr}$ --- to the classical critical energy
of the regularity-chaos transition.}
\end{figure}

We can also use the above introduced analog of thermodynamic entropy
$S_k$ to estimate the degree of saturation in the basis.
Fig.\ref{fig25} gives $S_k$ values for the states of $A_1$ type
for the dimension of sub-matrices $408$ and $690$. Increasing of the
basis dimension does not lead to significant changes of $S_k$ values
for the states with energy up to the saddle one $E_S$.

\begin{figure}
\includegraphics[width=\textwidth]{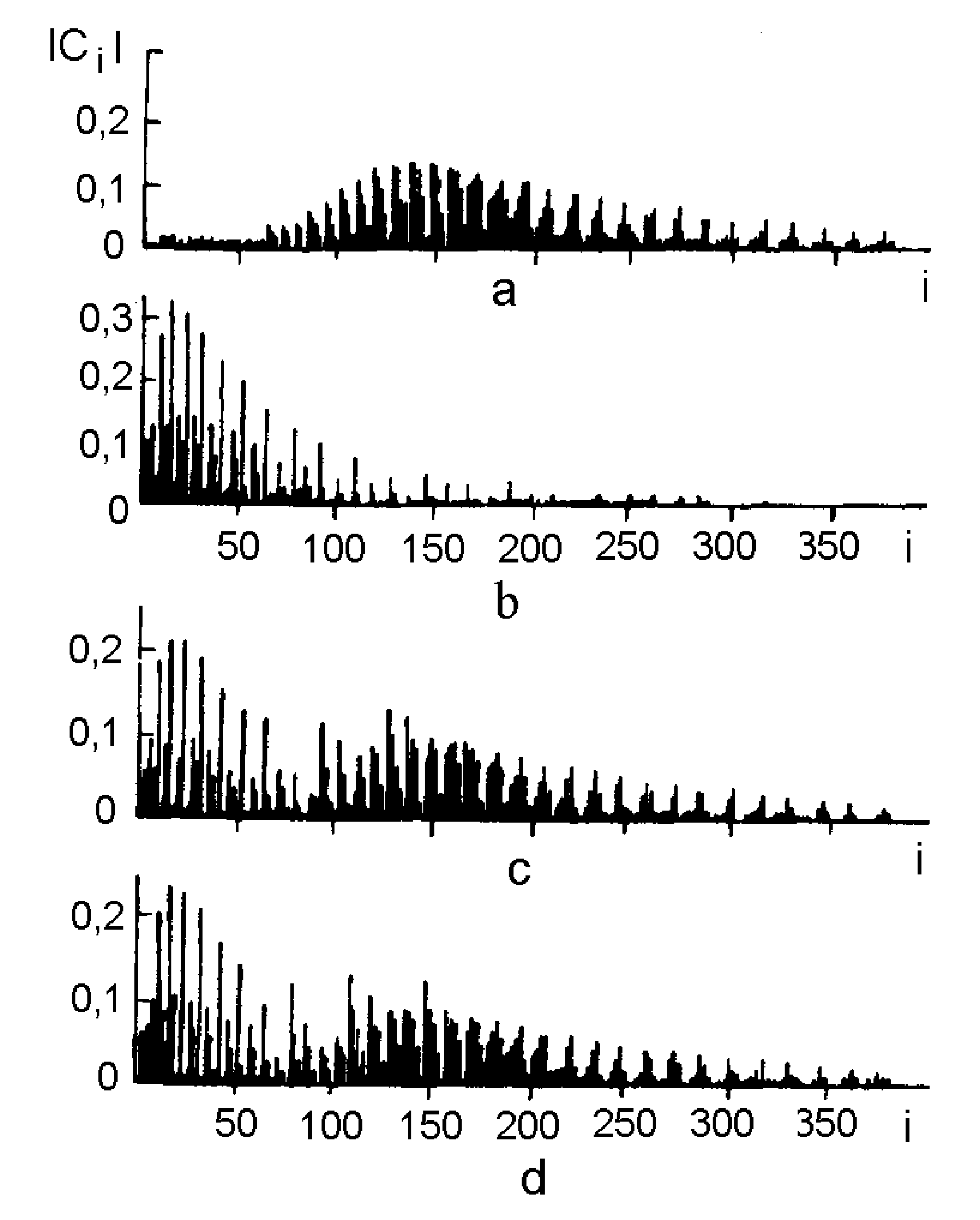}
\caption{\label{fig26} Distribution of coefficients $C_i^k$ by
number of basis state $i=(NLj)$ ($E$ - type) for localized
states with $k=3$ (a) and $k=4$ (b) at $b=0.17379$ ($W=17.8$) and
for the delocalized states with $k=3$ (c) and $k=4$ (d) in the point
of quasi-crossing $b=0.1737924$ ($W=17.8$).}
\end{figure}

The states localized in the central or in the lateral minima have
essentially different distributivity of the coefficients $C_i^k$,
$\{i\equiv NLj\}$ (see Fig.\ref{fig26}a,b) and thus different
entropies: states localized in the central minimum have less
entropy. In the neighborhood of the points of level
quasi-crossings, delocalization of wave functions corresponding
to these levels takes place; these wave functions have close
distributivity of the coefficients $C_i^k$ (see Fig.\ref{fig26}c,d).

\begin{figure}
\includegraphics[width=\textwidth]{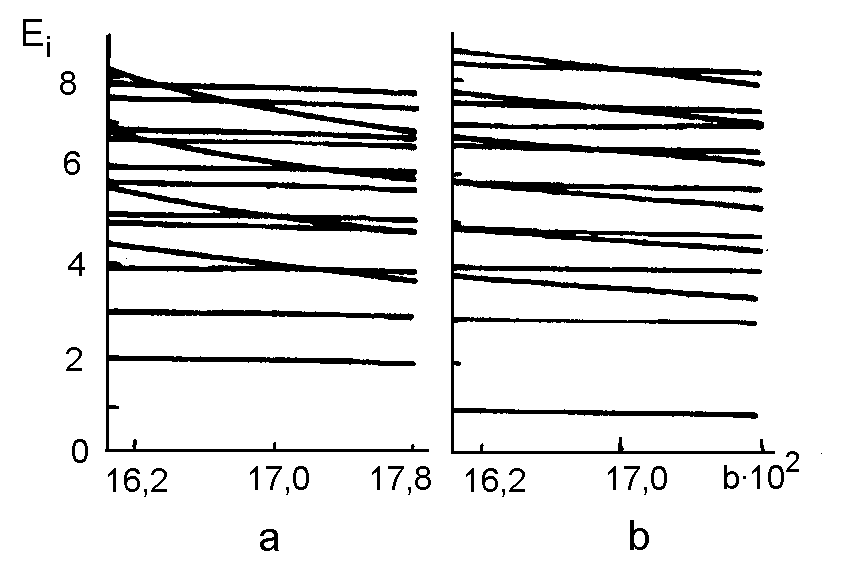}
\caption{\label{fig27}Dependence of energy levels $E_i$ of the $QO$
Hamiltonian (\ref{qo_ham}) on the parameter $b$ for $W=17.8$: a) ---
spectrum of $E$-type, b) --- spectrum of   $A_1$-type.}
\end{figure}

\begin{figure}
\includegraphics[width=\textwidth]{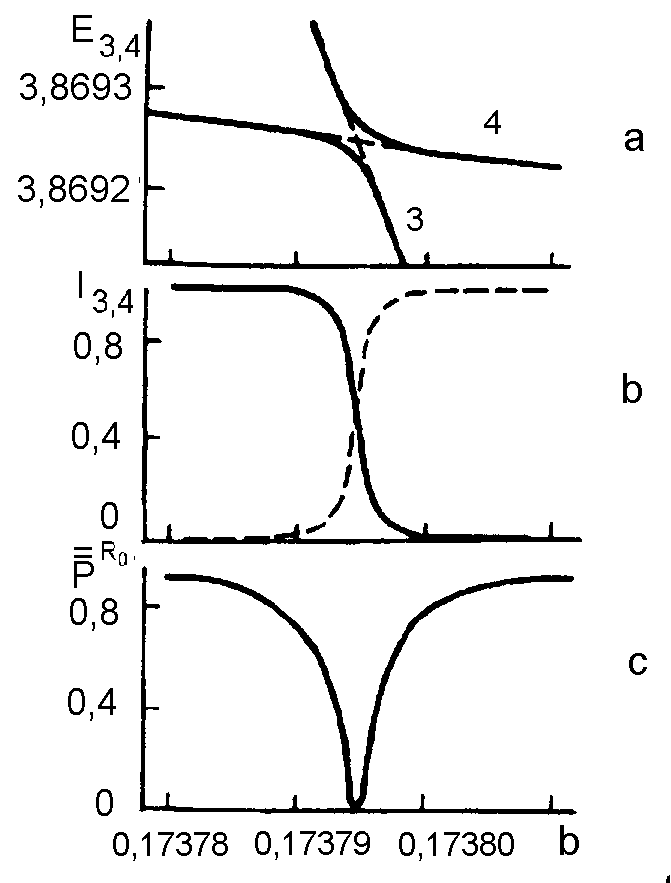}
\caption{\label{fig28}a) - Quasi-crossing of energy levels of
$E$-type with $k=3,4$ for $QO$ Hamiltonian (\ref{qo_ham}); b -
localization of wave functions of states with $k=3,4$ in the central
well at different $b$; c - dependence $\bar{p}^R(b)$.}
\end{figure}

Fig.\ref{fig27} shows the sub-barrier part of the energy spectrum
obtained by the diagonalization. As is easy to see, the tunneling
of the wave packet, composed of the sub-barrier wave functions, can
be described in a two-level approximation. Indeed, there are
approximately $10$ level quasi-crossings of $A_1$ and $E$-types,
where the nonlinearity parameter $b$ changes are of the order of
$10^{-2}$, while the effective half-width of the overlap integral in
(3.66) is about $10^{-5}$ (see Fig.\ref{fig28}). Hence, all
non-diagonal elements will be close to zero with the overwhelming
probability in the matrix
\[\alpha_{mn}=\int\limits_R \psi^*_m(\mathbf{r})\psi_n(\mathbf{r}) d\mathbf{r}\]
of an arbitrary nonlinearity parameter (e.g., $b$). Two essential
non-zero out-of-diagonal matrix elements (two-level approximation)
appear only in the vicinity of quasi-crossings. The probability of
double quasi-crossings at a fixed nonlinearity parameter is almost
excluded. This probability is by two or three orders lower than that
of rather rare $\sim10^{-3}$ single quasi-crossings.

So now we can give the answer to the question we have posed above.
The stringiest correlations between level quasi-crossings,
delocalization of wave functions and resonant tunneling across the
potential barrier take place in the two-dimensional case (and, most
likely, in the multi-dimensional one too).

The existence of the mixed state for many-well potentials must
be clearly manifest in the dynamics of quantum wave packets. The
pre-exponential factor of the tunnel amplitude depends on the type
of classical motion and consequently we expect to observe asymmetry of the effective barrier penetration in the mixed state.
This purely classical effect may be observed only if the uncertainty
in level energy is comparable with the average distance between
levels, and the system does not yet "feels" that the spectrum is
discrete. This determines the time scale for which observation of
the effect is possible. It is the same time scale on which
transition from classical diffusive evolution with linearly
increasing energy to the quasi-periodical quantum one is observed
\cite{casati_chirikov}.

\sat\section{Dynamical tunneling}\sat

Transition from integrability to chaos in classical dynamics not
only significantly modifies the tunneling process, but moreover it
leads to the appearance of principally new scenarios of the tunneling
effect. The simplest example of such a scenario is the so-called
dynamical tunneling \cite{davis}. Dynamical tunneling appears in systems where the phase space contains such regions that transition
between them is forbidden in classical mechanics, but this
restriction is not due to the potential barrier. Of course such an
effect is possible in systems with two or more degrees of freedom,
where additional to the energy integrals of motion are sources of
the corresponding restrictions. This new type of tunneling is more
complicated than the traditional (barrier) tunneling. The
complication is connected with the fact that simple consideration of the potential energy surface does not reveal the conditions of
the restriction. Instead of static potential surfaces one should
consider dynamical behavior of the trajectories.

In order to understand the origin of dynamical tunneling in a
bounded system, let us recall the semiclassical solution of the
one-dimensional symmetric double well problem. If we quantize the
system, considering each well separately, we obtain a spectrum
composed from exactly degenerate doublets. Only by taking into account
the interaction between the wells via overlap of exponentially small
sub-barrier tails of the wave functions, we can obtain the correct
result --- almost degenerate pairs of levels with the well-known
splitting. The magnitude of the latter determines the tunneling rate.

A situation similar to those in the one-dimensional symmetric
double well can exist  also in a multi-dimensional potential in
the absence of energy barrier. Let us consider
\cite{bohigas93,tomsovic,doron} a dynamical system with reflective
symmetry of the phase space $T$. Let there be two non-connected
regions $A_1$ and $A_2$ in the phase space, each of them being
invariant with respect to classical dynamics, mapping one onto
another with the symmetry transformation $A_2=TA_1$. Let the
classical motion in $A_{1,2}$ be regular, i.e. those regions
represent the stability islands embedded in the chaotic sea. In that
case as we have seen above the regions $A_{1,2}$ represent invariant tori in classical phase space. We make an additional
assumption that in the semiclassical limit there is a set of states
$\psi(\mathbf{q})$ mostly localized in the regions $A_1$ and $A_2$.
Using the standard procedure we can quantize motion in each region
independently and construct degenerate wave functions
$\psi^{(1)}(\mathbf{q})$ and
$\psi^{(2)}(\mathbf{q})=\psi^{(1)}(T\mathbf{q})$, sometimes called
{\it quasi-modes}. Taking into account the interaction between the
regions, the quasi-modes $\psi^{(1,2)}(\mathbf{q})$, as in the one-dimensional case, must be replaced by their combinations ---
symmetric and antisymmetric:
\[\psi^{(\pm)}(\mathbf{q})=\frac{1}{\sqrt2}\left(\psi^{(1)}(\mathbf{q})\pm\psi^{(2)}(\mathbf{q})\right)\]

Energy degeneration between the functions is removed due to the
tunneling processes. However, in contrast to the one-dimensional case, the
invariant tori $A_{1,2}$ are not always separated by the energy
barrier in the configuration space. Transitions $A_1\leftrightarrow
A_2$ may be classically forbidden by conservation of the constant of
motion distinct of the energy.

Like in the one-dimensional case, a wave packed constructed from linear
combinations of the wave functions
\begin{equation}\label{wp12}\Psi^{(1,2)}(\mathbf{q})=\psi^{(+)}(\mathbf{q})\pm\psi^{(-)}(\mathbf{q})\end{equation}
will mimic a classical particle, initially localized in one of the
regions $A_{1,2}$. The wave packet (\ref{wp12}) will make the
transitions between regions with the frequency determined by the
energy splitting of states $\psi^{(+)}(\mathbf{q})$ and
$\psi^{(-)}(\mathbf{q})$. A natural question arises: if in the phase
of the system there are two symmetric stability islands separated by chaotic sea, then how that sea affects the dynamical tunneling
between the islands? Semiclassical reasoning points out that the
probability distribution associated with the wave function of the
quantized torus decays exponentially outside the torus. The small
overlap in the classically forbidden region of the decaying
distributions, centered on the two congruent quantized tori, will
lead to the tunneling splitting. If there is no chaotic region
between the tori, then the overlap will be very small. However if
a chaotic region exists between the tori, then the wave functions
corresponding to the tori overlap initially with the chaotic
states. Due to the ergodic nature of chaotic wave functions (uniform
distribution of the probability density), the connection between the
two tori appears to be more efficient that in the case of a regular
intermediate state. Therefore we can expect that the tunneling will
be assisted by the presence of the chaotic region.

All these qualitative considerations are confirmed by the analysis
\cite{bohigas93} of the splitting magnitude in the tunneling doublet
as a function of the parameter $\sigma$ --- the measure of
chaoticity of the system. The splitting undergoes considerable
fluctuations, which are connected with quasi-crossing of the
"outside" chaotic level and the considered regular tunneling
doublet. It is important to note that because there is no dynamical
fragmentation of the chaotic region on separated symmetric parts as
happens in the regular part of the phase space, we have no reason to
expect the appearance of chaotic doublets. So every chaotic state
has fixed parity.

Peculiarities of such a spectral region can be described in the
three-level model \cite{tomsovic} defined by the Hamiltonian
\begin{equation}\label{tunnel_ham}\hat{H}=\left(\begin{array}{ccc}
E+\varepsilon & 0 & 0\\
0 & E-\varepsilon & \nu \\
0 & \nu & E^C
\end{array}\right)\end{equation}
Here $E\pm\varepsilon$ are energies of the regular quasi-doublet,
composed of the symmetric and antisymmetric combinations of the
corresponding quasi-modes. The model assumes that the even chaotic state
$|C\rangle$ with energy $E^C$ is connected with the $|+\rangle$
state with energy $E-\varepsilon$ by the coupling constant
$\nu$. In practice one of the constants $\nu$ or $\varepsilon$
dominates. If $\varepsilon$ dominates then we turn back to the above
considered two-level model. Therefore let us assume that the
dominating constant is the coupling between the $|C\rangle$  and
$|+\rangle$ states and let us take $\varepsilon=0$. In that case the
doublet splitting $\Delta E$ will be governed by the energy shift
$E_+$ due to the coupling $\nu$. Diagonalization of the Hamiltonian
(\ref{tunnel_ham}) leads to
\[\Delta E=\left\{\begin{array}{cc}
\frac{\nu^2}{E-E^C} & E-E^C\gg\nu\\
|\nu| & E-E^C\ll\nu
\end{array}\right.\]
It means that with the variation of $\sigma$ in the tunneling doublets
splitting we will observe peaks with magnitude $|\nu|$ when
$E^C(\sigma)$ crosses $E_+(\sigma)$.

\begin{figure}
\includegraphics[width=\textwidth]{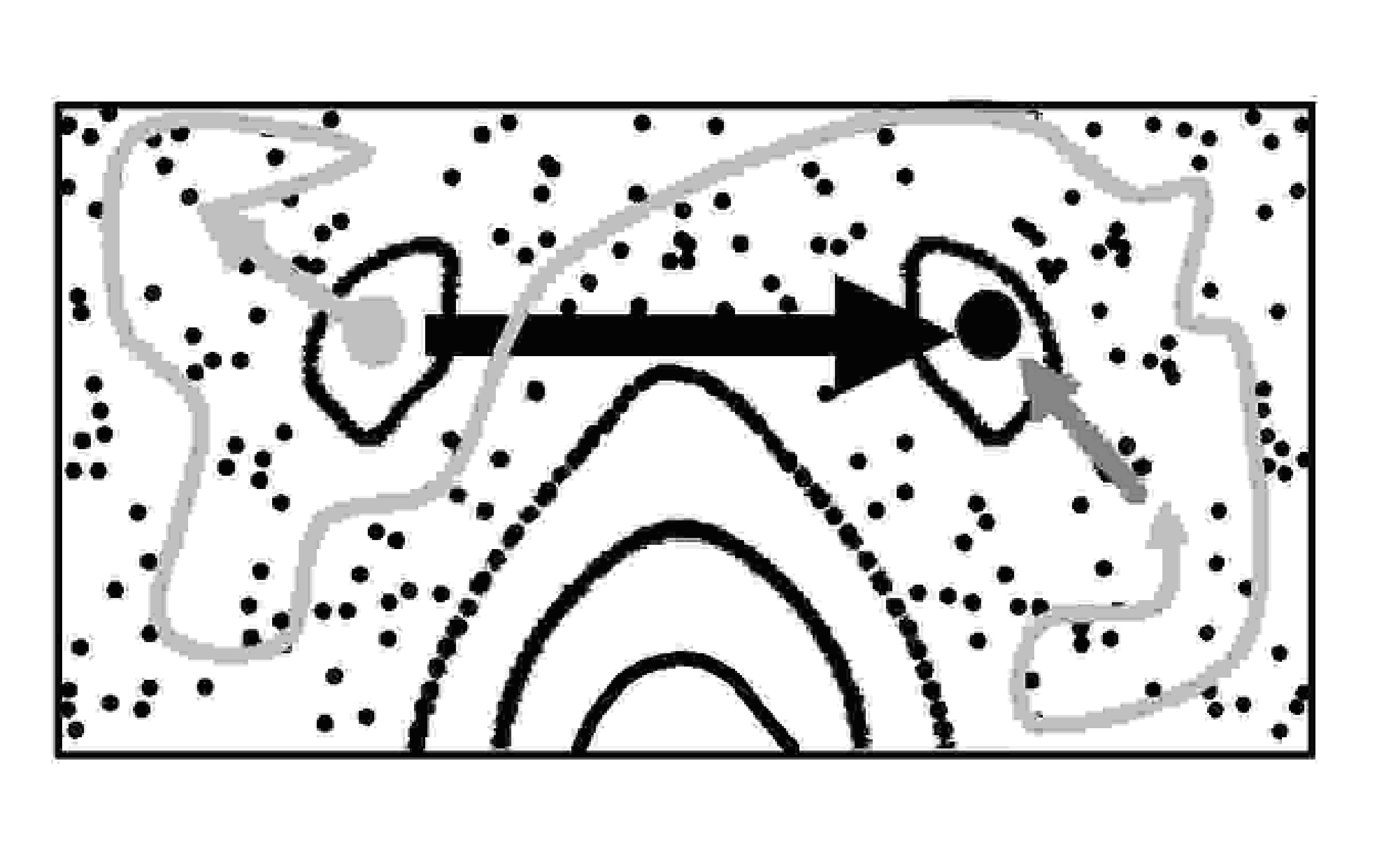}
\caption{\label{tunnel} A schematic representation of direct and
chaos assisted tunneling \cite{tureci}}
\includegraphics[width=0.3\textwidth]{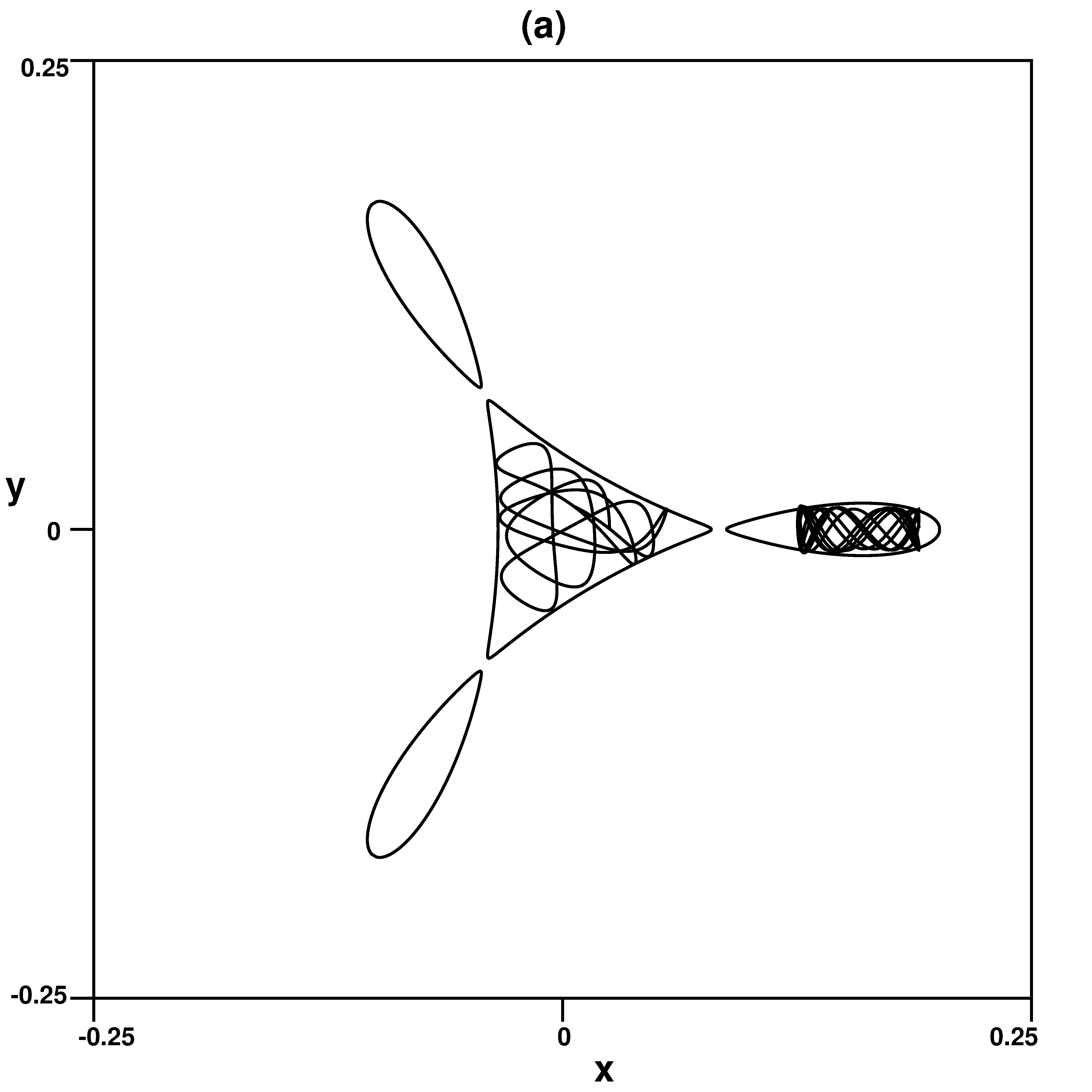}
\includegraphics[width=0.3\textwidth,height=0.3\textwidth]{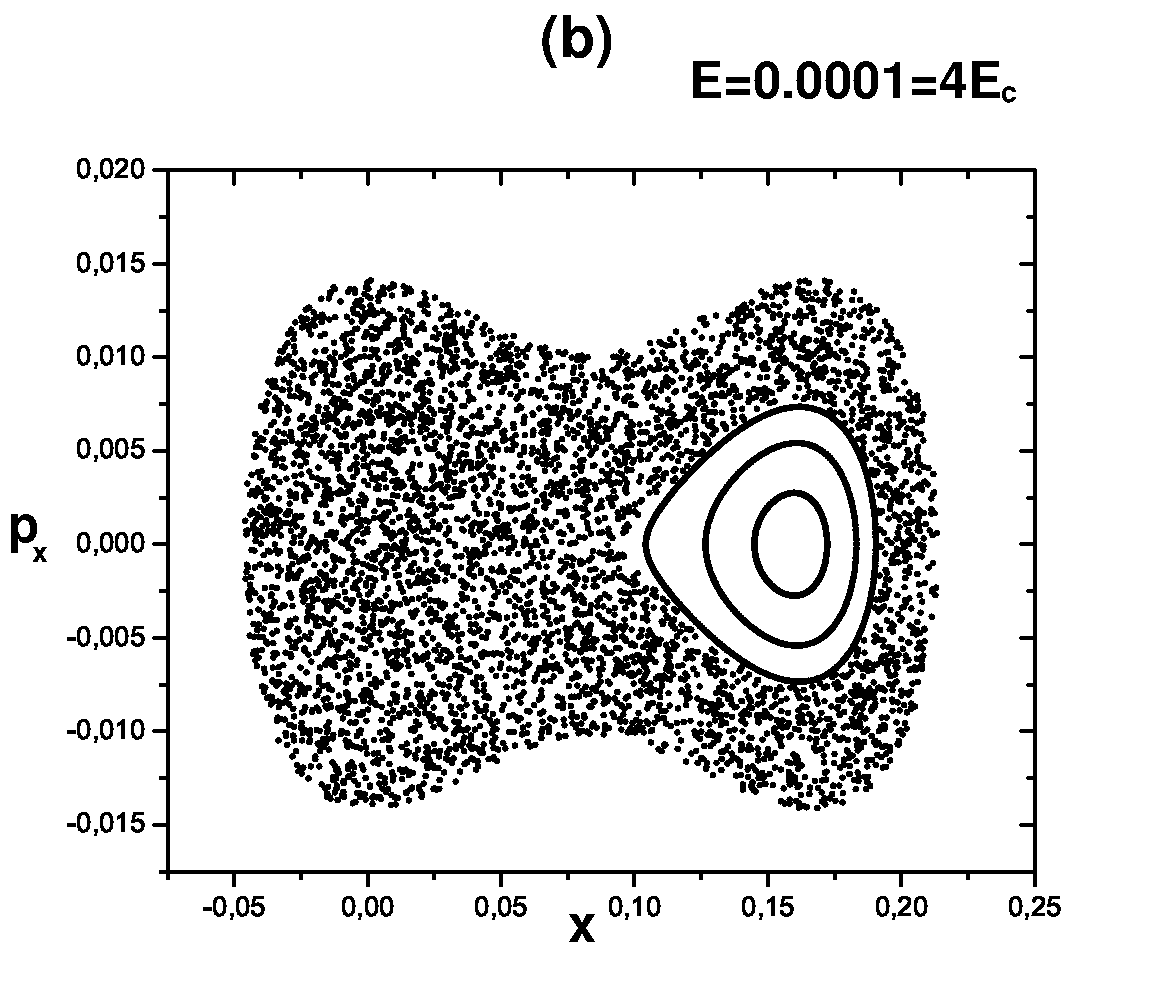}
\includegraphics[width=0.3\textwidth]{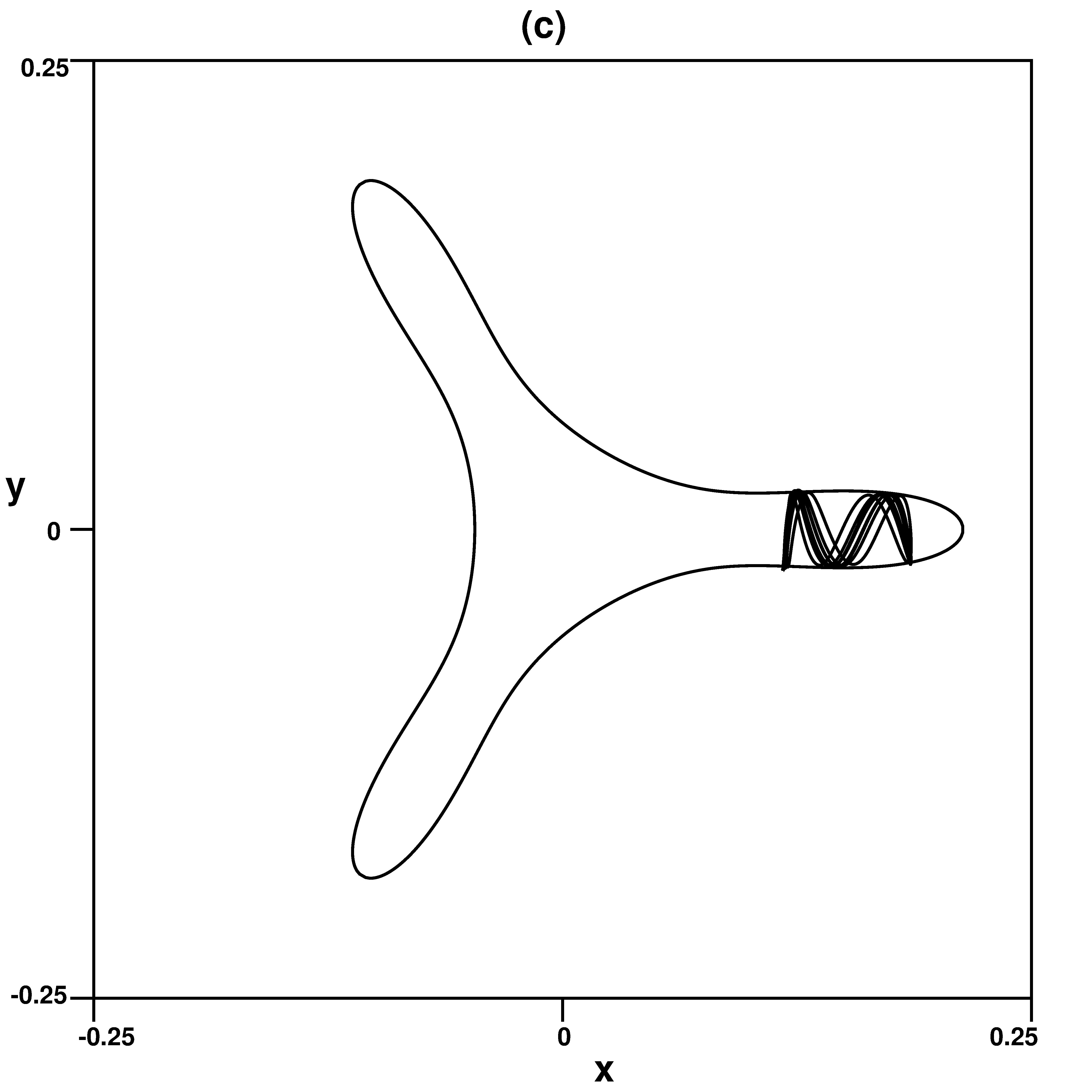}
\caption{\label{traj} Regular and chaotic trajectories for
$E\lesssim E_S$ (a), PSS (b) and regular quasiperiodic trajectory
(c) for $E=2E_S$ in the $QO$ potential (\ref{u_qo}).}
\end{figure}

Now we draw some conclusions. As distinct from the integrable case, systems with dynamics of mixed type, where the phase space
contains both regular and chaotic regions, demonstrate a new
tunneling mechanism called chaos assisted tunneling
\cite{bohigas93}. The doublets splitting that determines the
tunneling rate in two-level approximation in systems of mixed
type, is as a rule several orders of magnitude higher than for
similar integrable systems. In contrast to direct processes
when a particle tunnels directly from one state to another, chaos assisted tunneling corresponds to the following three-step
process:
\begin{enumerate}
\item tunneling from a periodic orbit to the closest point in the
chaotic sea;
\item classical propagation in the chaotic region of the phase space
towards the vicinity of another periodic orbit;
\item tunneling from the chaotic sea onto the latter periodic orbit.
\end{enumerate}
In other words doublet splitting due to reflective symmetry
occurs not directly but through the compound-process of wave
function destruction, piece by piece, near one regular region, and
then chaotic transport through the chaotic sea towards the neighboring
symmetry conjugated regular region, and at last restoration of the
initial state image. Schematically this process is presented on
Fig.\ref{tunnel}. Let us note that chaos assisted tunneling is a
formally higher perturbative order process than direct
tunneling. But the corresponding matrix elements for chaos
assisted tunneling are much larger than for the direct process.
Intuitively it can be pictured as the following: as the main part of
the distance (through the chaotic sea) represents a classically
allowed transition, then we can expect that those indirect
trajectories will make a larger contribution to the tunneling flow
than the direct ones. In the case of direct tunneling all
sub-barrier trajectory represents a classically forbidden process.

Let us now consider the dynamics of Gaussian wave packets in $QO$
potential (\ref{u_qo}) with $W=18$, when PES have four local minima
--- the central one with chaotic motion and three symmetric
peripheral with regular motion for sub-barrier energies (see
fig.\ref{traj}a).

At considerably super-barrier energies of motion, chaotic
dynamics is observed in all local minima; however in the peripheral
one, a large stability island survives even for $E=2E_S$
(fig.\ref{traj}b), which correspond to quasiperiodic trajectories
localized in those minima (fig.\ref{traj}c).

\begin{figure}
\includegraphics[width=0.3\textwidth]{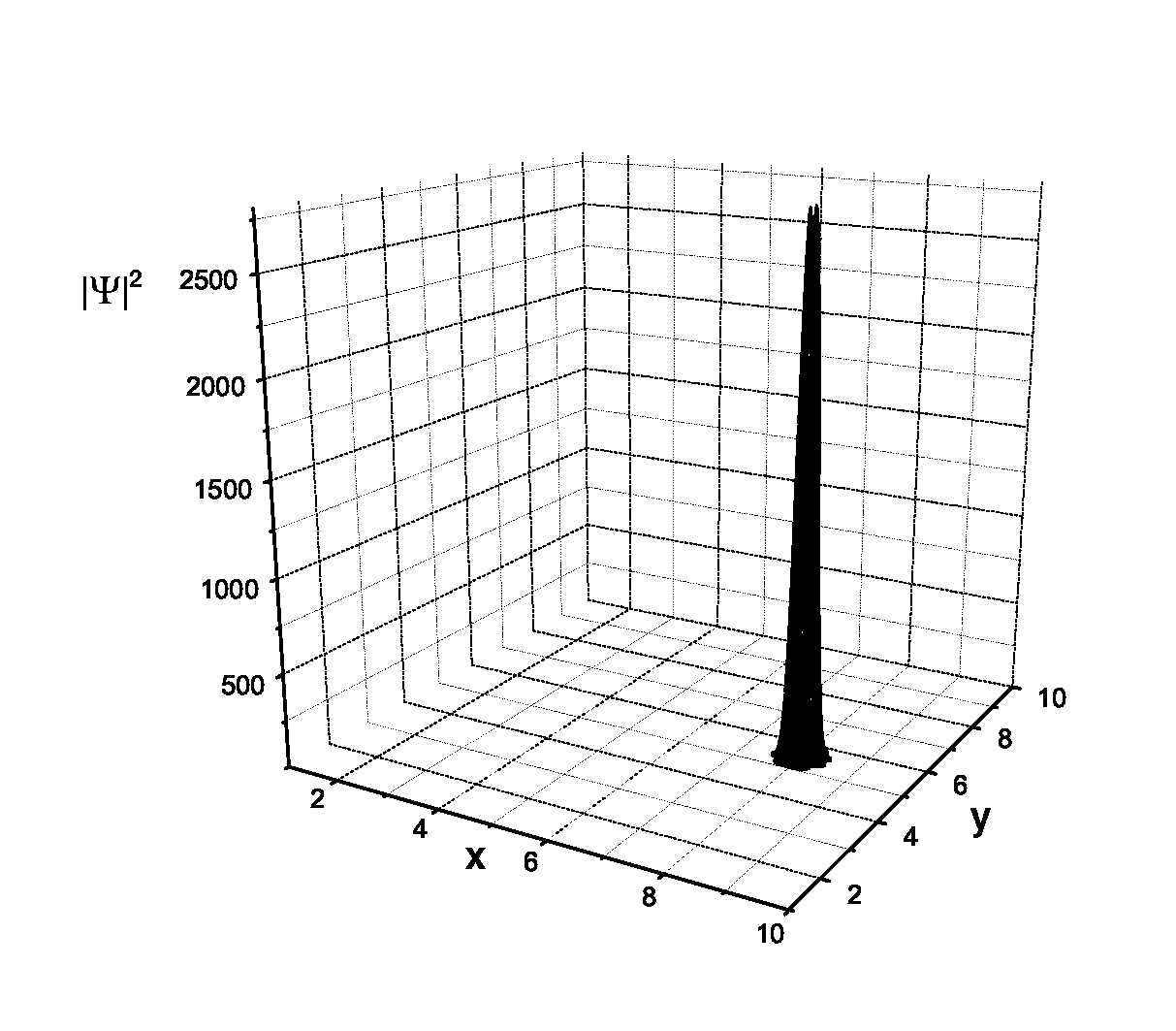}
\includegraphics[width=0.3\textwidth]{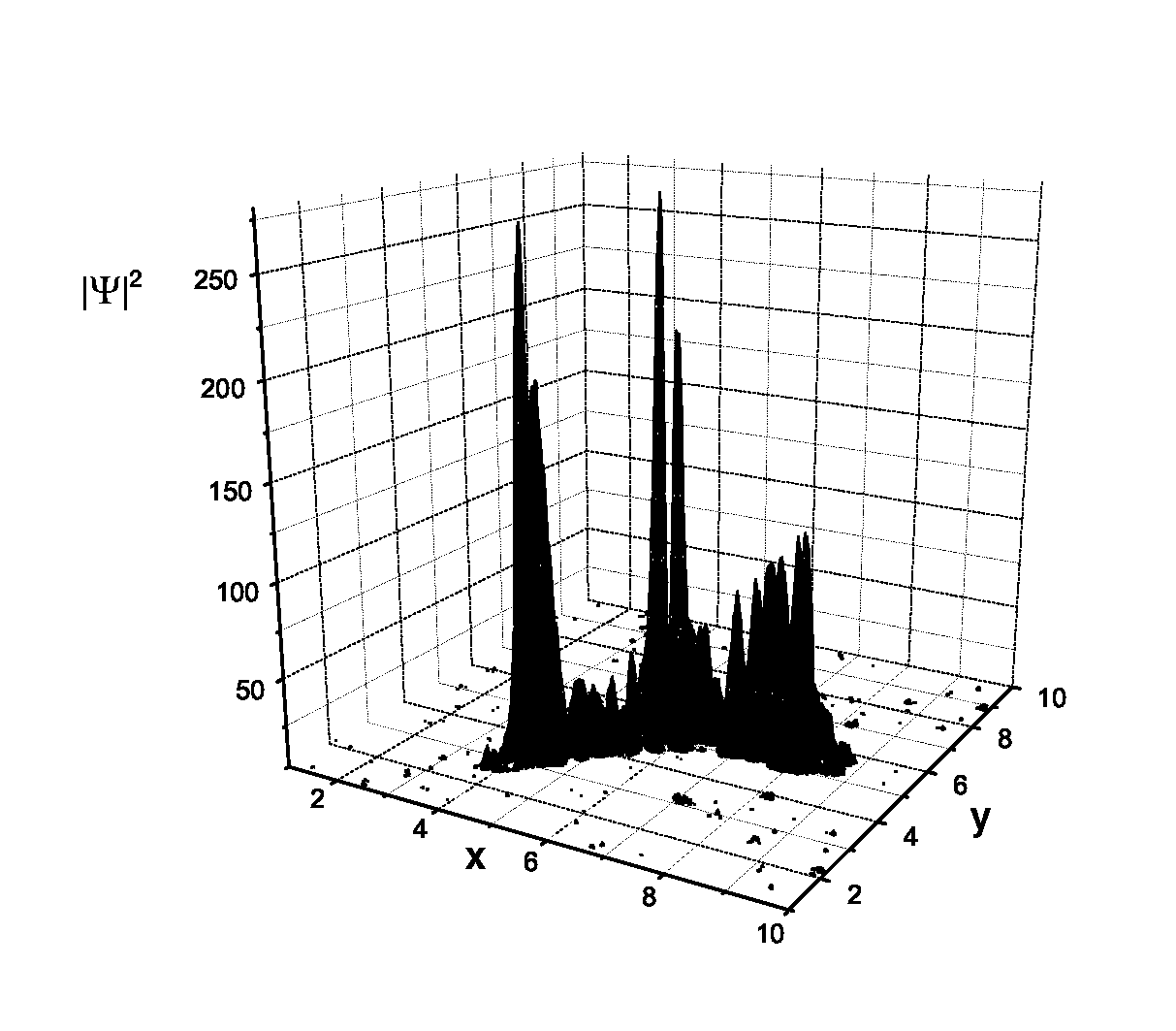}
\includegraphics[width=0.3\textwidth]{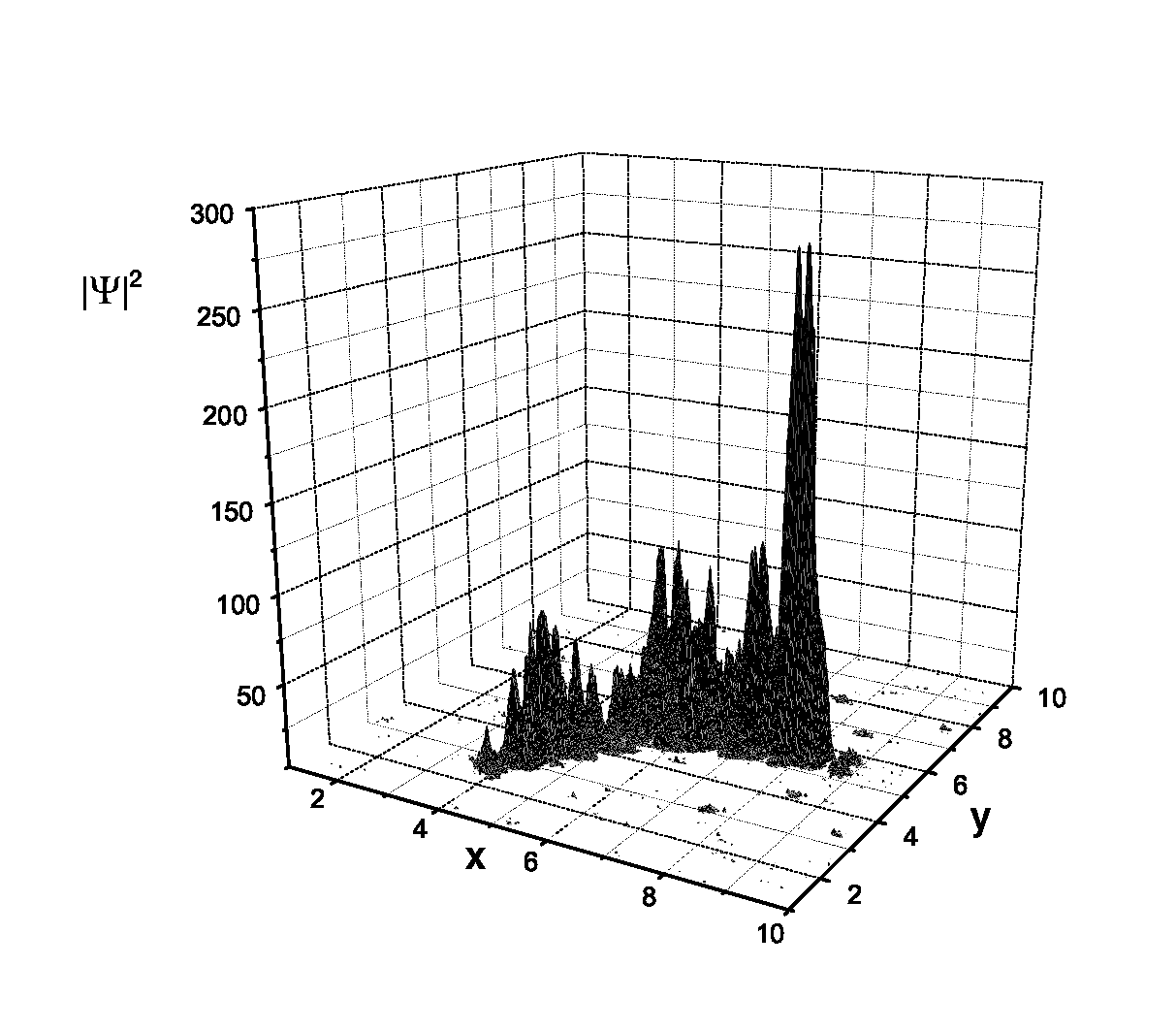}
\caption{\label{psit} Temporal evolution of gaussian wave packet
(\ref{gwp}) under dynamical tunneling in the $QO$ potential
(\ref{u_qo}).}
\includegraphics[width=0.5\textwidth]{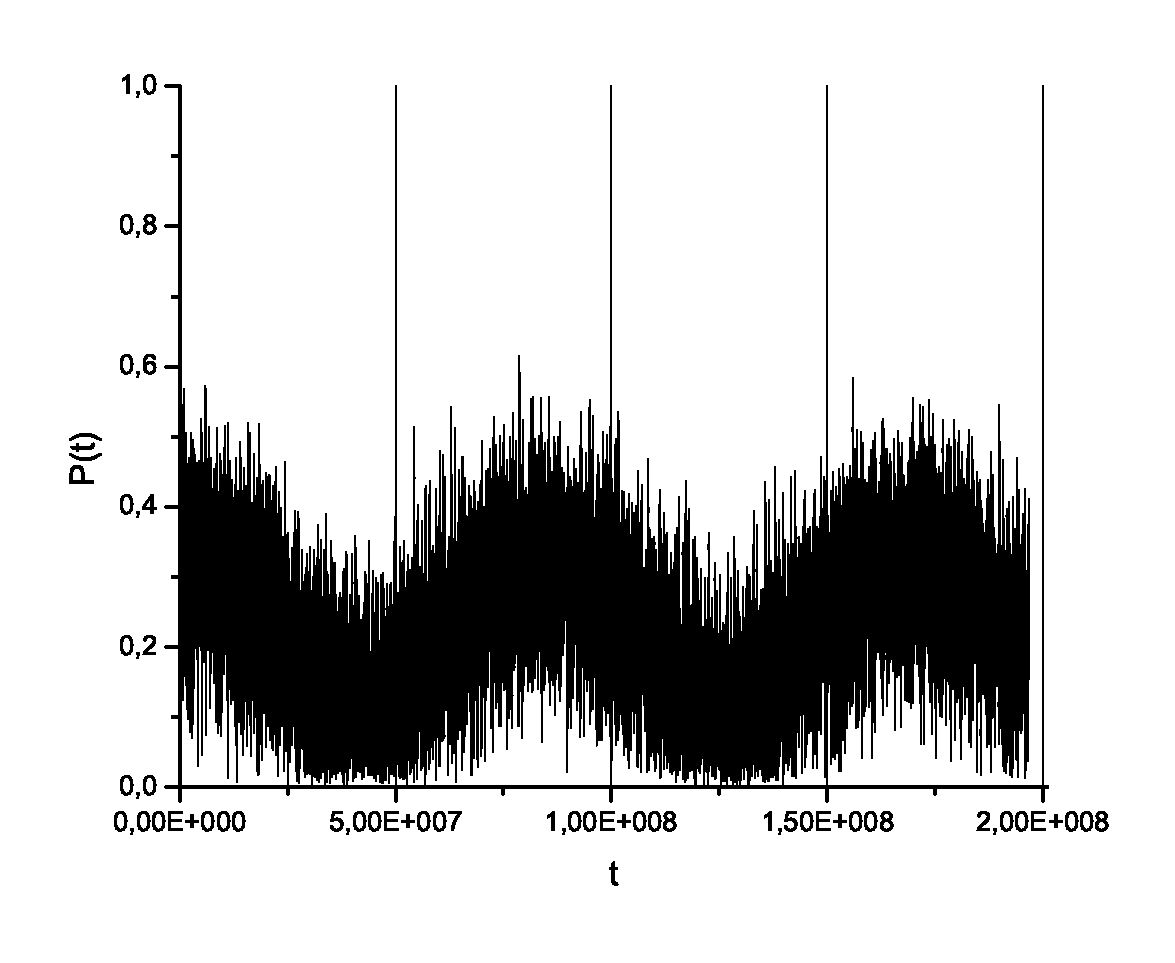}
\includegraphics[width=0.5\textwidth]{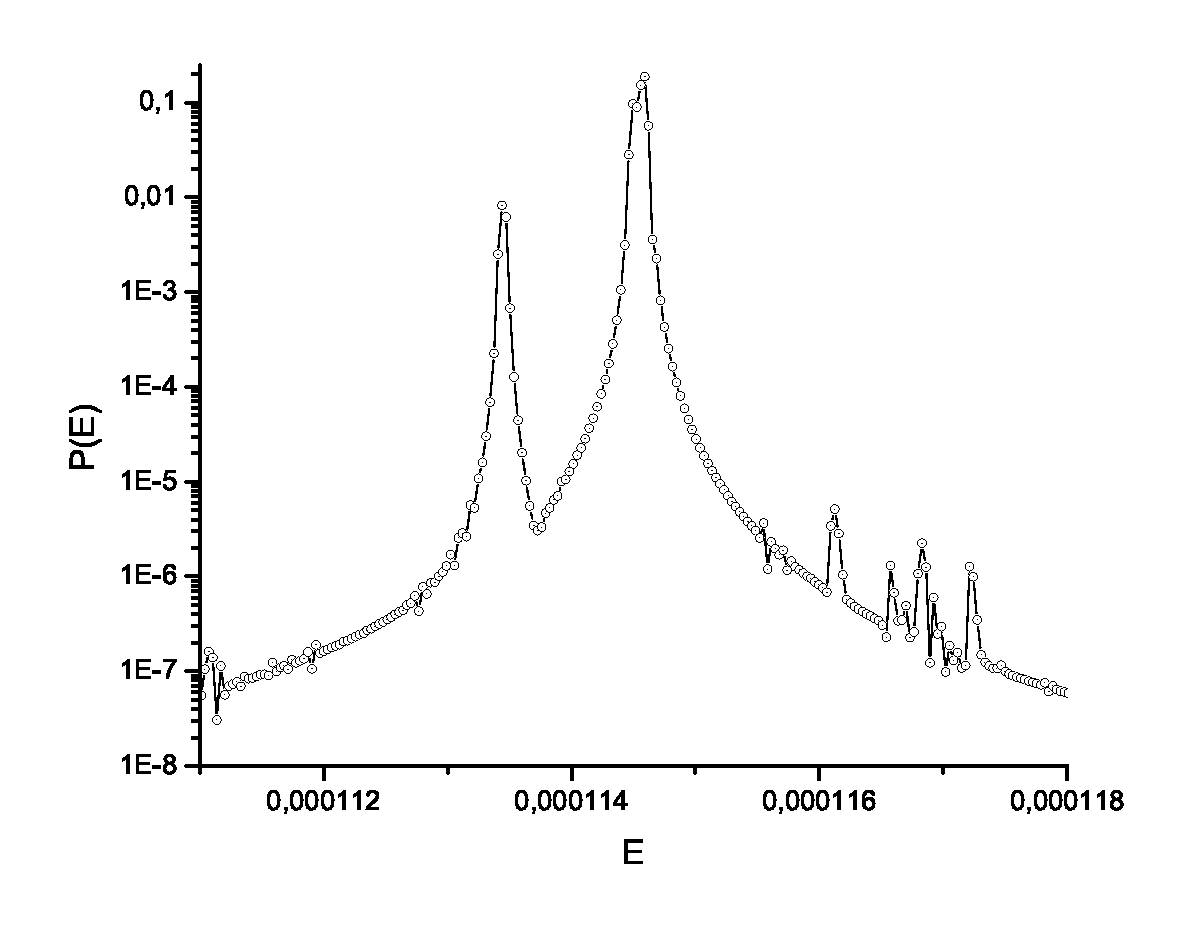}
\caption{\label{ptpe} The auto correlation function (\ref{autocor})
(left) and its Fourier transform (right) for Gaussian wave packet
(\ref{gwp}) tunneling dynamics in the $QO$ potential (\ref{u_qo}).}
\includegraphics[width=0.3\textwidth]{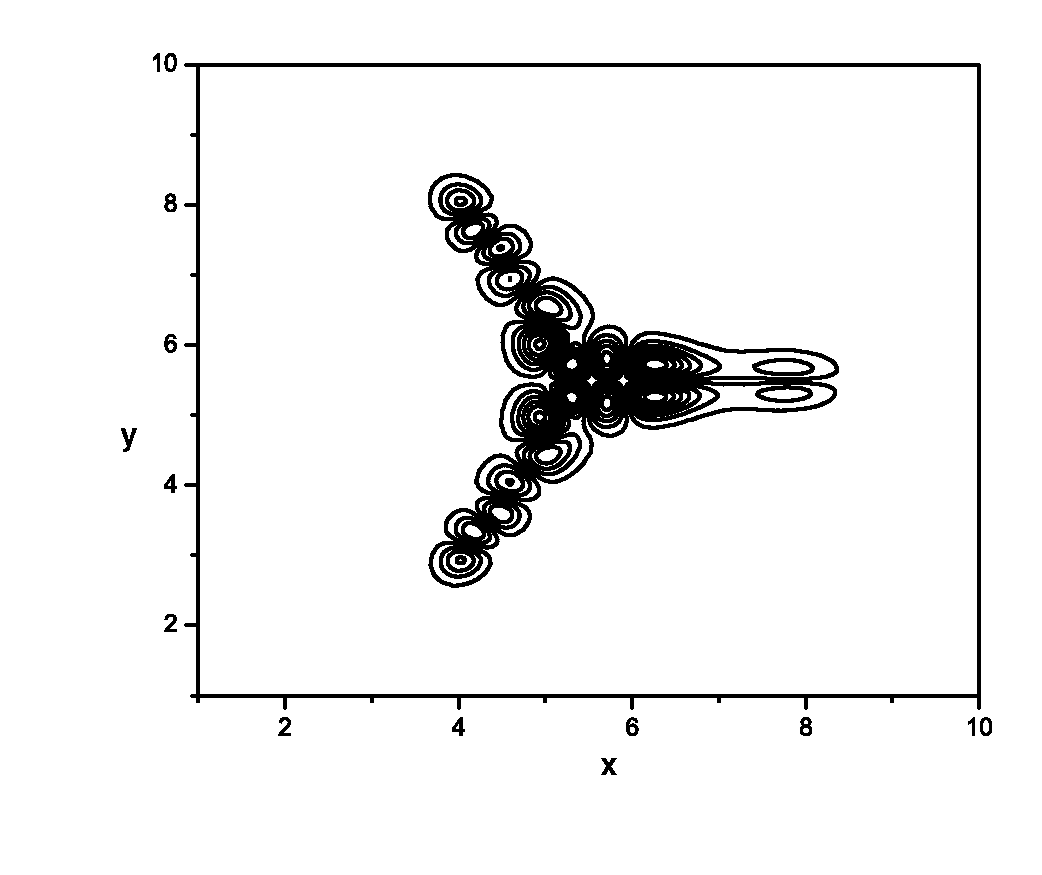}
\includegraphics[width=0.3\textwidth]{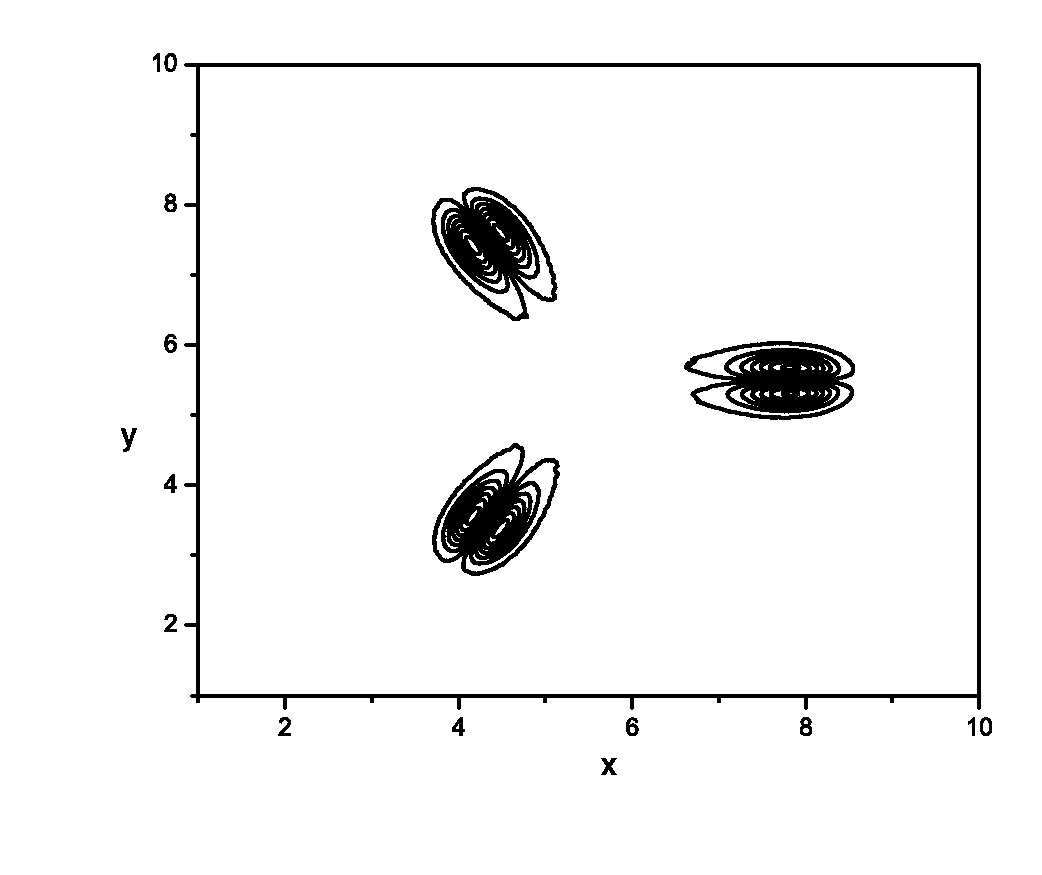}
\includegraphics[width=0.3\textwidth]{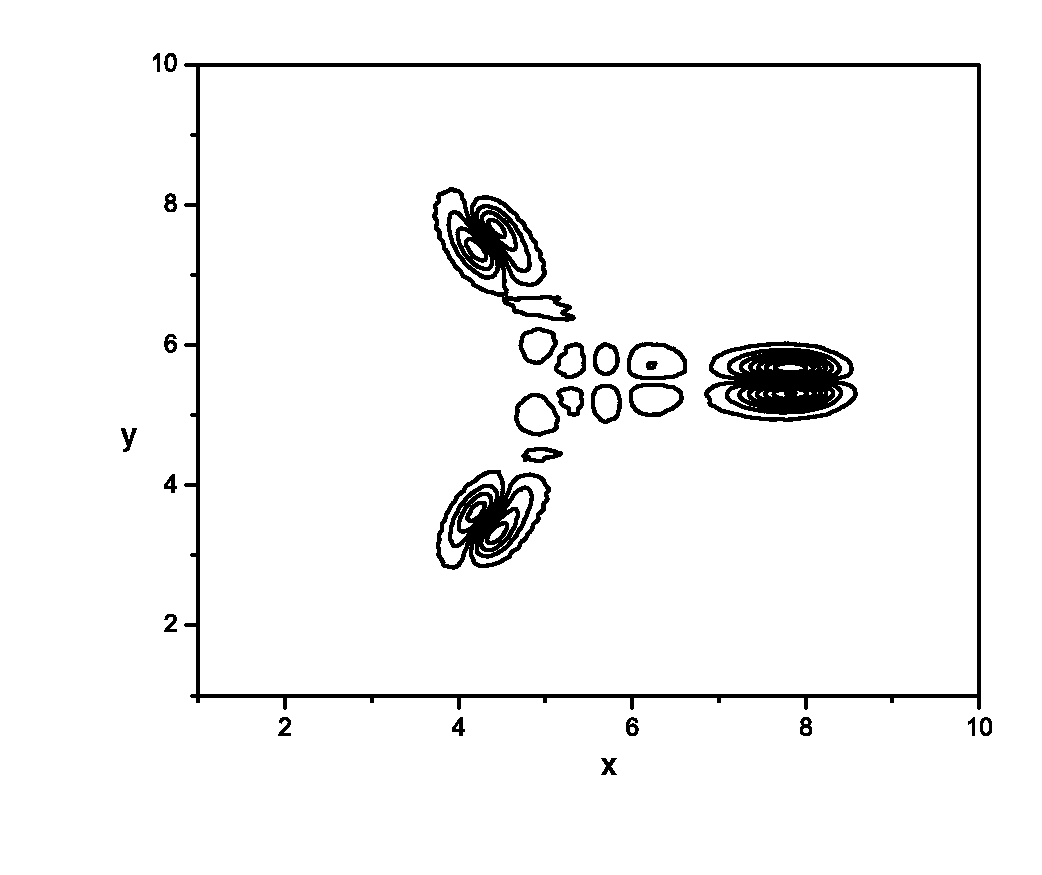}
\caption{\label{tunnel3} Level lines for $|\psi(x,y)|^2$
distribution for the tunneling triplet wave functions (the energy
grows from left to right).}
\end{figure}

With the spectral method it is easy to perform a numerical
simulation of the dynamical tunneling process for super-barrier motion
in $QO$ potential (\ref{u_qo}). In order to do that we prepare the initial state $\psi_0(x,y)$ in the form of a Gaussian minimum
uncertainty wave packet (\ref{gwp}) with such parameters
$(x_0,y_0,p_{x0},p_{y0})$ that correspond to classical initial
conditions for a trapped regular trajectory (fig.\ref{traj}c). For
some time the quantum dynamics of such a wave packet will imitate classical motion on the semiclassical trajectory. After a sufficiently
long period of time, the dynamical tunneling effect starts to be
visible --- the wave packet leaves the local minimum of its initial
localization and appears in two others --- a process forbidden in
classical mechanics. After the same time period the wave packet more
or less restores in its initial states (fig.\ref{psit}).

The oscillating nature of dynamical tunneling is clearly described in
terms of the autocorrelation function (\ref{autocor}) (fig.\ref{ptpe} on
the left). Its Fourier transform (fig.\ref{ptpe} on the right)
demonstrates three dominant states in the wave packet composition,
that form the tunneling triplet (the right maximum on the $P(E)$ diagram
is actually a very tiny tunneling doublet). Figure \ref{tunnel3}
shows the probability distribution $|\psi(x,y)|^2$ for all three
members of the tunneling triplet in order of energy increasing: the
intermediate state is purely regular and forms a tunneling doublet
with the state of highest energy which is almost regular with a small
admixture of chaotic modes. The lowest lying state is predominately
chaotic and overlaps with only one of the tunneling doublet partners
(the higher one). Repulsion of those states considerably broadens
the level splitting in the tunneling doublet and therefore increases
the tunneling rate substantially, which is the essence of the chaos
assisted tunneling effect.

\sat\chapter{The mixed state and concrete physical effects}\sat
\sat\section{Decay of mixed states}\sat

The escape of trajectories (particles) from localized regions of
phase or configuration space has been an important topic in
dynamics, because it describes the decay phenomena of metastable
states in many branches of physics: chemical and nuclear reactions,
atomic ionization, nuclear fusion and so on. This problem has a rich
history. Almost a century ago, Sabine \cite{sabine} considered the
decay of sound in concert halls. Legrand and Sornette \cite{legrand}
have shown that this problem is equivalent to the escape one: a
small opening of width $\Delta$ for escape must be identified with
$\int\alpha(S)ds$, where $\alpha(s)$ is the absorption coefficient
at position $S$ of the container (billiard) boundary, $\alpha(s)=1$
over the width of window and $\alpha(s)=0$ elsewhere. Szepfalusy and
Tel \cite{tel} connected the escape problem with the problem of
chaotic scattering.

Exponential decay is a common property expected in strongly chaotic
classical systems \cite{bauer,alt,nemes}. Let us consider as an
example \cite{bauer} point particles in a rectangular box bouncing
elastically off the walls. We allow our system to decay by providing
a small window in one of the box walls through which particles are
allowed to escape. As is well known, the motion of particle in a
rectangular billiard is regular: two independent integrals of motion
are the absolute values of momentum projection on the billiard
walls. The trajectories of particles become chaotic if a circular
scattering center is placed somewhere inside the box.

For the chaotic case, simple consideration leads to exponential
decay. The number of particles leaving per time interval is given by
\begin{equation}\label{n_t}\frac{dN}{dt}=\Delta\rho(t)\int d^2p\
\mathbf{pe}_n=-2\Delta\rho(t)p^2\delta p\end{equation} Here $p$ is
the absolute value of momentum, $\mathbf{e}_n$ is a unit vector
normal to the opening in the surface, and integration in momentum
space is taken over a circular ring with radius $p$ and
infinitesimal width $\delta p$. Function $\rho(t)$ is the phase
space density, which in the ergodic motion is only a function of
time. In our case
\begin{equation}\label{rho_t}\rho(t)=\frac{N(t)}{2\pi p\delta
pA_c}\end{equation} where $A_c$ is total coordinate space area
available. Inserting (\ref{rho_t}) into (\ref{n_t}) yields
\begin{equation}\label{ntb}N(t)=N(0)e^{-\alpha t};\ \alpha=\frac{p\Delta}{\pi A_c}\end{equation}
Analytically calculated decay constant $\alpha$ is in a good
agreement with the graphically extracted value.

Exponential law at extremely long times turns into the power law
typical for decay of regular systems. One possible mechanism for
generation of power tails is the effect of "sticking" of the chaotic
orbits to outer boundaries of stability islands \cite{karney}, or a
very similar effect connected with the existence of marginally
stable periodic or "bouncing ball orbits". Although some qualitative
models, which show how the algebraic tail emerges, were introduced
in \cite{bunimovich}, no critical conditions for the distinct decay
laws were formulated in terms of the billiard geometrical
constrains. Experimental escape of cold atoms from a laser trap of
billiard type with a hole was studied in \cite{milner,friedman}.

Transition from the billiard escape problem to escape from potential
wells from one hand substantially broadens the number of possible
applications, and on the other significantly complicates the
problem. Of course, the one-well case is a simpler one. Zhao and Du
\cite{zhao_du} reported a study on the escape rates near the
threshold of Henon-Heiles potential (\ref{u_mu}). Simulations
performed by the authors show that the escape from a Henon-Heiles
system at energy slightly exceeding the saddle one, follows an
exponential law similar to chaotic billiard systems. They derived an
analytic formula for the escape rate as a function of energy. The
derivation is based on the fact that the phase space considered
potential (as well as the billiards) is practically homogeneous near
the saddle points. It should be noted that in such a case all
trajectories with energy higher than the saddle one leave the
potential well in finite time. The only problem to solve is to
determine the probability of a particle to escape from the well in
unit time interval.

We intend to study the particles escape from the local minima in the
case when the phase space contains macroscopically significant
components of regular as well as of chaotic type \cite{mi}. As we
have seen, such a possibility is realized in potentials with several
local minima in the form of a mixed state. The problem of particles
escape from a potential well in the presence of the mixed state has
an essential distinction from the one considered in \cite{zhao_du}.
Detailed analysis of the Poincar\'e sections for the potentials
$D_5$ and $QO$, made in chapter \ref{ms}, shows that in the
"regular" well at energy higher than the saddle, chaotic
trajectories appear and their measure increases with energy growth.
At the same time the regular trajectories remain mostly trapped in
the well. The phase space of such a system contains macroscopic
regions of both the regular and chaotic motion. Therefore the
ensemble of particles initially situated in the well, divides on two
components. So the ultimate problem is to determine the probability
to escape per unit time for the chaotic component, and the relative
measure of non-escaping trajectories trapped in the well.

\begin{figure}
\includegraphics[width=0.45\textwidth]{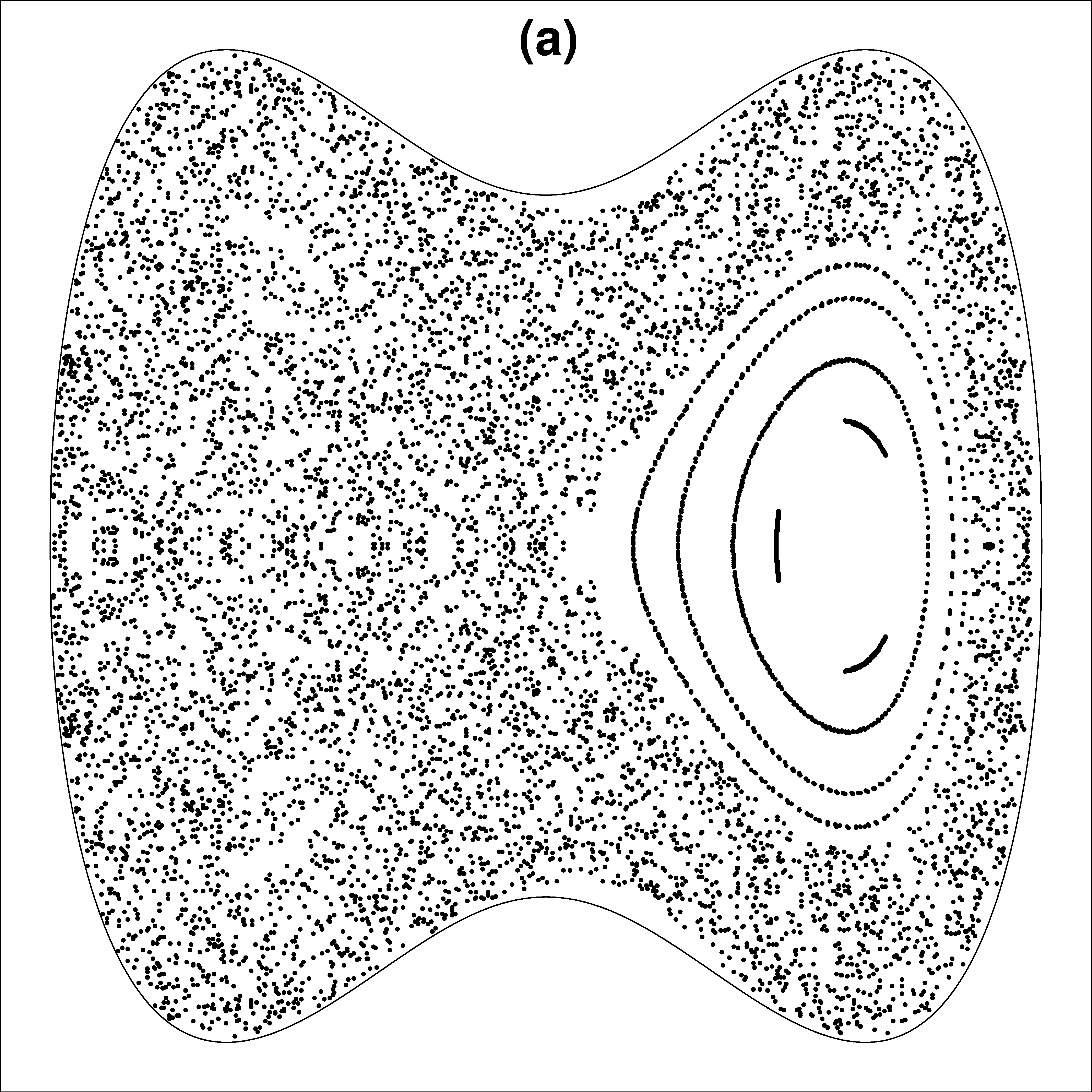}
\includegraphics[width=0.45\textwidth]{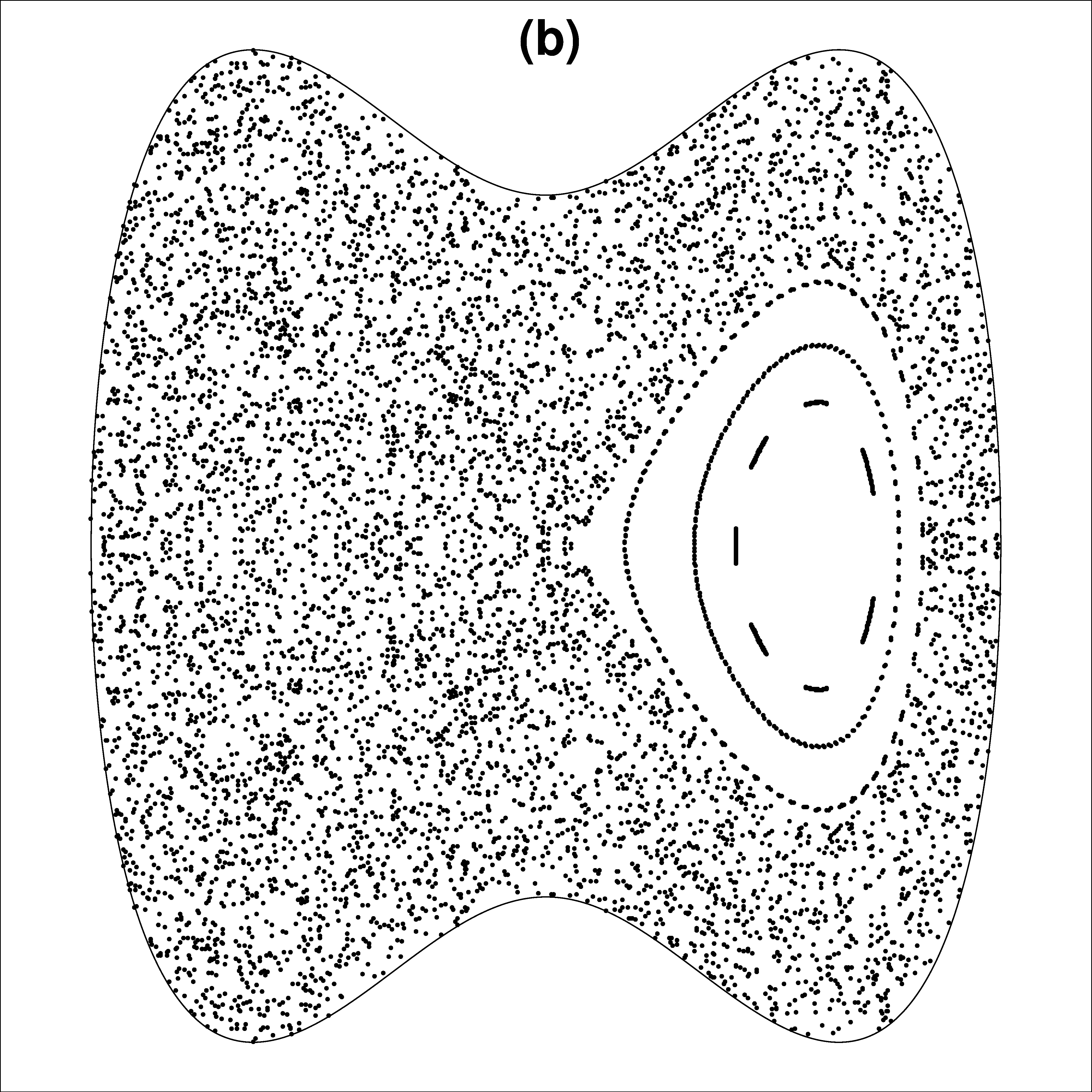}
\caption{Poincar\'e surfaces of section $y=0$ in the $(x,p_x)$ plane
for the potentials $D_5$ (a) and $QO$ (b) for $E=2E_S$
\label{d5qopss}}
\end{figure}

Let us address again the example to above considered in details
potentials of quadrupole oscillations (\ref{u_qo}) and umbilic
catastrophe $D_5$ (\ref{u_d5}). In the first as well as in the
second case, in some local minima up to the saddle energies the
motion remains absolutely regular. Moreover, at energies
significantly higher than the saddle energy (see fig.\ref{d5qopss})
the phase space structure preserves division on chaotic and regular
components. The latter is localized in the part of configuration
space which corresponds to regular motion at energies below than
saddle energies.

This phase space structure leads to the fact that escape from such
local minima has all the above mentioned properties of decay of
chaotic systems, and also a diversity of principally new features,
representing an interesting topic for conceptual understanding of
chaotic dynamics, and for application as well. We are interested
only in the "first passage" effects, leaving aside the problem of
dynamical equilibrium setup for the finite motion (for example, in
$QO$  potential). It is important to stress that though we study the
process of escape from a concrete local minimum, the over-barrier in
the case of a mixed state has a specific memory: general phase space
structure at super-saddle energies is determined by the
characteristics of motion in all other local minima.

We carried out numerical simulation and analytical estimates
\cite{mi} of trajectories escape in potentials $D_5$ and $QO$
through the hole over the saddle point. Results of the escape
problem for systems with multi-component phase volume (regular and
chaotic components) essentially depends on choice of initial
ensembles for dynamical variables. Fig.\ref{n_t2} presents the
normalized particle number $N(t)/N(t=0)$ for $10^6$ initial
conditions, uniformly distributed inside the right minimum in the
potential $D_5\ (x>0)$, and peripheral minimum in the $QO$ potential
$(x>1/12)$, together with the typical trajectories and Poincar\'e
sections. The results for different potentials are evidently similar
and have such characteristic features:
\begin{figure}
\includegraphics[width=\textwidth,draft=false]{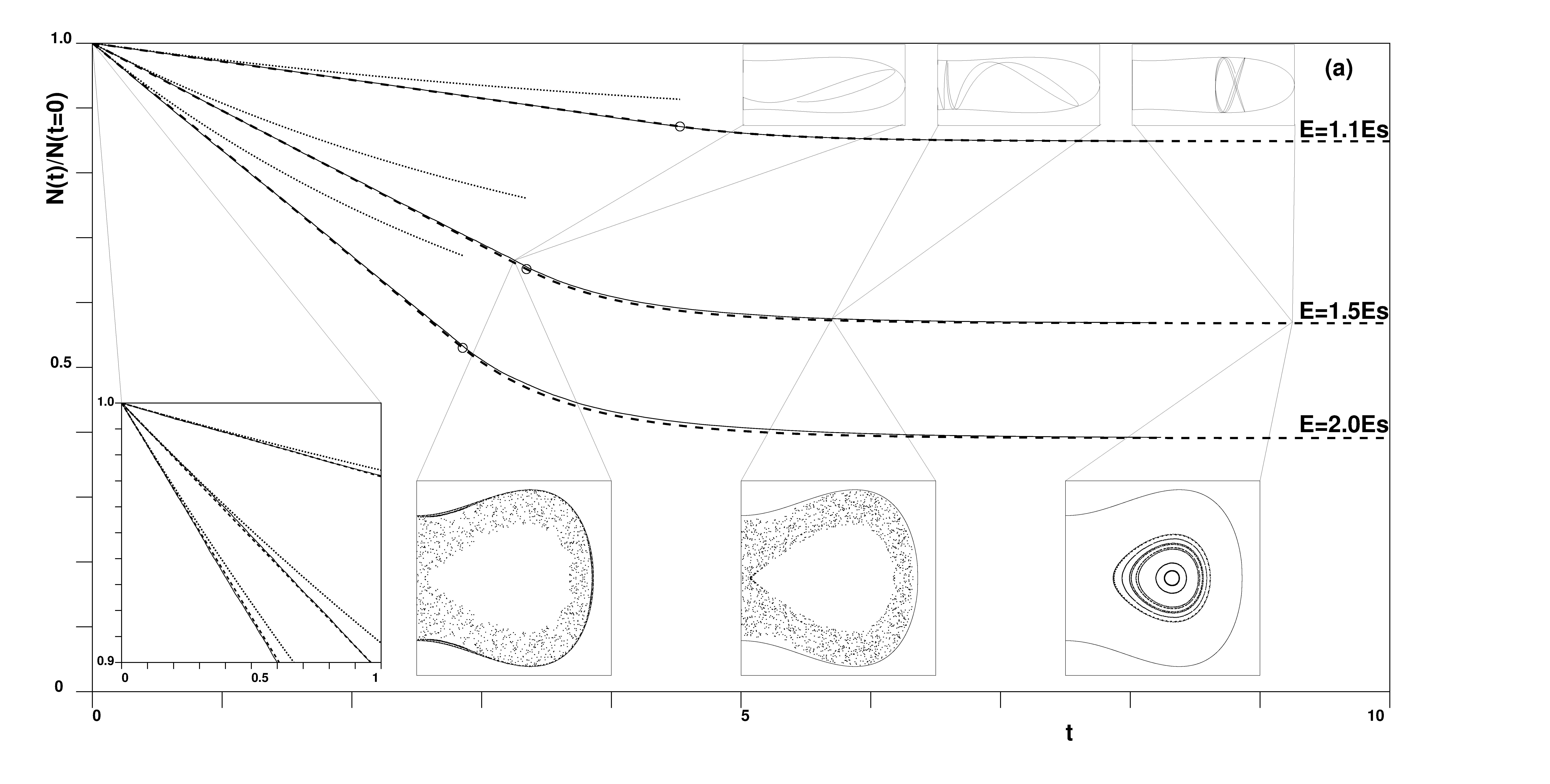}
\includegraphics[width=\textwidth,draft=false]{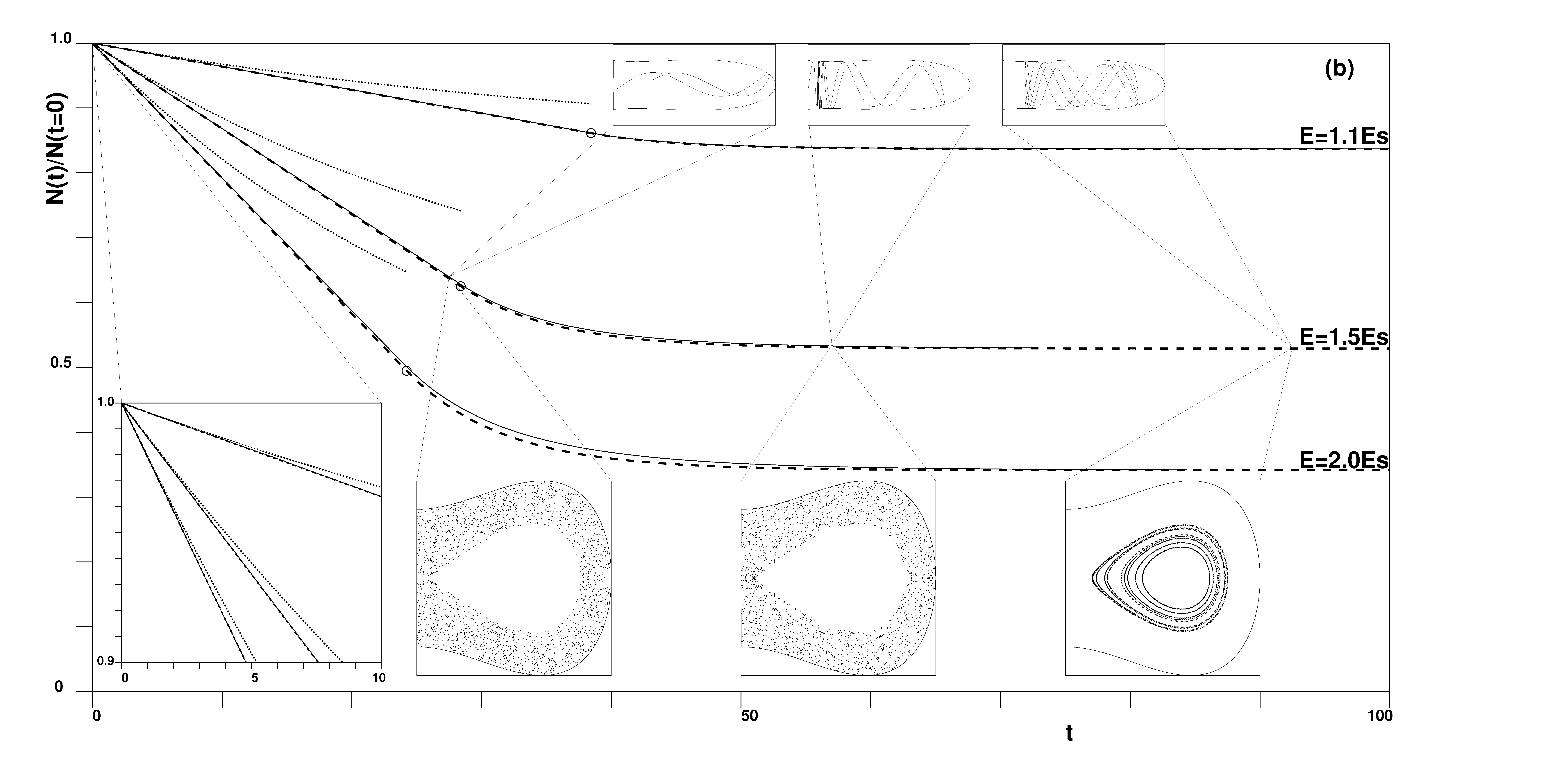}
\caption{\label{n_t2} Decay law for mixed states in the $D_5$ (a) and
$QO$ (b) potentials. Solid lines --- numerical simulation for $E/E_S
= 1.1, 1.5, 2.0$; dotted and dashed lines
--- theoretically obtained exponential and linear decay laws
respectively, zoomed on the inset figure in the lower left corners.
Other inset figures represent the typical trajectories and
Poincar\'e sections for the three different types of initial
conditions: linearly escaping, exponentially escaping and
non-escaping. Circles show the joining points between the linear and
the exponential decay laws at critical time $t=\tau$.}
\end{figure}
\begin{itemize}
\item  At times $t\rightarrow\infty$ decay law saturates at \[N(t\rightarrow\infty)=\rho^{(ne)}N_0\] where $\rho^{(ne)}$ is
equal to the relative phase volume of "never-escaping" trajectories,
which represents the regular trajectories completely localized
inside the considered minimum. All such trajectories, therefore,
have infinite escaping times.
\item For $t>\tau(E)$ the decay law
has exponential form \begin{equation}\label{exp} N(t)/N_0 =
\rho^{(ne)} + \rho^{(e)}e^{- \alpha^{(e)} (t-\tau)}\end{equation}
where $\rho^{(e)}$ represents the relative number of exponentially
escaping particles.
\item $t<\tau(E)$ For the decay law is linear
\begin{equation}\label{lin}N(t)/N_0 = 1 - \alpha^{(l)}
t.\end{equation}
\end{itemize}

We should stress, that (\ref{lin}) is in no way a linear
approximation of (\ref{exp}) for small $t$: in general
$\rho^{(ne)}+\rho^{(e)}e^{-\alpha^{(e)}\tau}\ne1$ and
$\alpha^{(e)}\rho^{(e)}e^{-\alpha^{(e)}\tau}\ne\alpha^{(l)}$.
Instead, from the condition of smooth joining of curves (\ref{lin})
and (\ref{exp}) in the transition point $t=\tau$, we obtain
\[\begin{array}{c}
\alpha^{(e)} = \alpha^{(l)}/\rho^{(e)}\\
\rho^{(e)} = 1 - \rho^{(ne)} - \rho^{(l)}\end{array}\] where
$\rho^{(l)}=\alpha^{(l)}\tau$ is the relative number of linearly
escaping particles. Moreover, already on time scales
$t\lesssim\tau(E)$ linear decay law (\ref{lin}) is apparently
different from its exponential analogue \[N(t)/N(t=0)=\rho^{(ne)} +
(1-\rho^{(ne)})e^{(- \alpha^{(l)} t)}\] (see the inset on
fig.\ref{n_t2}).

As we can see from the inset Poincar\'e section on fig.\ref{n_t2}
both the chaotic and regular trajectories contribute to linear
escaping regime (\ref{lin}), because for sufficiently small times
$t<\tau$ chaotic and regular motions are not yet distinguishable. Up
to transient time $t=\tau$ all quasi-one-dimensional regular
trajectories, oriented along the $x$-axis, have already escaped and
for $t>\tau$ the escape of the remaining chaotic particles follows
exponential law (\ref{exp}). The particles escaping the last show
already mentioned sticking phenomena (see fig.\ref{n_t2}).

The transient time $\tau(E)$ in fact coincides with the passage time
of the longest one-dimensional path from the opening to the opposite
wall of the potential well and back (see fig.\ref{n_t2}). For the
potentials $D_5$ and $QO$ corresponding theoretical estimates read
(we assumed $m=1$) \begin{equation*}\begin{array}{c} \tau_{D_5}(E) =
2\int\limits_{0}^{\sqrt{2(1+\sqrt{E})}}\frac{dx}{|p|}=
\frac{\sqrt{2}}{E^{\frac 1 4}} K\left(\sqrt{\frac{1 + \frac{1}{\sqrt E}}{2}}\right)\\
\tau_{QO}(E) = 12\left(\frac{E_S}{E}\right)^{\frac 1 4}
K\left(\sqrt{\frac{1 + \sqrt\frac{E_S}{E}}{2}}\right) =
6\sqrt{2}\tau_{D_5}(\frac{E}{E_S})
\end{array}\end{equation*} where $K(k)$ is the complete elliptic integral of
the first kind and $E_S=1/12^4$ is the saddle energy in the $QO$
potential for $W=18$ (for the $D_5$ potential $E_S=1$).

Theoretical estimates for the escape rate were obtained by averaging
the escape probability over the opening \cite{zhao_du}:
\begin{equation*} \alpha(E) = \rho(E)\int\limits_{x=x_S} dy
\int\limits_{-\frac \pi 2}^{\frac \pi 2}d\theta |p|\cos\theta,
\end{equation*}
where $x_S$ is the coordinate of saddle point and $\rho(E)$ is the
normalized particle density: \begin{equation*} \rho(E)=\frac{1}{2\pi
A(E)},
\end{equation*}
where $A(E)$ denotes area of the classically allowed region inside
the well: \[A(E)=\int\limits_{x>x_S} dx dy
\Theta\left(E-U(x,y)\right).\] Such $\rho(E)$ corresponds to uniform
distribution of particles on the energy surface $H(\mathbf{p,q})=E$.

For the potentials $D_5$ and $QO$ the explicit formulae are the
following:
\begin{eqnarray*}A_{D_5}(E)=2\int\limits_0^{\sqrt{2(1+\sqrt{E})}} dx
\sqrt{\frac{E-\left(\frac{x^2}{2}-1\right)^2}{x+2a}}\\
A_{QO}(E)=\frac{1}{72}\int\limits_0^{\sqrt{1+\sqrt{\frac{E}{E_S}}}}
d\xi \sqrt{(\xi+4)^2-7} \times \\
\times \sqrt{\sqrt{1+\frac{\frac{E}{E_S}
-\left(\xi^2-1\right)^2}{[(\xi+4)^2-7]^2}}-1}\end{eqnarray*} where
$\xi=\sqrt[4]{E_S}(x-x_S)$.

Finally, the general expression for the escape rate is
\begin{equation*} \alpha(E) = \frac{\langle p\rangle}{\pi A(E)}.
\end{equation*}
For our case $\langle p\rangle=\int\limits_{x=x_S} dy |p|$. In the
case of billiards with small opening $p=const\Rightarrow\langle
p\rangle=p\Delta$, and we recover expression (\ref{ntb}). For the
potentials $D_5$ and $QO$ we get the results in closed form:
\begin{equation*} \alpha_{D_5}(E) =
\frac{E-1}{2\sqrt{a}A_{D_5}(E)}\end{equation*}
\begin{equation*}
\alpha_{QO}(E) = \frac{\sqrt[4]{\varepsilon}}{12\pi A_{QO}(E)}
\left\{\left(16\sqrt{\varepsilon}+1\right)K\left(\sqrt{\frac{1-\frac{1}{16\sqrt{\varepsilon}}}{2}}\right)
-2E\left(\sqrt{\frac{1-\frac{1}{16\sqrt{\varepsilon}}}{2}}\right)\right\}
\end{equation*}
where $\varepsilon=E-E_S+1/256$, $K(k)$ and $E(k)$ are the complete
elliptic integral of the first and second kind respectively.

In order to obtain $\alpha^{(l)}(E)$  we correct $A(E)$ subtracting
the relative phase space occupied by the non-escaping particles
\begin{equation*}A^{(l)}(E)=A(E)(1-\rho^{(ne)})\Rightarrow
\alpha^{(l)}(E)=\frac{\alpha(E)}{1-\rho^{(ne)}}\end{equation*}
Fig.\ref{n_t2} demonstrates good agreement between our theoretical
and numerical results for a wide energy range.

\begin{figure}
\includegraphics[width=\textwidth,draft=false]{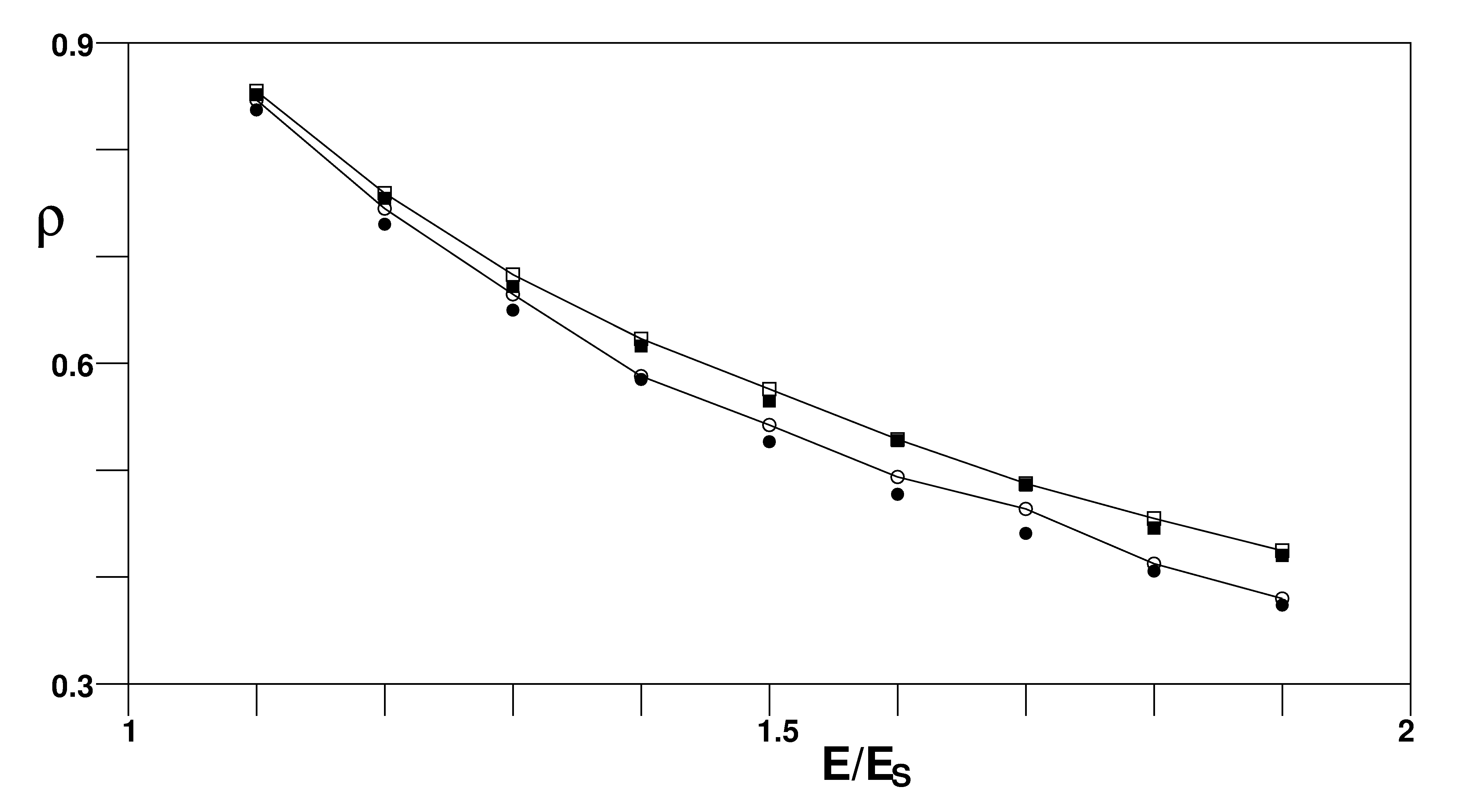}
\caption{\label{rho_e_f} Correlations between relative area of the
stability island $\rho^{(si)}(E)$ and non-escaping particles ratio
$\rho^{(ne)}$: empty squares -- $\rho^{(si)}_{D_5}(E)$, filled
squares -- $\rho^{(ne)}_{D_5}(E)$, empty circles --
$\rho^{(si)}_{QO}(E)$, filled circles -- $\rho^{(ne)}_{QO}(E)$.}
\end{figure}

The fraction of the non-escaping particles $\rho^{(ne)}$ coincides
with the relative phase space volume of trajectories, localized in
the regular minimum, which may be well estimated by the relative
area of the stability island $\rho^{(si)}$ on the Poincar\'e section
(fig.\ref{n_t2}). Calculation of the relative area of regular island
in the Poincar\'e section was performed by the following scheme. By
numerical integration of equations of motion the island boundary was
determined and then the interior area was calculated. Further the
obtained area was divided on the entire area of classically allowed
motion, defined by the conditions $x>0$ and $p^2>0$. While the phase
volume itself is 4-dimensional, the stability island in Poincar\'e
section is 2-dimensional, and so we cannot expect absolute
coincidence of the corresponding measures. However, the calculations
show very close correspondence between them (see fig.\ref{rho_e_f}).
Therefore, numerical analysis of Poincar\'e sections together with
our theoretical results gives all the information necessary to predict
the escape dynamics by an independent method.

In summary, we have considered classical escape from separated local
minima in two representative $2D$ multi-well potentials, realizing
the mixed state. We have found that escape from regular minima
contains a number of new features. The most important among them are
the following: i) decay law saturates at long time ranges; ii) on
small time scales  there exists a linear segment, which is not
connected with linear approximation to the exponential decay law
observed in chaotic systems with homogeneous phase space. The
fraction of particles, remaining in the well is determined by the
relative phase volume of the regular component, which in its turn
monotonically decreases with growth of energy. It was shown that the
linear segment of the decay law is generated by the
quasi-one-dimensional trajectories, oriented perpendicular to the
opening, and the transient time of the linear-to-exponential regime
lies in perfect agreement with the analytical estimates.

We should note that we devote our main attention to escape from the
regular local minima, because the specifics of the mixed state
manifests only in them. However let us recall that in the case of
mixed state the phase space structure at super-saddle energies is
determined by dynamical characteristics in different local minima of
the whole potential energy surface.

The above mentioned peculiarities of the escape problem may have
found practical application for extraction of the required particle
number from atomic traps. Changing the energy of particles trapped
inside the regular minimum, we can extract from the trap any
required number of particles. The problem of particle energy
changing in the potential well may be solved by the introduction of
small dissipation. The obtained results may be of interest also for
the description of induced nuclear fission in the case of
double-humped fission barrier. Revealed peculiarities must manifest
also in over-barrier dynamics of wave packets, initially localized
in the regular minima.

\sat\section{Chaos assisted tunneling in superdeformed nuclei}\sat

Since its discovery the atomic nucleus has been constantly used for
verifying new physical ideas such as superfluids, superconductivity,
supersymmetry and dynamical chaos. The aim of this section is to
demonstrate the possibility of observing the above mentioned
features of chaotic multi-well dynamics in concrete physical effect
--- the decay of the superdeformed states of atomic nuclei. In
contrast to the classical decay of the mixed state considered above,
this effect is purely quantum, but nevertheless dependent on the
structure of the classical phase space.

The superdeformed (SD) nucleus has an ellipsoid shape with axes
ratio $a/b\sim2$. Such extreme nuclei shapes are the result of the
quantum nature of particles that compose the nucleus and fill the
discrete energy levels (shells). In large deformations, the energy
interval (gap) between filled and next free shells could be very
large. In combination with other effects this energy gap could
provide stability of the nucleus in such large deformations.
Superdeformed nuclei were found more than twenty years ago
\cite{twin}, and today they have been observed in several mass
regions around $A=20,40,80,130,150,165,190$ and $240$ \cite{singh}.

Study of SD nuclei offers a new way of understanding nuclear
structure. This is a good illustration of the above mentioned fact
that nuclei are a laboratory for research of the general physical
effects such as tunneling, chaos and phase transitions. On the one
hand, nuclei provide interesting data for the study of these general
phenomena. On the other hand, we obtain new information about
nuclear structure.

\begin{figure}
\includegraphics[width=\textwidth]{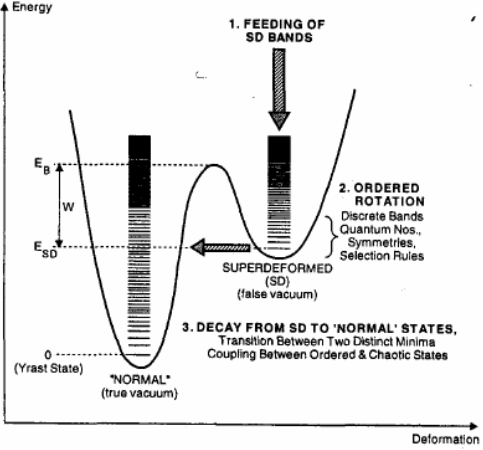}
\caption{\label{sd3steps}The three stages in the "life" of a
superdeformed nucleus together with some characteristic properties \cite{khoo}.}
\end{figure}

SD nuclei are produced in reactions with heavy ions. Initially
accelerated heavy ions, when colliding with nuclei of target
material, produce highly excited and fast rotating compound nuclei.
These nuclei release part of the excitation energy by the emission
of light particles (neutrons, protons, alpha particles) and photons.
In all observations of fast rotating SD nuclei, rotation breaks off
(already at low momentum), when nucleus suddenly changes shape and
decays to the state that corresponds to lower deformations. Three
stages of the transition from SD state to the normal deformed (ND)
one are presented on the fig.\ref{sd3steps} \cite{khoo}: feeding of
superdeformed bands, ordered rotation and decay from superdeformed
to normal states.

The lowest states in SD local minimum correspond to high excitations
in the main minimum. As a result, the lowest SD states (cold states)
are characterized by good quantum numbers and symmetries, while
their decays to the main minimum are controlled by strict rules of
selection. In contrast, ND states at the same energy could
correspond to chaotic motion in the semi-classical limit.
Statistical analysis confirms this idea: energy spectrum in the
region $E_{ext}\sim 6-7 Mev$ demonstrates all the signs of chaos,
while yrast region is regular \cite{aberg97}. This means that in the
considered region of nuclei excitation for which deformation
potential allows for the existence of second minimum, there is
realized a mixed state, considered in detail through the present
report. In that case a natural question arises: can the chaos
assisted tunneling mechanism be used for description of the
transition from the SD state to the ND one? We should stress that we
can consider dynamical tunneling because the transition takes place
in multi-dimensional space: nuclear shape is characterized by at
least two parameters.

Aberg \cite{aberg99} estimated chaos assisted tunneling in
perturbation theory. In the limit of no mixing between the ND states
(completely regular ND system) the wave function of the SD state
mixes with the doorway state only. In this case the tunneling
probability is given by
\[T^{regular}\approx\left(\frac{V_t}{\Delta E}\right)^2\]
where $V_t$ is tunneling coupling between SD and doorway states, and
$2\Delta E$ is the energy distance between the doorway states.

In the other limit situation (quantum chaos) the tunneling strength
is spread out over all ND states. Because tunneling probability in
the chaotic case is
\[T^{chaotic}\approx\left(\frac{V_t}{\Delta E}\right)^2 N.\]
Consequently
\[\frac{T^{chaotic}}{T^{regular}}\approx N\approx\frac{\rho_{ND}(E)}{\rho_{doorway}}\]

For $\rho_{doorway}\approx1\ Mev^{-1}$, and relative excitation
energy between the SD and ND state $E=3-5\ Mev$
$\rho_{ND}/\rho_{doorway}\approx10^4-10^6$. Therefore we expect the
tunneling probability to be enhanced by $10^4-10^6$ times if the ND
states are chaotic. In other words, the tunneling process connected
to the decay out of superdeformed states is strongly enhanced by
chaotic properties of the ND states.
\sat\chapter{Conclusions}\sat

Autonomous Hamiltonian systems with many local minima describe a
variety of physical processes, including chemical nuclear reactions,
nuclear fission, phase transitions etc. Nevertheless, the potentials
of non-trivial topology still remain far aside of attention, from
both the technical and ideological difficulties. The present work is
one of the first studies of regular and chaotic classical and
quantum dynamics in multi-well potentials.

The most general type of classical motion in potential with two and
more local minima represents the so-called mixed state, which was
first observed for simple polynomial 2D potentials. We proposed
quantum mechanical treatment of the mixed state, and demonstrate
signatures of quantum chaos in the energy levels statistical
properties, stationary wave function structure and Gaussian wave
packets dynamics. In particular, we have presented a new approach to
the investigation of QMCS in wave function structure, which can be
realized in potentials with two or more local minima. The main
advantage of the proposed approach is the possibility to detect QMCS
in comparison not different wave functions, but different parts of
the same wave functions. Efficiency of the approach was demonstrated
for the deformational potential of surface quadrupole oscillations
of nuclei and lower umbilic catastrophe $D_5$.

In this work we study the energy spectra statistical properties in
the case of the mixed state. Spectral series in the energy ranges,
where the mixed state is realized, open new possibilities for
studies of intermediate statistics. At these energies chaotic and
regular components are separated not in phase space (as usually the
case), but in the configuration space. A priori the nearest
neighbors spacing distribution function in the mixed state does not
obviously reduce to the properly weighted superposition of Poisson
and Wigner distributions. In the mixed state we deal not with
statistics of a mixture of two level systems with different nearest
neighbors spacing distribution functions, but with statistics of a
level system, where each level does not belong to definite
statistics. Statistical properties of such systems still wait
investigation, however such systems correspond to the common case
situation.

The solution of Schr\"odinger equation for two-dimensional potential systems of general form imposes high requirements on
efficiency of numerical methods used. The matrix diagonalization
method is efficient for one-well potentials. However, this numerical
procedure becomes less attractive at the transition to multi-well
potentials. We have shown that, in this case an attractive
alternative to the matrix diagonalization may become the spectral
method. The advantages of the spectral method compared to the
diagonalization method are demonstrated in the analysis of quantum
chaos problems for two-dimensional potentials with complicated
geometry.

The report presents also some concrete realizations of physical
effects, connected with the mixed state. Having considered classical
escape from 2D multi-well Hamiltonian system, realizing the mixed
state, we show that escape from local minima has a diversity of
principally new features, representing an interesting topic for
conceptual understanding of chaotic dynamics and applications.

In conclusion, let us dwell on open problems. Configuration space
for multi-well potentials breaks up naturally into different regions
(separate local minima); in each of them different dynamical regimes
are realized. Thus, there is a need for a calculating scheme that
will enable us to use the partial information about each isolated
region with the aim of obtaining a solution of the full problem. We
believe that the path decomposition expansion of Auerbach and
Kivelson \cite{auerbach} will prove to be useful for describing the
mixed state. This formalism allows the expression of the full time
evolution operator as a time convolution and surface integrations of
products of restricted Green functions, each of which involves the
sum over paths that are limited to different regions of
configuration space. Even for complicated nonseparable potentials
qualitative behavior is readily inferred and quantitative solutions
can be obtained from knowledge of the classical dynamics.

In the late 1990s Zaslavsky and Edelman
\cite{zaslavsky_edelman} considered a model of a billiard-type
system, which consists of two chambers connected through a hole. One
chamber has a circle-shaped scatterer inside, and the other one has
a Cassini oval with a concave border. As was shown, the
corresponding distribution function does not reach equilibrium even
during an anomalous large time. We want to emphasize, that the mixed
state, at energy a little exceeding the saddle one, can serve as a
more realistic model for studies of anomalous kinetics.

\end{document}